\documentclass[11pt]{book}
\usepackage{graphicx}                  
\usepackage{amssymb}
\usepackage{amsmath}
\usepackage{bm}
\numberwithin{equation}{section}
\usepackage{multicol}
\usepackage{array}
\newcolumntype{C}[1]{>{\centering\arraybackslash}m{#1}}
\usepackage{epstopdf}
\usepackage{caption}
\usepackage{subcaption}
\usepackage{mathrsfs}
\usepackage{grffile}
\usepackage{floatrow}
\usepackage[amssymb]{SIunits}
\usepackage{floatflt} 
\usepackage[toc,page]{appendix}
\usepackage[listings,theorems]{tcolorbox}
\usepackage[section]{placeins}
\usepackage{wrapfig}
\usepackage{amsthm}
\usepackage[utf8]{inputenc} 
\usepackage{dsfont}
\usepackage{geometry}
\usepackage{setspace}
\usepackage{physics}
\usepackage{authblk}
\usepackage[round]{natbib}
\usepackage{dsfont}
\usepackage[bookmarks]{hyperref}
\usepackage{titlesec}
\usepackage[font=small,labelfont=bf]{caption}
\DeclareMathOperator\supp{supp}
\titleformat*{\section}{\large\bfseries}
\titleformat*{\subsection}{\normalfont\bfseries}
\titleformat*{\subsubsection}{\normalfont\bfseries}

\newtheorem*{theorem*}{Theorem}

\newtheorem{definition}{Definition}
\usepackage{breqn}
\usepackage{thumbpdf}

\DeclareMathAlphabet{\mymathbb}{U}{BOONDOX-ds}{m}{n}
\DeclareMathOperator{\spn}{span}

\title{\textbf{Predictability, indeterminacy, and observability in quantum and post-quantum physics}}

\author[1]{Johannes Fankhauser\thanks{\href{mailto:johannes.j.fankhauser@gmail.com}{\it johannes.j.fankhauser@gmail.com}}, \it University of Oxford.}

\date{(\today)} 

\begin{document}

\begin{titlepage}
	\centering
	
	{\scshape\Large dphil thesis for the degree in philosophy\par}
	\vspace{3cm}
	{\huge \textbf{Observability and Predictability in Quantum and Post-Quantum Physics}
		\par}
	\vspace{0.2cm}
	\vspace{1.3cm}
	Author\par
	{\Large\itshape Johannes Fankhauser\par}
	\vspace{1.2cm}
	\begin{figure}[H]
		\centering
		\includegraphics[width=0.3\linewidth]{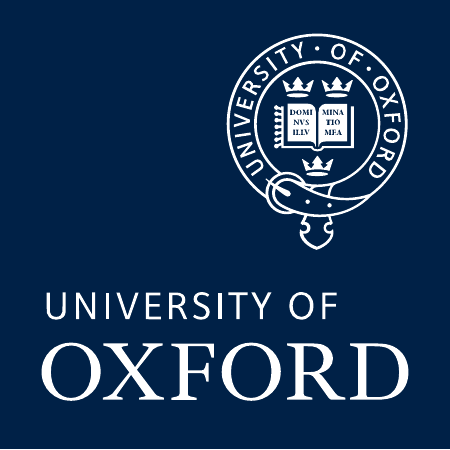}
	\end{figure}
	\vfill
	Advisors: \par
	Christopher G. \textsc{Timpson}\\
	Owen J. E. \textsc{Maroney}\\
	\vspace{2cm}
	{\Large \textbf{University of Oxford}\par}
		\vspace{0.1cm}
	{\Large\itshape St Edmund Hall\par}
	\vspace{0.8cm}
    A thesis submitted for the degree of \par
	\vspace{0.1cm}
	\textit{Doctor of Philosophy}\par
	\vspace{0.8cm}
	Michaelmas Term 2022 \par
	\vfill
\end{titlepage}
\hspace{5cm}

\section*{Papers and Publications}

\begin{itemize}
\item Fankhauser, J. (2019). Taming the Delayed Choice Quantum Eraser. \textit{Quanta}, 8(1):44-56. \\
The paper is based on and closely follows thesis Chapter \ref{section:Taming the delayd choice quantum eraser}.
\item Fankhauser, J. and Dürr, P. M. (2021). How (not) to understand weak measurements of velocities. \textit{Studies in History and Philosophy of Science Part A}, 85:16-29. \\
In collaboration with Patrick M. Dürr. Both authors provided critical feedback and helped shape the manuscript; the main ideas and analysis are by Johannes Fankhauser. It closely follows thesis Chapter \ref{section:Weak measurements}.
\item Fankhauser, J. Quantum Predictive Advantage and Pilot Wave Theory. \\
\textit{(In preparation)}
\item Fankhauser J. Quantum Predictability and Wigner’s friend.\\
\textit{(In preparation)}
\end{itemize}

\chapter*{Abstract}
	
I introduce a framework to distinguish two domains of physics --- the manifest (i.e. the directly observable empirical records in terms of manifest configurations) and the non-manifest domain of physics (i.e. the things that the manifest configurations signify according to a physical theory). I show that many quantum `paradoxes' rest on ambiguous reasoning about the two domains. More concretely, I study so-called `surrealistic' trajectories, the `delayed choice quantum eraser', and weak measurements. Finally, I show how the alleged puzzles resolve in the framework provided. 

I then formally define and address the question of whether quantum uncertainty could be fundamental or whether post-quantum theories could have predictive advantage whilst conforming to the Born rule on average. This notion of what I call `empirical completeness' refers to actual prediction-making beyond the Born probabilities, and thus delineates the operational notion of predictability from a `hidden variable' programme in quantum theory. 

I study how empirical completeness connects to signal-locality, and argue that a partial proof for the impossibility of predictive advantage can be established for bi-partite quantum systems. The relevant results demonstrate signal-locality as a sufficient principle that might explain the fundamental chanciness in present and future quantum theories and, in turn, reconciles us to many quantum features as aspects of limits on Nature's predictability.
\chapter*{Acknowledgements}

I am grateful for the immense guidance provided by my supervisors Christopher G. Timpson and Owen J. E. Maroney. Their mentorship has been invaluable, and I am truly thankful for the stimulating discussions and the freedom they provided me in pursuing my ideas.

I am indebted to my examiners Jonathan Barrett and Guido Bacciagaluppi for carefully reviewing my thesis and offering constructive feedback and remarks.

I would also like to express my gratitude to the Swiss Study Foundation, the University of Oxford, Magdalen College, St Edmund Hall College, ETH Zurich, the University of Lausanne with Michael Esfeld, my mother, and my friends Sivert and Martin for their financial support over the past few years.

I thank my extended Oxford family and the Philosophy of Physics group for enriching my time in this inspiring and diverse place. I would like to specifically thank Christopher Timpson, Owen Maroney, Harvey Brown, James Read, Tushar Menon, Simon Saunders, Oliver Pooley, and Adam Caulton.

The past four and a half years have been more than just a PhD, they have significantly contributed to my personal development. I want to thank my friends Sivert, Moritz, Jonas, Felix, Adam, Philomene, Fabian, Rosaria, Kathi, Hamza, Kate, and many more good souls for being part of this journey.

I also want to mention Magdalen College, St Edmund Hall College, the Austrian, German, and Swiss Society for their support and communal spirit.

I am grateful to Sivan, Valentine, and Amanda for teaching me valuable life lessons, and a special thanks to Sivert for being my support during the tough pandemic times and for cheering me up during my lonely times in the Austrian mountains. To all the people in my life who made this journey a special experience, I thank you.

Finally, I thank Mother Nature and the gifts she offers to experience ourselves and the world around us.

\tableofcontents

\chapter*{Introduction}
\addcontentsline{toc}{chapter}{Introduction}

As I'm writing this thesis, Alain Aspect, John F. Clauser, and Anton Zeilinger are awarded the 2022 Nobel Prize in physics for their pioneering work in quantum information science. In a private conversation, Zeilinger once said, `Quantum mechanics has told us that the only thing that’s real is its randomness'. A study of this claim being one of the aims of this thesis, I argue that quantum indeterminacy may be fundamental (at least to some extent) in future physical theories.

One of quantum theory's salient lessons is its inherent indeterminacy. Insofar as the formalism is deemed valid, its predictions are generally probabilistic. That is, generic physical states imply uncertainty for the outcomes of measurements. At face value, the quantum formalism suggests that its probabilism is unsurmountable, and so quantum indeterminacy has become one of the characteristics of the theory. Unlike in pre-quantum theories, where the physical state usually determines the values of all measurable observables, physical states in quantum physics imply uncertainty for the outcomes of measurements. This feature is quite peculiar. Why does the quantum formalism suggest that there is a limit to the predictability of an event? Is it because Nature is fundamentally indeterministic? Is it because the relevant degrees of freedom are uncontrollable? Or is quantum indeterminacy the consequence of a more fundamental physical principle?

I introduce a framework to study what I call the `empirical completeness problem' of quantum mechanics, i.e. the question of whether alternative theories could have a predictive advantage over the standard quantum probabilities and thus reduce the theory's indeterminacy whilst recovering the quantum predictions on average. To do this, I distinguish the manifest from the non-manifest domain in physics to delineate empirical completeness from metaphysical completeness. 

This thesis is a contribution to two related topics in the foundations of quantum theory. In the first part, I argue that absent an account of what quantum mechanical measurements signify, many alleged quantum `paradoxes' don't come off the ground. By a careful analysis of the physical arguments involved and being clear on what the terms in question mean, the paradoxes are resolved in the framework laid out. 

The main aim of the second part is to give an analysis of where we stand towards answering the empirical completeness problem, discuss the relevant results and arguments available to date, clarify them and offer more vital and general interpretations, and explore the possible avenues towards a generalised understanding of predictability in quantum and post-quantum theories. 

I shall proceed as follows. In Chapter \ref{section:manifest-non-manifest-domains}, I set the stage by providing a framework to distinguish the manifest domain from the non-manifest domain in physics and prediction-making. This will lay the ground to account for several quantum puzzles related to the problem of what quantum mechanical outcomes are supposed to signify (Chapter \ref{section:Quantum paradoxes}). After introducing the main issue in Section \ref{section:What do quantum measurements signify}, I begin with an illustration thereof in Section \ref{section:Surrealistic Bohmian trajectories} in the context of so-called `surrealistic' trajectories. 
In Chapter \ref{section:Taming the delayd choice quantum eraser}, I will turn to the `delayed choice quantum eraser' experiment. Chapter \ref{section:Weak measurements} contains an account of weak velocity measurements in the context of de Broglie-Bohm theory. A brief discussion on measurement disturbance will be the subject of Chapter \ref{section:measurement disturbance}. In Chapter \ref{section:the empirical completeness problem} I make the relevant definitions to introduce the empirical completeness problem of quantum theory. Chapter \ref{section:de Broglie-Bohm theory and predictive advantage} comprises an account of empirical completeness in pilot wave theory. It is shown under what assumptions the Bohmian framework can have a predictive advantage over quantum probabilities. The relationship between non-locality, signal-locality and quantum indeterminacy is examined in Chapter \ref{section:Non-locality and quantum predictability}. I argue that the signal-locality assumption leads to a partial proof of quantum empirical completeness. That is, there exist fundamental limits to what can be predicted independently of the quantum formalism. More concretely, relativistic locality appears as a sufficient principle to rule out predictive advantage for the probabilities of arbitrarily entangled bi-partite systems. My analysis leaves open whether a general proof is possible for the empirical completeness of arbitrary quantum systems, particularly individual ones. If there do not exist such principles which rule-out predictive advantage, there remains the question of why it does not seem to be observed in Nature --- why the Born rule seems accurate and complete. This may be a brute fact, or it may be that there is a dynamical explanation why --- as a kind of equilibrium --- a universe which begins as one in which improved predictability is possible will tend to one in which quantum theory is empirically complete. Finally, I shall comment on some ideas for future work in Section \ref{section:outlook}. The findings are summarised in the conclusions.

\chapter[The Manifest and Non-Manifest Domains of Physics]{The Manifest and Non-Manifest Domains of Physics}
\chaptermark{The Manifest and Non-Manifest}
\label{section:manifest-non-manifest-domains}

This is how I see three things connected --- the phenomena of the world, the elements of a physical theory, and what may be called the world's ontology: The physical world can be decomposed into two domains. 

One contains the empirical data, i.e. the manifest elements like outcomes of measurements and records in the environment. It comprises what is directly observable as outputs, settings and physical data --- the bare phenomenological facts.

It is worth noting that all of what exists in the manifest domain is in terms of positions of objects, events, or configurations in 3 or 4-dimensional space. Every measurement result and observable fact is ultimately encoded in spatial degrees of freedom like ink on paper, points on a screen or the position of a pointer. Hence, this empirical domain will be called the manifest domain of physics or the manifest domain of the world. Let it be labelled with $\mathcal{M}$.\footnote{Note that this description is of course already theory-laden. It serves as a minimal and natural assumption on the operational terms involved, one there is arguably a very solid consensus about.} 

The manifest domain alone doesn't entail any \textit{a priori} commitment to what is real in the physical world beyond itself, i.e. what the elements in the manifest domain may or may not signify. For instance, a configuration of dots on a screen does not --- without further theorising --- imply the existence of a particle in this very spatial location. Theories about what is contained in the non-manifest domain will be informed by what is observed as the manifest domain. But the point is that there exists no \textit{a priori} connection between the two. More theorising and metaphysical assumptions will be necessary to explain what the manifest domain ultimately instantiates or what lies beneath it. The manifest domain is the empirical reality only. There is, moreover, no \textit{a priori} commitment as to whether the manifest configurations persist in time. For instance, an ink spot on paper or a dot on a screen do, in fact, appear and disappear. An underlying theory may ultimately tell a story about persisting ink particles that went from a pen onto the paper or persisting electrons that locally excited a photomultiplier screen, but this is part of a theory and not the bare manifest domain.    

Then there is the second domain, that which contains the elements, objects and facts that \textit{aren't} directly observable. These will be the things that exist to which a physical theory relates the manifest elements. For example, it will tell when a manifest element, e.g. a position measurement record, actually refers to a non-manifest element, e.g. the position of a particle. In physical theorising, the standard criteria of theory choice --- such as simplicity and elegance --- usually determine a minimal set of properties that may exist in the non-manifest domain. The real world (according to our best understanding) is built from the domain of non-manifest elements that feature in our physical theories of what the real world is. We may label those elements with the familiar attribute of being `ontological' and refer to it as the ontological sector of the non-manifest domain.  

The non-manifest domain, furthermore, contains another category, which I term the nomological sector. These are the laws of physics that determine the dynamics of the ontological elements (and so it is also assumed that one can make sense of a notion of time). Whether one includes the laws of physics in the ontology is up for grabs, but I will distinguish the ontological from the nomological to better understand how the physical world can be divided. Note that there is no claim to be made that the laws that actually govern the behaviour of the world (if they exist) are represented by the laws that are used in practice to make predictions. Hence, I tend to think about the non-manifest domain first as a mathematical device for prediction making, and only in a second (philosophically more difficult) step as ontic and nomic. 

The important point is that all the contingent facts in the non-manifest domain are epistemologically derivative of the manifest domain. It's also well known that there is generally always a many-to-one relationship between physical laws and the patterns in the manifest domain.\footnote{Jumping ahead, it is a trivial observation that the nomology can, in principle, be chosen to be deterministic by introducing enough information in the initial state of a system. However, the resulting form of determinism comes at the cost of predictability and will therefore be rather uninteresting. Hence, determinism ought to be carefully distinguished from predictability.} 

An instrumentalist might worry about realist commitments acquired by introducing the non-manifest domain. One way to accommodate this worry is to think of the properties of the non-manifest domain as merely mathematical objects. This allows one to assign states, make predictions, and talk about the details of a theory without resorting to questions about what is or isn't real. As opposed to the instrumentalist, who may take all of the non-manifest domain or at least parts of it as mathematical convenience, the realist attributes ontological status to the non-manifest domain. And this is just a matter of interpretation which hasn't any consequence on the nature of prediction-making in a physical theory (though it may have for explanation and understanding).

Nevertheless, in the face of quantum theory, many seeming paradoxes arise from a lack of stringency in delineating the empirical from the possibly non-empirical. An instrumentalist reading of the paradoxes below often says nothing about the physical significance thereof. Most paradoxes don’t occur since an honest instrumentalist can only make statements about empirical data sets. In this view, a measurement outcome is not a measurement of anything real. This isn't to say a commitment to instrumentalism must solve the said paradoxes. Instead, a realist wants to be in a position to say more than an instrumentalist about what the empirical data may refer to and, importantly, can resolve the alleged paradoxes when sticking to a clear-cut distinction between the manifest and non-manifest domains. 

Distinguishing the two domains is essential for at least two reasons. One is that the nature of the non-manifest domain cannot be inferred na\"ively from the manifest domain. Physical theorising and metaphysical considerations are necessary to draw connections between the domains. Moreover, what goes into the non-manifest domain is strongly informed by what can coherently explain the empirical data. Discussions on the meaning of modern physics theories have demonstrated the relevance of making the manifest/non-manifest distinction. As discussed in Chapter \ref{section:Quantum paradoxes} on paradoxes, quantum mechanics is notoriously susceptible to confusion. A number of seeming quantum paradoxes are due to misunderstandings about what measurement outcomes signify in quantum mechanics. The theory doesn't come with a natural interpretation of how the outcomes, i.e. manifest configurations, relate to or inform us about the nature of the non-manifest domain. Since the theory leaves plenty of room for filling in the non-manifest framework, inconsistencies often arise.

Quantum theory is a fascinating case, for it manages to be an enormously successful theory without making satisfactory reference to the non-manifest domain. That is, the theory tells us how things are observed, and predictions are made (inside the manifest regime) but not what is observed (lacking a description of the non-manifest domain). In this sense, we are dealing with an operational, or instrumentalist theory. The motivation to resort to purely operational terms stems from the interpretational challenges inherent in theories like quantum mechanics. 


The second reason why the distinction is necessary involves insights from the hidden variable programme. For instance, quantum non-locality and contextuality show that what holds on a macroscopic, operational level doesn't necessarily carry over to the microscopic level. The distinction between microscopic versus macroscopic doesn't fully align with the non-manifest versus manifest distinction, but the point is that the relationships between things in one domain are no guarantee for the relationships between things in the other domain. A similar conclusion holds for concepts like determinism and predictability. Thus, in a rather trivial sense, the non-manifest domain allows virtually unlimited room to introduce all sorts of characteristics (hidden from the manifest domain). Therefore, our job will be to assess the physical principles that prevent these characteristics from being carried over to the manifest domain. 

In quantum mechanics, one common approach to studying the theory and pertinent puzzles more rigorously is resorting to the ontological models framework, as introduced by \cite{Spekkens-ontological-models}. An ontological model assumes there exists some set of ontic states in some state space $\lambda \in \Lambda$ for a physical system providing a \textit{complete} characterisation of its physical properties. Every preparation of a quantum state is thus said to correspond to preparation of some particular ontic state $\lambda$. But since the preparation may not fully control the ontic state, it describes a probability distribution $\mu(\lambda)$ over the state space $\Lambda$. Moreover, a measurement of the system is supposed to be a probabilistic response to a particular ontic state prepared. For this thesis, the manifest/non-manifest distinction can be understood to build on the ontological models framework in the following manner. First, the stipulation of the manifest domain as localised elements in three or four-dimensional space makes a direct and unambiguous claim to what is observable, i.e. --- the manifest configurations. Whereas measurements in the ontological models framework can, in a sense, be `incomplete'. Although the latter explains how measurement results come about by virtue of ontic states, it remains agnostic about how those results are ultimately represented as manifest configurations.\footnote{This makes all the difference for, e.g. spin degrees of freedom as we will see, for instance, in Section \ref{section:Surrealistic Bohmian trajectories} on so-called `surrealistic' trajectories. There, the outcome of a spin measurement attains coherent meaning if it is represented by some manifest degree of freedom.} At any rate, \textit{some} starting commitments have to be made as to the concrete nature of these domains, and manifest configurations as introduced here appear as a natural choice on which empirical experiences supervene. Secondly, when studying predictive advantage, it may be desirable to invoke variables that do not represent the complete ontic state of a quantum system. For instance, some manifest configurations could refine the quantum probabilities but might effectively only correspond to a coarse-graining of ontic states. The variables of an ontological model are, therefore, not suited to represent the variables of hypothetical predictive refinements of quantum theory. More concretely, in an ontological model one would have to specify what variables are accessible on the manifest domain, whereas the additional variables in a theory with predictive advantage are always assumed to be accessible.      

In sum, we first should be clear that there exists no \textit{a priori} entailment from the manifest to the non-manifest. And second, be wary about the possibility that what may apply in one domain may not in the other. 

\begin{figure}[h]
	\centering
	\includegraphics[width=1\linewidth]{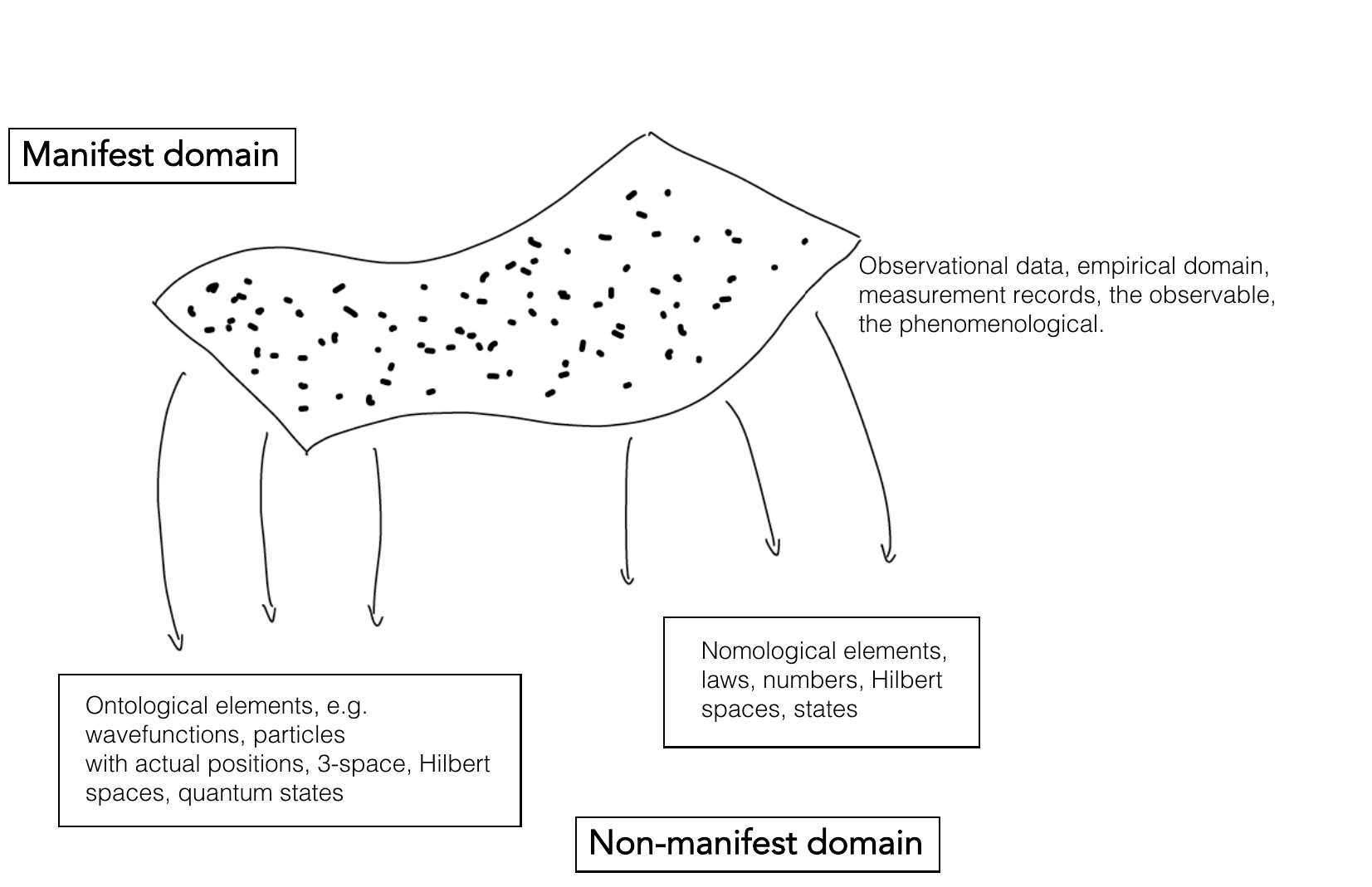}
	\caption{The set of empirical data (the manifest domain) informs the commitments on the non-manifest domain of physics, i.e. the ontological states and the physical laws.}
	\label{fig:manifest-nonmanifest}
\end{figure}

The framework introduced gives a clear picture of reality and distinguishes the empirical and observable, i.e. the manifest domain, from the non-manifest and not directly observable.\footnote{In a related but distinct fashion, Sellars distinguishes between the `manifest image' and the `scientific image' of the world (cf. \citealt{Sellars}). For a critique of this dichotomy see \cite{vanFraasen}.} The latter further decomposes into an ontological and a nomological category. Figure \ref{fig:manifest-nonmanifest} depicts the relationship. 

The manifest/non-manifest domains serve several purposes. One is giving a clear-cut framework and scope in which we can formulate the notion of predictive advantage and empirical completeness of a physical theory (see Chapter \ref{section:the empirical completeness problem}). The other is to provide the grounds for analysing a number of alleged quantum paradoxes, which are resolved given an account in the framework put forward (see Chapter \ref{section:Quantum paradoxes}). I wish to highlight that quantum mechanics is a compelling example showing that the connection between the manifest and non-manifest domain is anything but straightforward. In fact, the theory's peculiarities are arguably the primary motivation for making this distinction. In most pre-quantum theories like General Relativity Theory and Electromagnetism, the non-manifest elements --- e.g. classical matter, charged test particles --- connect more naturally to the observed manifest configurations and the non-manifest domain is usually taken for granted or implied implicitly.

Moreover, the emphasis on the manifest domain reflects how physics is commonly done. Instead of postulating an extensive set of non-manifest structures and theories to then strip away the superfluous elements, we start from what's directly observable, i.e. the manifest configurations, and then formulate theories by introducing variables that more or less directly derive from the configurations. For instance, the existence of particles may be postulated, which is natural given localised manifest position outcomes. But it may turn out that a theory like this is insufficient to explain all phenomena and the non-manifest domain may have to be extended. The point is that theory construction in physics usually follows the top-down rather than bottom-up approach.\footnote{I will set aside the familiar problems of operationalism (see, for instance, \citealt{sep-operationalism}). Note that it isn't necessary to be committed to a pure operationalist starting point, but we want to look at a quasi-operationalist starting point for methodological purposes.}

\section{Physical Theories and Prediction Making}

Despite the lack of agreement on whether the purpose of physics is to provide explanations just about empirical facts, or perhaps even just about maths, there seems general consensus that physics at least ought to include making predictions. To formalise the general idea of a prediction according to a physical theory, I introduce the notion of a target system and measurement apparatus, or pointer device. Ultimately, both systems are represented by subsets of configurations in the manifest domain. That is, the target system and the apparatus can be defined through variables that comprise localised patches of the manifest domain at some particular time $t$. 

For example, we may relate the target system to a small set of localised configurations at a time. Thus, any operational description of a target system will be determined by such configurations. Similarly, the properties of the apparatus may derive from a different subset of configurations in the manifest domain. A physical theory will then assign states derived from those chosen configurations.\footnote{Technically, the framework is not restricted to Markovian theories since the state variables could, in principle, contain information about the system at a given time \textit{and} its past.}   

In the present context, this means we get a prediction by the theory about the probabilities of the configurations of the manifest domain at a later time $t+\Delta t$ given the configuration of the manifest domain at an earlier time $t$. 

By definition, the manifest domain is idealised as containing a collection of single points in $\mathbb{R}^3$ for each space-like Cauchy surface in $\mathbb{R}^4$ labelled by $t$. These are the coordinates of the localised configurations accessible in the world, i.e. the empirical content. Formally, each space-like slice $\mathcal{M}_t$ of the manifest domain $\mathcal{M}$ can be represented as 

\begin{equation}
\mathcal{M}_t\in \bigcup\limits_{n\in\mathbb{N}}\mathbb{R}^{3n},
\end{equation} with $\mathbb{R}^0:=\emptyset$. The definition of the manifest domain thus accounts for the fact that the number of manifest points in each slice may vary over time. 

Physical predictions on the manifest domain are a family of probability maps

 \begin{equation}
		\left\{f_{t,\Delta t}: \bigcup\limits_{n\in\mathbb{N}}\mathbb{R}^{3n} \rightarrow \mathcal{D}\left(\bigcup\limits_{n\in\mathbb{N}}\mathbb{R}^{3n}\right)\right\}_{t\leq t+\Delta t},
\end{equation} where $\mathcal{D}(\cdot)$ denotes the space of distributions. That is, each slice of the manifest domain fixes a probability for each possible future slice, i.e. $p(\mathcal{M}_{t+\Delta t}|\mathcal{M}_{t})$, where $\mathcal{M}_{t+\Delta t}, \mathcal{M}_{t} \in \bigcup\limits_{n\in\mathbb{N}}\mathbb{R}^{3n}.$

\begin{figure}[h]
	\centering
	\includegraphics[width=0.8\linewidth]{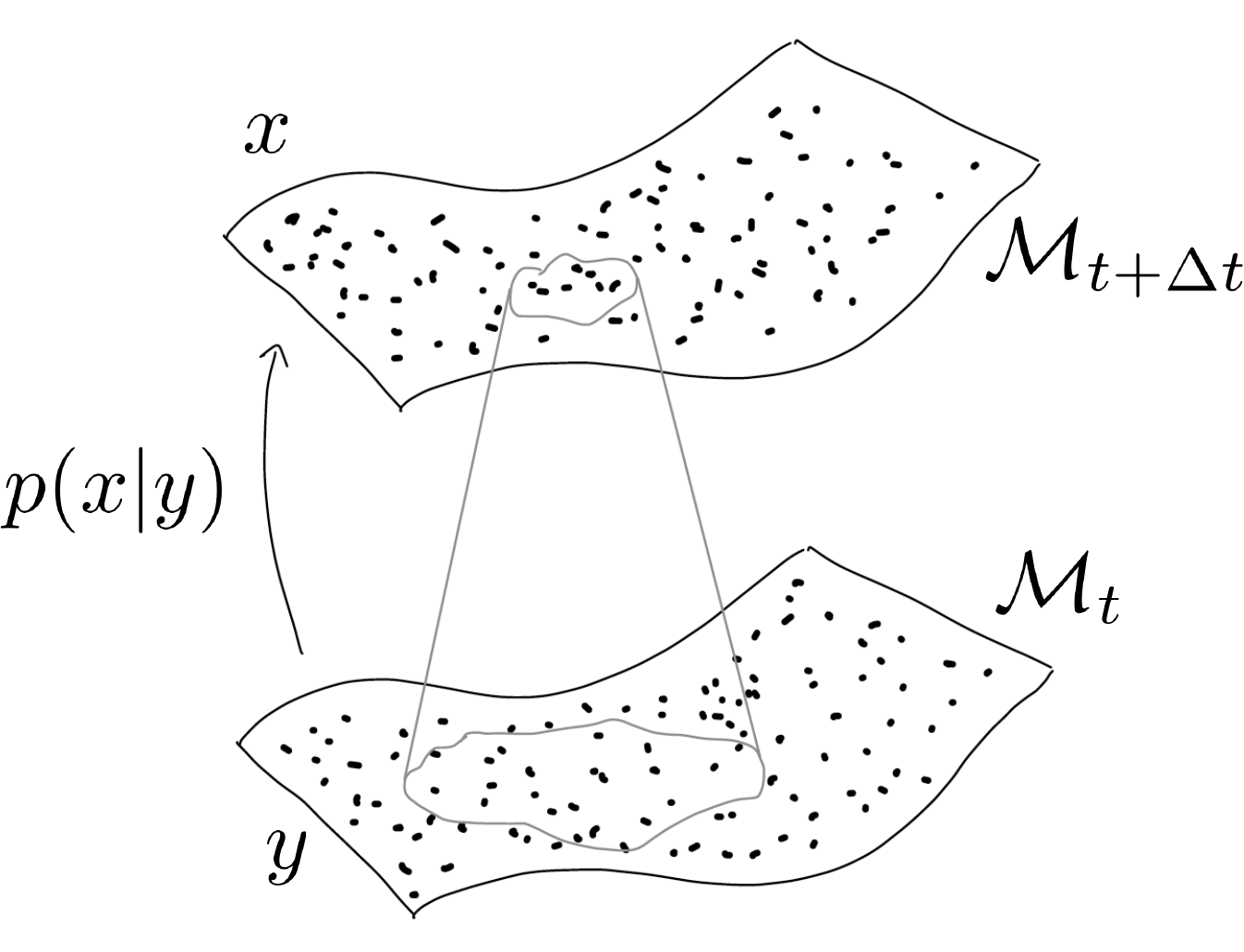}
	\caption{A slice of the manifest domain $\mathcal{M}_t$ at time $t$ and another at time $t+\Delta t$. A prediction, i.e. probability assignment $p(x|y)$, to manifest outcomes $x$ is based on manifest configurations $y$.}
	\label{fig:manifest-prediction}
\end{figure}

For all practical purposes a prediction is usually not made for an entire slice $\mathcal{M}_{t+\Delta t}$ of the manifest domain. Moreover, the variables on which a prediction is based do usually also not contain the entirety of $\mathcal{M}_{t}$. We consider, therefore, variables that pick out only localised patches in the slices. For instance, a preparation of a target system at time $t$ may be represented by a collection of points $y=(y_1, y_2,..., y_k)\in\mathbb{R}^{3k}$ which contains only the points in the tuple $\mathcal{M}_{t}$ for which $|y_i-y_j|^2\leq R$ is smaller than some radius $R\in \mathbb{R}$ for all $i\neq j$. Similarly, the outcomes $x=(x_1, x_2,..., x_{k'})\in\mathbb{R}^{3k'}$ of an apparatus may be localised as well, i.e. containing only the points in the tuple $\mathcal{M}_{t+\Delta t}$ for which $|x_i-x_j|^2\leq R'$ for all $i\neq j$, and radius $R'\in \mathbb{R}$. A prediction can then simply be expressed as $p(x|y)$ (see Figure \ref{fig:manifest-prediction}). Individual points in the space of the manifest domain can also be thought of as ordinary spacetime \textit{events}.

When a physical theory identifies configurations over time, i.e. in different manifest slices, the temporal evolution of systems can be defined. This then allows to talk about relativistic locality on the manifest domain. It will be required that every theory formulated on a world like this will satisfy a locality condition for the pertinent variables (cf. Chapter \ref{section:Non-locality and quantum predictability} for the definition of signal-locality). 

A physical theory requires an operational procedure for how measurement results are obtained, yielding its empirical content. That is, a theory $T$ assigns states to the target system and the measurement apparatus from which the statistical predictions are derived. Both the formal states of the target system and apparatus are derived from configurations in $\mathcal{M}$, but are usually states in the \textit{non-manifest} domain. For example, possible states may contain variables such as actual positions of particles and their momenta, local intensities of fields, or Hilbert space vectors.

Let $\lambda\in\Lambda$ be the state of the target system and $\mu\in \Lambda$ the state of the apparatus, with $\Lambda$ the state space of the non-manifest domain. For a given observable $Q$ to be measured, the initial apparatus state is denoted as $\mu_Q$ and assumed to be fixed for all preparations. The variable $Q$ refers to the setting of the apparatus. But, again, also $Q$ is just part of the apparatus whose states are a function of some configurations in $\mathcal{M}$. By choosing some setting that is supposed to measure the physical feature $Q$, we pick a manifest configuration of the apparatus. No claim is made as to what extent these states are ontological, i.e. elements of the non-manifest domain. For simplicity, it suffices to assume that they are just mathematical objects that describe the physical system. 

The target system and apparatus' joint system is denoted by non-manifest states $\lambda\times\mu_Q$. However, the structure of states on the joint system need not be represented by the structure of the Cartesian product. What's important is that there is a way to compound individual systems. Hence, $`\lambda\times\mu_Q$' is to be understood as merely a placeholder for a state in that bigger space, whatever its structure. The theory $T$ contains a statistical algorithm that maps to each target system-cum-apparatus state a statistical distribution of outcomes $p(x|\lambda\times\mu_Q)$ for observable $Q$, with mean value $\langle Q\rangle_T$. A theory is called \textit{deterministic} if $p(x)\in \{0,1\}$ for all outcomes $x$. An ensemble of target systems is called homogeneous when each element in the ensemble is assigned the same target system state $\lambda$. The apparatus states are assumed to be fixed. When it is clear how the physical states are obtained --- derived from the manifest configurations --- and what observable is measured, we may again abbreviate the probability distribution as $p(x|y)$ with $x$ and $y$ defined as above. For every outcome $x$ to occur given the configuration $y$, the theory specifies conditional probabilities $p(x|y)$.\footnote{This implies that prediction-making amounts to a Cauchy initial value problem. A set of (dynamical) equations determines the predicted probability distributions given a set of initial data. Spacetimes that are not globally hyperbolic (e.g. contain closed time-like curves) are thus ruled out by such a requirement. I wish to emphasise that there is, in principle, no assumption about how the laws are supposed to be formulated on the non-manifest domain. The variables on the non-manifest domain may as well allow a Cauchy initial value formulation, but they needn't. That is, the laws aren't necessarily informed by the traditional time evolution paradigm but can be more general, such as non-dynamical equations, constraints and consistency conditions. For example, there may be other non-standard notions of determinism where measurement outcomes are determined by more than just the information on a space-like hyper-surface or where the formulation doesn't allow a time evolution altogether (see, e.g., \citealt{Adlam-determinism-beyond-time} for a current account). But importantly, operational prediction-making is based on current manifest configurations to infer what some future manifest configurations will be. In a nutshell, the point is that the manifest-non-manifest distinction ought not to pose any restrictions on physical law-hood. However, any theory's prediction-making algorithm must be formulated in the way described above. For instance, assume that an all-at-once-law exists where the manifest outcomes are determined by a non-Markovian condition rather than an initial state and a governing equation. Then, loosely speaking, for all practical purposes, the prediction of the weather tomorrow can still only be informed by the manifest configurations of today. The point is that for an observer like us, actual prediction-making seems to be possible only in the way outlined here, i.e. in terms of a probability mapping from \textit{past} observed configurations to \textit{future} configurations.} The mapping from manifest configurations $y\in \mathcal{M}$ to target system states $\lambda$, however, may not be isomorphic, i.e. one $y$ does not uniquely select the state $\lambda$. In this case, each preparation procedure creates an ensemble of states with some probability distribution $\mu(\lambda|y)$. The ensemble average then yields the individual outcome probabilities 

\begin{equation}
	p(x|y)=\int_{\Lambda} p(x|\lambda\times\mu_Q)\mu(\lambda|y)d\lambda.
\end{equation}

Operationally, the conditional probabilities are used to predict the likelihood of some future outcome $x$ to be observed given the configuration $y$. If an observer does not have enough information of a system to determine $p(x|y)$ completely, a prediction can subjectively contain more uncertainty due to the presence of ignorance.

The assignment of states $\lambda\times\mu_Q$ used for prediction-making must lead to unique predictions $p(x)$. Otherwise, a theory could make ambiguous predictions about future configurations of the outcomes of measurements. This seems an obvious requirement, but as I will later argue, the theory of quantum mechanics allows for situations to be constructed where it is unclear what predictions should be made (see Sections \ref{section:predictability and measurement problem} on internal observations). Therefore, in this case, a precise assessment of the theory’s empirical content remains out of reach unless a more filled-in framework for the standard quantum formalism is introduced. Furthermore, I shall conclude below that statements about empirical completeness are contingent on solutions to the notorious measurement problem in quantum mechanics.  

The aforementioned formal ingredients allowed a characterisation of the notion of a prediction in purely operational terms by referring to configurations $x, y$ in the manifest domain only. 

\chapter[Quantum Paradoxes and the Manifest/Non-Manifest Domains Framework]{Quantum Paradoxes and the Manifest/Non-Manifest Domains Framework}
\chaptermark{Quantum Paradoxes}
\label{section:Quantum paradoxes}

\section{What do Quantum Measurements Signify?}
\label{section:What do quantum measurements signify}

Let's jump straight into the deep end. What does quantum mechanics tell us about what is \textit{real}? Unfortunately, about a hundred years after the theory's advent, the answer still is: not much! At best, we have a multitude of proposals in the form of quantum interpretations, some of which try to answer this question, some of which don't. But all come with their problems, and the quantum community is deeply split on which one, if any, to opt for. At worst, up to the present day, the quantum keeps being fuelled by confusion about what quantum measurement results supposedly signify. In modern literature, it remains customary to conflate the manifest with the non-manifest domain. This supports the need for a clear distinction between the manifest and non-manifest domains. Confusion about the latter may muddy the waters while assessing the former.  

The main strength of the traditional quantum formalism is its empirically well-established predictions. That is, the theory tells a story about how the statistics come about that are observed upon measurement of a quantum system. And this story is the vast empirical success of the theory. But the formalism also tells a good deal more. For example, the kinematics and dynamical laws explain a wide range of physical phenomena, such as the stability of matter, the nature of spin, or field quantisation, and lead to many novel predictions in most branches of physics. 

But here is one thing the theory doesn't tell us: \textit{What} can be inferred from empirical evidence about the physical system that is measured? In other words, what meaning is to be assigned to a particular measurement result, i.e. manifest pointer configuration? And what (if any) information does it convey about the physical system?   

Some argue that the key lies in altogether giving up the relationship between the manifest and non-manifest domains. Hence, resorting to operationalism or some sort of instrumentalism (see, for instance \citealt{sep-science-theory-observation}, instrumentalism in \citealt{sep-scientific-realism}, and references therein). Others attempt to propose a coherent underpinning by appealing to more ontological commitments (see, for instance, \citealt{allori2013primitive, sep-scientific-realism}, and references therein). But, unsurprisingly, a half-hearted mid-way or toing and froing between the two will invariably cause trouble. 

As it turns out, quantum mechanics teaches us that some measurement scenarios don't reliably reveal specific properties of a system once na\"ively thought to be measurable. Thus, the term quantum \textit{observable} ought to be taken with a grain of salt. For what it is that's `observed' is not \textit{a priori} entailed by the use of it. But this may not be so surprising. As Einstein aptly noted: what's observable is contingent on the theory (cf., for example, \citealt{Heisenberg1958-physics-and-philosophy}).

The manifest/non-manifest framework introduced in this thesis shall provide the grounds to identify where claims are unwarranted about the significance of quantum measurement results. A number of alleged quantum paradoxes can be resolved by consistently applying the manifest/non-manifest distinction. Or by pointing out that often the assumptions on which a paradox rests originate from sloppy thinking about the two domains. Absent an account of this crucial distinction, most paradoxes aren't genuine puzzles. 

I discuss some of them in said framework and identify the assumptions that lead to the alleged paradoxical conclusions. I show that many arise in the presence of inconsistent reasoning about quantum measurements. 

I shall first investigate a thought experiment where particular measurement results were deemed to give rise to observation of `surreal' behaviour. That is, quantum scenarios where certain combinations of observations allegedly prove that systems can possess paradoxical, conflicting properties. This analysis is motivated by the concrete case of so-called `surrealistic trajectories' in the context of de Broglie-Bohm theory, on which I shall comment. 

I then discuss the `Delayed Choice Quantum Eraser' experiment and argue that its paradoxical implications can be eliminated by taking into proper account the role of measurements in the setup. Finally, I give a straightforward description in standard quantum mechanics and show that the puzzle partly rests on misleading assumptions about what quantum measurement records signify.

Furthermore, I examine so-called `weak values' and `weak measurements' with regard to observation of the non-manifest. The alleged detectability of Bohmian particle trajectories is discussed in light of weak velocity measurements. Here as well, clarity on the meaning of observation resolves alleged paradoxes. Some general remarks are made on the ontological significance of weak values.  

To conclude, I shall say some words on disturbance upon quantum measurement. The question of what (if anything) is disturbed in a quantum interaction is intimately bound up with answers to the quest of what measurement results signify. 

What the examples show, ultimately, is that we have to free ourselves from a direct connection between the observable manifest phenomena of the world and what brings about these phenomena. 

\section[Surrealistic Observables and the Non-Manifest Domain]{Surrealistic Observables and the Non-Manifest Domain}
\sectionmark{Surrealistic Observables}
\label{section:Surrealistic Bohmian trajectories}

The detection procedure leading to what has (somewhat inappropriately) been coined \textit{`surrealistic'} trajectories presents a particularly illustrative quantum example where a na\"ive conflation of the manifest and non-manifest leads to inconsistencies. The term `surrealistic' is misleading as there is nothing in the standard quantum theory that one would refer to as being \textit{real}. As I alluded to before, the standard framework doesn't reveal any statements about the ontological sector of the non-manifest domain. So-called `surrealistic' trajectories or, more generally, `surrealistic' observables are thus meaningless terms. For the theory lacks a foundation for when a trajectory is supposed to be \textit{measurable}. This, in fact, blocks the paradox right from the outset because any claim that some `path of a particle' was detected in an experiment is unwarranted. 

\subsection{Introduction}

The physics of bad weather is quite interesting. All sorts of things fall from the sky, invisible forces carry around objects on the ground, and the night sometimes becomes some light show. But technically, a lightning bolt may, in fact, not be located where it was seen. Perhaps it was observed through a reflection in the window or a lake. Or, the radiation of the lightning is scattered such that some of it appears somewhere else from where it emerged. And then the lightning is followed by the thunder, which, as we know, is often not anywhere `close' to the lighting when it is experienced. Thus, even in classical physics --- here classical electrodynamics --- the directly observable, i.e. looking in the sky, or listening to the noise of thunder, are no direct and reliable sources of their underlying phenomena. It is hence no surprise that in quantum mechanics similar stories will be told (even though, their nature is, arguably, even more bizarre). This turns out to be the case for so-called `surrealistic' trajectories where the click of a detector must not be na\"ively identified with the position of a particle.


There is no need to rehearse the many treatments of surrealistic trajectories in the literature. Instead, the point is to present surrealistic trajectories as a case where a consistent underpinning of quantum theory leads to the fact that not all position measurements are genuine measurements of a particle's actual location. 


\subsection{The Surrealistic Trajectories Experiment}

\begin{figure}[h]
	\centering
	\includegraphics[width=0.4\linewidth]{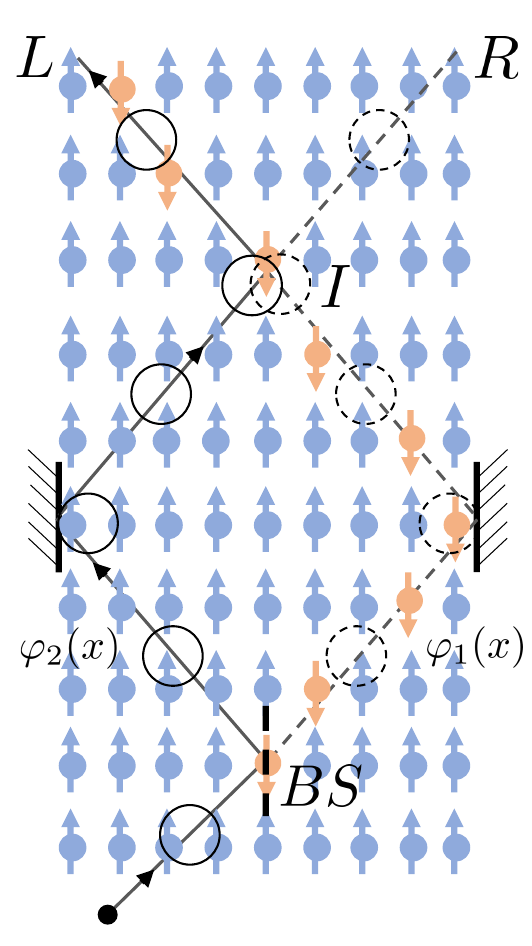}
	\caption{A particle's trajectory is `detected' by a grid of spins that flip when the particle travels by them. Although both theories produce the same outcomes for the spin flips, as opposed to standard quantum mechanics (chain of flipped spins in orange), pilot wave theory predicts the particle to traverse a path that isn't indicated by the series of spin flips (the solid black $L$-path to the left).}
	\label{fig:spin-flip-surreal}
\end{figure}

In the surreal trajectory experiment --- an instructive variant of which I discuss here\footnote{Another simple account is discussed in \cite{barrett2000persistence}, the original experiment was presented in \cite{EnglertScullySussmannWalther}.} --- a Mach-Zehnder interferometer is supplemented with a grid of spin systems interacting with the particle. The interaction is designed such that when the particle traversing the interferometer hits, the corresponding spins flip from their ready state $\ket{\uparrow}$ to the state $\ket{\downarrow}$. This allows, so the story goes, to track the particle's path. After the experiment is performed, the spin flips are read off, indicating the trajectory the particle took through the experiment (see the setup in Figure \ref{fig:spin-flip-surreal}).  

Assume that before the interaction the experiment consists of $n$ unperturbed spins $\ket{\uparrow}^{\otimes n}$. Without loss of generality, we assign the spins in the orange path $L$ to the first $k$ slots of the total state, and the spins in the alternative path $R$ to the next $k$ slots, respectively. Hence, the state is represented as

\begin{equation}
\ket{\uparrow}^{\otimes n}:=\ket{\uparrow_{L_1}} \cdots \ket{\uparrow_{L_k}}\otimes\ket{\uparrow_{R_1}}\cdots \ket{\uparrow_{R_k}}\otimes\ket{\uparrow}^{\otimes (n-2k)}.
\end{equation}

The quantum description of the experiment looks like so: When the particle travels through the experiment, the total initial state of particle and spin grid  
\begin{equation}
	\label{eqn:initial state surreal trajectory}
	\frac{1}{\sqrt{2}}(\ket{\varphi_1(x)}+\ket{\varphi_2(x)})\ket{\uparrow}^{\otimes n},
\end{equation} evolves into the final post-interaction state

\begin{equation}
	\label{eqn:post measurement spin flips surreal trajectory}
	\frac{1}{\sqrt{2}}(\ket{\varphi_1'(x)}\ket{\downarrow_{L_1}} \cdots \ket{\downarrow_{L_k}}\ket{\uparrow}^{\otimes (n-k)}+\ket{\varphi_2'(x)}\ket{\uparrow}^{\otimes k}\ket{\downarrow_{R_1}}\cdots \ket{\downarrow_{R_k}}\ket{\uparrow}^{\otimes (n-2k)}),
\end{equation} where $\ket{\varphi_1(x)}, \ket{\varphi_2(x)}$ are the two disjoint wave packets of the particle emerging from the beam-splitter $BS$, and $\ket{\varphi_1'(x)}, \ket{\varphi_2'(x)}$ the two disjoint wave packets of the particle at the end of the experiment (top region in Figure \ref{fig:spin-flip-surreal}). 

The two branches of the wave function are reflected at two walls or mirrors $M$ and subsequently overlap in the region $I$ before they travel further on their path. In light of the standard quantum description it is thus clear what is going on in terms of how the trace of spin-flips arises. Each branch of the particle's wave function causes a series of adjacent spin particles to switch states like a game of dominoes, creating a path a single particle would be expected to take. Since the initial quantum state (Equation \ref{eqn:initial state surreal trajectory}) is a superposition of those two branches, each track occurs with equal probability. Admittedly, this looks daringly close to the detection of a particle's trajectory. Nothing mysterious so far. 

But pilot wave theory (committed to actual particle positions) tells a different story of where the particle went. To compute the Bohmian prediction, the total wave function of all involved particles is rewritten in position space, and spin degrees of freedom are used in spinor form. The trajectory is then calculated from Bohm's guiding equation.\footnote{For details, see the use of the guiding equation for spin in Chapter \ref{section:de Broglie-Bohm theory and predictive advantage}.}  

The total quantum state of an individual spin particle on the grid then reads

\begin{equation}
	\Phi(y)\begin{pmatrix}
		1 \\
		0		
	\end{pmatrix},
\end{equation} for spin up, and 
\begin{equation}
	\Phi(y) \begin{pmatrix}
		0 \\
		1		
	\end{pmatrix},
\end{equation} for spin down, respectively. Here $\Phi(y)$ is a single spin particle's spatial wave function centred at the position of the particle. Notice, importantly, that the wave function of the position degree of freedom is identical for both spin values. 

The total final state, thus, is found to be 

\begin{dmath}
	\frac{1}{\sqrt{2}}\left(\ket{\varphi_1'(x)}\Phi_{L_1}(y-y_{L_1})\begin{pmatrix}
		0 \\
		1		
	\end{pmatrix} \cdots \Phi_{L_k}(y-y_{L_k})\begin{pmatrix}
		0 \\
		1		
	\end{pmatrix}\otimes
\Phi_{R_1}(y-y_{R_1})\begin{pmatrix}
	1 \\
	0		
\end{pmatrix} \cdots \Phi_{R_1}(y-y_{R_k})\begin{pmatrix}
	1 \\
	0		
\end{pmatrix}\otimes\ket{\uparrow}^{\otimes (n-2k)} +\ket{\varphi_2'(x)}\Phi_{L_1}(y-y_{L_1})\begin{pmatrix}
		1 \\
		0		
	\end{pmatrix} \cdots \Phi_{L_k}(y-y_{L_k})\begin{pmatrix}
		1 \\
		0		
	\end{pmatrix}
\otimes
\Phi_{R_1}(y-y_{R_1})\begin{pmatrix}
	0\\
	1		
\end{pmatrix} \cdots \Phi_{R_1}(y-y_{R_k})\begin{pmatrix}
	0 \\
	1		
\end{pmatrix}\otimes\ket{\uparrow}^{\otimes (n-2k)}
\right)
\end{dmath}, with $y_{L_i}, y_{R_i}$ the Bohmian positions of the spin particles.  Nothing mysterious so far here either. But what happens in the region $I$ according to the Bohmian description? 

Throughout the experiment the spatial degrees of freedom of the spin particles remain untouched since only their spin state flips. In other words, the spin measurement is incomplete since no record was created in terms of manifest configurations that would displace the spin particles (or further entangle them to spatial degrees of freedom of an apparatus). 

Therefore, without a manifest read-out, each particle's wave function is the same irrespective of whether that particle's spin has flipped or not, i.e. $\Phi_{L_i}(y-y_{L_i})=\Phi_{R_i}(y-y_{R_i})=:\Phi$, such that the final state can be rearranged as

\begin{equation}
	\frac{1}{\sqrt{2}}\Phi^n\left(\ket{\varphi_1'(x)}\begin{pmatrix}
		0 \\
		1		
	\end{pmatrix}^{\otimes k}\otimes\ket{\uparrow}^{\otimes (n-k)}+\ket{\varphi_2'(x)}\ket{\uparrow}^{\otimes k}\begin{pmatrix}
	0 \\
	1		
\end{pmatrix}^{\otimes k}\ket{\uparrow}^{\otimes (n-2k)}\right).
\end{equation} 

This makes it possible for the two branches of the target particle's wave function to interfere coherently in region $I$. Plugging the state in $I$ into the Bohmian guiding equation for spinor-valued wave functions reveals that the target particle's velocity has contributions from both branches of its wave function, i.e.

\begin{equation}
	\label{eqn:surreal wave function}
	m v_x= \hbar\Im \frac{\Psi^*\nabla_i \Psi}{\Psi^*\Psi}=(|\Phi|^2)^{n}((\varphi^I_1(x))^*\nabla\varphi_1^I(x)+(\varphi^I_2(x))^*\nabla\varphi_2^I(x)).
\end{equation}

The guiding potential of particle $x$ is independent of all $y$ degrees of freedom! The interaction is designed such that only the spin state evolves when a particle passes by, and the spin particles don't move upon that (otherwise the effect wouldn't occur). If one of the spin particles were to move during the interaction, one of the two contributions would be suppressed. That is, although the two spin states (the left and right branches) of an individual spin particle are orthogonal, i.e. $\bra{\Phi(y)}\ket{\Phi(y)}\bra{\uparrow}\ket{\downarrow}=0$, they may still overlap in configuration space. Hence, the overlapping spin states in configuration space will affect the target particle's motion as Equation \ref{eqn:surreal wave function} suggests.\footnote{This was highlighted in \citep{maroney-phd-thesis-2004} by distinguishing orthogonality of quantum states from `super-orthogonality', where the latter refers to the feature that two quantum wave functions don't overlap in configurations space.}  

As Equation \ref{eqn:surreal wave function} shows, the particle will move straight up when the two branches are made to overlap since the total velocity is the sum of the two contributions coming from the left and right moving wave packet (i.e. horizontal velocity component cancels). When the particle reaches the interference area $I$, the two wave packets can interfere before they continue to pass through each other and separate again. Moreover, this causes the particle to be reflected from point $I$ and continue travelling on either the experiment's left or right half.   

In the last step, a measurement of the final position of the particle is performed, and the spin states are recorded (e.g. with a series of Stern-Gerlach apparatuses). Note that if the spatial wave functions and positions $y_i$ of the spin particles on the grid were to change, they couldn't be pulled out as an overall factor in Equation \ref{eqn:surreal wave function}, and would therefore effectively decohere the target particle's wave function into its two branches.

As a result, the standard quantum framework predicts a chain of spin flips to emerge that \textit{doesn't} coincide with the particle trajectory as predicted by the Bohmian guiding equation. If we believe that the spin track in the experiment reliably indicates the particle's path, one could elicit a puzzle: The observed path (the flipped spins) is the one on the right, but the according to de Broglie-Bohm theory, the particle never was there. That is, the Bohmian trajectory is `surreal', and therefore the theory can't be right.\footnote{In the same vein, there exists, in fact, an even more straightforward puzzle concerning the measurement of velocity or momentum: The Bohmian momentum $mv_x$ given by the guidance equation \textit{also} doesn't correspond to the value observed by a measurement of the momentum operator $\hat{p}$. It's not even preserved over time.} 

We first observe that, of course, concerning the state of the spin, whether flipped or unflipped, the hypothetical emergence of a chain of flipped spins is not an actual measurement of anything yet. Although one may claim that some spin state must have changed upon interaction with the particle flying through the experiments, no configurational record exists in the manifest domain indicating a measurement result. This turned out to be crucial. The argument relies on the fact that none of the spin particle's configurations $y_i$ change during the experiment. Otherwise, the guiding equation wouldn't give rise to the described behaviour of the target particle. But this also means that during the experiment, the spin chain must not be observed in terms of manifest configurations. After the interference region $I$ was passed, subsequent measurements may create records thereof. 

The existence of so-called surrealistic trajectories was supposed to be evidence for testable false predictions of pilot wave theory (see \citealt{EnglertScullySussmannWalther}). The trajectories predicted, so the story goes, contradict what an actual measurement of the particle trajectory would yield. So they were used to claim that the Bohmian theory gives wrong predictions for the \textit{actual} position of a particle. If this were true, it would falsify the theory.

However, one recalls that first, the predictions of de Broglie-Bohm theory for the measurement outcomes of this thought experiment coincide with the ones of standard quantum mechanics. Second, in the standard theory, no meaning is assigned to the trajectory of a particle, so that it remains unclear what such a measurement would even have to do with the dynamics of a particle (cf. also, for instance, \cite{durr1993commentsurreal, lazarovici2019position, bricmont2016making, daumer1996naive}). Classical particles do not exist in standard quantum theory, let alone have trajectories. As I alluded, `surrealistic' trajectories result from an unjustified relationship between the manifest and non-manifest domains. The observed spin track is na\"ively identified with a particle's actual trajectory. Since this claim would go beyond the postulates of the standard quantum framework, it is unclear whether that assumption can be consistently made. Furthermore, a theory where a claim about the existence of actual particle trajectories \textit{can} consistently be made is pilot wave theory. But as the example shows, the trajectory cannot coincide with the one indicated by the chain of spin flips.\footnote{The surrealistic trajectories experiment crucially relies on \textit{whether} and \textit{when} the spin degrees of freedom are made manifest as pointer configurations. The pertinent assumption was that all experience supervenes on manifest configurations. Having said that, it would still be interesting to devise alternative descriptions where this claim is false and where experiences supervene on other variables than manifest configurations. It would also be interesting whether surrealistic measurements arise in all hidden variable theories.}  

But maybe the existence of surrealistic trajectories are just a deficiency of how the standard Bohmian framework is defined. Is there a way to tell a less puzzling story about the paths of particles where (some version of) a Bohmian theory isn't in conflict with the observed spin track?\footnote{Either by choosing a different guiding equation or other alternative Bohmian laws for the motion of particles?}  However, surrealistic quantum properties aren't just an artefact inherent to Bohmian commitments. 

Another strategy to present the discrepancy is by resorting to \textit{weak} velocity measurements. I shall discuss this in much more detail in Chapter \ref{section:Weak measurements} on weak measurements. Performing a weak measurement of the particle's trajectory in this experiment is yet another way to detect a particle's path. In fact, what's interesting, weak velocity measurements seem a plausible and intuitive \textit{standard} quantum prediction about where a particle went and didn't include any Bohmian commitments about actual particle positions. Anticipating the conclusion, one encounters two measurement scenarios with an intuitive grasp that contradict each other. So which to believe? The weak measurements, the chain of observed spin flips, the Bohmian prediction, or something completely different? The answer is, as mentioned, that this depends on the details of the theory. This again shows the importance of assigning a coherent meaning to the measurement results. 

Moreover, there are other analogous instances of `surreal' quantum observables. For example, as Brown et al. indicate in their \citeyear{brown1995bohm}, other properties besides spin, such as mass, charge, and magnetic moment are all inconsistent with what they call the thesis of `localised particle properties'. 

Another example of `surreal' quantum properties worth mentioning is the Cheshire cat experiment, where a cat appears to be `separated' from its grin. In particular, the experiment is designed so that a photon is detected in one arm of a Michelson-Morley interferometer, but its angular polarisation is detected in the other \citep{Aharonov_2013_Cheshire}. Similarly to surrealistic Bohmian trajectories, the thought experiment, in effect, shows an apparent discrepancy between a particle's `location' and its `location' observed by interaction with its spin degree of freedom (the photon's polarisation). 

%
%
%
%
%

\chapter[Taming the Delayed Choice Quantum Eraser]{Taming the Delayed Choice Quantum Eraser}
\chaptermark{Delayed Choice Quantum Eraser}\label{section:Taming the delayd choice quantum eraser}

I discuss the delayed choice quantum eraser experiment (DCQE) by drawing an analogy to a Bell-type measurement and giving a straightforward account in standard quantum mechanics. The delayed choice quantum eraser experiment turns out to resemble a Bell-type scenario in which the paradox's resolution is rather trivial, so there really is no mystery. At first glance, the experiment suggests that measurements on one part of an entangled photon pair (the idler) can be employed to control whether the measurement outcome of the other part of the photon pair (the signal) produces interference fringes at a screen after being sent through a double slit. Significantly, the choice of whether there is interference can be made long after the signal photon encounters the screen. The results of the experiment have been alleged to invoke some sort of `backwards in time influence'. I argue that issue can be eliminated by taking into proper account the role of the signal photon. I show that the alleged paradox partly rests on misleading assumptions about what quantum measurement records signify. Moreover, in the de Broglie-Bohm picture, the particle's trajectories can be given a well-defined description at any instant of time during the experiment. Thus, it is again clear that there is no need to resort to any kind of `backwards in time influence'.  

\section{Introduction}

Delayed choice scenarios in slit experiments as found in \cite{wheeler1978past}, and earlier in \cite{von1941deutung} and \cite{bohr1996discussion}, have formed a rich area of theoretical and experimental research, as evidenced in the literature (\cite{eichmann1993young}, \cite{englert2000quantitative},  \cite{doi:10.1119/1.19257}, \cite{doi:10.1119/1.19258}, \cite{Kim1999}, \cite{walborn2002double}, \cite{kwiat2004science}, \cite{aharonov2005time}, \cite{Peres2000}, \cite{Egg2013}, to name a few). 
From the results of the original delayed choice experiment, Wheeler concluded that  `no phenomenon is a phenomenon until it is an observed phenomenon', and `the past has no existence except as it is recorded in the present' (ibid.). Others have also been inclined to conclude that such experiments entail some kind of backwards in time influence or another (e.g. \cite{doi:10.1119/1.19258}). I shall discuss a modified version of Wheeler's delayed choice experiment, which was first proposed by  \cite{scully1982quantum} and later realised in the experiments of \cite{Kim1999}. It is termed the Delayed Choice Quantum Eraser. I show that the actual evolution of the quantum state in the experiment and a novel analysis in terms of a Bell-type scenario prove Wheeler's conclusions and other conclusions about backwards in time influence unwarranted. The puzzlement about delayed choice experiments emerges from misinterpreting and ignoring the symmetry of time-ordered measurement events. The general analysis applies to all cases of quantum erasure where two systems become entangled. For a 3-slit quantum eraser experiment, see, for example, \cite{shah2017quantum}.

\section{Delayed Choice in a Bell-type Scenario}
\label{Bell}

\begin{figure}[H]
	\centering
	\includegraphics[width=0.7\linewidth]{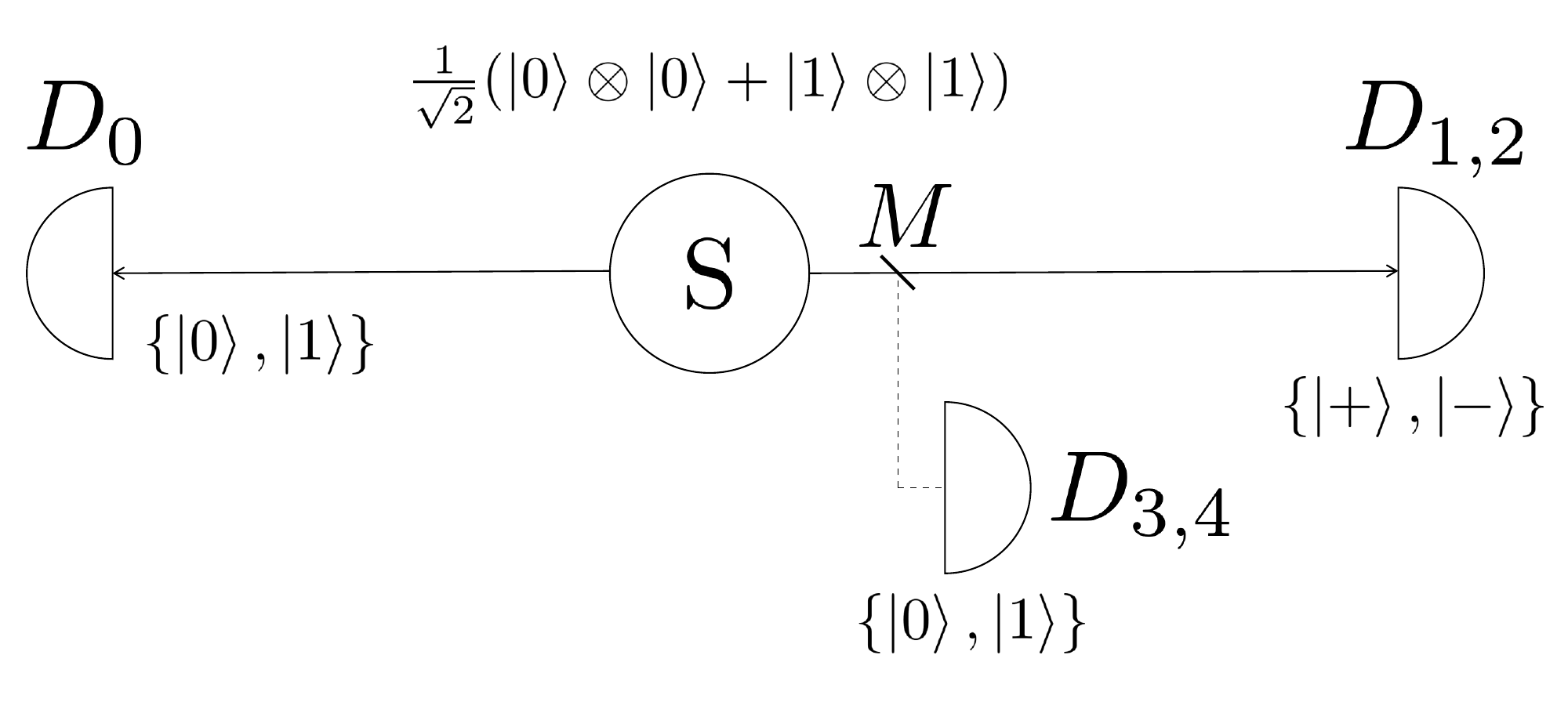}
	\caption{A Bell-type experiment resembles the delayed choice quantum eraser experiment.}
	\label{fig:BellD}
\end{figure}

Let us begin by considering a simple and familiar case, which will nonetheless provide the key to illuminating the DCQE. Imagine a source $S$ emitting photons. Both Alice (detector $D_0$) and Bob (detector $D_{1,2}$ and $D_{3,4}$) receive one particle of an entangled photon pair in the Bell state
\begin{equation}
	\label{state}
	\ket{\psi}=\frac{1}{\sqrt{2}}(\ket{0}\otimes\ket{0}+\ket{1}\otimes\ket{1}).
\end{equation} 
The states of the photons are taken to be qubit states. Let us, for ease of comparison with the DCQE later, call Alice's photon the signal photon and Bob's photon the idler. Figure \ref{fig:BellD} depicts the experiment.

$M$ denotes a mirror that can be used to reflect the idler photon into detector $D_{3,4}$. In Bob's arm, two measurements can be performed: Either he chooses to use mirror $M$ to measure the idler photon in the computational basis $\{\ket{0},\ket{1}\}$ or the mirror is removed, and the photon travels to detector $D_{1,2}$ where Bob performs a measurement in the diagonal basis  $\{\ket{+},\ket{-}\}$, where $\ket{+}=\frac{1}{\sqrt{2}}(\ket{0}+\ket{1})$ and $\ket{-}=\frac{1}{\sqrt{2}}(\ket{0}-\ket{1})$. 

The quantum predictions for the experimental outcomes are familiar and relatively simple: The probabilities for the outcomes $0$ and $1$ at detector $D_0$ both are $0.5$ as quickly seen from state $\ket{\psi}$ in Equation \ref{state}. And the same holds for the case where Bob measures in the diagonal basis at detector $D_{1,2}$. For this latter case, the state is rewritten as   

\begin{align}
	\label{eraser state}
	\ket{\psi}&=\frac{1}{2}(\ket{0}\otimes(\ket{+}+\ket{-})+\ket{1}\otimes(\ket{+}-\ket{-}))\\ \nonumber 
	&=\frac{1}{2}((\ket{0}+\ket{1})\otimes\ket{+}+(\ket{0}-\ket{1})\otimes\ket{-}). \nonumber
\end{align}

The statistics at $D_0$ are independent of which measurement Bob performs, as expected. Furthermore, by conditioning on Bob's outcomes, we can make the following statements:

(1) Assuming Bob's measurements come before Alice's (for, say, the rest frame of the laboratory), if he measured $\ket{0}$ at detector $D_{3,4}$, we know that Alice is going to measure $\ket{0}$ at $D_0$. Likewise, if the outcome was $\ket{1}$, she is going to measure $\ket{1}$. For the sake of comparison with the DCQE let us call this a `which-path measurement' since Bob's measurement will tell us with certainty which result Alice will subsequently observe. By contrast, if Bob decides to perform a measurement at detector $D_{12}$, conditioned on outcome $\ket{+}$ the state arriving at $D_0$ is going to be $\frac{1}{\sqrt{2}}(\ket{0}+\ket{1})=\ket{+}$. Thus, Alice will equally likely observe outcomes $0$ and $1$. Conversely, conditioned on Bob measuring $\ket{-}$ Alice will receive the phase shifted state $\frac{1}{\sqrt{2}}(\ket{0}-\ket{1})=\ket{-}$ and she would again detect outcomes $0$ and $1$ with probability $0.5$. This directly follows from Equation \ref{eraser state}. We shall call this an `interference measurement' on Alice's side since Alice will detect a spread of results. 

(2) The case when Bob's measurements happen after Alice's is similar: When Alice obtains $0$, Bob is going to see $0$ as well in the $D_{3,4}$ measurement (the same holds for outcome $1$). In this case, just as previously, therefore, conditioning on Bob's recording a 0(1) outcome, Alice will have recorded a 0(1) outcome. In the case of the measurement at detector $D_{1,2}$, i.e. in the diagonal basis, Bob expects state $\frac{1}{\sqrt{2}}(\ket{+}+\ket{-})$ if Alice's outcome was $0$ and $\frac{1}{\sqrt{2}}(\ket{+}-\ket{+})$ otherwise. In both cases the outcomes $\ket{+}$ and $\ket{-}$ show up with probability $\frac{1}{2}$. Again, therefore, we see that the conditional statistics are the same as in the previous scenario; that is, the statistics are the same as those which would be recorded if Bob's measurement had come first rather than Alice's: conditioned on Bob recording a +(--) result, Alice's will record 50\% outcome 0 and 50\% outcome 1. Indeed, they have to be the same since any measurement on Alice's side commutes with any measurement on Bob's side. 

It is important to note that on an operational view in terms of statistics of outcomes, nothing is puzzling about any of this, for there are no ontological commitments made other than the existence of conditional probabilities, which could just be understood as relative frequencies for certain pre- and post-selected subensembles. But as an exercise, we could elicit a puzzle: We can now argue as follows that there must be retrocausal action. Bob is free to perform his measurements at any time. In particular, he can decide to perform a `which-path measurement' or 'interference measurement' well after the signal photon has reached Alice's detector. Since Bob, by choosing to measure with detector $D_{1,2}$ or $D_{3,4}$ can decide to create either a computational basis state $\ket{0}$, $\ket{1}$ or a superposed state at Alice's detector ($\frac{1}{\sqrt{2}}(\ket{0}+\ket{1})$ or $\frac{1}{\sqrt{2}}(\ket{0}-\ket{1})$), the state of the signal photon that hit Alice's detector had to change retroactively in order to get the outcomes expected. In other words, Bob's measurement determines the state of the signal photon, but since that photon has already been measured, it must have done so by acting on the past of it. Moreover, we might argue that one `has' to reason in this way since the statistics that we obtain for Alice's outcomes when we condition on Bob's later outcome are `exactly those' which are generated when Alice receives a quantum state produced as a result of Bob's measurement.

The error in this naive argument for `action into the past' applied to the Bell-type scenario is apparent immediately. It is true that conditioned on Bob's outcome (if it happens first), we can infer the signal photon's state, but if Alice's measurement happens first, we need to condition Bob's state as well! Thus, story (2) is to be told. In a nutshell, the puzzle arises from ignoring the role of the photon that hits detector $D_0$ conditioned on whose outcome explains the behaviour at Bob's site. If the outcome of measuring the idler at $D_{1,2}$ is, say, $\ket{+}$, would we expect the measurement to have changed the past of the other particle to $\frac{1}{\sqrt{2}}(\ket{0}+\ket{1})$? Certainly not. Only when the signal photon has not yet encountered detector $D_0$ would we say it evolved to $\frac{1}{\sqrt{2}}(\ket{0}+\ket{1})$ given that the state of the idler photon yielded $\ket{+}$. Otherwise, the outcome of Alice's measurement will first give $\ket{0}$ or $\ket{1}$ and, as a result, leave the state of the idler photon in a superposition of $\ket{+}$ and $\ket{-}$.

Two comments are in order. For one, the puzzle only has any grip since the outcome statistics of Alice measuring first and Bob after equals the outcome statistics of Bob measuring first and Alice measuring after. Obviously, the time order in which measurement happens first does not matter as is clear from the system's wave function and is enforced by the no-signalling theorem of quantum mechanics. 
This allows the post-selection to be done after the outcomes have occurred. Violation of this condition would indeed lead to a genuine paradox, let alone allow for signalling. However, noticing this symmetry does not dispel the paradox but is the reason for why people can get confused about what is going on in the experiment. 

For the other, assuming the signal photon to be in one of the basis states before it or the idler photon were measured puts one into the business of hidden variable theories (after all, non-locality has it that the actual state of Alice indeed changes when Bob performs his measurement; see Section \ref{Bohm}). This introduces an ontological commitment to the value definiteness of states prior to measurement. For example, Egg and Ellerman correctly point out that if a detector can only detect one collapsed eigenstate, this does not mean that the photon was already in that state prior to that measurement (\cite{Egg2013,ellerman2011very,Ellerman2015}). That's why one might want to avoid phrases like `which-path information' as there isn't any information about which-path since no path was ever uniquely taken. Strictly speaking, without an account of what quantum measurements are to signify, i.e. a clear-cut distinction between the manifest and non-manifest (cf. Chapter \ref{section:manifest-non-manifest-domains}), the term `erasure' is a misnomer. It suggests that \textit{something} is supposedly erased in the experiment. But such a statement is empty, for it's unclear of what sort the thing is that is erased. Equally, if it means that some information is erased, information of what?

Most importantly, one can tell a coherent story about the states in the experiment without resorting to `retrocausal action into the past'. In light of this analysis, the puzzle seems trivial. 

In the double slit DCQE experiment (Section \ref{Scully}), the paradox is more disguised by the details of the experiment, but we will see that the same story can be told as in the Bell scenario.

\section{The Double Slit Delayed Choice Quantum Eraser}
\label{Scully}

The setup employed by Kim et al. uses double slit interference of photons and raises a conceptual problem, which, according to Wheeler and others, would allegedly imply that there was a change in the behaviour from `acting like a particle' to `acting like a wave', or vice versa, well after the particle entered the double slit.

\begin{figure}[h]
	\centering
	\includegraphics[width=0.8\linewidth]{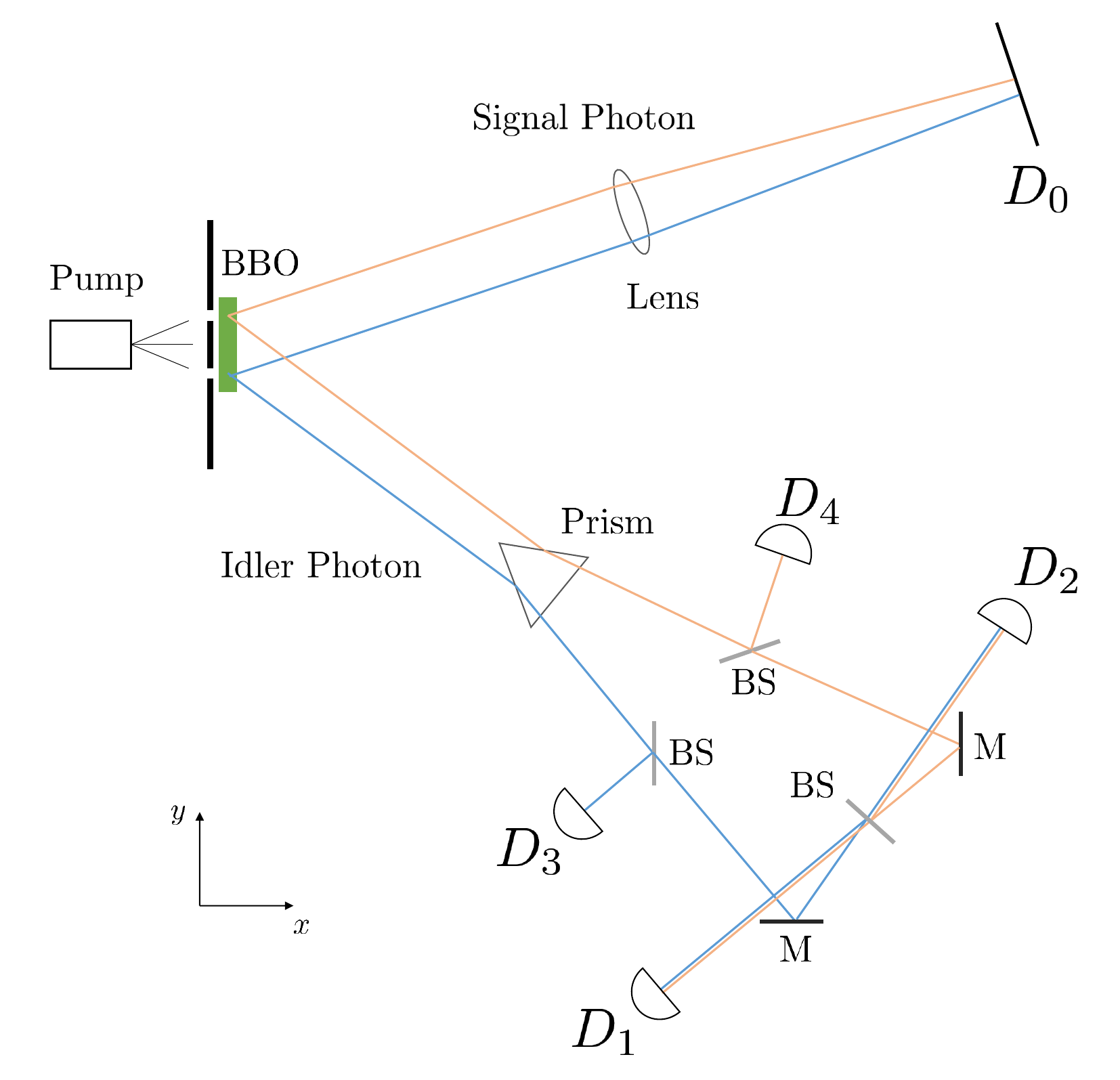}
	\caption[]{A delayed choice quantum eraser experiment. A laser beam aims photons at a double slit. After a photon passes the slits, it impinges on a Barium borate (BBO) crystal placed behind the double slit. The optical crystal destroys the incoming photon and creates an entangled pair of photons via spontaneous parametric down conversion at the spot where it hit. Thus, if one of the photons of the entangled pair can later be identified by which slit it went through, one also knows whether its entangled counterpart went through one or the other side of the crystal. Whether which-path information about the signal photon arriving at detector $D_0$ is obtained or erased is decided by manipulating the idler photon well after the signal photon has been registered.}
	\label{fig:Kim}
\end{figure}

In the old days of quantum mechanics, it was believed that the loss of interference in double slit experiments was due to Heisenberg's uncertainty principle, for no measurement device could be so fancy as not to perturb the system observed and destroy coherence. Such a perturbation leads to so-called `which-path information' that `collapses the wave function', making interference effects disappear. More concretely, when an interaction of the quantum particle with the measurement device occurs, the terms in the quantum state that led to interference are coupled with states of the measurement device, and those states of the combined system are orthogonal for a suitable measurement device. In the DCQE case, the which-path information of the photon is obtained by entanglement with an auxiliary photon without disturbing the wave function (cf.~Einstein's move in the EPR experiment \cite[p. 779]{PhysRev.47.777}). Significantly, the which-path information can be `erased' long after the photon encounters the double slit. This is possible by further measurement procedures on the entangled photon. The interference pattern, as a result, reappears. This was deemed inconceivable in the old picture since the state was believed to have been irrevocably disturbed as soon as a measurement happened. Figure \ref{fig:Kim} illustrates the experimental setup. 

The familiar story of how the DCQE works goes like this: A laser beam (pump) aims photons at a double slit. After a photon passes the slits, it impinges on a Barium borate (BBO) crystal placed behind the double slit. The optical crystal destroys the incoming photon and creates an entangled pair of photons via spontaneous parametric down conversion at the spot where it hit. Thus, if one of the photons of the entangled pair can later be identified by which slit it went through, we will also know whether its entangled counterpart went through one or the other side of the crystal.   
By contrast, we will have no which-path information if we cannot later identify where either of the photons came from. Even though the entangled photons created at the crystal are now correlated, the experiment can manipulate them differently. We call one photon of the pair the signal photon (sent toward detector $D_0$) and the other one the idler photon (sent toward the prism). The naming is a matter of convention. The lens in front of detector $D_0$ is inserted to achieve the far-field limit at the detector and, at the same time, keep the distance small between slits and the detector. The prism helps to increase the displacement between paths. Nothing about these parts gives which-path information, and detector $D_0$ can not distinguish between a photon coming from one slit or the other. At this point, we might expect interference fringes to appear at $D_0$ if we were to ignore that the signal photon and idler photon are entangled. Considering the signal photon alone we might think that the parts of the wave function originating at either slit should interfere and produce the well-known pattern of a double slit experiment. On the other hand, quantum mechanics would predict a typical clump pattern if by taking into account the idler photon, which-path information were available.

After the prism has bent the idler photon's path, the particle heads off to one of the 50-50 beamsplitters \textit{BS}. The photon is reflected into the detector $D_3$ a random 50\% of the time when it travels on the lower path, or reflected into the detector $D_4$ a random 50\% of the time when it is travelling on the upper path. If one of the detectors $D_3$ or $D_4$ clicks, a photon is detected with which-path information. That is, we know at which slit both photons of the entangled pair were generated. In that case, the formalism of quantum mechanics predicts no interference at $D_0$. In all of the other cases, the photon passes through the beamsplitter and continues toward one of the mirrors $M$. Importantly, it does not matter if the choice of whether the photon is reflected into the which-path detectors $D_3$ or $D_4$ is made by beamsplitters. This is because the original experiment uses beamsplitters, so it is randomly decided which measurement is performed. But we could equally replace the beamsplitters with moveable mirrors. In that way, the experimenter can decide which-path information is available by either keeping the mirrors in place or removing them so that the photon can reach the eraser. 

After being reflected in one of the mirrors, the photon encounters another beamsplitter $BS$ --- the quantum eraser. This beamsplitter brings the photon into a superposition of being reflected and transmitted. To that end, for an idler photon coming from the lower mirror, the beamsplitter either transmits the photon into detector $D_2$ or reflects it into detector $D_1$. Likewise, for an idler photon coming from the upper mirror, the beamsplitter either transmits it into detector $D_1$ or reflects it into detector $D_2$. If one of the detectors $D_1$ or $D_2$ clicks, it is impossible to tell which slit the photon came from. 
To summarise the above, detectors $D_1$ and $D_2$ placed at the output of $BS$ erase the which-path information, whereas a click of detectors $D_3$ or $D_4$ provides which-path information about both the idler and the signal photon. 
Notably, when the photon initially hits $D_0$, there is no which-path information available, only later when the entangled idler photon is detected at $D_3$ or $D_4$.
\begin{figure}[H]
	\centering
	\includegraphics[width=1\linewidth]{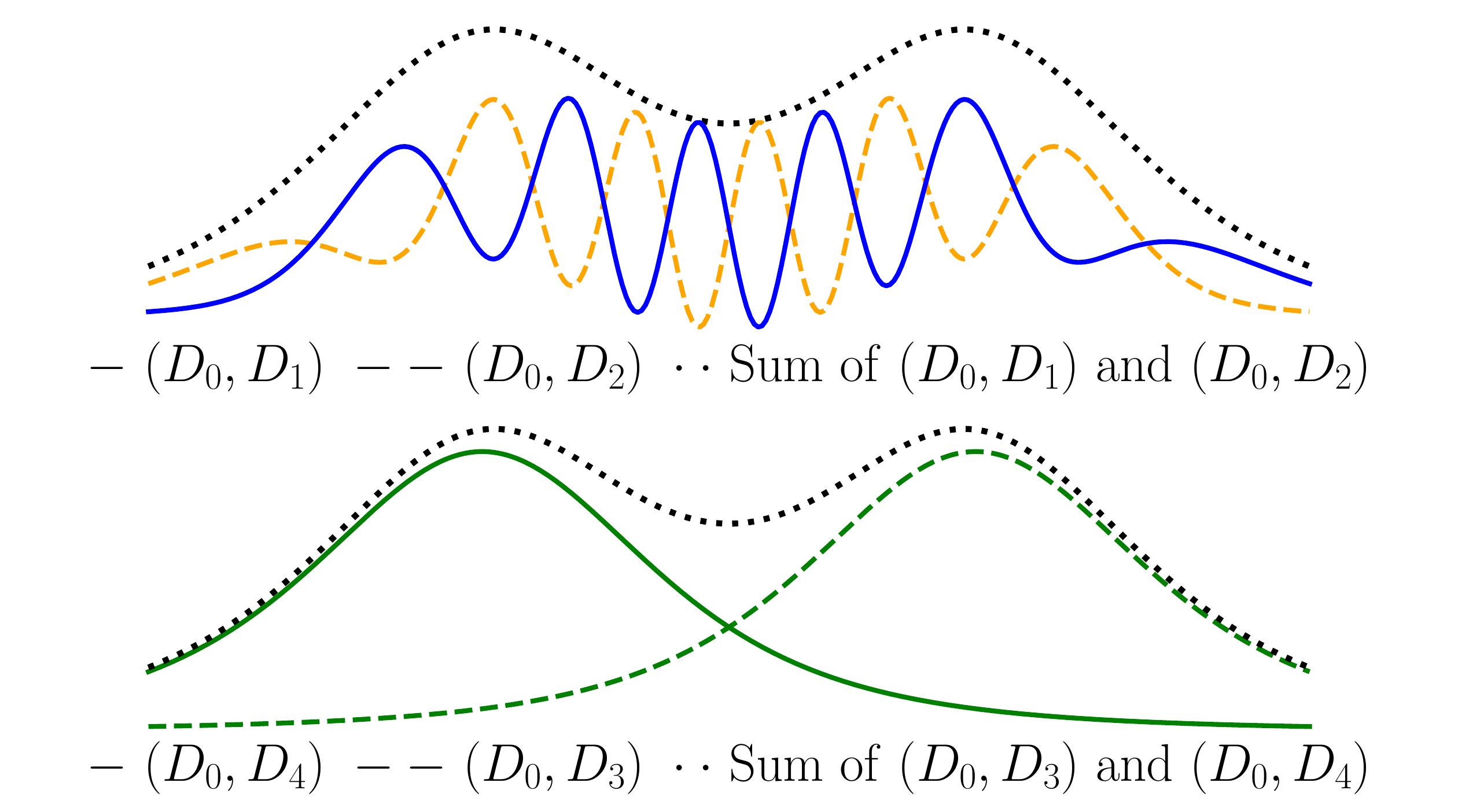}
	\caption{Joint detection events at detector $D_0$ and detectors $D_1$-$D_4$. The figure shows a plot of the bits of Equation \ref{expression}. Records of $D_0$ and $D_1$ ($D_2$) show interference fringes. On the contrary, records of $D_0$ and $D_3$ ($D_4$) show a clump pattern. If added together, the distribution on the screen in both cases (with and without the eraser) is always two clumps.}
	\label{fig:patterns}
\end{figure} 
This is key. The setup ensures that the which-path information is only erased or provided, respectively, after $D_0$ has detected the signal photon. We, therefore, say the choice is delayed.  
For each incoming photon from the laser beam, there will be a joint detection of the signal photon at $D_0$ and the idler photon at $D_1$--$D_4$. Figure \ref{fig:patterns} shows the expected results.\footnote{The results in \cite{Kim1999} show a single clump as opposed to two clumps in Figure \ref{fig:patterns}. This is simply due to the close distance between the slits Kim et al. chose for their experiments.} When which-path information is provided, a clump pattern appears, but interference fringes appear when no which-path information is available. Therefore, the two interference patterns corresponding to correlation with $D_1$ and $D_2$ are out of phase. The reason for that will become clear in the next sections. 

Those inclined to instrumentalism might be satisfied at this point, for the predictions of standard quantum mechanics give the desired results to confirm the experimental observation. Then, however, the philosopher might start to worry about what is happening here. 

\section{Backwards in Time Influence?}
Indeed, it may be tempting to interpret these results as instances of future measurements influencing past events. Seemingly, there is something odd going on in the experiment. The appearance on the screen (either one that shows interference or one that shows a clump pattern) of the signal photon is determined by how the idler photon is measured --- a choice that occurs after the signal photon has already been detected. Can a later, distant measurement cause an entangled particle to alter its wave function retroactively? It seems the detection of the idler photon and thus the choice of which-path information affects the behaviour of the signal photon in the past. Is this a process that reverses causality? Wheeler comments on his original \textit{Gedankenexperiment} as follows:
\begin{quote}
	\textit{`Does this result mean that present choice influences past dynamics, in contravention of every formulation of causality? Or does it mean calculating pedantically and do not ask questions? Neither; the lesson presents itself rather like this, that the past has no existence except as it is recorded in the present.'} \cite{wheeler1978past}
\end{quote}
Thus, Wheeler accepts that the past could be created \textit{a posteriori} by happenings in the future. In comparison, Bohr concludes that understanding the quantum behaviour of particles is confused by giving pictures that try to maintain conceptions of classical physics. He states that a sharp separation of the quantum system and the observing measurement device is impossible \cite{bohr1961atomic}. In his view, there is no point in visualising the process as a path a particle takes when not in a well-defined state. In more recent work, Brian Greene argues that delayed choice quantum eraser scenarios may not alter the past, but future measurements certainly determine the story we tell about a particle's past behaviour. His account, although, is too vague to reach a satisfactory resolution \cite{brian2004fabric}[pp. 194-199]. 

One should not expect the formalism of quantum mechanics to provide clear images of what could be actually going on, for at the moment, it is a framework with different interpretations. Only if one is to adopt an interpretation can a conclusion be meaningful. Many physicists and philosophers did not accept the views of Wheeler or Bohr and had been continuing to debate the delayed choice experiment to seek possibilities that account for physical intuition. 

\section{Delayed Choice in Standard Quantum Mechanics}

The first significant point is that there never appears an interference pattern at $D_0$ without conditioning on whether we choose which-path information to be available or erased.\footnote{Note that in the experiment of Kim et al., the decision is made randomly by the beamsplitters next to the prism, but as I mentioned, they can be replaced with mirrors and allow the experimenter to make this choice.} Technically, by conditioning, we mean to constrain the already observed measurement results to the subset of coincidence detections of the signal photon with the idler photon in a chosen detector $D_1$--$D_4$. Moreover,  the two interference patterns from the joint detection events of $D_0$ and $D_1$ or $D_2$, respectively, obtain a relative phase shift of $\pi$ and cancel when added together. This feature is often left out but is crucial, as we shall see.    

I shall analyse the experiment proposed by Kim et al. using standard quantum mechanics.\footnote{Egg discussed Scully's delayed choice version involving a cavity to distinguish between which-path measurement and interference measurements \cite{Egg2013}. He gives a straightforward account of the experiment in standard quantum mechanics. However, he presents some insights on metaphysical claims on entanglement realism that are taken into question by delayed choice scenarios. But, in my view, there is too little emphasis on the importance of the point that the paradox results from misinterpreting the pre/post measurement symmetry in quantum mechanics and the significance of detecting the signal photon and condition on its outcome. Having said that, it might not have been Egg's endeavour to resolve the paradox in the DCQE in the way I am describing here since I think the main point for him was to argue against non-realist arguments (e.g. in \cite{healey2012quantum}) that seem to undermine the physical reality of entanglement utilising delayed choice scenarios.} Schrödinger's equation describes the wave functions involved, which strictly applies to massive particles. For a rigorous treatment with photons, we need to avail ourselves of quantum field theory.

Nevertheless, we can straightforwardly replace photons with electrons for the sake of a \textit{Gedankenexperiment}. The interference phenomena qualitatively remain the same. However, I shall retain the term `photons' in the derivation throughout the paper for clarity. 

The incoming laser beam can be described as a plain wave 
\begin{equation}
	\label{laser}
	\psi=e^{ik_xx}
\end{equation} impinging on the double slit, where $k_x$ is the wave vector.\footnote{For the sake of simplicity, we can suppress time dependence of the wave function since it does not affect the argument. I omit normalisation factors where not stated explicitly.} After the slits, the wave function can be decomposed into two interfering parts as
\begin{equation}
	\psi= \frac{1}{\sqrt{2}}(\psi_1+\psi_2).
\end{equation} Wave function $\psi_1$ belongs to the part of the wave function emerging from the upper slit and $\psi_2$ to the part of the wave function emerging from the lower slit. We may assume waves of the form
\begin{equation}
	\psi_i=\frac{e^{ikr_i}}{r_i},
\end{equation} where $r_i$ is the distance from the slit $i$. 
These give the well-known two-slit interference fringes. The crystal then creates an entangled pair of photons with opposite momenta in the $y$-direction such that 

\begin{equation}
	\label{pair}
	\psi=\frac{1}{\sqrt{2}}(\psi_1\otimes \psi_1'+\psi_2\otimes \psi_2'),
\end{equation} where unprimed wave functions correspond to the signal photon and primed to the idler photon. The signal photon sent to detector $D_0$ is now entangled with the idler photon. This affects the probability amplitudes at $D_0$, and interference between $\psi_1$ and $\psi_2$ vanishes since $\psi_1\otimes \psi_1'$ and  $\psi_2\otimes \psi_2'$ are orthogonal states. Note that $\psi_1'$ and $\psi_2'$ are thought to be non-overlapping, and thereby the inner product vanishes. Note also that, in general, the orthogonality of two states does not imply zero overlap of the states in a position basis. The squared norm of the wave function yields

\begin{equation}
	|\psi|^2= \frac{1}{2}(|\psi_1|^2|\psi_1'|^2+|\psi_2|^2|\psi_2'|^2).
\end{equation} By integrating out the idler degrees of freedom, we find for the probability distribution of the signal photon on the screen
\begin{equation}
	\rho= \frac{1}{2}(|\psi_1|^2+|\psi_2|^2).
\end{equation} From this, it is clear that no interference will appear on the screen for this state. Assuming the signal has not yet reached $D_0$, if the idler gets reflected into detector $D_3$ the state would be $\psi_2\otimes \psi_2'$, and if reflected into $D_4$ it would be $\psi_1\otimes \psi_1'$. 
If the idler photon encounters the quantum eraser, the wave function undergoes another unitary evolution. The eraser puts the idler photon in a superposition of being transmitted to one detector or reflected to the other. At each reflection at a beamsplitter or mirror the wave function picks up a phase of $\frac{\pi}{2}$ (a multiplication of the wave function by $e^{i\frac{\pi}{2}}=i$) such that
\begin{align}
	\psi_1' &\mapsto \ ~ i\psi_{D1}-\psi_{D2}\nonumber\\
	\psi_2' &\mapsto  -\psi_{D1}+i\psi_{D2}.
\end{align} The joint wave function then turns into
\begin{align}
	\label{expression}
	\psi &= \frac{1}{2}(\psi_1\otimes (i\psi_{D1}-\psi_{D2})+\psi_2\otimes (-\psi_{D1}+i\psi_{D2})) \nonumber \\ &=\frac{1}{2}((i\psi_{1}-\psi_{2})\otimes\psi_{D1} + (-\psi_{1}+i\psi_{2})\otimes\psi_{D2})
\end{align} 
once the idler photon has passed the quantum eraser. Indices in $\psi_{D_1}$, $\psi_{D_2}$ refer to which detector the part of the wave function is reflected into. In this form, state \ref{expression} makes it clear that when detector $D_1$ clicks, conditioned on this event, the state of the signal photon is $i\psi_{1}-\psi_{2}$, yielding a probability distribution of interference fringes, 
\begin{align}
	|\psi_{D_0, D_1}|^2 &= (i\psi_{1}-\psi_{2})\overline{(i\psi_{1}-\psi_{2})}\nonumber \\
	&=|\psi_1|^2+|\psi_2|^2-2\Im(\overline{\psi_1}\psi_2).
\end{align} In the case in which $D_2$ clicks, conditioned on that event the state is $-\psi_{1}+i\psi_{2}$ and yields a distribution showing shifted anti-fringes:
\begin{align}
	|\psi_{D_0, D_2}|^2 &= (-\psi_{1}+i\psi_{2})\overline{(-\psi_{1}+i\psi_{2})} \nonumber 
	\\ &=|\psi_1|^2+|\psi_2|^2-2\Im(\psi_1\overline{\psi_2}) \nonumber
	\\ &=|\psi_1|^2+|\psi_2|^2+2\Im(\overline{\psi_1}\psi_2).
\end{align} In both of the cases, there is a path on which the idler is reflected twice, and a path on which it is reflected once.

So far, there is no puzzle. The experiment of Kim et al., however, is designed such that the choice of whether the state produces interference fringes or a clump pattern happens after the signal photon has been detected at $D_0$. We, therefore, say the choice is delayed. In the setup of \cite{Kim1999} the optical length of the idler photon is about $8\ \nano\second$ longer than that of the signal photon. If we accepted that the causal story to be told about what is going on with the signal photon at any time is purely determined by what happens to the idler photon, then this would suggest backwards action from the future since the measurement of the signal photon has already occurred. With all this in mind, must we conclude that a measurement in the present retroactively changes the past to make it agree with the measurement outcomes? 

Crucially, at detector $D_0$ there never appears an interference pattern, regardless of whether the idler photon reaches the quantum eraser. This can readily be seen by adding up the distributions:
\begin{equation}
	|\psi_{D_0, D_1}|^2 + |\psi_{D_0, D_2}|^2 = |\psi_1|^2+|\psi_2|^2.
\end{equation} The interference terms cancel out when added together which effectively leads to a clump pattern. Each sub-case shows an interference pattern, but the overall statistics add up to two clumps. Note that there is no way to avoid the phase difference in the interference fringes since any additional device would act symmetrically on both paths. Insert, for instance, a $\lambda/4$-plate into the paths of the idler photon, and it will affect both superposed paths reflected into the detectors. Thus, the effect of the plate would cancel out.

Of course, the fact that at detector $D_0$, interference fringes never occur guarantees consistency with signal-locality between $D_0$ and the other detectors. That is to say, it is not possible to decide what distribution (either an interference pattern or a clump pattern) appears at the detector $D_0$ by choice of whether the idler photon will trigger the which-path detectors $D_3$ and $D_4$ and thus communicate information. However, as I noted above, this choice can be realised by replacing the former two beamsplitters with mirrors which can be inserted \textit{ad libitum} by the experimenter (compare no-signalling in \textit{EPR}).

The apparent retroactive action vanishes if a click in $D_0$ is regarded to condition the overall wave function too, not only a click in the detectors $D_1$--$D_4$ (think back to the discussion of the Bell-type experiment in Section \ref{Bell}). In the standard explanation, if the idler photon's detection happens before the signal photon's detection at $D_0$, the detectors $D_1$--$D_4$ determine the state of the signal photons end up in. But similarly, in the case when the signal photon is detected at the moment in time preceding the observation of the idler photon, the detected position of the signal photon determines the state of the idler photons, which will go on to trigger one of the detectors $D_1$--$D_4$. 

We can see this by rewriting state \ref{pair} in the position basis of the signal photon. Let's first work out the state for the which-path measurement:
\begin{align}
	\psi&=\frac{1}{\sqrt{2}}\int(\psi_1(x)\ket{x}\otimes \ket{\psi_1'}+\psi_2(x)\ket{x}\otimes \ket{\psi_2'})dx\\ \nonumber
	&= \frac{1}{\sqrt{2}}\int\ket{x}\otimes(\psi_1(x) \ket{\psi_1'}+\psi_2(x)\ket{\psi_2'})dx
\end{align}
From this, we can see that if the signal photon hits the screen at position $x$, the probability for a click in $D_3$ is (roughly, since the two lumps slightly overlap) $|\psi_2(x)|^2$. And for detector $D_4$ it is roughly $|\psi_1(x)|^2$.  In other words, we can be almost sure which of the two detectors $D_3$ or $D_4$ will fire by looking at what lump on the screen the signal photon ends up in. Note that, vice versa, conditioned on a click in the respective detectors, we previously found  $\psi_2(x)$ or $\psi_1(x)$ to be the state determining the distribution of signal photon hits on the screen. Thus, the two causal stories are consistent with the observations, and the same correlations between signal and idler photon arise as expected. Note that for a quantum eraser, the wavepackets of the signal photon need to overlap sufficiently in position space in order to give rise to interference that can be `erased'. This is true at the screen and in the overlap region before the screen. Where the wavepackets do not overlap, there is a set of quantum particles that do not interfere (also compare with \cite{Quanta87}). In the idealised case of circular waves considered here, the waves, in fact, overlap at all points in space between the source and screen. 

Conversely, by rewriting Equation \ref{expression} for the interference measurement after the idler passed the quantum eraser, we obtain
\begin{align}
	\psi &= \frac{1}{2}\int[\psi_1(x)\ket{x}\otimes (i\ket{\psi_{D1}}-\ket{\psi_{D2}})+\psi_2(x)\ket{x}\otimes (-\ket{\psi_{D1}}+i\ket{\psi_{D2}})]dx \nonumber \\ &=\frac{1}{2}\int\ket{x}\otimes[(i\psi_1(x)-\psi_{2}(x))\ket{\psi_{D1}}+(-\psi_{1}(x)+i\psi_{2}(x))\ket{\psi_{D2}}]dx.
\end{align} The probability for $D_1$ to fire is $|i\psi_1(x)-\psi_{2}(x)|^2$ and for $D_2$ it is $|-\psi_1(x)+i\psi_{2}(x)|^2$ if the signal photon was detected at position $x$ on the screen. Again, as before, we recover that the probabilities are consistent with the time-reversed story where the idler photon is detected first. That is, state $i\psi_1(x)-\psi_{2}(x)$ determines the outcomes on the screen for conditioning on $D_1$ and state $-\psi_1(x)+i\psi_{2}(x)$ for conditioning on $D_2$.


One needs clarification if one is to stubbornly stick to the notion that a measurement of the idler photon determines the probability distribution at $D_0$ for the signal photon. Observation of individual subsystems of entangled pairs never changes the probability distribution of the remote particle. After all, the conditional probabilities of signal and idler photon measurement outcomes are spatio-temporally symmetric. Gaasbeek tried to reason in his \citeyear{Gaasbeek2010} that this idea alone were to demystify the paradox. Though, I wish to emphasise again that the symmetry in the time-ordering is crucial to realising why the alleged paradox arose in the first place. If the two causal stories I just outlined above gave \textit{different} predictions for the probabilities of outcomes depending on which photon is detected first, a causal direction can be established since from the patterns on the screen one could tell which measurement happened first. However, as the probabilities are invariant under the time-order of measurements on the signal and idler photon, one can get confused as to whether the idler photon could retroactively determine the patterns on the screen. 


This tells us that no matter how the idler photon gets manipulated, the probability distribution on $D_0$ is a clump pattern. Still, when we condition on the outcome of the detectors, which either give which-path information or not, we find correlations as expected and, most importantly, the same correlations arise when the conditioning on the outcome of the signal photon. The quantum eraser does not influence the past of the signal photon; rather, it reveals the correlations of an entangled photon pair in just another way. This only is puzzling because the probabilities conditioned on the post-selection are time-symmetric in the pre and post-selected states. 

To reinforce the point, compare the situation with the Bell scenario in Section \ref{Bell}: The source $S$ of an entangled pair of photons can be identified with the laser beam, the double slit, and the BBO crystal. $M$ denotes a mirror that can be used to reflect the idler photon into $D_{3,4}$. In the Bell scenario detectors $D_3$ and $D_4$ were concatenated into one detector, where an outcome $\ket{0}$ would correspond to detection at $D_3$ and an outcome $\ket{1}$ to detection at $D_4$. We stipulate that the signal photon is sent towards the lens and the idler photon to the prism. If we are to perform a which-path experiment, we measure the idler photon in the computational basis $\{\ket{0},\ket{1}\}$ at $D_{3,4}$. Detector $D_0$ measures the signal photon in the computational basis, which corresponds to an interference measurement if the state of the signal photon, for instance, is one of the states of the diagonal basis $\{\ket{+},\ket{-}\}$. The measurement on the idler photon in the diagonal basis (at $D_{1,2}$) acts as the quantum eraser, i.e. a measurement of the idler photon in the diagonal basis is consistent with the signal photon being in a superposition of $\ket{0}$ and $\ket{1}$. The results of the detectors $D_0$ conditioned on the outcome of $D_{1,2}$ show familiar correlations when compared. 

When a photon has passed the BBO crystal, the quantum state ends up entangled. After rewriting the second slot of the state in the diagonal basis, we recover a wave function that is qualitatively identical to Equation \ref{expression}. Thus, the paradox in the double slit case resolves in the same ways as the one we alleged to the Bell-type scenario in Section \ref{Bell}.


\section{Delayed Choice in de Broglie–Bohm Theory}
\label{Bohm}

In pilot wave theory, Bohmian Mechanics or de Broglie-Bohm theory, it is supposed that particles follow definite trajectories at all times. Thus, working out the DCQE in such a framework is particularly appealing when it comes to verifying whether past trajectories could in any way depend on future measurements. As it turns out, as well as in the standard quantum treatment, in the hidden variable approach, the puzzle resolves. Although, for instance, Hiley and Callaghan treated a double slit version of the delayed choice quantum eraser in Bohmian mechanics in \cite{hiley2006erased} it is instructive to treat the DCQE version described here. Hiley and Callaghan analyse a case in which which-path information is acquired by a cavity wherein atoms get excited. This leads to interference fringes and anti-fringes that do not appear in the setup employed by Kim et al. Therefore, I shall also work out the ongoings of Kim's setup within de Broglie-Bohm's theory in the following. 

I will use the term `de Broglie-Bohm theory' to stand for the interpretation discussed by \cite{bohm2006undivided}. A detailed account of the theory is given in Appendix \ref{Appendix}. Here it is assumed that a particle always travels on only one path. The wave function is considered as a pilot wave and used in its polar form, i.e. 

\begin{equation}
\psi(\vec{r},t)=R(\vec{r},t)e^{iS(\vec{r},t)/\hbar}.
\end{equation} The dynamics of the pilot wave obey the Schrödinger equation
\begin{equation}
i\hbar\partial_t\psi=H\psi
\end{equation} and the particle's trajectory is determined by

\begin{equation}
\vec{v}\left(t\right)=\dot{\vec{x}}(t)=\frac{1}{m}\nabla S(\vec{r},t)|_{\vec{r}=\vec{x}}
\end{equation} where $m$ is the mass of the particle. For simplicity, I will set $\hbar=1$ for the remainder.

Now let us turn to consider how particles behave according to de Broglie-Bohm in this experiment. First, we construct a set of possible trajectories, each individually corresponding to one initial value of the position of the particle within the incident beam. Supposedly, de Broglie-Bohm theory should reveal whether present observations influence the past since it assumes a well-defined path of the particles at all times. Note that the de Broglie-Bohm interpretation does allow us to illustrate such a process and reproduce all the known experimental results in tension with Wheeler's and Bohr's conclusions about these phenomena.

The wave function of the incoming laser beam \ref{laser} is already in polar form, and the trajectories in this region are straight lines. First, we consider the case without the eraser. To work out what happens, we must write the final wavefunction in Equation \ref{pair} in the form\footnote{For simplicity, I suppress normalisation factors.}

\begin{equation}
\psi(r,r')= R(r,r')e^{iS(r,r')}.
\end{equation} The wave function is evaluated at the positions of the signal photon $r$ and the idler photon $r'$. It decomposes as 
\begin{align}
\psi(r,r')&=R_1(r)e^{iS_1(r)}
R_1'(r')e^{iS_1'(r')}\nonumber\\
&+R_2(r)e^{iS_2(r)} R_2'(r')e^{iS_2'(r')}.
\end{align} Again, primed variables correspond to the idler photon. For the final amplitude $R$ and the phase $S$ we find
\begin{equation}
R^2 =(R_1R_1')^2 + (R_2R_2')^2 + 2R_1R_1'R_2R_2'\cos\Delta \phi,
\end{equation} by the law of cosines, where $\Delta\phi=(S_2+S_2')-(S_1+S_1')$. Also, 
\begin{equation}
\tan S = \frac{R_1R_1'\sin(S_1+S_1')+R_2R_2'\sin(S_2+S_2')}{R_1R_1'\cos(S_1+S_1')+R_2R_2'\cos(S_2+S_2')}.
\end{equation}
We need to evaluate this term for each trajectory. For the photon travelling through the upper slit, the entangled pair is created at this slit, and since the probability of creating an entangled pair at the lower slit is zero when the photon does not pass through it, $R_2'=0$ (since $R_2'$ has no support in the upper slit). Importantly, $R_2\neq 0$ at points where $R_1$ has support. Vanishing $R_2'$ on this trajectory cancels out overlapping terms so that $R^2=(R_1R_1')^2$ and interference vanishes. Recall that the wave function is evaluated at the positions of all the particles involved. Likewise, if the photon's path goes through the lower slit, $R_1'=0$. Thus, $R^2=(R_2R_2')^2$ and interference vanishes as before. The guiding phase in the former case yields
\begin{equation}
S= S_1(r)+S_1'(r').
\end{equation} That means that the guidance equation for the signal photon becomes independent of $S_2$ and $S_2'$:

\begin{equation}
p_1=\nabla_{r}S= \nabla_{r}S_1(r),
\end{equation} with $p_1$ the particle's momentum.\footnote{Again, we should talk about massive particles for the guidance equation to make sense.} The idler photon then continues to travel to detector $D_4$ or $D_1$. Similarly, in the latter case, the signal photon is independent of $S_1$ and $S_1'$. The idler photon then continues to travel to detector $D_3$ or $D_2$. The gradients $\nabla S_1, \ \nabla S_2$ (and consequently the momentum) point in the radial direction away from the slits. The interference terms cancel whatever paths the photons follow (i.e., `the signal photon has passed the upper slit' or `the idler photon follows a path towards detector $D_4$'). Then, all of the other potential states do not contribute to the guidance equation, so the interference term cancels.

I will now turn to the situation where the quantum eraser is present, but we remove the two beamsplitters reflecting the idler photons into the which-path detectors. The question is whether the trajectories change when we consider the wave function of the eraser.
Recall the wave function of the system when the idler photon has passed the eraser:
\begin{align}
\label{eraser}
\psi &= \frac{1}{2}(\psi_1\otimes (i\psi_{D1}-\psi_{D2})+\psi_2\otimes (-\psi_{D1}+i\psi_{D2})) \nonumber \\
&= \frac{1}{2}((i\psi_{1}-\psi_{2})\otimes\psi_{D1} + (-\psi_{1}+i\psi_{2})\otimes\psi_{D2}).
\end{align} Or in polar form
\begin{align}
\label{bohmeraser}
\psi&= R_1(r)e^{iS_1(r)}(R_{D_1}(r')e^{iS_{D_1}(r')+i\frac{\pi}{2}}\\
&-R_{D_2}(r')e^{iS_{D_2}(r')})\nonumber \\
&+ R_2(r)e^{iS_2(r)}(-R_{D_1}(r')e^{iS_{D_2}(r')}\nonumber\\
&+R_{D_2}(r')e^{iS_{D_2}(r')+i\frac{\pi}{2}})\nonumber.
\end{align} Consequently, unlike in the case without the eraser, the signal photon is guided by a potential with contributions both from $R_1$ and $R_2$. Indeed, assume the idler photon to end in the path leading to detector $D_1$. That means $R_{D_2}=0$, and the trajectory of the signal photon is determined by
\begin{align}
& R_1(r)e^{iS_1(r)}R_{D_1}(r')e^{iS_{D_1}(r')+i\frac{\pi}{2}} \nonumber \\
-& R_2(r)e^{iS_2(r)}R_{D_1}(r')e^{iS_{D_2}(r')},
\end{align} and vice versa by
\begin{align}
&- R_1(r)e^{iS_1(r)}R_{D_2}(r')e^{iS_{D_2}(r')} \nonumber
\\ &+R_2(r)e^{iS_2(r)}R_{D_2}(r')e^{iS_{D_2}(r')+i\frac{\pi}{2}}
\end{align} if the idler photon travels toward detector $D_2$. In both cases, the paths are those wiggly trajectories which photons take in the usual double slit experiment (up to a phase shift). These trajectories produce the same interference patterns we came across in Figure \ref{fig:patterns}. Bear in mind that if added, they produce a clump pattern.    

\begin{figure}[H]
\centering
\begin{subfigure}[b]{0.3\textwidth}
	\includegraphics[width=\textwidth]{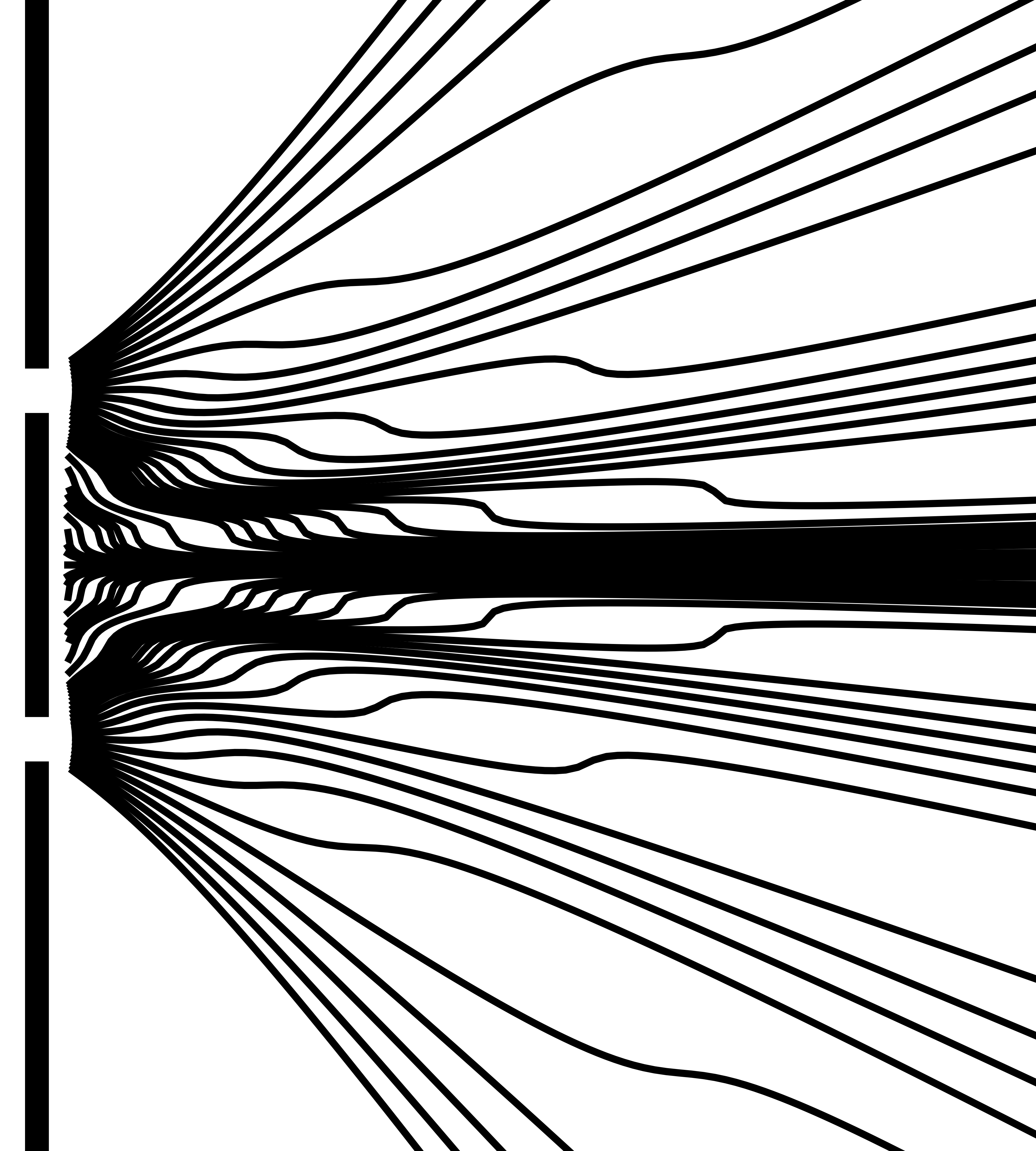}
	\caption{}
	\label{fig:s1}
\end{subfigure}
~
\begin{subfigure}[b]{0.3\textwidth}
	\includegraphics[width=\textwidth]{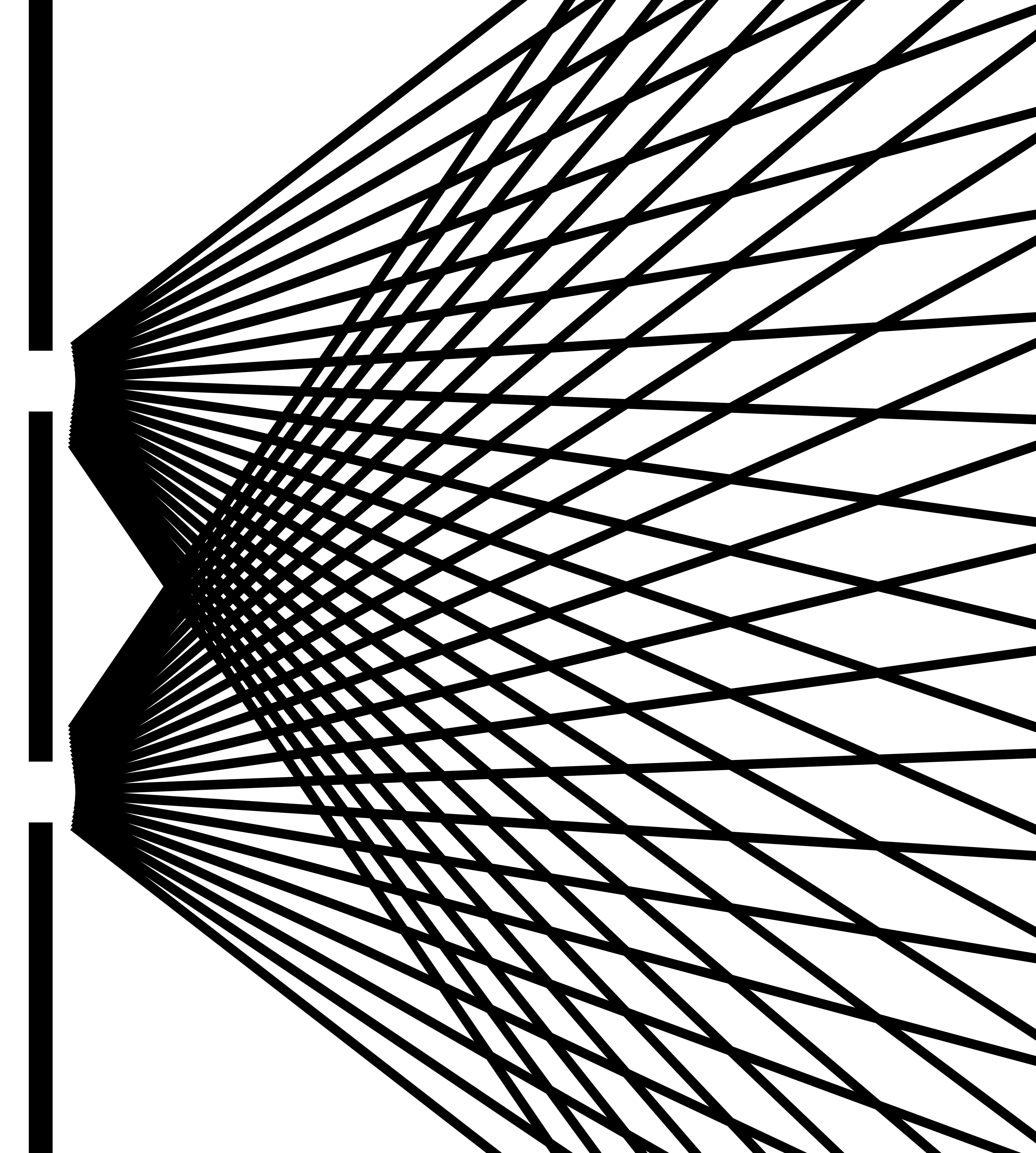}
	\caption{}
	\label{fig:s2}
\end{subfigure}
~
\begin{subfigure}[b]{0.3\textwidth}
	\includegraphics[width=\textwidth]{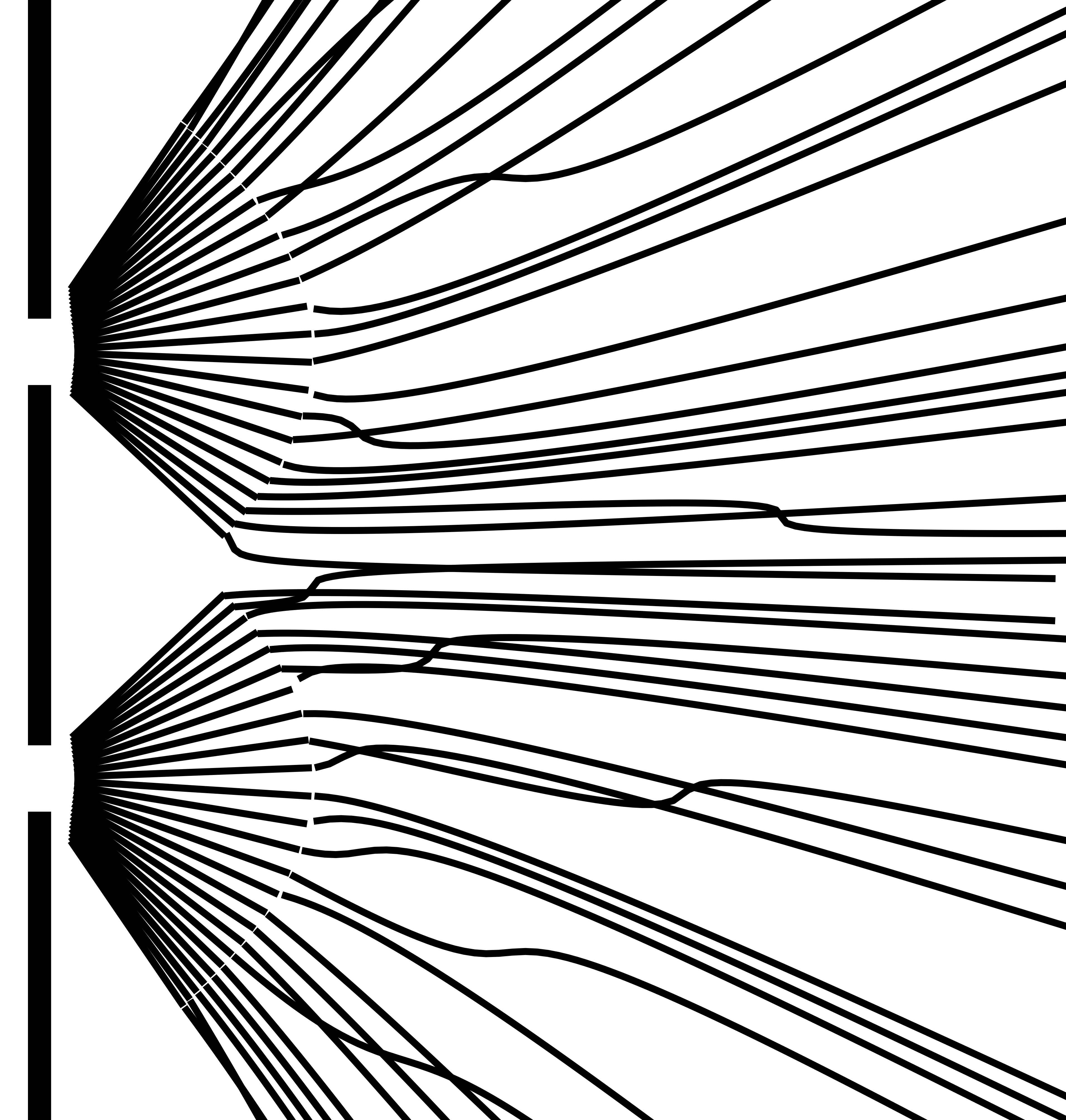}
	\caption{}
	\label{fig:s3}
\end{subfigure}

\caption{The signal photon follows different trajectories depending on when the idler photon encounters the quantum eraser. \textbf{(a)} The well-known wiggly trajectories that lead to an interference pattern in a usual double slit experiment. \textbf{(b)} In the case where the idler photon hits the quantum eraser after the signal photon arrives at the screen (which is how the experiment is set up in \cite{Kim1999}), the signal photon moves on straight lines.  \textbf{(c)} Before the idler photon has encountered the quantum eraser, the signal photon follows straight lines. When the idler photon travels to detector $D_1$ or $D_2$, a jump in the guidance relation happens, leading to trajectories as in the interfering case.}
\label{fig:Doppelspalt}	
\end{figure}

The eraser drastically changes the wave function, but at the same time, the signal photon's past trajectory is not influenced by the change. Depending on when the idler photon enters the region between the eraser beamsplitter and detectors $D_1$ or $D_2$, the signal photon jumps from moving on straight lines to following wavy trajectories typical for interference. This is striking, for the effects on the signal photon are mediated superluminally, in conflict with special relativity. On the other hand, this should not be surprising, for non-locality is one of the features of a hidden variable theory like de Broglie-Bohm's. In the experiment of \cite{Kim1999}, the moment when the idler photon encounters the eraser is always after the signal photon hits the detector. Therefore, the Bohmian trajectories, in that case, look like the straight lines in Figure \ref{fig:s2}. This shows that it is possible to observe interference effects without wavy trajectories in de Broglie-Bohm theory. After the post-selection of the correlated sub-ensembles of signal and idler photon, the two phase-shifted interference patterns are recovered (even without the guidance of a superposition state!). If one adjusted the delay and shortened the optical length of the idler photon such that it passes through the eraser during the signal photon travelling toward $D_0$, the trajectories would look like those in Figure \ref{fig:s3}.

Let us recap. There are two ways in which interference fringes can emerge at the detector $D_0$. When the idler photon arrives at the eraser during the flight of the signal photon, then the signal photon continues to move on wiggly lines giving rise to fringes. There is no change of the past whatsoever. When the idler photon arrives after the signal photon encounters $D_0$, the trajectories are straight lines (see Figure \ref{fig:Doppelspalt}). In this case, selecting out interference patterns by conditioning on $D_1$ and $D_2$ does not change trajectories of the past. The reason we can extract interference fringes is that one subset of the trajectories of the signal photon is consistent with the idler photon being detected at $D_1$ (interference fringes), and another subset is consistent with detection in $D_2$ (anti-fringes), and both add up to a clump pattern. This is the case in the experiments by Kim et al. and confuses if we do not consider conditioning on the signal photon, thus calling for the need for `backwards in time influence' to restore the interference outcomes. It also trivially follows from my analysis that there is no need to invoke `entanglement in time'. I do not use any non-standard features of standard quantum mechanics or de Broglie-Bohm theory. Pilot wave dynamics restores the conventional view of the world as particles having a definite trajectory and past. In Wheeler's view, the past comes into existence only after the measurement in the present, but my analysis gives an account that consistently attributes a past to the photon's trajectory.

\section{Conclusion}

The delayed choice quantum eraser experiment resembles a Bell-type experiment and thus is not more mysterious than that. 
There is no need to invoke a notion such as `the present action determines the past'. I have shown this to be fairly straightforward in the Bell framework. The original puzzle arises due to the symmetry property that the time ordering of which party measures first is irrelevant for the statistics of the outcomes and this allows for an alternative explanation in terms of `action into the past'. But such is unwarranted. One can treat the same phenomenon in two different ways given this property, but one of them looks as if something would have to change in the past.  

We can consistently derive the probabilities for different measurement outcomes in the delayed choice quantum eraser experiment from standard quantum mechanics. Whenever the idler photon is manipulated in a way that provides which-path information about the signal photon, detector $D_0$ does not show interference, even if conditioned on the idler photon's specific measurement results. On the other hand, if the idler photon is detected such that the measurement irrevocably erases which-path information about the signal photon, then the interference patterns reappear. Those distributions are complementary because they add up to a clump pattern. Further, the patterns can be extracted only conditioned on the idler photon's detector outcomes.

I have shown that the experiment can be understood without invoking `backwards in time influence' in standard quantum mechanics and in the de Broglie-Bohm theory. Properly conditioning the system's state without neglecting the measurement of the signal photons explains why there is no paradox. The seemingly retroactive action disappears if the effect of measurement on the state of the signal photon is considered to also change the overall state. In the de Broglie-Bohm theory, the particle takes one definite trajectory and does not change its past during its motion. However, the idler photon may determine the pilot wave function of the signal photon depending on when the idler photon passes the quantum eraser. Most importantly, de Broglie-Bohm theory allows one to consistently construct the trajectories the particles have taken in the past. 

Alongside the double slit delayed choice experiment, there are further cases like delayed choice entanglement swapping or delayed choice Bell experiments (see, for example, \cite{Bacciagalupi-post-selection, Glick2019}). Are these experiments all of the same kind? I presume they all can be elucidated similarly to what I did in this paper. Considerations of delayed choice experiments could also affect ideas like entanglement realism. For example, what is the status of entanglement if quantum effects like interference, entanglement swapping, and violation of Bell inequalities can equally well be confirmed via post-selection in delayed choice scenarios?

\chapter[How (not) to Understand Weak Measurements of Velocity]{How (not) to Understand Weak Measurements of Velocity}
\chaptermark{Weak Velocity Measurements}
\label{section:Weak measurements}

To date, the most elaborate attempt to complete quantum mechanics by the addition of hidden variables is the de Broglie-Bohm theory or pilot wave theory. It endows particles with definite positions at all times. Deterministic dynamics govern their evolution. By construction, however, the individual particle trajectories generically defy detectability in principle. Of late, this lore might seem to have been called into question in light of so-called weak measurements. Due to their characteristic weak coupling between the measurement device and the system under study, they permit the experimental probing of quantum systems without essentially disturbing them. It's natural, therefore, to think that weak measurements of velocity, in particular, offer the opportunity to observe the particle trajectories. If true, such a claim would not only experimentally demonstrate the incompleteness of quantum mechanics: it would provide support for de Broglie-Bohm theory in its standard form, singling it out from an infinitude of empirically equivalent alternative choices for the particle dynamics. Here I examine this possibility. The result is deflationary: weak velocity measurements constitute no new arguments, let alone empirical evidence, in favour of standard de Broglie-Bohm theory. One mustn't na\"ively identify weak and actual positions. Instead, weak velocity measurements admit a straightforward standard quantum mechanical interpretation independent of any commitment to particle trajectories and velocities. This is revealed by a careful reconstruction of the physical arguments on which the description of weak velocity measurements rests. In turn, they present another case where a careful analysis of the relationship between the manifest and non-manifest proves essential. Moreover, for weak velocity measurements to be reliable, one must already presuppose de Broglie-Bohm theory in its standard form: in this sense, they can provide no new argument, empirical or otherwise, for de Broglie-Bohm theory and its standard guidance equation. 

\section{Introduction}

Since its inception, Quantum Mechanics (QM) has faced three major interpretative conundrums (see, e.g. \citealt{lewis2016quantum, myrvold2018-sep-qt-issues}). The first is the so-called Measurement Problem (see, e.g. \citealt{maudlin1995three}): how are we to make sense of the superpositions of states that the formalism of QM (if assumed to be universally valid) appears to attribute to objects? The second pertains to the interpretation of Heisenberg’s uncertainty relations (see, e.g. \citealt{sep-qt-uncertainty}): do they circumscribe an absolute limit of simultaneous knowledge of, say, a particle’s momentum and position? Or does it reflect an \textit{ontological} indeterminacy? Finally, how should one understand entanglement (see, e.g. \citealt{ney2013wave}) --- the fact that generically, composite systems appear to defy an unambiguous description of their individual constituent parts? 

These three puzzles culminate in the so-called EPR paradox (see, e.g. \citealt[Chapter 3]{redhead1987incompleteness} or \citealt{fine-sep-qt-epr}). Suppose one widely separates the partners of an entangled pair of particles. They can then no longer interact. Hence we may, according to Einstein, Podolsky and Rosen, `without in any way disturbing the system' perform (and expect a well-defined outcome of) a position measurement on one partner and a simultaneous momentum measurement on the other \citep[p.~777]{einstein1935can}. Prima facie, it looks as if thereby we can bypass the uncertainty relations. However, this raises the question of whether QM in its current form is complete: does every element of physical reality have a counterpart in the description of the QM formalism?  

Famously, Einstein thought otherwise (see, e.g. \citealt{lehner2014einstein}). He was `[...] firmly convinced that the essentially statistical character of contemporary quantum theory is solely to be ascribed to the fact that this [theory] operates with an incomplete description of physical systems' \citep[p.~666]{Schilpp1949}.  
To date, the most elaborate attempt to thus `complete' (cf. \citealt[Section~4]{goldstein-sep-qm-bohm}) QM dates back to \citet{bohm1952suggested, bohms1952suggested2} ---  `Bohmian Mechanics' or, in recognition of de Broglie's earlier proposal, `de Broglie-Bohm theory''.\footnote{There exist two \textit{distinct} variants of de Broglie-Bohm theory: the `quantum potential'  school (expounded, e.g. by \citealt{bohm2006undivided}, or \citealt{holland1995quantum}), and the `$1
	^{st}$-order formulation', canonised by Dürr, Goldstein, Zangh\`i and their collaborators. 
	
	The present treatment will only be concerned with the latter; `de Broglie-Bohm theory' will exclusively refer to this variant of de Broglie-Bohm theory.} (I’ll stick to the latter term throughout.) The theory underpins the statistical nature of quantum theory with deterministic trajectories.

But de Broglie-Bohm theory isn't free of problems. From its early days on, a principal objection to it\footnote{For subtleties in the early objections to de Broglie-Bohm theory, related to de Broglie-Bohm theory's unobservability, we refer to \citet{myrvold2003some}} targets the unobservability of its particle dynamics. By construction, in de Broglie-Bohm theory, the individual particle trajectories seem to be undetectable \textit{in principle}. Only their statistical averages are observable. They coincide with the standard quantum mechanical predictions. Thereby, standard de Broglie-Bohm theory achieves empirical equivalence with QM.\footnote{Here, we'll set aside possible subtleties, see \citet{arageorgis2017bohmian}.}

Recently, this lore seems to have been called into question in light of a novel type of measurement --- so-called weak measurements \citep{aharonov1988result}. These denote setups in which some observable is measured without significantly disturbing the state.

Inspired by \citet{wiseman2007grounding}, eminent advocates of standard de Broglie-Bohm theory have advocated such weak measurements as a means of actually observing individual trajectories in standard de Broglie-Bohm theory (e.g. \citealt[Section~4]{goldstein-sep-qm-bohm}). Moreover, they point to already performed experiments (e.g. \citealt{kocsis2011observing, mahler2016experimental}) that appear to corroborate de Broglie-Bohm theory's predictions and claim to show the particle trajectories. 

The present chapter will critically examine those claims. Should they hold up to scrutiny, they would not only establish the incompleteness of QM. But, almost more spectacularly, they would also furnish the remedy: they would vindicate de Broglie-Bohm theory in its standard form.

Those claims are mistaken: weak measurements constitute no new arguments, let alone empirical evidence in favour of de Broglie-Bohm theory's guidance equation. To show this, I'll carefully reconstruct the physical arguments on which the description of weak measurement rests. de Broglie-Bohm theory is entirely dispensable for a coherent treatment and interpretation of weak velocity measurements; they receive a natural interpretation within standard QM as observational manifestations of the gradient of the wave function's phase. For weak velocity measurements to disclose the particles' actual velocities, one must presuppose the prior existence of deterministic (and differentiable) trajectories and the specific form of standard de Broglie-Bohm theory's particle dynamics. We contest Dürr et al.'s suggestion of a legitimate sense in which weak velocity measurements allow a genuine measurement of particle trajectories.  

I'll proceed as follows. Section \ref{section underdetermination} will revisit de Broglie-Bohm theory's empirical underdetermination. Then, I'll turn to weak velocity values in Section \ref{de Broglie-Bohm theoryweak}. Section \ref{wisevel} will introduce Wiseman's measurement protocol for so-called weak velocity measurements. I'll subsequently illustrate it in the double-slit experiment (Section \ref{doubleslit}). The primary analysis of the significance of weak measurements for velocities in de Broglie-Bohm theory will form the subject of Section \ref{weak measurements are not genuine}. I'll first elaborate on when actual velocities and weak ones (as ascertained in Wiseman's measurement protocol) coincide (Section \ref{secton when do weak and actual velocities coincide}). This will enable a critical evaluation of both Dürr et al.'s claim that weak velocity measurements are in some sense genuine (Section \ref{DGZgenuine}), and as well as the idea that they provide \textit{non}-empirical support for standard de Broglie-Bohm theory (Section \ref{grounding}). The findings will be summarised in Section \ref{conclusion}. A mathematical appendix (Section \ref{Appendix}) contains a concise review of weak interactions within the von Neumann measurement scheme, as well as of post-selection and the two-vector-formalism.

\section{Underdetermination in de Broglie-Bohm Theory}
\label{section underdetermination}

The Bohmian framework supplements the QM formalism with deterministic but manifestly non-local dynamics for particles. At all times, they occupy determinate positions, evolving continuously in time according to a guidance equation. Only the particles' initial exact distribution is unknown. Due to this, QM emerges from de Broglie-Bohm theory in a manner `approximately analogous [...] to the statistical mechanics within the framework of classical mechanics' ---  as Einstein (ibid) had hoped. For a detailed account of the theory's postulates, the reader is referred to the Appendix \ref{appendix:de Broglie-Bohm theory}. 

Empirically, the guidance equation \ref{standardguidancelaw} isn't the only option, though. More precisely, it isn't necessary for empirical equivalence with QM. Infinitely many different choices 
\begin{equation}
	\label{altguidanceequation}
	\bm{v}^{\Psi}\mapsto \bm{v}^{\Psi}+ |\Psi|^{-2}\bm{j}
\end{equation} are equally possible for otherwise arbitrary vector fields $\bm{j}$ whose divergence vanishes, $\nabla \cdot \bm{j}=0$. They yield coherent alternative dynamics with distinct particle trajectories whilst leaving the predictive-statistical content unaltered \citep{deotto1998bohmian}. 

One needn't even restrict oneself to deterministic dynamics (an option expressly countenanced by, e.g. \citealt[Chapter~1.2]{teufel2009bohmian}): a stochastic dynamics, with $|\Psi|^{-2}\bm{j}$ corresponding to a suitable random variable can also be introduced. As a result, the particles would perform random walks, with the r.h.s. of the integral equation
\begin{equation}
	\nonumber
	\bm{Q}(t)-\bm{Q}(t_0)=\int\limits_{t_0}^{t}\bm{v}^{\Psi}d\tau
\end{equation} containing a diffusion term. A proposal of this type is Nelson Stochastics (see, e.g. \citealt{goldstein1987stochastic}; \citealt{Bacciagaluppi2005intronelson}). In short, de Broglie-Bohm theory's individual particle trajectories are observationally inaccessible by construction.

In consequence, de Broglie-Bohm theory is vastly underdetermined by empirical data: all versions of de Broglie-Bohm theory with guidance equations of the type \ref{altguidanceequation} are experimentally indistinguishable. Yet, the worlds described by them clearly differ. (This is illustrated in Figure \ref{fig:trac}.) 

\newsavebox{\smlmat}
\savebox{\smlmat}{$\bm{j}:=\frac{1}{x^2+y^2}\left(\begin{array}{c} 
		x\\
		-y 
	\end{array}\right)$}

\begin{figure}[ht]
	\centering
	\begin{subfigure}[b]{0.4\textwidth}
		\includegraphics[width=\textwidth]{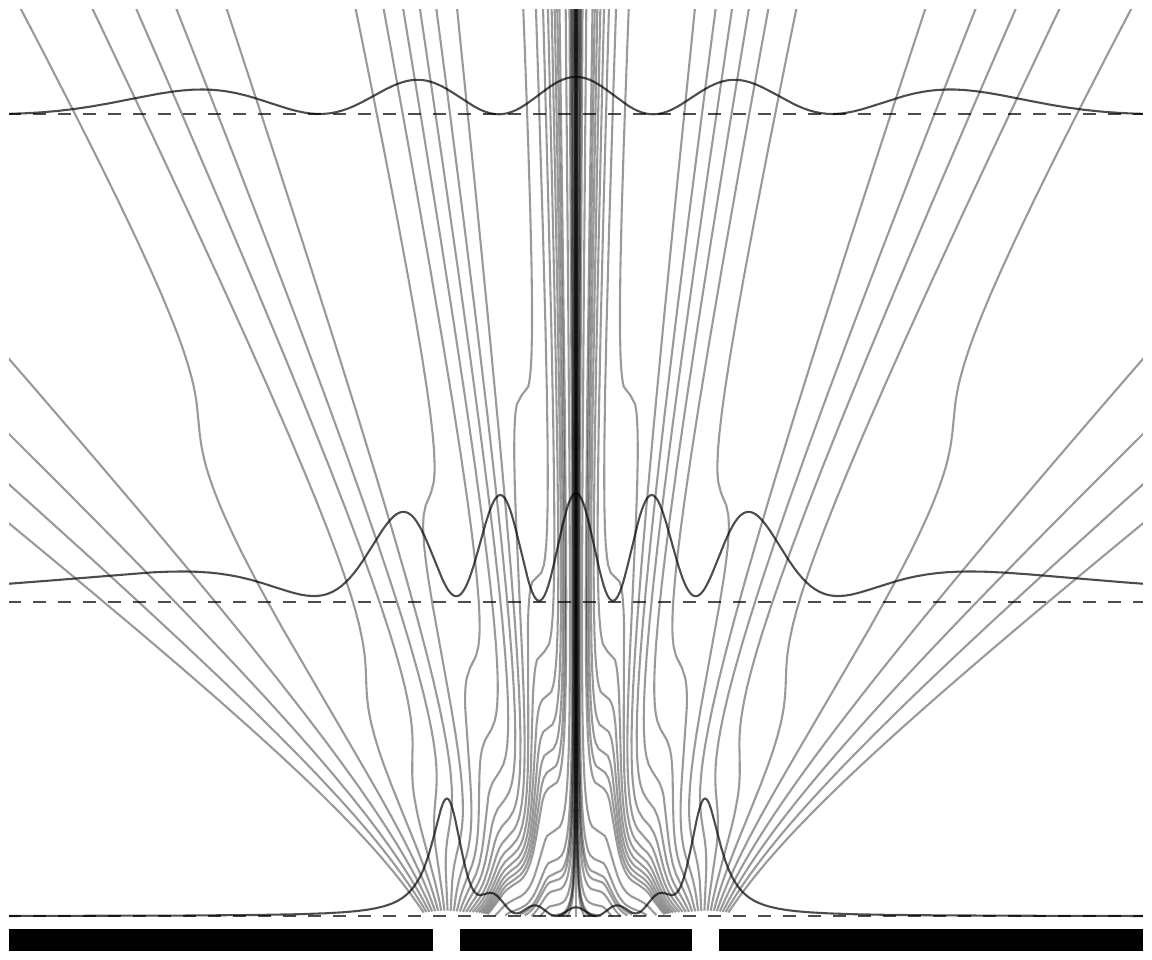}
		\caption{}
		
	\end{subfigure}
	~
	\begin{subfigure}[b]{0.4\textwidth}
		\includegraphics[width=\textwidth]{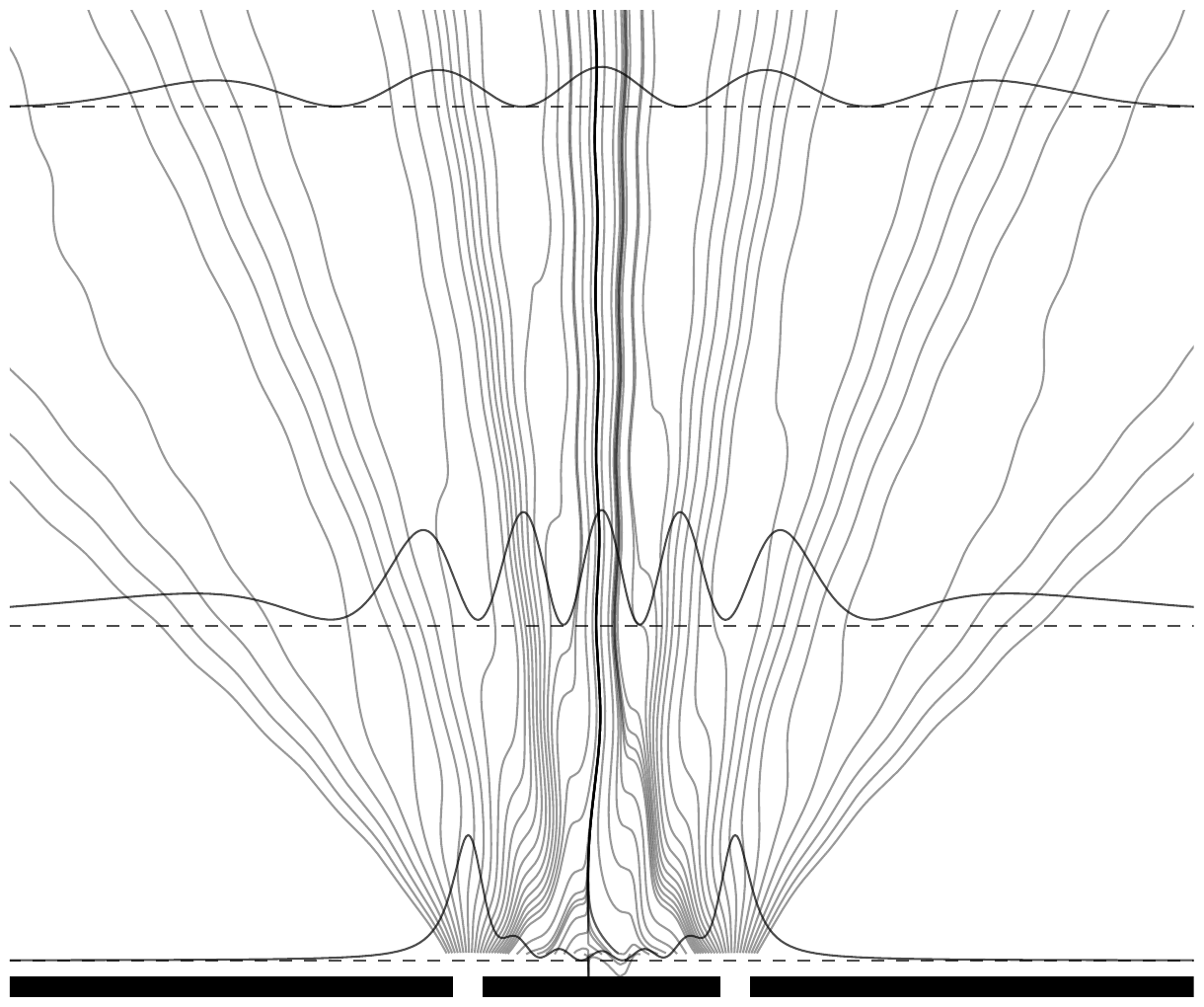}
		\caption{}
		
	\end{subfigure}
	~
	\begin{subfigure}[b]{0.4\textwidth}
		\includegraphics[width=\textwidth]{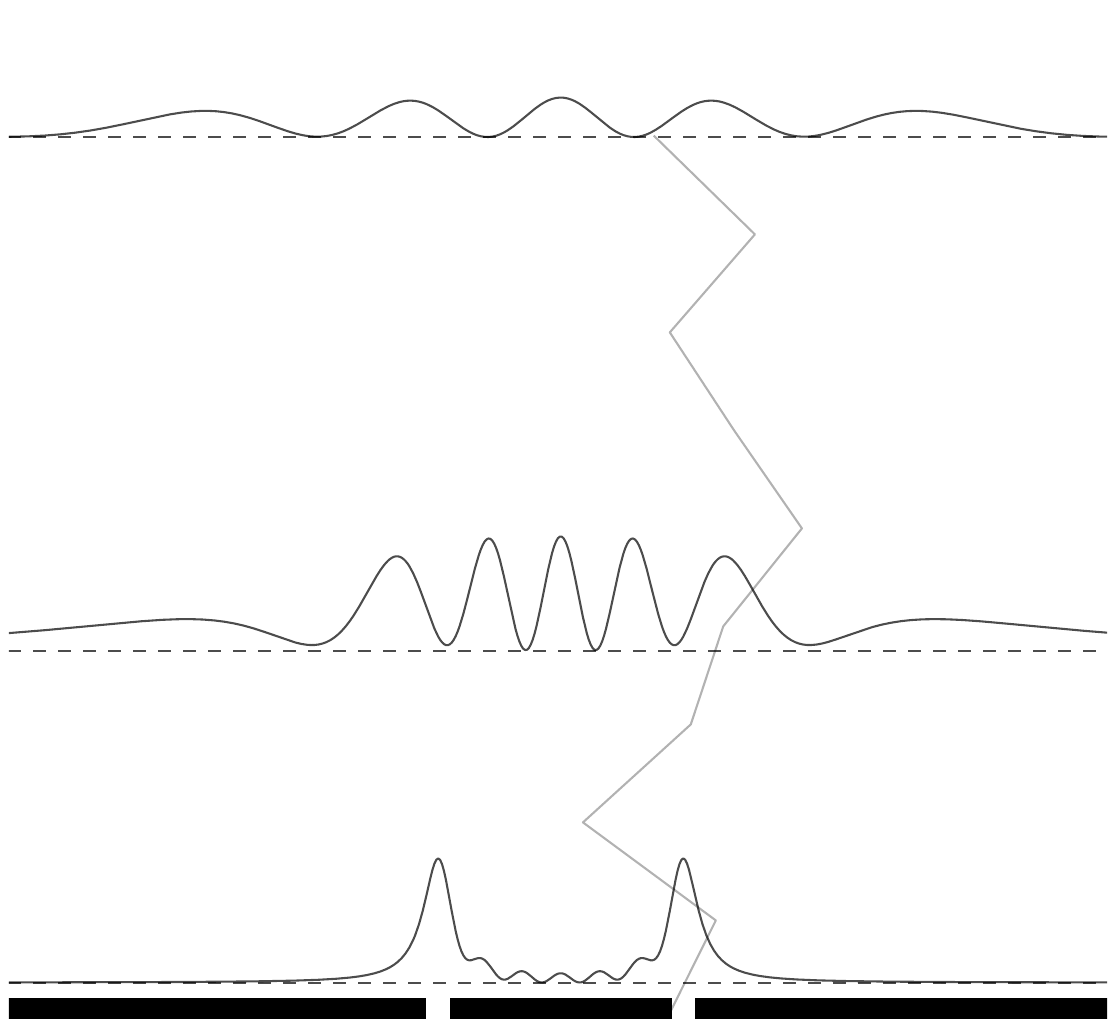}
		\caption{}
		
	\end{subfigure}
	
	\caption{A particle follows different trajectories corresponding to different/non-standard guidance equations.
		\textbf{(a)} The familiar wiggly deterministic trajectories that lead to the interference pattern in a double-slit experiment determined by the standard guidance equation. 
		\textbf{(b)} Alternative trajectories obtained from adding a divergence-free vector field~\usebox{\smlmat} to the standard Bohmian velocity field.  
		\textbf{(c)} A single stochastic trajectory generated by a random variable sampled according to $|\psi|^2$. For illustration, the probability density $|\psi|^2$ is shown at three different snapshots in time.
		All choices of the dynamics (i.e. (a), (b), (c)) are observationally indiscernible: The resulting measurable distributions at the screen at the top of each figure are the same.}
	\label{fig:trac}	
\end{figure} 

This underdetermination poses a challenge to a realist understanding of de Broglie-Bohm theory (cf. for example \citealt[Chapter~3.2]{Stanford-sep-scientific-underdetermination})

For the purposes of this analysis, I'll confine the class of considered choices to the family of de Broglie-Bohmian-like theories (cf. \citealt[Section 3.4]{durr2017probabilities}). It encompasses, e.g. the `identity-based Bohmian Mechanics' \citep{goldstein2005all} or `Newtonian QM' \citep{sebens2015quantum}. Let's whittle down the list of candidate theories to deterministic variants of de Broglie-Bohm theory with differentiable paths, i.e., variants of de Broglie-Bohm theory that differ only with respect to their vector field of the type in Equation \ref{altguidanceequation}. Still, the underdetermination persists; its severity is scarcely diminished: how to justify the particular choice for the standard guidance equation amongst the uncountably infinite alternatives?

An argument frequently cited in response is a result of \citet[p.~852]{durr1992quantum}: `The standard guidance equation is the simplest first-order equation that respects Galilei covariance and time-reversal invariance.'

But this is not decisive. First, individually neither desideratum of Dürr et al.'s theorem seems compelling --- unless one is already guided by intuitions shaped by either classical physics or standard de Broglie-Bohm theory. In particular, one may reject the ab initio requirement of Galilei covariance as implausible: Galilei covariance is the symmetry group of \textit{Classical} Mechanics.\footnote{Not even this is entirely obvious. On the one hand, at least once one incorporates Newtonian gravity, the most perspicuous spacetime setting is no longer Galilei spacetime (e.g. \citealt{pooley_subst_2013}). On the other hand, one may be attracted to the idea of a theory of classical mechanics that incorporates the Leibniz Group as its symmetry group, such as in the Barbour-Bertotti theory (ibid., Section 6.2). (Notice that recent attempts have indeed been made, e.g. by \citet{vassallo2015can}, to combine Barbourian shape dynamics with de Broglie-Bohm theory.) In either case, the symmetry groups would be larger than the Galilei group.}

Why impose it on a more fundamental theory --- de Broglie-Bohm theory --- which is supposed to \textit{supersede} Classical Mechanics?


Secondly, Dürr et al.'s argument rests on an assumption about how the Galilei group acts on the wave function. As \citet{skow2010symmetry} has argued, such an assumption is essentially unwarranted.

Thirdly, Dürr et al.'s argument turns on mathematical simplicity. As a super-empirical criterion, simplicity may well be felt a dubious indicator of truth (see, e.g. \citealt[Chapter~4.4]{van1980scientific}; \citealt{norton2000nature}, \citealt[Chapter~5-7]{norton2018material}; \citealt{ivanova2014there,ivanova_forthcoming}). At best, it may be regarded as a pragmatic criterion. 

Finally, Valentini has argued that the spacetime of pilot-wave theory is Aristotelian, not Galilean \citep{Valentini-Aristotelian}.

This context --- underdetermination --- renders weak value measurements particularly interesting. By (prima facie) allowing measurements of individual particle trajectories, they appear to directly overcome de Broglie-Bohm theory's underdetermination. But wouldn't that contradict the empirical inaccessibility of the trajectories? Let us see.

\section{Weak Velocity Values}
\label{de Broglie-Bohm theoryweak}

This section will offer a concise review of so-called weak values. I'll first outline how they are harnessed in Wiseman's measurement protocol for weak velocity measurements (Section \ref{wisevel}). An application to the double-slit experiment will further illustrate the salient points (Section \ref{doubleslit}). This will pave the way for our subsequent discussion in
(Section \ref{weak measurements are not genuine}).

\subsection{Wiseman's Measurement Protocol for Weak Velocity Measurements}
\label{wisevel}

Following \citealt{aharonov1988result}, weak measurements are measurement processes (modelled via the von Neumann scheme, see Appendix \ref{strong measurements appendix}) in which the interaction between the measurement apparatus (`pointer device') and the particle (`system') is weak: it disturbs the wave function only slightly. As a result, one can combine a weak measurement of one quantity (say, initial momenta) and a subsequent ordinary `strong' (or projective) measurement (say, positions).

More precisely: after a weak interaction (say, at $t=0$), the pointer states aren't unambiguously correlated with eigenstates of the system under investigation. In contradistinction to strong measurements, the system doesn't (effectively) `collapse' onto eigenstates; the particles can’t be (say) located very precisely in a single run of an experiment. This apparent shortcoming is compensated for when combined with a strong measurement a tiny bit \textit{after} the weak interaction: the experimenter is then able not only to ascertain the individual particle’s precise location (via the strong measurement); for a sufficiently large ensemble of identically prepared particles with initial state $\psi_{in}$ (viz. Gaussian wave packets with a large spread), she can also gain statistical access to the probability amplitude of all sub-ensembles whose final states --- the so-called `post-selected' state --- have been detected (in the strong measurement) to be $\psi_{fin}$:

\begin{equation}
	\langle \hat{x}\rangle _ w:=\frac{\langle \psi_{fin}|\hat{x}|\psi_{in}\rangle}{\langle\psi_{fin}|\psi_{in}\rangle}
\end{equation}

This quantity is called the `weak position value' (for the position operator $\hat{x}$). (The concept is straightforwardly applied to other operators, mutatis mutandis.) It can be shown (see the Appendix \ref{weak values appendix}) that after many runs, the pointer’s average position will have shifted by $\langle \hat{x} \rangle_w$.
Specifically, if we characterise the final/post-selected state via position eigenstates $|x\rangle$, determined in a strong position measurement and unitary evolution of the initial state, we obtain 

\begin{equation}
	\langle \hat{x} (\tau)\rangle _ w=\Re\left(\frac{\langle x|\hat{U}(\tau)\hat{x}|\psi_{in} \rangle}{\langle x|\hat{U}(\tau)|\psi_{in} \rangle}\right),
\end{equation}

where $\hat{U}(\tau)$ denotes the unitary time evolution operator during the time interval $[0;\tau]$. Following \citealt{wiseman2007grounding}, it’s suggestive of construing $\langle \hat{x}(\tau) \rangle _w$ as the mean displacement of particles whose position was found (in a strong position measurement at $t=\tau$) to be at $x$. From this displacement, a natural definition of a velocity field ensues:

\begin{equation}
	\label{operational velocity}
	\textbf{v}(\textbf{x},t)= \lim\limits_{\tau \rightarrow 0}\frac{1}{\tau}(\textbf{x}-\langle\hat{x}_w\rangle).
\end{equation} 

Note that all three quantities entering this velocity field --- $\tau$, $x$ and $\langle \hat{x} (\tau)\rangle _w$ --- are experimentally accessible. In this sense, the velocity field is `defined operationally' (Wiseman). In what follows, I’ll refer to the application of this measurement scheme --- a strong position measurement in short succession upon a particle’s weak interaction with the pointer --- for the associated `operationally defined' velocity field as `Wiseman’s measurement protocol for weak velocity measurements', or simply \textit{`weak velocity measurements'}.

For a better grasp of its salient points, let’s now spell out such weak velocity measurements in the context of the double-slit experiment.

\subsection{Weak Measurements in the Double-Slit Experiment}
\label{doubleslit} 

Consider the standard double-slit experiment with, say, electrons hitting a screen. It enables the detection of the electrons' positions. This constitutes a strong position measurement. Accordingly, I'll dub this screen the \textit{strong screen}. Let a weak measurement of position be performed between the strong screen and the two slits from which the particles emerge. Let this be called the \textit{weak screen}. The two screens can be moved to measure various distances from the double-slit. Suppose it takes the particles some time $\tau>0$ to travel from the weak to the strong screen. 

After passing through the slits, the electron will be described by the wave function $\ket{\psi}=\int \psi(x,t)\ket{x}dx$. This leads to the familiar double-slit interference fringes. I assume that the weak screen, i.e. the pointer variable, is in a Gaussian ready state with width $\sigma$, peaked around some initial position. After the particles have interacted with the measurement device (at time $t=0$), the composite wave function $\ket{\Psi(0)}$ of particle-\textit{cum}-weak screen is 

\begin{equation}
	\label{state}
	\ket{\Psi(0)}=\int \psi(x,t)\ket{x}\otimes \varphi(y-x)\ket{y}dxdy.
\end{equation} 

Here, $\ket{\varphi}$ denotes the wave function of the weak screen, and $y$ its free variable  (e.g. the position of some pointer device). The wave function then evolves unitarily for some time $\tau$, according to the particle Hamiltonian $\hat{H}$: 

\begin{equation}
	\ket{\Psi(\tau)}=\hat{U}(\tau)\ket{\Psi(t)}=e^{-\frac{i}{\hbar}\hat{H}\tau}\ket{\Psi(0)}.
\end{equation} 

After weakly interacting, the particle and pointer are entangled. Hence, only the composite wave function --- \textit{not} the reduced state of the pointer --- evolves unitarily during time $\tau$. The unitary operator $\hat{U}(\tau):e^{-\frac{i}{\hbar}\hat{H}\tau}$ only acts on $x$ (not on $y$). Due to this evolution, the post-selected position $\textbf{x}$ on the strong screen will, in general, differ from the weak value $\langle\hat{x}_w\rangle$, obtained from averaging the conditional distribution of the pointer of the weak screen. The procedure is depicted in Section \ref{secton when do weak and actual velocities coincide} below in Figure \ref{fig:correspondence}. 

On both screens, the wave function is slightly washed out. It differs from an undisturbed state (i.e. in the absence of the weak screen). To obtain the two position values --- the weak and the strong one --- strong measurements are now performed both at the weak and the strong screen (i.e. on the pointer variable and the target system). For each position outcome $x$ at the strong screen, let's select a sub-ensemble. We then read out the statistical distribution of the position measurement outcomes at the weak screen for any such sub-ensemble. 

We have thus assembled all three observable quantities needed for Wiseman's operationally defined velocity \ref{wisevel}: the time $\tau$ that elapsed between the two measurements, the positions $x$ (obtained as values at the strong screen), and the average value of all positions of the sub-ensemble, associated with (i.e. post-selected for) $x$.  
This may now be done for different positions $x$ on the strong screen: move the screens to different locations; there repeat the same measurement procedure. 

With this method, one can eventually map the velocity field for a sufficiently large number of measurements. The discussion of how to construe this result is deferred to the next section.

\begin{figure}[h]
	\centering
	\includegraphics[width=0.9\textwidth]{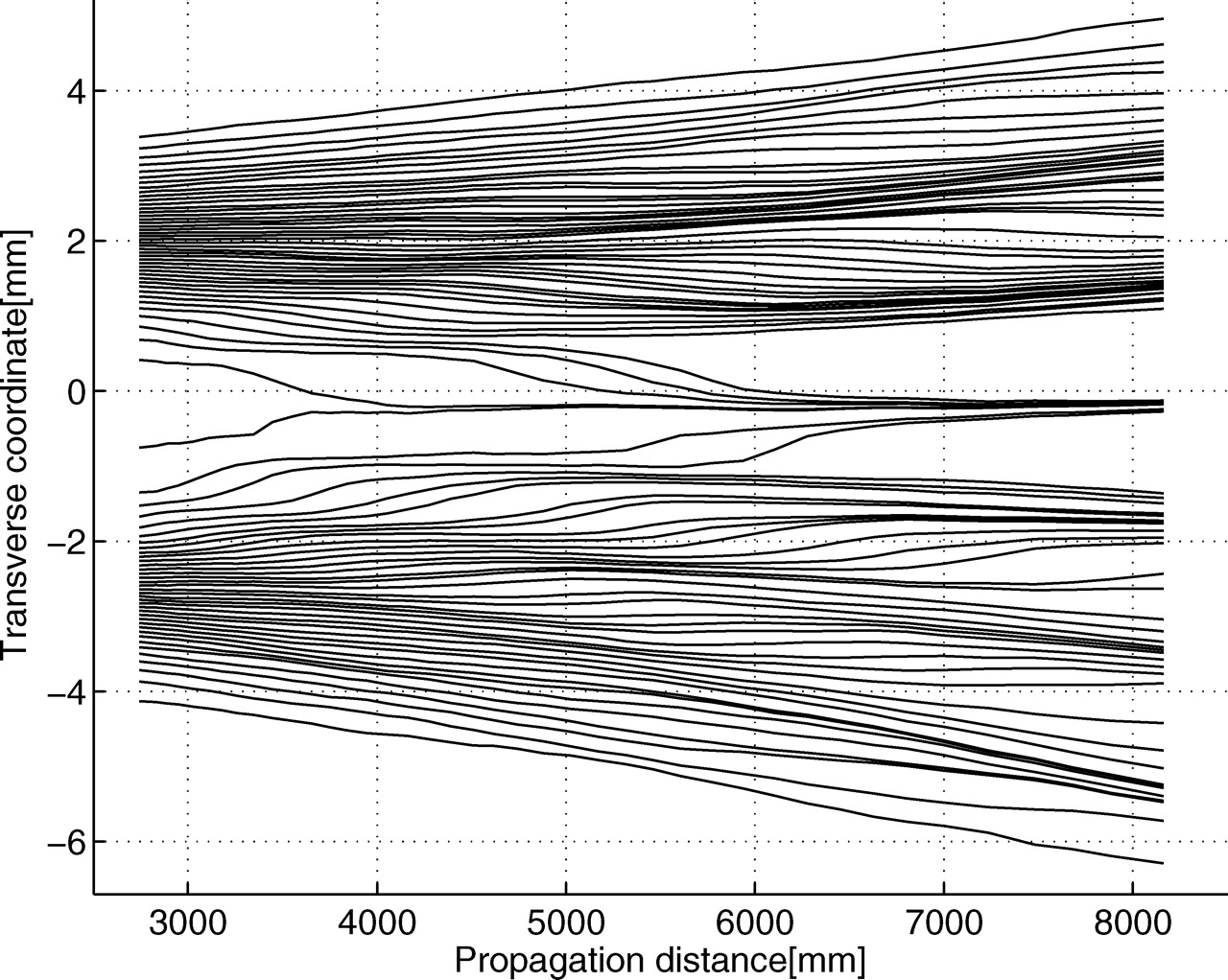}
	\caption{A weak velocity measurement for photons allows the reconstructions of trajectories, qualitatively identical to those of particles in standard de Broglie-Bohm theory. Particle trajectories in a double-slit experiment performed by \citealt{kocsis2011observing}.}
	\label{fig:Kocsis}
\end{figure}

Kocsis et al. have indeed performed an experiment of a similar kind, using weak measurements of momentum. Their result, depicted in Figure \ref{fig:Kocsis}, qualitatively reproduces the trajectories of standard de Broglie-Bohm theory. (I'll return to this experiment and how to understand it in Section \ref{weak measurements are not genuine}. Here, I mention it primarily to convey an impression of the qualitative features of Wiseman's operational velocity when experimentally realised.) Moreover, it can be shown (cf. Section \ref{weak velocity and the gradient of the phase}) that weak velocity measurements are measurements of the gradient of the phase of the wave function. Thus, they coincide with the definition of standard Bohmian velocities in the guidance equation

\begin{equation}
	v= \frac{\hbar}{m}\nabla S, 
\end{equation} where $S$ is the gradient of the phase of the wave function, $\psi(x)=|\psi(x)|e^{i S(x)}$. 

Notice that only the standard quantum mechanical formalism has been utilised for this. Therefore, we may conclude that --- based solely on standard QM --- weak velocity measurements permit experimental access to the gradient of the wave function's phase. 

\section{Why Weak Velocity Measurements do not Measure Velocities}
\sectionmark{Weak and actual velocities}
\label{weak measurements are not genuine}

Suggestive as these results are, I will now show that such measurements could not provide direct experimental evidence displaying the shape of particle trajectories, even if it is assumed that some deterministic particle trajectories exist. They cannot, that is, go anyway to experimentally resolving the underdetermination in putative de Broglie-Bohm theory guidance equations mentioned previously. I will first analyse the relationship between Wiseman's operationally defined velocity Equation \ref{operational velocity} and the particle's actual velocity. In particular, I'll show that a strong assumption is required that would render it question-begging to employ weak velocity measurements in order to infer the particles' actual velocities. This analysis will subsequently allow a critical evaluation of two stances regarding the significance of weak velocity values for de Broglie-Bohm theory --- Dürr et al.'s portrayal of weak velocity measurements as allegedly `genuine' measurements, and a view of weak velocity measurements as \textit{non}-empirical support of standard de Broglie-Bohm theory.

\subsection{When do Weak and Actual Velocities Coincide?}
\label{secton when do weak and actual velocities coincide}
Here, I'll address the question of whether --- or rather: \textit{when} --- weak velocities coincide with the particles' actual velocities, assuming that they exist. That is, I'll explicate the conditions under which weak velocity measurements count as reliable. That, I argue, turns out to \textit{presuppose} standard de Broglie-Bohm theory.  

In the following, $x$ and $y$ will denote the position variables of the individual particles to be measured and the measurement apparatus, respectively. For simplicity, consider one dimension only. Let the particles be prepared in the initial state 
\begin{equation}
	\ket{\psi}=\int dx ~\psi(x)\ket{x}.
\end{equation}  

Furthermore, let the pointer device (i.e. the  weak screen of the double-slit version of weak measurements in §\ref{doubleslit}) be prepared in the initial state given by a Gaussian with large spread $\sigma$, centred around $0$: 
\begin{equation}
	\ket{\varphi}=\int dy ~\varphi(y)\ket{y}=N\int dy e^{-\frac{y^2}{4\sigma^2}}\ket{y},
\end{equation} where $N$ is a suitable normalization factor. Together, the particle and the pointer form the compound system with the joint initial state

\begin{equation}
	\ket{\psi}\otimes\ket{\varphi}= \int dxdy ~\psi(x)\varphi(y)\ket{x}\otimes\ket{y}.
\end{equation}

Now consider the first --- the weak --- measurement process. It consists of an interaction between the particle and the pointer. Upon completion of this process (say at $t=0$), the compound system ends up in the entangled state

\begin{equation}
	\ket{\Psi(x,y,0)}= \int dxdy ~\psi(x)\varphi(y-x)\ket{x}\otimes\ket{y}.
\end{equation}

The probability distribution for the pointer variable $y$, \textit{given} some position $X$ of the particle, is therefore:

\begin{equation}
	\rho_X(y)=\frac{|\Psi_0(X,y)|^2}{|\psi(X)|^2}=|\varphi(y-X)|^2.
\end{equation} This probability density determines the expectation value 

\begin{equation}
	\label{expectation value at t=0}
	\mathbb{E}(y|x=X)=\int dy~y \rho_X(y)=X.    
\end{equation} That is, the mean value of the pointer distribution, conditional on the particle occupying position $X$, coincides with that position. This underwrites the following counterfactual:

\begin{itemize}
	\label{C_0}
	\item[$\mathbf{(C_0)}$] \textit{If one were to perform an ordinary (strong) position measurement on the particles \textit{immediately after} the weak interaction, the expectation value would yield the actual position of the particle.}
\end{itemize} 

Via $\mathbb{E}(y|x=X)$, the particle position thus is empirically accessible through the statistics of large ensembles of identically prepared particles from which we pick the post-selected outcomes $x=X$.  

This thought is further exploited in the final steps of Wiseman's procedure. In the preceding considerations, the strong measurement was performed immediately upon the weak one. Instead, we'll now allow for a small delay. After the particle and the pointer have (weakly) interacted, the total system evolves freely for some small but finite time $\tau$. Its state, then, is    

\begin{equation}
	\ket{\Psi(x,y,\tau)}= e^{-\frac{i}{\hbar}\tau\hat{H}_0}\ket{\Psi(x,y,0)},
\end{equation} where $\hat{H}_0$ denotes the system’s free Hamiltonian. 

Eventually, we perform a strong measurement of the particle’s position $X_{\tau}$ at $t=\tau$. (The strong coupling between the measurement device and the particle enables precise detection of the latter's actual position.) We thus get the expectation value for the pointer variable, conditional on the particle occupying the position $X_{\tau}$ at $t=\tau$:

\begin{equation}
	\label{weak position value}
	\mathbb{E}(y|x=X_{\tau})=\int dy~y |\Psi(X_{\tau},y,\tau)|^2.
\end{equation}

Through the statistics of a sub-ensemble of particles whose strong position measurements yielded $X_{\tau}$, this expectation value is empirically accessible. 

In \textit{analogy} to Equation \ref{expectation value at t=0}, let's define the position: 

\begin{equation}
	\label{position}
	X_0:=\mathbb{E}(y|x=X_{\tau}).
\end{equation}

Combined with the particle position $X_\tau$, obtained from the strong measurement at $t=\tau$, it thus appears as if we have access to particle positions at two successive moments. Using Equation \ref{position}, the associated displacement is

\begin{equation}
	\label{displacement}
	X_{\tau}-X_0=X_{\tau}-\mathbb{E}(y|x=X_{\tau})
\end{equation}

Let's grant one can make it plausible that the particles' trajectories are differentiable. Then, the displacement (Equation \ref{displacement}) gives rise to the velocity field 

\begin{equation}
	\label{velocity0}
	v(X_0):= \lim\limits_{\tau \rightarrow 0}\frac{1}{\tau} (X_{\tau} - \mathbb{E}(y|x=X_{\tau})).
\end{equation}

Note that all terms on the right-hand side of Equation \ref{velocity0} are observable. (Hence, presumably, Wiseman's labelling \ref{velocity0} as an `operational definition'.) In conclusion, it seems, via the statistics of an experimental setup implementing Wiseman’s procedure, we can empirically probe this velocity field. 

But what does this velocity field signify? It's tempting to identify it with the particles' actual velocities. That is, should this be true, the flow of Equation \ref{velocity0} generates the particles' trajectories (assumed to be deterministic and differentiable). Is this identification justified?

By \textit{defining} an $X_0 \overset{def}{=}\mathbb{E}(y|x=X_{\tau})$ via Equation \ref{position}, our notation certainly suggests so. Let’s assume that this is correct and dub this the `Correspondence Assumption' (COR). That is, suppose that the actual particle position $X_{\tau}$ at $t=\tau$ is connected with the earlier particle position $x(0)=X_0=\hat{T}_{-\tau}X_{\tau}$ at $t=0$, where $\hat{T}_{-\tau}$ denotes the shift operator that backwards-evolves particle positions by $\tau$. (In other words: for arbitrary initial positions, $\hat{T}_{\tau}$ supplies the full trajectory.) Then, according to (COR), the expectation value (\ref{position}) corresponds to the particles' position at $t=0$. For post-selection of sub-ensembles with $x(\tau)=X_{\tau}$ , (COR) thus takes the form (in the limit of large spread $\sigma$):

\begin{equation}
	\textbf{\text{(COR)}}~  \mathbb{E}(y|x(\tau)=X_{\tau})=\hat{T}_{-\tau}X_{\tau}.
\end{equation}

In other words, (COR) implies the counterfactual:

\begin{itemize}
	\label{C_t}
	\item[$\mathbf{(C_t)}$] \textit{If one were to perform a strong position measurement at $t=\tau$ (with the weak interaction taking place at $t=0$), yielding the particles' position at $x(\tau)=X_{\tau}$, the weak value would be directly correlated with the particles' earlier position $\hat{T}_{-\tau}X_{\tau}$. That is, upon a strong measurement at $t=\tau$, the expectation value would reveal the particles' true positions:
		\begin{equation}
			\mathbb{E}(y|x(\tau)=X_{\tau})=\hat{T}_{-\tau}X_{\tau}.
	\end{equation}}
\end{itemize} 

On (COR), the weak value thus gives the particle’s \textit{actual} position at the weak screen: the expectation value on the l.h.s. is reliably correlated with the particle's earlier positions. But most importantly, this is an \textit{if and only if condition}: If (COR) is satisfied, then we recover the actual position, but if it is not, we don't. As a result, one ought to assume that (COR) is valid for weak position measurements to yield actual particle positions. 

Thereby, any set of data compatible with QM appears to corroborate standard de Broglie-Bohm theory: \textit{given (COR)}, weak velocity measurements yield standard de Broglie-Bohm theory's velocity field. It thus seems as if standard de Broglie-Bohm theory’s empirical underdetermination has been overcome.

Such an apparent possibility of confirming standard de Broglie-Bohm theory would be remarkable. It crucially hinges, however, on the soundness of (COR). So why believe that it’s true? As I'll show (COR) is generically false, and there is no plausible argument for why it should hold. This will eventually be illustrated with a simple example.
Prima facie, (COR) looks like a plausible extrapolation of a strong measurement immediately after the weak interaction (i.e. at $t=0$). This idea may be developed in three steps. First, (COR) holds in the limit $\tau\rightarrow 0^+$. Next, in a deterministic world, it would seem that

\begin{equation}
	\mathbb{E}(y|x(\tau)=\hat{T}_{\tau}\kappa)=\mathbb{E}(y|x(0)=\kappa),
\end{equation} where $\kappa\in\mathbb{R}$ denotes a position. 

By appeal to $C_0$, this would then yield 

\begin{equation}
	\mathbb{E}(y|x(\tau)=\hat{T}_{\tau}\kappa)=\mathbb{E}(y|x(0)=\kappa)=\kappa,
\end{equation} as desired.

On the face of it, this argument looks watertight. Its first step results from QM's standard rules (see Equation \ref{expectation value at t=0}). Its third step, too, seems innocuous: a few lines earlier, $(C_0)$ was derived from the standard QM formalism. Let’s, therefore, throw a closer glance at the second step. It’s convenient to cast it in terms of the probability densities associated with the expectation values:

\begin{equation}
	\mathbb{P}(y|x(\tau)=\hat{T}_{\tau}\kappa)=\mathbb{P}(y|x(0)=\kappa).
\end{equation}

Prima facie, given determinism, this identity stands to reason. One might think the events $\{(x(\tau),y)\in \mathbb{R}\times\mathbb{R}:x(\tau)=\hat{T}_{\tau}\kappa\}$ and $\{(x(0),y)\in \mathbb{R}\times\mathbb{R}:x(0)=\kappa\}$ refer to the same events of our probability space (i.e. the same diachronically identical configurations, and \textit{therefore} are assigned the same probability measure.

Yet, this inference is illicit. While it’s true that $\{(x(\tau),y)\in \mathbb{R}\times\mathbb{R}:x(\tau)=\hat{T}_{\tau}\kappa\}$ and $\{(x(0),y)\in \mathbb{R}\times\mathbb{R}:x(0)=\kappa\}$ contains the same pointer configurations, this \textit{doesn't} imply that $\mathbb{P}(y|x(\tau)=\hat{T}_{\tau}\kappa)=\mathbb{P}(y|x(0)=\kappa)$. For this to hold, the conditional probabilities --- as defined via post-selection --- on both sides must be well-defined. That is, 
\begin{equation}
	\label{conditional probabilities}
	\frac{\mathbb{P}(y \& x(\tau)=\hat{T}_{\tau}\kappa)}{\mathbb{P}(x(\tau)=\hat{T}_{\tau}\kappa)} \ \text{and} \ \frac{\mathbb{P}(y \& x(0)=\kappa)}{\mathbb{P}(x(0)=\kappa)}
\end{equation} must exist (and coincide). 

In classical Statistical Mechanics, one may take this for granted. However, in a \textit{quantum} context, entanglement complicates the situation: it compromises the ascription of probability measures to certain events. One must heed the time with respect to which the assigned probability measure is defined. This is the case with weak velocity measurements. Recall that Wiseman’s measurement protocol only performs the strong measurement at $t=\tau$. This precludes defining the second term in \ref{conditional probabilities}. That is, no strong measurement is performed --- and no attendant `effective collapse' of the wave function occurs --- at an \textit{earlier} time (viz. at $t=0$). As a result, at the time of the weak interaction ($t=0$), the wave function of the pointer and particles is entangled. That means, however, that we \textit{can't} na\"ively assign the event of any particular particle position at $t=0$ an objective, individual probability measure.

In the present context, one can’t simply assign the particles a probability measure --- that of the reduced density matrix --- \textit{per stipulation}: it must be deduced from the probability measure of the composite pointer-device system --- using only the other axioms. The quantum operation of a partial trace, implementing the transition to the reduced density matrix, transcends those fundamental axioms (see, e.g. \citet[Section~6]{Durr2004}); that would require post-selection at $t=0$. Only the entangled pointer-\textit{cum}-particle system as a whole has a physically grounded, objective probability measure.

This follows from the fact that $\mathbb{P}(x(0)=\kappa)$ is obtained from the pointer-\textit{cum}-particle system’s reduced density matrix (i.e. by partially tracing out the pointer’s degrees of freedom). But this transition from the density matrix of a pure state to the reduced density matrix of an `improper mixture' \citep[Chapter~7]{espagnat2018conceptual} lacks objective-physical justification (see, e.g., \citealt[Chapter~3-4]{mittelstaedt2004interpretation}). Contrast that with the situation of $\frac{\mathbb{P}(y \& x(\tau)=X_{\tau})}{\mathbb{P}(x(\tau)=X_{\tau})}$: this \textit{is} well-defined via post-selection. That is, due to the `effective collapse' (see, e.g., \citealt[Chapter 9.2]{teufel2009bohmian}), induced by the strong measurement at $t=\tau$, the event $x(\tau)=X_{\tau}$ \textit{can} be assigned a well-defined probability measure. In d'Espagnat's terminology, we are dealing with a `proper mixture'. In short: Owing to the pointer’s entanglement with the particle, determinism \textit{doesn't} imply $\mathbb{E}(y|x(\tau)=\hat{T}_{\tau}\kappa)=\mathbb{E}(y|x(0)=\kappa)$. The initially promising argument for (COR) therefore fails.

From its failure, we also gain a wider-reaching insight: unless (at $t=0$) the strong measurement is \textit{actually} performed (unlike in Wiseman’s measurement protocol), the conditional probabilities $\mathbb{P}(y|x(0)=\kappa)$ (or equivalently: their associated expectation values) aren't objectively defined --- \textit{if} one adopts their usual definition in terms of post-selection. Strictly speaking, the \textit{un}realised measurement renders $\mathbb{P}(y|x(0)=\kappa)$, thus defined, meaningless.

\begin{figure}[h]
	\centering
	\includegraphics[width=1\textwidth]{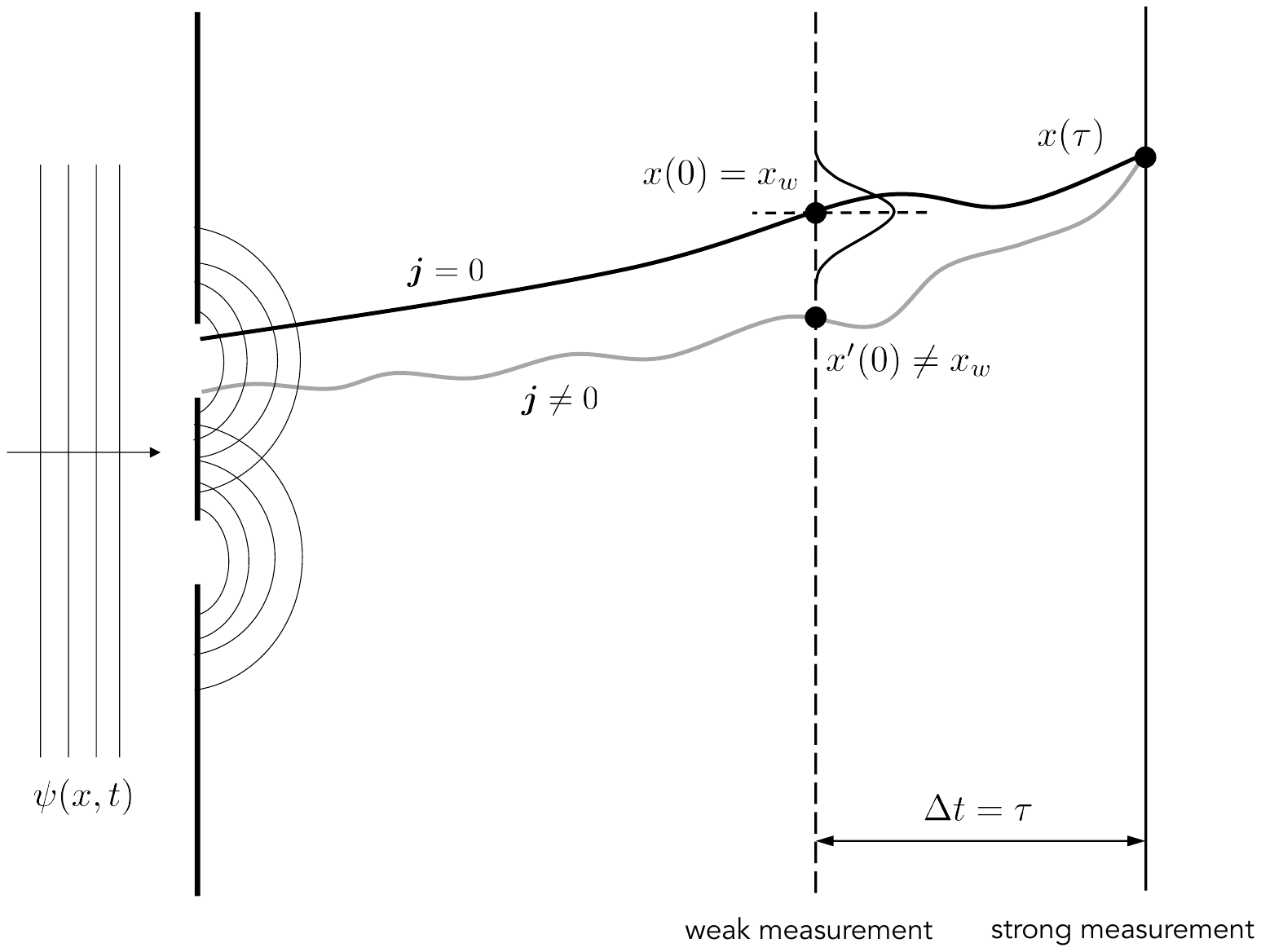}
	\caption{The weak measurement procedure for a given post-selected state $x(\tau)=X_{\tau}$. The weak value is obtained from the distribution on the weak screen. When the velocity field is that of standard de Broglie-Bohm theory ($\bm{j}=0$), the actual position of the particle $x(0)$ matches the weak value $x_w$. For an alternative guidance equation ($\bm{j}\neq 0$), it doesn't: the particle crosses the weak screen at a point $x'(0)$, other than the weak value. This shows that depending on which guidance equation one chooses, the weak value needn't yield the actual position of the particle at time $0$.}
	\label{fig:correspondence}
\end{figure}

No \textit{independent} reasons have been given so far for believing that (COR) is true, though.
(Conversely, the lack of independent reasons for standard de Broglie-Bohm theory (rather than any of its non-standard variants), especially in light of its empirical underdetermination, was our major motivation for applying weak velocities in the context of de Broglie-Bohmian theories.)  Consequently, counterexamples to (COR) abound --- and are perfectly familiar: \textit{any} non-standard variant of de Broglie-Bohm theory of the type of Equation \ref{altguidanceequation} (i.e. with non-vanishing, divergence-free vector field $\mathbf{j}$). In them, the particle's trajectory generically crosses the weak screen at a point \textit{distinct} from what the weak velocity measurements would make us believe. Figure \ref{fig:correspondence} illustrates this.

The outcome cannot be regarded as representing the actual position of the particle at time $t$. It's just unknown: it could have traversed \textit{any} location within the support of the Gaussian wave function, centred around the weak value. Still, the operationally defined velocity (obtained from averaging) wouldn't change: as long as the Born Rule and the validity of the Schrödinger Equation --- the only prerequisites for deriving the result Equation \ref{gradient weak velocity} --- hold, its value remains the same. (In this sense, any guidance equation of the type of Equation \ref{altguidanceequation}, is compatible with Wiseman's operationally defined velocity.)  

Absent an independent argument for the correspondence between weakly measured and actual positions (i.e. COR), it remains unclear what --- if anything --- Wiseman's operational velocity \ref{operational velocity} signifies non-manifestly, i.e. ontologically.

By na\"ively generalising $C_0$ to $C_t$, one neglects the relevance of time in the present setup: it matters both when the weak measurement interaction occurs, and \textit{when one post-selects}. If both happen simultaneously, the weak position value corresponds to the particle's actual position at time $t=0$. If, however,  some time $\tau$ elapses between interaction and post-selection, generically this is no longer the case. 

It's instructive to rephrase this result. The assumption $C_t$, necessary for the correspondence of weak and actual velocities, is, in fact, equivalent --- in virtue solely of the quantum mechanical formalism and the supposition of deterministic differentiable particle trajectories --- to standard de Broglie-Bohm theory. (First, suppose that $C_t$ is true. Then, the weak velocity measurement yields the actual particle velocities. Wiseman's operationally defined velocity \ref{operational velocity} uniquely picks out a guidance equation --- that of standard de Broglie-Bohm theory.

Conversely, suppose standard de Broglie-Bohm theory to be true. A weak velocity measurement then discloses the actual particle velocities. Thus, $C_t$ holds.) 

In conclusion, I argued that a particle's weak velocity coincides with its actual velocity (provided one is willing to attribute deterministic, differentiable paths to the particles), if and only if standard de Broglie-Bohm theory is true. But this coincidence is a sine qua non for deploying weak velocity measurements in \textit{support} of standard de Broglie-Bohm theory. To attempt to do so --- absent independent arguments for the reliability of weak velocity measurements --- would one thus incur circularity.  

\subsection{Weak Measurements as Genuine Measurements?}
\label{DGZgenuine}

The previous analysis sheds light on a recent claim by Dürr, Goldstein, and Zangh\`i (\citeyear{Durr2009}). These authors (henceforth abbreviated as `DGZ') propose that Wiseman’s measurement protocol for weak velocities allows `in a reasonable sense, a \textit{genuine} measurement of velocity' in standard de Broglie-Bohm theory (ibid., pp.~1025, DGZ’s emphasis). Such a statement is misleading. DGZ themselves identify a condition as crucial for their claim. This identification, too, we deem the source of further potential confusion. The crucial --- but in DGZ's account \textit{tacit} --- condition for weak velocity measurements to be reliable, as we saw in the previous sections, is (COR). But (COR) is equivalent to assuming the standard form of the guidance equation. The essential equivalence between (COR) and de Broglie-Bohm theory's standard guidance equation impinges upon the significance of weak measurements for de Broglie-Bohm theory. Whether we regard weak velocity measurements as enabling genuine measurements of the particle's actual velocity is essentially equivalent to an \textit{antecedent} commitment to standard de Broglie-Bohm theory. Pace DGZ, this curtails the significance of weak velocities as genuine. Yet, albeit misplaced in the context of weak measurements, DGZ's (misleadingly) identified crucial condition might open up a potentially illuminating perspective on standard de Broglie-Bohm theory.

DGZ assert that weak velocity measurements, as realised by Wiseman’s measurement protocol, constitute real measurements in standard de Broglie-Bohm theory (cf. \citealt[Section 3.7]{Durr2004}). What is more, in his review of de Broglie-Bohm theory \cite[Section~4]{goldstein-sep-qm-bohm} writes: ``In fact, quite recently \cite{kocsis2011observing} have used weak measurements to reconstruct the trajectories for single photons `as they undergo two-slit interference, finding those predicted in the Bohm-de Broglie interpretation of quantum mechanics' '' (cf. \citealt[pp.~142]{durr2018verstandliche} for a similar statement). Of course, DGZ are aware of the fact that such a claim needs an additional assumption; however, as I'll show, they seem to misidentify that `crucial condition' \citep[p.~1026, 1030]{Durr2009}.  

Figure \ref{fig:Kocsis} displays the weak velocities measurements ascertained in Kocsis et al.'s double-slit experiment. Indeed, they qualitatively tally with the trajectories of standard de Broglie-Bohm theory (cf., for instance, Figure 5.7 in \citealt[p.~184]{holland1995quantum}). Still, nothing immediately follows from that regarding the status of standard de Broglie-Bohm theory (see also \citealt{flack2014weak,flack2016weak}; \citealt[p.~181]{bricmont2016making}).  Kocsis et al.'s experiment has been performed for (massless) \textit{photons}. However, standard de Broglie-Bohm theory is a non-relativistic quantum theory for massive particles: it can't handle photons.\footnote{The treatment of photons within field-theoretic \textit{extensions} of de Broglie-Bohm theory, capable of dealing with photons (or bosons, more generally), is a delicate matter, outside the present Chapter's scope. I refer the interested reader to e.g. \citealt[Chapter~11]{holland1995quantum} and \citealt[Chapter~10]{durr2012quantum} (also for further references).} Kocsis et al.'s experiment hence has no direct bearing on de Broglie-Bohm theory's status.\footnote{Rather than the trajectories of \textit{individual} photons, \citealt{flack2014weak} and \citealt{flack2016weak} have argued that Kocsis et al.'s experiments measure mean momentum flow lines.} 

This interpretation has a counterpart in weak velocity measurements of the electrons of the present setup: per se, the weak velocity measurements only allow experimental access to the gradient of the wave function's phase.

Now to DGZ's main claim: for a coherent application of weak velocity measurements to the Bohmian framework as reliable velocity measurements, an assumption on the disturbance of actual velocities is needed. Only standard de Broglie-Bohm theory, so the story goes, has this feature. In turn, it appears that weak velocity measurements can constitute genuine measurements of the particle's actual velocities only in standard de Broglie-Bohm theory.

DGZ’s considerations seem to \textit{start from} the reliability of weak velocity measurements; they are predicated on (COR). DGZ (correctly) state that only standard de Broglie-Bohm theory is consistent with that. As the `crucial condition' responsible for that result, they identify a characteristic feature of standard de Broglie-Bohm theory's velocity field.

\begin{itemize}
	\item[\textbf{(SPE)}]  
	\textit{Whenever the particle-\textit{cum}-pointer compound system has the form $\psi(x)\otimes\phi(y-x)$, the particle’s velocity field $v$ (conceived of as a function of the compound system’s wave function $\psi\otimes\phi$) is supposed to depend only on the particle’s wave function $\phi$:}
	$v[\psi\otimes\phi]=v[\phi]$.
\end{itemize}

Let's dub this condition `separability of particle evolution' (SPE). It uniquely singles out standard de Broglie-Bohm theory \citep[Section~4]{Durr2009}.

DGZ's mathematical proof of this latter claim is sound. Their identification of (SPE) as a \textit{physically} essential condition, however, is wrong-headed: (SPE), in fact, plays no obvious role in the attempt to exploit weak velocity measurements for standard de Broglie-Bohm theory (see §\ref{de Broglie-Bohm theoryweak} and §\ref{secton when do weak and actual velocities coincide}): nowhere is it invoked explicitly. Moreover, it remains elusive how (SPE) \textit{could} enter that analysis: (SPE) is an \textit{exact} equality, postulated to hold whenever the composite particle-pointer wave function is factorisable. By contrast, DGZ‘s decisive equations (viz. (21) and (22) in their paper) are only approximations, valid at $t=\tau$. Their terms linear in $\tau$ \textit{don't} take a factorisable form (nor do they vanish). Not even at $t=0$ is the pointer-particle wave function factorisable. Hence, (SPE) doesn't seem to be applicable from the outset. To call (SPE) `crucial' --- understood as \textit{directly} responsible --- for the reliability of weak velocity measurements in de Broglie-Bohm theory muddies the waters: \textit{it‘s solely in virtue of (SPE)'s essential equivalence with standard de Broglie-Bohm theory} that (SPE) is relevant at all. That (SPE) singles out standard de Broglie-Bohm theory is salient for the (mathematical) form of the standard guidance equation: the latter is uniquely characterised by the factorisation of velocities at $t=0$, as asserted by (SPE). 

As a result, only because (COR) presupposes standard de Broglie-Bohm theory, \textit{and} because the latter is essentially equivalent to (SPE) (recall the remark at the end of Section \ref{secton when do weak and actual velocities coincide}), is (SPE) `crucial' --- in the sense of necessarily satisfied for (COR) to hold. In short: (COR), (SPE) and standard de Broglie-Bohm theory’s guidance equation are essentially equivalent. That is:
\begin{align}
	(COR) \wedge (DIF) \wedge (DET) \nonumber& \Longleftrightarrow \text{standard guidance equation} \\\nonumber
	& \Longleftrightarrow (SPE) \wedge (DIF) \wedge (DET), \nonumber
\end{align} where (DET) and (DIF) denote the assumption of deterministic and differentiable particle trajectories, respectively. 

For weak velocity measurements to reveal the particles' actual trajectories (assuming determinism and differentiability, that is) --- i.e. for weak velocity measurements to be reliable --- (COR) not (SPE) --- is the crucial condition that must be satisfied: without it, the counterfactual $C_t$ no longer holds (recall \ref{secton when do weak and actual velocities coincide}); the particle's later positions can't be inferred from the weak measurements. In particular, given (COR)'s essential equivalence with standard de Broglie-Bohm theory or (SPE), this means that if weak velocity measurements are reliable, (SPE) needn't be assumed separately: it's implied by (COR). Unless independent reasons for (SPE), (COR) or standard de Broglie-Bohm theory are forthcoming, weak velocity measurements lack epistemic significance for gauging the status of de Broglie-Bohm theory. The analysis of weak measurements in a de Broglie-Bohmian framework \textit{doesn't} rely on (SPE). DGZ are right, however, in observing that if standard de Broglie-Bohm theory is true, weak measurements are reliable (i.e. weak position values and actual position values coincide).  

DGZ’s purely mathematical result --- the equivalence of (SPE) and standard de Broglie-Bohm theory --- hints at an alluring possibility (completely independently of weak measurements): it \textit{might} serve as a prima facie interesting avenue for justifying (or, at least, motivating) standard de Broglie-Bohm theory. Underlying (SPE) seems to be the hunch that for particle-pointer systems with separable (factorisable) quantum states, the particle is supposed to be guided exclusively by the particle’s wave function --- not by that of the pointer. More generally, due to (SPE), whenever a quantum system is prepared as separable, the dynamics for the particles of one subsystem doesn't depend on the quantum state of other subsystem(s).\footnote{This is somewhat reminiscent of (but not equivalent to) the so-called Preparation Independence, a key assumption in the Pusey-Barrett-Rudolph Theorem (see, e.g. \citealt[sect. 7]{Leifer2014extended}; specifically for the theorem in the context of standard de Broglie-Bohm theory, see \citealt{drezet2014pbr}). Preparation Independence asserts that for two systems with states $\rho_A$ and $\rho_B$ such that the compound state is the product $\rho_A\otimes\rho_B$, and ontic state spaces $\Lambda_A$, $\Lambda_B$, with preparations $\mu_A$, $\mu_B$, respectively, there exists an ontological model for the joint state with state space $\Omega=\Lambda_A\times\Lambda_B$ such that the product measure 
	
\begin{equation}
\mu_A\times\mu_B(\Omega)=\int\limits_{\Lambda_B} \mu_A(\Omega_{\lambda_B})d\mu_B(\lambda_B),
\end{equation} reproduces the quantum predictions for the separable product state, where $\Omega_{\lambda_B}=\{\lambda_A\in\Lambda_A|(\lambda_A,\lambda_B)\in\Omega\}$. For hidden-variable theories, this looks like a natural desideratum: it expresses how the separable systems' independence at the quantum level (cf., e.g. \citealt{howard1989holism,howard1992locality}) constrains the level of the more fundamental level of the hidden-variables (cf. \citealt[sect. 7.3]{Leifer2014extended}).
	
There exists a critical difference between Preparation Independence and (SPE): the former makes claims about the hidden-variable \textit{states} (in the present case: the particles' positions) and their statistically independent distribution; SPE, by contradistinction, claims their \textit{dynamics} --- its independence of other subsystems' quantum state (see main text). Consequently, one ought to expect justifications for either to differ.
	
Preparation Independence doesn't imply (SPE): all variants of de Broglie-Bohm theory respect the former (due to the Born Rule giving the distribution of the particles' actual positions, see, e.g. \citealt{Gao2019pbr}) --- but only standard de Broglie-Bohm theory satisfies (SPE).}

As a desideratum, (SPE) implements the expectation that the statistical independence on the manifest level translates to the level of the variables in the non-manifest level: whenever the outcomes of quantum states of a composite system $A\&B$ are independent, the dynamics of the particles constituting $A$ shouldn't be affected by $B$’s quantum state. One may deem this a plausible heuristic principle for the construction of hidden-variable theories: it aligns the statistical independence of the known (empirically accessible) realm of the quantum formalism (for separable quantum states) and the independence of the unknown  (empirically inaccessible) domain of the (putatively) more fundamental 'hidden variables' dynamics. A dynamics respecting this alignment then `naturally' explains the statistical independence at the coarse-grained quantum level. However, as I showed repeatedly before in this thesis, this desideratum often turns out to be unwarranted. I leave the prospects of (SPE) as a potentially promising motivation for standard de Broglie-Bohm theory to future inquiry.

	

This section afforded two main lessons. First, standard de Broglie-Bohm theory is mathematically uniquely characterised by a factorisation condition on the velocity field. But DGZ's identification of that condition as `crucial' for the reliability of weak measurements is misleading. Secondly, weak velocities coincide with the particle’s actual velocities if and only if standard de Broglie-Bohm theory is true. It thus remains questionable what argument (if any) weak velocity measurements provide in support of standard Bohmian trajectories or any other Bohmian theory. 

On their own, weak velocity measurements thus don't provide any empirical support for standard de Broglie-Bohm theory. What about \textit{non}-empirical inferential support, though?

\subsection{Non-Empirical Support for de Broglie-Bohm Theory?}
\label{grounding}

The main result of Wiseman's original paper can be read as a conditional claim: \textit{if} one adopts his operationally defined velocity, \textit{and} assumes deterministic, differential particle trajectories, the latter is uniquely determined as that of standard de Broglie-Bohm theory; on this reading, Wiseman remains neutral --- whether they are plausibly satisfied (or not). More exciting, however, would be the prospect of learning something novel about the status of standard de Broglie-Bohm theory from weak measurements (granting certain background assumptions). I'll now examine such a stronger interpretation of Wiseman's result --- as a non-empirical justification of standard de Broglie-Bohm theory.    

The starting point of the envisioned reasoning will be two tenets explicitly endorsed by Wiseman:
\begin{itemize}
	\item[(1)] One should construe the weak value in Wiseman’s weak measurement protocol of §\ref{operational velocity} as the average velocity of a large ensemble of particles \citep[sect. 3]{wiseman2007grounding}.
	\item[(2)] Albeit not per se referring to individual particles, this \textit{statistical} property provides a `justification for [standard de Broglie-Bohm theory's] law of motion [i.e. the standard guidance equation]' (ibid., p.~2).
\end{itemize}

According to tenet (1), the weak value, obtained in Wiseman’s setup, by itself corresponds to a real property only of an ensemble of particles --- rather than one na\"ively ascribable to the individual particles:  `Thus strictly the weak value [...] should be interpreted [...] only as the mean velocity in configuration space --- this noise could be masking variations in the velocity between individual systems that have the same Bohmian configuration $x$ at time $t$.' (\citealt[p.~5]{wiseman2007grounding}). 

Tenet (2) purports that \textit{in virtue of this statistical (ensemble) property} de Broglie-Bohm theory’s standard form `is preferred over all other on physical grounds' (\citealt[p.~12]{wiseman2007grounding}). That is, although other velocity fields generate the same (statistically-empirically accessible) mean velocity, we ought to believe that the standard velocity field is true --- rather than any of its alternatives: for Wiseman, (2) serves as a non-empirical rule of inference, `justifying [de Broglie-Bohm theory's] foundations' (ibid., p.~12). 

As Wiseman reiterates, no experiment can discriminate between de Broglie-Bohm theory’s standard velocity field and alternative choices. How, then, is the envisaged non-empirical justification supposed to work? 

One strategy turns on the allegedly natural character of his proposal to operationally define velocities via weak values: `(Standard de Broglie-Bohm theory) delivers thus the most natural explanation of the experiments described'(\citealt[p.~145]{durr2018verstandliche}, my translation). 

Quite generally, in Section \ref{secton when do weak and actual velocities coincide} and Section \ref{DGZgenuine}, we have seen that in the present case, the allegedly natural explanation would, at any rate, be deceitful: one mustn't \textit{na\"ively} take it for granted that they reveal the actual particle positions. \citet{leavens2005weak} draws attention to the fact that under certain experimental circumstances, `[...] there is no possibility of the weak value [...] reliably corresponding in general, even on average, to a successfully post-selected particle being found near (the weak value) at time $t=0$ when the impulsive weak position measurement begins and being found near (the post-selected value) an instant after it ends' (p. 477). 

The perils of na\"ive (i.e. literal) realism about weak position values are demonstrated in the so-called Three-Box-Paradox (\citealt{aharonov1991complete}; \citealt[Chapter~16.5]{aharanov2005quantum}; \citealt{maroney2017measurements}). Here we encounter another illustrative case where the identification of the manifest with the non-manifest is more complex than one would hope.

Imagine a particle and three boxes labelled $A$,$B$, and $C$. Let the particle‘s initial state be 

\begin{equation}
	\ket{\psi_i}= \frac{1}{\sqrt{3}}(\ket{A}+\ket{B}+\ket{C}),
\end{equation}

where $\ket{A}$ denotes the state in which the particle is in box $A$, and similarly, $\ket{B}$ and $\ket{C}$. For its final state, on which we’ll post-select, choose

\begin{equation}
	\ket{\psi_f}= \frac{1}{\sqrt{3}}(\ket{A}+\ket{B}-\ket{C}).
\end{equation}

Via the definition of weak values (see \ref{weak values appendix}), one then obtains the resulting weak values for the projectors onto state $i\in {A,B,C}$, $\hat{P}_i:=\ket{i}\bra{i}$:

\begin{align}
	\langle\hat{P}_A\rangle_w&= 1\nonumber\\
	\langle\hat{P}_B\rangle_w&= 1\nonumber\\
	\langle\hat{P}_C\rangle_w&= -1.
\end{align} 

If one were to believe that weak values invariably reveal the real positions of particles, one would have to conclude that box $C$ contains $-1$ particles. However, in de Broglie-Bohm theory's ontology (in any of its variants), this is an absurd conclusion: particles in de Broglie-Bohm theory either occupy a position or don’t; the respective position projectors take values only in $\{0,1\}$.

Consequently, adherents of de Broglie-Bohm theory must be wary of interpreting weak values as real position values without qualification. The reliability of weak position (or velocity) measurements is a non-trivial (and generically \textit{false}) assumption. Unqualified realism about weak position values inevitably conflicts with de Broglie-Bohm theory's default ontology. 

We are thus left with, at best, a considerably weaker position, one close to Bricmont's (\citeyear[p.~136]{bricmont2016making}): `[Weak velocity measurements via Wiseman’s protocol] (are) not meant to `prove' that the de-Broglie-Bohm theory is correct', because other theories will make the same predictions. Still, the result is nevertheless suggestive because the predictions made here by the de Broglie-Bohm theory are very natural within that theory [...].'

\section{Conclusion}
\label{conclusion}

Let's recapitulate the findings. I started from the empirical underdetermination of de Broglie-Bohm theory’s guidance equation. It impedes insouciant realism about the particles' trajectories, postulated by standard de Broglie-Bohm theory. Next, I scrutinised whether Wiseman’s weak velocity measurement protocol can remedy this underdetermination by empirical or non-empirical means. The result was negative. I elaborated that the reliability of weak velocities --- that they coincide with the particles' real velocities --- presupposes standard de Broglie-Bohm theory. For non-standard versions of de Broglie-Bohm theory, its presumption is generically false. Hence, weak velocity measurements don’t qualify as evidence or confirmation in favour of the velocity field, postulated by standard de Broglie-Bohm theory. Weak velocity measurements, thus, don't allow for genuine measurements in any robust sense (at least given the present knowledge). Finally, I critiqued an interpretation of Wiseman's measurement protocol as a non-empirical argument for standard de Broglie-Bohm theory regarding alleged theoretical virtues. The 3-Box Paradox demonstrated the dangers of \textit{any} na\"ive realism about weak \textit{position} values. 

In conclusion, I hope to have elucidated the status of weak velocity measurements in two regards. On the one hand, they are an interesting application of standard quantum mechanics in a novel experimental regime (viz., that of weak pointer-system couplings). They allow us to empirically probe the gradient of the system's wave function --- irrespective of any particular interpretation of the quantum formalism. 
On the other hand, however, with respect to the significance of weak velocity measurements, weak velocity measurements shed no light on the status of standard de Broglie-Bohm theory. In particular, on their own, they don’t provide any convincing support --- empirical or non-empirical --- for standard de Broglie-Bohm theory over any of its alternative versions. 

\section{Appendix: Weak Measurements and Weak Values}

\label{Appendix}
Methods of weak measurement have opened up a flourishing new field of theoretical and experimental developments  (see, e.g. \citealt{aharanov2005quantum}; \citealt{TamirCohen2013}; \citealt{svensson2013pedagogical}; \citealt{dressel2014colloquium}. Broadly speaking, weak measurements generalise strong measurements in that the final states of measured systems need no longer be eigenstates (and are thus a particular case of POVM measurements). This appendix will first provide a concise overview of weak measurements (Section\ref{strong measurements appendix}). In particular, I'll expound on how they differ from the more familiar strong ones. Then, in Section \ref{weak values appendix}, I'll introduce the notion of a weak value. 

\subsection{Strong versus weak}
\label{strong measurements appendix}

Strong or ideal measurements are closely related to the conventional interpretation of the Born Rule. Consider a quantum system $\mathcal{S}$ and a measuring device $\mathcal{M}$ with Hilbert spaces $\mathcal{H_S}$ and $\mathcal{H_M}$, respectively. The Hilbert space of the total system is $\mathcal{H}= \mathcal{H_S}\otimes\mathcal{H_M}$. The system be in a normalised state $\ket{\psi}$ before the measurement. We are interested in measuring an observable $A$ represented by the self-adjoint operator $\hat{A}$, which has a complete and orthonormal eigenbasis $\{\ket{c_i}\}$. In that basis the system's state reads $\psi=\sum\limits_i\alpha_i\ket{c_i}$ for some $\alpha_i$. Furthermore, we assume for simplicity the eigenstates are non-degenerate, i.e. have distinct eigenvalues. The only possible outcome of a strong measurement on this system is one of the eigenstates $\ket{c_i}$. The corresponding probabilities to observe $\ket{c_i}$ are
\begin{equation}
	p_i=|\bra{c_i}\ket{\psi}|^2=|\alpha_i|^2.
\end{equation} 
After the measurement was performed, the system ends up in the final state $\ket{c_i}$. This procedure is known as the \textit{von Neumann measurement} (cf., for example, see the reprint \citep{von2018mathematical}).  

In a weak measurement, the interaction of system and measurement device is modelled quantum mechanically with the pointer device as an ancillary system on which a strong measurement is performed after the interaction. That is, assume that system and pointer interact via a von Neumann Hamiltonian 
\begin{equation}
	\hat{H}=g(t)\hat{A}\otimes \hat{P}_M,
\end{equation} where $\hat{P}_M$ is conjugate to the pointer variable $\hat{X}_M$, i.e. $[\hat{X}_M, \hat{P}_M]=i\hbar$. As before $\hat{A}$ is the quantum operator of the observable to be measured, and $g(t)$ a coupling constant satisfying $\int\limits_0^T g(t) dt =1$. For simplicity, take a single qubit prepared in initial state
\begin{equation}
	\ket{\psi}= \sum\limits_i \alpha_i\ket{c_i}= \alpha\ket{0}+\beta\ket{1}.
\end{equation} We stipulate the eigenvalues of $\hat{A}$ are $\hat{A}\ket{0}=\ket{0}$ and $\hat{A}\ket{1}=-\ket{1}$. Suppose that the pointer that is to be coupled weakly to the qubit will initially be in a Gaussian ready state with spread $\sigma$ peaked around $0$, i.e.
\begin{equation}
	\ket{\varphi}=\int\varphi(x)\ket{x}dx=\int N e^{-\left(\frac{x}{2\sigma}\right)^2}\ket{x}dx,
\end{equation} with $N$ a normalization factor. During the unitary interaction, the total initial state $\ket{\Phi(0)}=\ket{\psi}\otimes\ket{\varphi}$ of system and pointer evolves according to Schrödinger's equation: 
\begin{align}
	\label{qubit evolution}
	\ket{\Phi(T)}&= e^{-\frac{i}{\hbar}\int\limits_0^T \hat{H} dt}\ket{\Phi(0)}\\ \nonumber
	&=  e^{-\frac{i}{\hbar}\hat{A}\otimes \hat{P}_M}\ket{\Phi(0)}\\\nonumber
	&= e^{-\frac{i}{\hbar}\hat{A}\otimes \hat{P}_M}\left(\alpha\ket{0}+\beta\ket{1}\right)\otimes \int\varphi(x)\ket{x}dx\\ \nonumber
	&= N\int \left(\alpha e^{-\left(\frac{x-1}{2\sigma}\right)^2}\ket{0}+\beta e^{-\left(\frac{x+1}{2\sigma}\right)^2}\ket{1}\right)\otimes\ket{x}dx.
\end{align} Recall that the momentum operator acts as a shift operator ($e^{-\frac{i}{\hbar}aP_M}\varphi(x)=\varphi(x-a)$). If the Gaussian peaks are narrowly localised and non-overlapping (to a good approximation), one can infer the system's state from the pointer measurement. However, for weak measurements, the Gaussians are assumed to spread over the pointer variable widely. The measurement outcome of the pointer is, therefore, consistent with the system being in states that are not eigenstates of the operator. This is read off from Equation \ref{qubit evolution}. If, say, the pointer ends up at position $0$, we recover the initial state $\ket{\psi}$ up to an overall factor. The two Gaussian amplitudes reduce to the same value. 

For arbitrary systems with finite-dimensional Hilbert space, the interaction generalises to 
\begin{equation}
	\label{endstate}
	\ket{\Phi(T)} = \sum\limits_i \alpha_i\ket{c_i}\otimes\int\varphi(x-a_i)\ket{x}dx,
\end{equation} where $a_i$ are the eigenvalues of the measurement operator $\hat{A}$. For simplicity, the free evolution Hamiltonian of system and pointer has been omitted; it would only give rise to additional total phases.  

So far, the measurement scheme has been standard. In Equation \ref{endstate}, no weakness is involved in particular. However, it becomes a weak one if the initial state of the pointer variable $X_M$ has a large spread $\sigma$. That is, the result of (strong) measurement on the pointer is not a projection onto eigenstates of the system. 

\subsection{Post-selection and two-vector-formalism}
\label{weak values appendix}

We may now introduce the notion of a weak value. A weak value of an observable $\hat{A}$ is the result of an effective interaction with the system in the limit of weak coupling and a subsequent post-selection. Coming back to the simple case of the qubit, if the state in Equation $\ref{qubit evolution}$ is post-selected on $\ket{0}$, for instance, the pointer ends up in a Gaussian lump centred around $1$. Similarly, conditioned on $\ket{1}$, the pointer is centred around $-1$, as one would expect from a strong measurement. However, depending on the post-selected state's choice, the pointer states are `reshuffled' and can be concentrated around mean values that can be far away from the eigenvalues of the observable $\hat{A}$. In the limit of large standard deviation $\sigma$, the distribution is again Gaussian, though. For post-selecting $\ket{+}:=\frac{1}{\sqrt{2}}(\ket{0}+\ket{1})$, for example, the distribution of the measurement device peaks around 
\begin{equation}
	a_w= \frac{\alpha-\beta}{\alpha+\beta}.
\end{equation} This is easily obtained by observing
\begin{equation}
	\ket{+}\bra{+}\otimes  \mathds{1}  \ket{\Phi(T)}= \ket{+}\otimes\frac{N}{\sqrt{2}}\int\left(\alpha e^{-\left(\frac{x-1}{2\sigma}\right)^2}+\beta e^{-\left(\frac{x+1}{2\sigma}\right)^2}\right)\ket{x}dx.
\end{equation} In the weak limit $\sigma\gg 1$ this gives
\begin{equation}
	\approx \ket{+}\otimes\frac{N}{\sqrt{2}}\int(\alpha+\beta) e^{-\left(\frac{x-\frac{\alpha-\beta}{\alpha+\beta}}{2\sigma}\right)^2}\ket{x}dx.
\end{equation}

Notably, the measurements on the pointer and the ones to find a post-selected state are \textit{strong measurements} in the sense defined above. For arbitrary post-selection on a final state $\ket{\psi_f}$ the state of the total system evolves according to

\begin{equation}
	\ket{\psi_f}\bra{\psi_f}e^{-\frac{i}{\hbar}\int\limits_0^T \hat{H} dt}\ket{\psi_i}\otimes\ket{\varphi}.
\end{equation} Since the spread $\sigma$ is large, the interaction Hamiltonian, which produces a shift in the pointer's wave function, can be effectively approximated by $e^{-\frac{i}{\hbar}\int\limits_0^T \hat{H} dt} \approx 1- \frac{i}{\hbar}\hat{A}\otimes\nobreak\hat{P}_M T$. Thus, the final state reads

\begin{align}
	&\approx \ket{\psi_f}\otimes\bra{\psi_f}\ket{\psi_i}\left(1- \frac{i}{\hbar}a_w\hat{P}_M T\right)\ket{\varphi} \nonumber\\
	&\approx \ket{\psi_f}\otimes\bra{\psi_f}\ket{\psi_i}e^{-\frac{i}{\hbar}a_w\hat{P}_M}\ket{\varphi},
\end{align} where

\begin{equation}
	\label{weak value}
	\langle\hat{A}_w\rangle:= a_w= \frac{\bra{\psi_f}\hat{A}\ket{\psi_i}}{\bra{\psi_f}\ket{\psi_i}}
\end{equation} the salient quantity of the weak value of the observable operator $\hat{A}$. That is, after many runs, the pointer's average position is $a_w$\footnote{There are cases in which $a_w$ is complex. Then, besides the position, the momentum is shifted too}. In other words, $\ket{\varphi}$ experiences the shift $\varphi(x) \mapsto \varphi(x-a_w)$. Note that the probability amplitude to obtain $\ket{\psi_f}$ in the post-selection is $p=|\bra{\psi_f}\ket{\psi_i}|^2$. If the initial and final state of $S$ are nearly orthogonal, the measurement may require many runs to find $a_w$ as the post-selected state occurs only rarely. If there is time evolution of the target system between the weak interaction and the final measurement of $\bra{\psi_f}$, then the expression would include $\bra{\psi_f}U$, where $U$ the unitary evolution operator:
\begin{equation}
	\langle\hat{A}_w\rangle:= a_w= \frac{\bra{\psi_f}U\hat{A}\ket{\psi_i}}{\bra{\psi_f}U\ket{\psi_i}}.
\end{equation} For a derivation, we refer the interested reader to literature.

\subsection{Weak velocity and the gradient of the phase}
\label{weak velocity and the gradient of the phase}
We can manipulate the definition of the operationally defined weak velocity to give us the velocity of the guidance equation of standard de Broglie-Bohm theory. That is, for the unitary evolution $\hat{U}(\tau)=e^{-i\hat{H}\tau/\hbar}$ during time $\tau$ (with the non-relativistic Hamiltonian of a massive particle $\hat{H}= \frac{\textbf{p}^2}{2m}+V(x)$), the expression for Wiseman's operationally defined velocity reduces to \citep[p.~5]{wiseman2007grounding}

\begin{align}
	\label{gradient weak velocity}
	\textbf{v}(\textbf{x},t)&= \lim\limits_{\tau \rightarrow \nonumber 0}\frac{1}{\tau}(\textbf{x}-\langle\hat{x}_w\rangle)\\\nonumber
	&= \lim\limits_{\tau \rightarrow 0} (\textbf{x}-\Re\frac{\bra{\mathbf{x}}\hat{U}(\tau)\mathbf{\hat{x}}\ket{\psi}}{\bra{\mathbf{x}}\hat{U}(\tau)\ket{\psi}})\\\nonumber
	&= \lim\limits_{\tau \rightarrow 0}\frac{1}{\tau}(\Re\frac{\bra{\mathbf{x}}\mathbf{\hat{x}}\hat{U}(\tau)\ket{\psi}-\bra{\mathbf{x}}\hat{U}(\tau)\mathbf{\hat{x}}\ket{\psi}}{\bra{\mathbf{x}}\hat{U}(\tau)\ket{\psi}})\\\nonumber
	&= \lim\limits_{\tau \rightarrow 0}\frac{1}{\tau}(\Re\frac{\bra{\mathbf{x}}[\mathbf{\hat{x}},\hat{U}(\tau)]\ket{\psi}}{\bra{\mathbf{x}}\hat{U}(\tau)\ket{\psi}})\\\nonumber
	&= \lim\limits_{\tau \rightarrow 0}\frac{1}{\tau}(\Re\frac{\bra{\mathbf{x}}[\mathbf{\hat{x}},\mathds{1}-\frac{i}{\hbar}\hat{H}\tau + \mathcal{O}(\tau^2)]\ket{\psi}}{\bra{\mathbf{x}}\mathds{1}-\frac{i}{\hbar}\hat{H}\tau + \mathcal{O}(\tau^2)\ket{\psi}})\\\nonumber
	&= \Re\frac{\bra{\mathbf{x}}[\mathbf{\hat{x}},-\frac{i}{\hbar}\frac{\hat{p}^2}{2m}]\ket{\psi}}{\psi(x)}\\\nonumber
	&= \Re\frac{\bra{\mathbf{x}}\frac{\hat{p}}{m}\ket{\psi}}{\psi(x)}\\ 
	&=\frac{\hbar}{m}\Im\frac{\nabla \psi(x)}{\psi(x)}=\frac{\hbar}{m}\nabla S(x),
\end{align} where $\nabla S(x)$ is the gradient of the phase of the wave function $\psi(x)$.

\chapter{What is Disturbed in a Measurement?}
\label{section:measurement disturbance}

In classical physics, so the story goes, a measurement allows one to determine any quantity of a system without disturbing it (at least in some ideal limit). Such statements are commonly made without proof or motivation and are mostly assumptions rather than established results. However, we shall see in this chapter that a quite general argument can be made in favour of arbitrarily high measurement accuracy. The details of the classical interaction between system and apparatus can be chosen to diminish the disturbance to arbitrary small degrees.  

In contrast, the story appears more complicated when it comes to quantum mechanics. Disturbance upon quantum measurements is often tacitly assumed to originate from the Heisenberg uncertainty relations and the finiteness of Planck's constant $h$: The interaction of an apparatus with a quantum system is supposed to introduce an unavoidable action upon the system bounded from below by $h$ \citep{Heisenberg-uncertainty-principles-1925}. Whatever pre-determines the individual result of a measurement (e.g. the position and momentum of a particle) is inevitably affected by the interaction with the apparatus. 

But of course, this early account of Heisenberg's uncertainty has to be overthrown in light of the insight that the variables allegedly being affected by a measurement do not even exist in the quantum formalism (cf. Section \ref{section:why are quantum predictions probabilistic} and see again Chapter \ref{section:manifest-non-manifest-domains}).  Thus, the question remains of what --- if anything --- is disturbed when a system is measured. 

In certain scenarios, quantum measurements don't seem to require any interaction at all. To mention a few examples: As we have seen before in Chapter \ref{section:Weak measurements}, weak measurements may fall in this category of interactions that do not (or only slightly) disturb a system; the negative result experiments first suggested by Schrödinger in his \citeyear{Schroedinger1934} assert that the lack of a particle detection also constitutes a measurement (cf. also \citealt{Renninger-negative-result}; in Hardy's paradox it is argued that an electron and positron can interact without annihilating each other \citep{hardy-paradox}; and the so-called Elitzur–Vaidman bomb tester supposedly can detect a bomb without setting it off \citep{Elitzur-Vadman-bomb}.  

I don't provide a comprehensive account of quantum measurement disturbance --- a task that arguably deserves a thesis on its own. Quantum theory doesn't give a sufficient framework that accounts for all variables that may determine a measurement result. Hence, a discussion of measurement disturbance in a quantum theory is intimately tied to the meaning of measurement results as well as the status of the quantum state. Thus, clearly delineating the manifest from the non-manifest will be necessary here too. Nevertheless, before moving on to empirical completeness in Chapter \ref{section:the empirical completeness problem}, a brief reflection on how disturbance might be linked to predictive advantage and empirical completeness is of benefit. 

I will start with a quick discussion of classical measurements and justify the claim that any quantity can be measured with arbitrary accuracy in classical mechanics. I then comment on epistemic states and argue that quantum measurements disturb the quantum state in a trivial sense, i.e. by updating a probability distribution. Finally, I conclude with some remarks on whether the non-manifest variables underpinning quantum theory are subject to disturbance. This, in turn, may serve as a motivation for why predictive advantage over quantum theory may be impossible.  

\section{Classical Measurements}
\label{classical measurements}

I shall first study the case where it's clear what's going on, i.e. classical mechanics. There, it turns out that predictions can be made with arbitrary accuracy and without disturbance.

A classical system is described by its configuration in phase space. For simplicity, we consider a single particle with position $x$ and momentum $p_x$. According to classical mechanics, all measured properties of the system are functions of its state $(x,p_x)$ in phase space. A quantity $A(x,p_x)$ may be obtained by letting the system interact with a pointer device via an interaction Hamiltonian\footnote{This interaction Hamiltonian is time dependent since it is `switched on' only for a (very short) finite amount of time. As a result, the conservation of momentum will be violated in general for such interactions. Note further that during the measurement, the interaction Hamiltonian dominates the total energy of the system and thus, the Hamiltonians of the free evolution are neglected.}

\begin{equation}
	H := g A(x,p_x) p_y,
\end{equation} where $g$ the interaction strength and $p_y$ the momentum of the pointer. Like the system, the pointer device's state is $(y,p_y)$, where $y$, for simplicity, can be thought of as the centre of mass of the pointer. 

We can readily find the change of the states by computing the Hamilton-Jacobi equations during the measurement interaction, i.e.

\begin{align}
	\label{Hamilton-Jacobi}
	\frac{dp_x}{dt}&=-\frac{\partial H}{\partial x}= -g\frac{\partial }{\partial x} A(x,p_x)p_y\\ \nonumber
	\frac{dx}{dt}&=\frac{\partial H}{\partial p_x}= g\frac{\partial }{\partial p_x} A(x,p_x)p_y\\ \nonumber
	\frac{dp_y}{dt}&=-\frac{\partial H}{\partial y}= 0\\ \nonumber
	\frac{dy}{dt}&=\frac{\partial H}{\partial p_y}= g A(x,p_x).\\ \nonumber
\end{align}

This shows that if the pointer's momentum $p_y$ is small, the system's state $(x,p_x)$ isn't affected. In fact, we can choose the initial momentum $p_y(0)$ to be arbitrarily small so that the measurement doesn't disturb the system. For this case, we can integrate the dynamical equations to find

\begin{align}
	p_x(t)&=p_x(0)\\
	x(t)&=x(0)\\
	p_y(t)&=p_y(0)\\
	y(t)&=y(0)+A(x(0),p_x(0))t.
\end{align}

After the measurement, the position of the pointer is correlated to the measured quantity $A(x,p_x)$. Therefore, this shows that in classical physics any property of the system can be measured without disturbance and to arbitrary accuracy. For example, a particle's initial position and momentum can be obtained by a measurement device with two pieces and interaction Hamiltonian $H=gxp_y +hp_xp_z$. Moreover, if the dynamics are deterministic, a non-invasive measurement lets one predict with certainty any future evolution of a system's state (cf. also \citealt{measurement-disturbance-zanghi}).  

\section{Epistemic States} 

The analysis can straightforwardly be generalised to cases where an epistemic state over position and momentum describes the system. The interest in studying this is that the quantum description of a physical system bears a resemblance to this situation since a quantum state defines a probability distribution via the Born rule.

For a classical system, when we do not know the precise ontic (non-manifest) state, i.e. the configuration position and momentum, an epistemic probability distribution over phase space usually represents the state. For example, when a particle's position and momentum are not known exactly, its behaviour is described only approximately. Assume that before measurement, these epistemic states are $f_0(x,p_x)$ and $g_0(y,p_y)$ for system and pointer device, respectively. Furthermore, let $h(x,y,p_x,p_y)$ be the compound epistemic state of system and pointer, and $h_0=f_0(x,p_x)g_0(y,p_y)$ the initial state. According to classical laws, for any initial point in phase space, the dynamics satisfy the Hamilton-Jacobi equations (\ref{Hamilton-Jacobi}) from above. If function $h$ is to be interpreted as an epistemic probability distribution over phase space, it has to satisfy a continuity equation, for the probability density has to be preserved over time. By evolving all the points in the phase space that the epistemic state $h$ supports, we arrive at Liouville's equation for its dynamics, i.e.

\begin{equation}
	\frac{\partial }{\partial t} h = \{h, H\} := \sum\limits_{s=x,y}\frac{\partial H}{\partial s}\frac{\partial h}{\partial p_s}- \frac{\partial H}{\partial p_s}\frac{\partial h}{\partial s}=: \hat{L}_H h,
\end{equation} where $\hat{L}_H$ is called the Liouville operator with Hamiltonian $H$.   

After writing out the equation for the classical interaction Hamiltonian and again assuming the momentum of the pointer to vanish approximately, i.e. $p_y\approx 0$, we arrive at

\begin{equation}
	\label{wave equation}
	\frac{\partial }{\partial t} h(x,y,p_x,p_y,t )= -\frac{\partial }{\partial y} h(x,y,p_x,p_y,t) \cdot x.
\end{equation}

Recall that Equation \ref{wave equation} describes a partial differential equation with wave-like solutions\footnote{Although Equation \ref{wave equation} differs from the familiar second order wave equation, its solutions still describe travelling waves.}, which allows making the ansatz 

\begin{equation}
	h= h_0(p_x,p_y)h_1(v(x)t-y), 
\end{equation} where $v(x)$ is an unknown function of $x$. Substituting into Equation \ref{wave equation} immediately implies

\begin{equation}
	v(x) = x. 
\end{equation} Moreover, for $t=0$ the epistemic state $h$ should coincide with $h_0=f_0(x,p_x)g_0(y,p_y)$. Thus, for the solution, we find 

\begin{equation}
	h(x,y,p_x,p_y,t )= f_0(x,p_x)g_0(v(x)t-y,p_y).
\end{equation} It is straightforward to repeat the derivation for arbitrary classical interaction Hamiltonians 

\begin{equation}
	H=\gamma A(x,p_x)p_y, 
\end{equation} designed to measure a more general property $A(x,p_x)$ of the system. What does the post-measurement state $h$ imply about the correlations of pointer outcome and this property of the system? After the measurement interaction at time $t=t_1$, a pointer reading $y_1$ is obtained that is then used to infer pre-measurement properties of the system.    

The first thing to note is the impossibility of measuring the system's initial distribution $f_0(x,p_x)$ with the readout variable $y(t_1)$. One single measurement outcome cannot tell what initial epistemic state for the system was prepared before the interaction. The individual pointer outcome is not correlated to the distribution $f_0(x,p_x)$. The claim that this cannot be is a consequence of the linearity of Liouville's equation. That is, for two initial preparations $f_1(x,p_x)$ and $f_2(x,p_x)$ the compound state $h_0(x,y,p_x,p_y) = (f_1(x,p_x)+f_2(x,p_x))g(y,p_y)$ evolves linearly via $\frac{\partial }{\partial t}h= \hat{L}_H h$. Thus, $h := h_1+h_2$ with 

\begin{equation}
	h_1= f_1(x,p_x)g(v_1(x)t-y,p_y),  h_2=f_2(x,p_x)g(v_2(x)t-y,p_y)
\end{equation} is a solution of the dynamics since $\frac{\partial }{\partial t}h= \frac{\partial }{\partial t}h_1+\frac{\partial }{\partial t}h_2= \hat{L}_H h_1+\hat{L}_H h_2 = \hat{L}_H h$, for arbitrary (non-linear) interaction Hamiltonians $H(x,p_x,y,p_y)$.  From this, it becomes clear that even if for the individual states $f_1$ and $f_2$, the final pointer position $y$ were correlated to the system's initial epistemic state, for an experiment with initial state $f_1+f_2$ the pointer variable $y$ could at best be correlated with \textit{either} of the states but not their sum. This is not a big surprise since for every individual run of the experiment, the system with ontic state $(x,p_x)$ contains no information about the subjective epistemic uncertainty on the particle's place in configuration space represented by $f_0(x,p_x)$. Nevertheless, a repeated measurement with identical initial preparation of the epistemic states can be used to gather statistics that allow tracing out $f_0(x,p_x)$. The epistemic state is defined as a statistical ensemble's probability density on phase space.

But what information, then, does the pointer contain about the system? Consider the general interaction Hamiltonian and the generic initial epistemic state $f_0(x,p_x)g_0(y,p_y)$ from before. After the measurement, the pointer will indicate some final position value $y_1(t_1)$. The final state for this outcome reads 

\begin{equation}
	h(x,y,p_x,p_y,t_1) = f_0(x,p_x)g_0(\gamma A(x,p_x)t_1-y_1,p_y).
\end{equation}  The quantity  $g_0(\gamma A(x,p_x)t_1-y_1,p_y)$ is interpreted as the probability of having found outcome $y_1$ at $t_1$. This post-measurement state limits the uncertainty about the value of $A(x,p_x)$. For a function $g_0$ that is highly peaked and centred around $y_1$, 

\begin{equation}
	\gamma A(x,p_x)t_1\approx y_1.
\end{equation} Hence, the system's property $A(x,p_x)\approx\frac{y_1}{\gamma t_1} $ can be determined up to the uncertainty of outcome $y_1$. For simplicity, assume without loss of generality that the function $g$ is so narrowly peaked that the pointer shows at most one outcome for $y$. That is, for fixed $(x,p_x)$, there is no uncertainty about the outcome of the pointer device. 

It is now asked what the final $h$ state entails for the post-measurement state of the system. The post-measurement probability density of the system we can write as 

\begin{equation}
	f_{t_1}(x,p)=\int g(\gamma A(x,p_x)t_1-y_1,p_y)f(x,p_x)dp_y.
\end{equation} Similarly to the uncertainty in pointer outcome $y$, the uncertainty of $(x,p_x)$ is limited by $g(\gamma A(x,p_x)t_1-y_1,p_y)$ since the final state $f_{t_1}(x,p_x)$ is peaked around values for which $A(x,p_x)\approx\frac{y_1}{\gamma t_1}$. For example, in the simple case where  $A(x,p_x)=x$, the system's final position $x$ is determined with high accuracy by $x\approx\frac{y_1}{\gamma t_1}$. Conversely, for values of $y_1$ that are located far away from $\gamma A(x,p_x)t_1$, the system's epistemic state remains approximately unchanged, i.e. $f_{t_1}(x,p_x)\approx f_0(x,p_x)$. In statistical mechanics described by epistemic probability states and subject to classical physics, we thus find that the system's property $A(x,p_x)$ is measurable to arbitrary accuracy. However, the epistemic state $f$ is not determinable in a single-shot measurement. But we would not have expected this to be true anyway since, as opposed to $(x,p_x)$, the $f$ states are not ontic. Epistemically, the pre-measurement probability state is viewed as having been disturbed by the measurement interaction. The initial probability density $f_0(x,p_x)$ evolved to the post-measurement state $f_{t_1}(x,p)$, which is peaked around $y_1$. But this sort of disturbance is of no further significance than a mere updating of the epistemic uncertainty contained in $f_0(x,p_x)$ for the system to possess property $A(x,p_x)$, to the decreased post-measurement uncertainty for  $A(x,p_x)\approx \frac{y_1}{\gamma t_1}$. 
As before, one can again introduce a measurement apparatus with more than one part, i.e. $H=\gamma_1 x p_x+\gamma_2 p_x p_y$ to simultaneously measure both position $x$ and momentum $p_x$ of the system. 

Moreover, even when the initially prepared state $f_0(x,p_x)$ involves uncertainty about the actual values of the ontic state, the fact that the evolution is deterministically fixed by the Hamilton-Jacobi equations still permits measurement of both $x$ and $p_x$ exactly: Knowledge of the interaction Hamiltonian and final pointer outcome $y_1$ can be used to work out the initial configuration of the system by simply reversing the deterministic dynamics. That is, the position of the pointer uniquely fixes the initial position and momentum of the system.  

The reason for invoking a statistical mechanics type of analysis is not coincidental. The quantum treatment of the situation is analogous in many ways. The conclusions about the epistemic states in statistical mechanics are also true for the quantum states insofar as they are epistemic \textit{qua} Born rule, but what is true for the ontic states in classical mechanics is generally violated by quantum mechanics. 

\section{Quantum Measurements}
\label{quantum measurements}

I will now comment on the disturbance in quantum measurements and how they differ from classical measurements. The nature of disturbance in quantum measurements is contingent on assumptions about what the variables in question \textit{are} that are supposedly disturbed. As mentioned before, this poses a problem to the analysis of measurement disturbance since quantum theory doesn't contain a full story of the manifest and non-manifest domains. Thus, it is unclear what could be disturbed in a quantum interaction. 

As a first relatively simple insight, on a purely operational level, i.e. when concerned with the manifest domain, only the conclusions about classical epistemic states directly translate to epistemic quantum states. Recall with the postulates of quantum theory that a quantum state is represented by a trace class density operator $\rho$ and via the Born rule is directly connected to the probability density $\tr(E_x\rho)$ with arbitrary observable $E_x$ (see a more detailed account of the theory's axioms in Section \ref{section:quantum theory and quantum predictions}). Thus, this probability density is an epistemic state on the uncertainty of the measurement outcome $x$, given that the measurement has in fact taken place (and the result not yet been observed). When an outcome $x$ obtains, in order to retain the requirement that immediately after the measurement interaction, it could be repeated and would show the same result, the post-measurement state must be reduced. Thus, a measurement reduces the initial state to a new state depending on the outcome. Analogously to classical epistemic states, the ``reduction'' of the state can be understood purely epistemically as an updating of a probability distribution given a definite outcome. I wish to emphasise the difference that in the classical case, the epistemic state is ignorance of pre-existing values (non-manifest domain), whereas in the quantum case, at this level, it is about measurement outcomes (manifest domain).

More interestingly, however, the claim can be directly stated as a direct consequence of the quantum formalism. Jumping ahead, in Section \ref{section: quantum decorrelation}, I introduce the quantum decorrelation principle, which states that the assignment of a pure state $\rho$ to a target system implies its decorrelation to the environment. For all outcomes $x$ of the target system and outcomes $y$ of its environment, the joint probability distribution factorises, i.e. $p(x,y)=p(x)p(y)$.
In other words, before the measurement, the maximum information of the target system encoded in the environment is the quantum state $\rho$. Therefore, a measurement will have to disturb that state to obtain more information on the system and create new records in the environment. Indeed, after the interaction, the environment is correlated to the system in an entangled state for which the reduced state cannot be pure anymore. This, in turn, presents a sense in which quantum states are disturbed during a measurement.\footnote{For a more formal and related argument, see the work of Busch on `no information gain without disturbance'. \citep{Busch2009}} 

In classical and quantum mechanics, the epistemic states are (trivially) disturbed upon measurement. However, as was shown, the classical theory also allows measurement interactions without disturbance of the outcome determining non-manifest variables, i.e. the position and momentum of particles. The interesting problem, thus, is whether similar claims hold in quantum theory. But in order to study such a question, one would need to know what the non-manifest variables are that ultimately determine a single outcome (deterministically or stochastically). 
Although the theory lacks a satisfactory non-manifest underpinning, a few things can be said.  

This is the question we are asking in terms of whether the relevant variables are disturbed in quantum measurement: Assume that some observable on a target system was measured that produced some record in its environment. Is it possible that upon further measurement (at least in some ideal limit) the target system behaves completely the same as if it hadn't been measured? If the quantum state was purely epistemic, understood as a probabilistic preparation of some non-manifest states, this will ideally hold. For all that is done in measurement would be revealing the actual value of a quantity the observer was uncertain of prior to measurement. Clearly, this must be false in quantum theory. The quantum decorrelation principle does a good deal more than just showing that the quantum state is updated epistemically. Not only can the target system's state not be pure after measurement correlations have been produced, but furthermore, the post-measurement \textit{mixed} state of the target system (which is entangled with the environment) is empirically distinct from the pre-measurement state. That is, there trivially exist observables that let us distinguish the pure state from the mixed state. A simple example is a projective measurement with the system's initial pure state in its basis. Before the entangling measurement interaction with the environment, the system is in an eigenstate of that observable's operator, whereas after, it isn't. In turn, unlike in classical mechanics, whatever determines a quantum outcome \textit{non-manifestly} must be disturbed in a measurement.\footnote{On a related note, this is akin to arguments on the quantum state having to be ontic (cf. \citealt{PBR-theorem}). But it isn't an argument for $\psi$-ontology since there the claim is that the ontic state is the full quantum wave function, and the present claim argues that whatever the ontic state is, it will be disturbed in non-trivial measurements.}   

In a similar vein, Maroney raises some interesting points on measurements and disturbance in the context of ontological models (see, \citealt{maroney2017measurements}). He distinguishes two kinds of disturbance: operational disturbance and ontic invasiveness. As before, the former effectively suggests assigning a joint probability distribution to the outcomes of two sequential measurements. Suppose $x$ and $y$ are the manifest outcomes of two observables obtained sequentially on a system prepared in some state $\rho$, then the first measurement is said to be operationally non-disturbing if the subsequent outcome can be described as a marginal probability, i.e.

\begin{equation}
	\label{eqn:operational non-disturbance}
	p(y|\rho)=\sum\limits_{x}p(x,y|\rho).
\end{equation} Ontic non-invasiveness, on the other hand, is the constraint that measurement does not affect the ontic state of the system. That is if $\lambda_1$ is the ontic state of the system before the first measurement and $\lambda_2$ the ontic state after that measurement and prior to the second measurement, then a measurement is said to be ontically non-invasive if these two ontic states coincide. Equivalently, the transition probability that $\lambda_2$ occurs conditioned on outcome $x$ and $\lambda_1$ is

\begin{equation}
	\label{eqn:ontic non-invasivness}
	p(\lambda_2|x,\lambda_1)=\delta(\lambda_1-\lambda_2),
\end{equation} (see also \citealt[pp. 4-5]{maroney2017measurements}).

I outlined above how, in general, a quantum measurement can't be operationally non-disturbing. Otherwise, the measurement is trivial, and no correlated records in the environment can exist --- the quantum decorrelation principle. Nevertheless, it doesn't rule out that, in some instances, a measurement is operationally \textit{non}-disturbing for a fixed set of observables. It depends on the final observable measured whether an intermediate interaction is operationally disturbing. More straightforwardly, every set of mutually commuting observables will lead to operational non-disturbance and thus satisfy Equation \ref{eqn:operational non-disturbance}. 

Maroney makes the interesting point, however, that in such scenarios where measurements are operationally non-disturbing, they might still be ontically invasive, i.e. violate Equation \ref{eqn:ontic non-invasivness}. He shows this explicitly for the so-called `three-box paradox' and generalises the claims to a wide range of measurements by drawing on \citeauthor{Legget-Garg-inequalities} inequalities (see also \citealt{Maroney-Timpson-on-macrorealism}). That is, while a quantum measurement may not create any effect on a system that can be detected empirically, it still holds that its ontic state is disturbed. What may classically be deemed as a non-disturbing interaction is usually not quantum mechanically. Therefore, unlike in classical physics, the inference from operational non-disturbance to ontic non-invasiveness, i.e. from \ref{eqn:operational non-disturbance} to \ref{eqn:ontic non-invasivness}, is generally incorrect. This, in turn, highlights a significant difference between the nature of measurement in classical and quantum physics. 

I conclude with a remark on how I think quantum measurement disturbance may be related to predictive advantage in future theories. The idea of connecting empirical completeness with measurement disturbance is the following. Assume there is a good theory of what and how quantum interactions disturb a system. This could then be elevated to a plausible principle of measurement disturbance based on which it may follow that any theory satisfying such a principle can't have a predictive advantage over quantum theory. Suppose it's, in fact, the case that a defining feature of quantum theory is the disturbance of any non-manifest variables that potentially pre-determine a measurement outcome (not necessarily deterministically). In that case, this could be taken as a principle to build post-quantum theories, which would have no predictive advantage. I leave it to future work to further explore this idea.

\chapter[The Empirical Completeness Problem of Quantum Mechanics]{The Empirical Completeness Problem of Quantum Mechanics}
\chaptermark{Quantum Empirical Completeness}
\label{section:the empirical completeness problem}


\section{Why are Quantum Predictions Probabilistic?} 
\label{section:why are quantum predictions probabilistic}

The tremendous success of quantum theory raises the question of whether its indeterminacy is a generic phenomenon of Nature or if alternative quantum theories exist with a predictive advantage over standard quantum mechanics. The quantum formalism --- if `empirically complete' --- suggests that its probabilism is fundamental to physics. 

The quantum uncertainty principles are usually taken to be the source of unpredictability. Heisenberg's ingenious move to introduce non-commuting operators into the quantum description led to the result that not all observables can be measured simultaneously \citep{Heisenberg-uncertainty-principles-1925}. For instance, when an experiment is set up to measure both the position and momentum of an electron (two conjugate, thus non-commuting observables), the quantum formalism implies a lower bound for the accuracy of those joint measurements, proportional to Planck's constant. 

However, Heisenberg's uncertainty inequalities have often falsely been interpreted as the claim that upon measurement, the relevant variables determining an outcome, i.e. the position and momentum in the example of the electron, would inadvertently be disturbed. Thus, making it impossible to predict measurement results with certainty. More formally, the product of the standard deviations of two observables is bounded from below by their commutator, i.e.

\begin{equation}
	\Delta \hat{A}\Delta\hat{B}\geq \frac{1}{2}|\langle[\hat{A},\hat{B}]\rangle|.
\end{equation}

But there is a problem with this view. It is too quick of a draw to think that predictions cannot be made better due to the disturbance of position or momentum --- properties that a particle \textit{allegedly} possesses. Quantum theory makes no \textit{a priori} statement about their existence other that the assignment of their corresponding measurement operators. As explained earlier, the basic theory is silent about what measurement results mean, i.e., what is contained in the non-manifest domain. The conditions that cause a single measurement result to occur are thus either undefined or do not exist (cf. also Heisenberg himself in his \citeyear{Heisenberg-physikalische-prinzipien-1930}, see also \citealt{Hermann1940-naturphilosophische-Uberlegungen} and in particular \citealt{Hermann_Determinism}).

An alternative, perhaps better formulation of quantum uncertainty is in terms of a system's decorrelation to its environment (see Section \ref{section: quantum decorrelation}). It states that whenever a (pure) quantum state is assigned to a target system, the only information that can exist in its environment is that of the (pure) state. If $x$ and $y$ are manifest configurations of target system and environment, respectively, signifying outcomes of arbitrary measurements, the total quantum state being $\rho$, and the target system's state $\rho_S$ being pure, then the quantum probabilities decouple

\begin{equation}
	p(x,y|\rho)= p(x|\rho_S)\cdot p(y|\rho_M), 
\end{equation} where $\rho_M$ is the reduced state of the environment. 

As a result, the question of empirical completeness becomes even more pressing. It is simply \textit{unknown} what the variables are (if they exist) that determine a particular measurement outcome. In this sense, empirical completeness would be tantamount to the claim that the quantum state --- by virtue of Born's rule --- is the only possible variable for making predictions. 

I am thus not concerned with the \textit{ontological} or \textit{metaphysical} status of the quantum state, but rather with its epistemic role as a means for making predictions. Thus, the focus is shifted away from the ontological underpinning of the theory to prediction-making. This, in turn, supports the need for a clear distinction between what's manifest, i.e. empirically accessible configurations, from what isn't, i.e. non-manifest variables. The variables that may determine a single measurement result don't necessarily coincide with the ones utilisable for predicting that result. Rather than to attempt restoring determinism in quantum mechanics, one of the main aims of my thesis is to follow a different train of thought: The question of whether quantum mechanics is \textit{empirically} complete. That is, I investigate the origin of the theory's limited predictability. Why the uncertainty? Whence the indeterminacy? Can there be theories with predictive advantage over quantum probabilities? Are there physical principles ruling out such a possibility? In other words, I'm concerned with a particular source for measurement uncertainty: Dispersion in the statistics of outcomes due to hypothetically undiscovered information about the system and its state. 

We can evade the problems of determinism by focusing on actual prediction-making, away from trying to make sense of what the measurement outcomes signify. And so the lesson of Heisenberg's uncertainty relations is not that the \textit{precise} position and momentum (na\"ively assumed to exist before the measurement takes place even though their meaning hasn't been defined) aren't determinable or controllable exactly. But merely that the predictions for observable manifest configurations are statistical in their nature. This paves the way for the possibility of an empirical extension. For the limits of predictability are not bound anymore by a commitment to the classical variables believed to determine the outcome of an experiment.\footnote{The sharp distinction between the notions of predictability and determinism is made clear by distinguishing metaphysical completeness from empirical completeness (cf. Section \ref{Section:metaphysical and empirical completeness}).
	
Furthermore, concrete quantum models have been formulated where the latter may be satisfied, but the former is wrong --- the Bohmian theory being a paradigm example. Determinism doesn't imply predictability.} Furthermore, it also opens up the possibility that prediction-making could involve more than the quantum states of an isolated target system.

Born is right to point out that to save determinism (and, as we shall see, any predictive advantage) from quantum theory, any extension or refinement will need to disprove some of its predictions empirically (see, for instance, \citealt{Born1983}). For if applied universally to all possible observables that will ever be discovered, the quantum predictions are subject to their inherent statistical uncertainty. This is laid out in detail in Chapter \ref{Section:quantum state preparation} by introducing the quantum decorrelation principle. Therefore, one immediate response to the problem of empirical completeness in quantum mechanics could be the following objection: Since quantum predictions would have to be violated in some regime for predictive advantage to exist, but quantum statistics are well established through empirical testing, why should we find predictive advantage in the first place? It must be noted that the way in which a new theory or new experimental evidence would lead to predictive advantage may leave the quantum theory untouched. Quantum statistical predictions are based on the variables that are known \textit{today}. However, there is no argument for why there shouldn't be further variables not accounted for in quantum theory, whose existence implies predictive advantage. For instance, post-quantum theories may involve novel properties of systems leading to refined predictability (cf. also Chapter \ref{section:the empirical completeness problem} for examples in classical physics). In fact, how non-quantumness would enter for predictive advantage is rather innocuous since indeterminism and unpredictability can be detached from the theory. For who is to impose the dictum that all variables of future theories are quantum observables and therefore bound by the current theory’s predictive limitations? And so the question remains: Why the quantum probabilism? 

Nevertheless, the quantum formalism suggests a fundamental limit to predictability. And that statement is independent of its interpretation and valid if the theory is deemed universal. Thus, quantum theory must be violated in some regimes for predictive advantage to exist. Therefore, the purpose of the present investigations is to determine when and how it would have to be violated and what physical principles could be resorted to, showing that it shouldn't be.

Quantum uncertainty was declared insurmountable by many authors. For instance, Feynman famously believed that quantum mechanics' indeterminism is a necessary component of nature. When it comes to the problem of `trying to predict exactly what will happen in a given circumstance' he states: `Yes! Physics has given up.'  \cite[Volume~1, Chapter~37, p.~10]{Feynman-lectures} But even if physics has given up, on what grounds? By accepting that quantum theory is supposed to be universally valid without further justification?

Feynman's sentiment, arguably, is reminiscent of the arguments on predictability by von Neumann and Dirac, who tried to provide an allegedly rigorous mathematical proof for `completeness'.

In his `Deductive Development of the Theory,' von Neumann claims to have demonstrated the mathematical necessity of quantum mechanics' indeterminism \citep{von2018mathematical}. His argument is supposed to rule out so-called `dispersion free', i.e. deterministic hidden variable theories. Such a result would spoil the possibility of (deterministic) metaphysical extensions and empirical extensions. If no variables exist, introducing more pre-determination of measurement outcomes than what is encoded in the quantum state, then quantum theory would be empirically complete. This proof is widely seen as an unsuccessful no-go argument for hidden variables. In a nutshell, von Neumann's inference from expectation values of measurement outcomes to fundamental properties of the hidden variables are readily shown to be wanting (see, for instance, \citealt{Bell-on-von-Neumann, Bub-on-von-Neumann-proof}). 

Similarly, \citeauthor{dirac1995collected}'s presentation of the quantum formalism puts indeterminism into the theory by hand. He assumes that the quantum amplitudes will always represent the outcome probabilities, whatever new `traits' a future description of a system is subject to (cf. \citealt{Hermann_Determinism}). Thus, the argument can't be a genuine empirical completeness theorem either. 

The theorems by Gleason and Kochen-Specker are more interesting, however. They dispense with the interpretations of von Neumann and Dirac, and none of the unwarranted assumptions of the earlier proofs of `indeterminism' are made. More specifically, Gleason proved that for Hilbert spaces with $2$ or more dimensions, all probability measures for projective measurements in quantum theory must be identical to the Born rule \citep{Gleason-theorem}. Thus, no deterministic hidden variable theory could exist since Born predictions exhibit indeterminism.   Notwithstanding, the result doesn't establish indeterminism either. A closer look reveals that Gleason's theorem rules out \textit{non-contextual} deterministic hidden variable theories (cf. \citealt{Bell-on-von-Neumann, Wuthrich-determinism}). For instance, de Broglie-Bohm theory is a famous example of a deterministic hidden variable theory which is contextual for all observables but position. The Kochen-Specker theorem explicitly shows that all deterministic hidden-variables theories must be contextual \citep{Kochen-Specker-theorem}. 

In conclusion, since quantum predictions are generally indeterministic for all variables to which the formalism is applied, it follows that whatever new variables may be discovered, they all exhibit the familiar indeterminism if they are assumed to conform to the standard quantum description. But clearly, this begs the question. The empirical completeness problem poses the question of whether \textit{non-quantum} variables could exist to overcome the theory's indeterminism or predictive limitations.










There are several options for how an answer to the empirical completeness problem could pan out. According to Hermann, there is only one sufficient reason why the search beyond quantum theory for the final `causes’ of physical events is doomed \citep{Hermann1940-naturphilosophische-Uberlegungen}: That quantum theory already contains all the causes that lead to a particular outcome. But if that is true, a more filled-in quantum theory comprising those causes has to tell a story for \textit{why} they are either incomprehensible or empirically inaccessible.\footnote{Of course, Hermann here anticipates what has later been coined a `hidden variable theory’ of quantum mechanics. I shall show below how pilot wave theory precisely realises her idea of a causal theory while stating how these causes are inaccessible in principle (under suitable assumptions).}  Or, the theory doesn't contain those causes, and then the question arises why in this case, any future investigations are prevented from finding them. Alternatively, quantum theory isn't empirically complete; thus, other causes exist to predict an outcome beyond the current possibilities of quantum mechanics. We should then embark on the endeavour to look for them and cast them into better theories. 

Another strategy is to look for law-like or methodological principles compatible with quantum theory from which empirical completeness follows. If nature is such that a set of basic physical principles implies quantum unpredictability, then insight is gained into indeterminacy independently of the formal structure of the theory. Such tenets may then explain in post-quantum theories why predictability is limited.  

Alternatively, quantum theory may be empirically \textit{in}complete. Then better theories exist with predictive advantage. But then one should have at least some epistemological motivations for why predictive advantage was \textit{hitherto} either not observed, rare, and where to look for its existence.

A further option may be motivated by dynamical reasoning. Perhaps quantum probabilities are only approximate, arising from a more fundamental temporal process. The question then reduces to what such a theory could look like and what properties of the time evolution would lead to the observed accuracy of quantum predictions. 

This thesis explores these possibilities by providing a generic framework to study the predictability in post-quantum theories and the empirical completeness problem of quantum theory. The aim is to comprehensively analyse where we stand towards answering the question, discuss the pertinent results and arguments available to date, clarify them and offer stronger, more general interpretations, and explore the possible avenues towards a generalised understanding of predictability in quantum and post-quantum theories. 

But why should we care whether quantum theory is predictively complete? One motivation is practical: What is the best possible prediction for physical phenomena that can be made? Is quantum mechanics (at least in principle) the most powerful prediction-making algorithm? What is the maximally achievable information of measurement outcomes encoded in the environment? Since quantum theory is probabilistic, empirical completeness would imply that there are fundamental limits to how accurate predictions can be in nature. 

Secondly, future experiments may demonstrate that quantum theory isn't applicable or valid in specific regimes. If that turns out to be accurate, it's desirable to have physical principles that can be used to make statements about the predictability in post-quantum theories. That is, an understanding of predictability could be reached independently of the quantum framework, which may also inform future theory construction.

Thirdly, there is a big question in the background: Are there objective probabilities in nature? If yes, are those the quantum probabilities? That is, are quantum predictions irreducible? 


\section{Metaphysical and Empirical Completeness}
\label{Section:metaphysical and empirical completeness}

I shall now define what it means for a theory to be empirically complete with respect to another theory. In turn, this allows a study of whether possible future quantum theories have predictive advantage over the traditional approach.

The completeness of a theory can mean many different things. It is often expressed as some kind of ontological completeness in the literature. For instance, a theory could be said to be complete if `it says all there is to say about nature' (cf. \citealt{shimony1983foundations} who expressed this intuition). Or, a theory could be said to be complete when `every element of physical reality must have a counterpart in the physical theory' (see \citealt{einstein1935can}). At the very least, both of those sentiments are concerned with a sense of ontological and predictive completeness. Here I discuss the specific notion of \textit{empirical} completeness, accounting for predictions in terms of the manifest configurations in the environment. The notion is an operational and relative one. That is, it compares the predictability of theories based on the possible variables in the manifest domain only. 

Thus, it is crucial to assess whether a theory has an empirical advantage to determine its empirical content — the actual results in terms of configurations in the manifest domain. It is possible for the physical states to be inaccessible to the manifest domain, i.e. the variables of the target system are generally (in part) unknown. Thus, were we to conditionalise instead on the general physical states $\lambda\times\mu_Q$, it is unclear to what extent the theory could make better predictions. The Bohmian theory is an instance thereof. There the physical states uniquely fix the outcome of a measurement, but nevertheless, those states can’t be prepared and therefore don’t lead to better predictions (at least in equilibrium pilot wave theory).  

I shall show below in more detail that one can always find a deterministic completion of a probabilistic theory without improving its predictive power. Therefore, I delineate what I call a \textit{metaphysical completion} from an \textit{empirical completion}. A metaphysical completion fills in the framework of a theory with additional variables producing a (more) deterministic theory. That is, conditionalising on the physical states of a system, including the additional variables, would lead to more refined predictions. However, the actual empirical content of the theory is preserved. In contrast, an empirical completion introduces variables or changes the theory such that its empirical content is different, and more refined predictions are possible. 

With the definitions introduced in Chapter \ref{section:manifest-non-manifest-domains}, consider two different theories $T_1$ and $T_2$, used to measure an observable $Q$ of a physical system. The same system consisting of target system and apparatus shall be described by state $\lambda^1\times\mu_Q^1$ according to theory $T_1$, and state $\lambda^2\times\mu_Q^2$ according to $T_2$, respectively. Consequently, the prediction algorithms of the two theories let us deduce the probabilities for all possible outcomes $x$ of observable $Q$:

\begin{align}
	p_1(x)&= p(x|\lambda^1\times\mu_Q^1),\\
	p_2(x)&= p(x|\lambda^2\times\mu_Q^2).
\end{align}

If the two probabilities do not coincide for some $x$, $Q$, or preparation $\lambda$, then we are dealing with two (at least) metaphysically distinct theories since the states $\lambda^1\times\mu_Q^1$ and $\lambda^2\times\mu_Q^2$ do give rise to different probabilities for the same target system. If, furthermore, the preparation of the target system produces homogeneous ensembles according to both $T_1$ and $T_2$, the two theories are empirically incompatible and could be falsified by actual measurements. The actual measurement results would then produce different outcome statistics. 

In order to state more precisely the first important notion of metaphysical completeness, I introduce the concept of a metaphysical extension. 

\begin{definition}{\textbf{Metaphysical extension.}}
	\label{def:metaphysical extension}
	If for two theories $T_1$ and $T_2$ the \textit{same} system is described by states $\lambda^1\times\mu_Q^1$ and $\lambda^2\times\mu_Q^2$, and it holds that $p(x|\lambda^1\times\mu_Q^1)\neq p(x|\lambda^2\times\mu_Q^2)$ for some outcomes $x$, then $T_2$ is said to be a metaphysical extension or metaphysical decomposition of $T_1$ if 
	\begin{equation}
		p(x|\lambda^1\times\mu_Q^1)=\int p(x|\lambda^2\times\mu_Q^2)\rho(\lambda^2|\lambda^1)d\lambda^2,
	\end{equation} where $\rho(\lambda^2|\lambda^1)$ is a probability distribution over the target system states of theory $T_2$ depending on the details of the preparation for the fixed state $\lambda^1$. 
\end{definition} 

\begin{definition}{\textbf{Metaphysical completeness.}}
	A theory $T$ is metaphysically complete if and only if it has no metaphysical extension. 
\end{definition}

Note again that the initial non-manifest states of the apparatus are assumed to be fixed. 

I shall point out below the result that only deterministic theories are metaphysically complete since one can always construct metaphysical extensions for probabilistic theories. 

We say a metaphysical extension of a theory is \textit{merely} metaphysical if and only if it is not an empirical extension of the theory. I shall turn now to empirical extensions and empirical completeness. 

Nothing has been said so far about whether one of the theories has some predictive advantage. For it needs to be clarified how the manifest configurations  --- on basis of which the states are prepared and selected --- relate to the physical states used for predictions. In brief, metaphysical completeness is silent about the predictive strength of a theory. For example, the states of a metaphysical extension of some theory could be uncontrollable or unpreparable while predictability is preserved. No predictive advantage compared to the original theory would be gained. One theory may make probabilistic predictions when conditionalising on its physical states, and the metaphysically extended theory may be fully deterministic.

Nevertheless, the deterministic theory could restrict the degree to which its variables can be accessed. For example, I will conclude below that this is the case for the theories of pilot wave theory and standard quantum mechanics. That is, equilibrium pilot wave theory is merely a metaphysical extension of standard quantum mechanics. 

Outcome probabilities must be conditionalised on configurations in the manifest domain utilisable for state preparation in order to compare the theories' predictability. To formalise the notion of empirical completeness, I define an empirical extension of a theory: Suppose the two theories $T_1$, and $T_2$ assign states $\lambda^1\times \mu_Q^1$ and $\lambda^2\times \mu_Q^2$ to target system plus apparatus. Furthermore, suppose the states are represented by the manifest configurations (or sets of configurations) $y_1, y_2\in \mathcal{M}$ being from theory $T_1$, and $T_2$ respectively. As opposed to metaphysical extensions, this implies that some experimental procedure can controllably prepare the states. For simplicity, let's identify the configurations $y_1$ and $y_2$ with the preparations of the corresponding states. Note also that a theory’s prediction algorithm should also be able to state what states are and aren't preparable experimentally. For instance, one could read off from the space of possible outcome results what information about the states the manifest configurations can contain in principle. An empirical extension can now be defined as:

\begin{definition}{\textbf{Empirical extension.}}
	\label{def:empirical extension}	
	If for two theories $T_1$ and $T_2$ the \textit{same} system is described by states represented by configurations $y_1\in \mathcal{M}$ being from $T_1$, $y_2\in \mathcal{M}$ being from $T_2$, and it holds that $p(x|y_1)\neq p(x|y_2)$ for some outcome $x$, then $T_2$ is said to be an empirical extension or empirical decomposition of $T_1$ if
	\begin{equation}
		p(x|y_1)=\int p(x|y_2)\rho(y_2|y_1)d y_2,
	\end{equation} where $\rho(y_2|y_1)$ is some probability distribution over the target system states of theory $T_2$ for fixed preparation $y_1$.
\end{definition} 

Analogously to metaphysical completeness, I define empirical completeness: 

\begin{definition}{\textbf{Empirical completeness 1.}}			
	\label{def:empirical completeness 1}
	When there exists no \textit{empirical} extension for a theory $T$, it is called \textit{empirically complete}.
\end{definition}


The domain of predictions of a particular theory $T$ may be restricted, but the manifest data on which its predictions may be refined ought to encompass the universal manifest domain, for the aim is to make the most general unrestricted claim on empirical completeness. 

Consider a classical thermodynamic system of a gas in a box to illustrate how empirical \textit{in}completeness may turn out. The box of gas may be described as a closed system with a constant temperature. Subject to kinetic gas theory, a measurement of the velocity $v$ of the gas particles in the ensemble is expected to give values distributed according to Maxwell-Boltzmann statistics $f_{MB}(v)$, peaked around some mean velocity. Thus, the velocity $v$ of an individual particle will be found with some likelihood $p(v)$. But let’s imagine the outcomes of many repeated measurements on the same gas yield doubly peaked distributions corresponding to two Maxwell-Boltzmann distributions of different mean velocities. And let’s assume the two peaks originate from the fact that there are two distinct gases in the box, namely nitrogen and oxygen. Since the two gases have different kinetic properties, this leads to separate mean velocities. But to an experimenter without a physical theory distinguishing the two molecules, only one gas is in the box. Thus, the experimenter’s original theory is empirically incomplete since the measurement outcomes will show an in-principle uncertainty resulting from treating the gas as one single substance. A more advanced theory that can distinguish nitrogen from oxygen will obtain more complete predictions. For this to be possible, an actual experiment, i.e. preparation, has to exist that distinguishes nitrogen from oxygen, such that, on average, the total distribution  $f_{MB}(v)$ is still recovered. This is crucial, for otherwise, the improved theory wouldn't be an empirical completion but rather a metaphysical completion. Thus, to extend the predictability of the original gas theory, a variable in the environment must be present to distinguish the two gases in an ensemble of preparations.

\begin{figure}[h]
	\centering
	\includegraphics[width=1\linewidth]{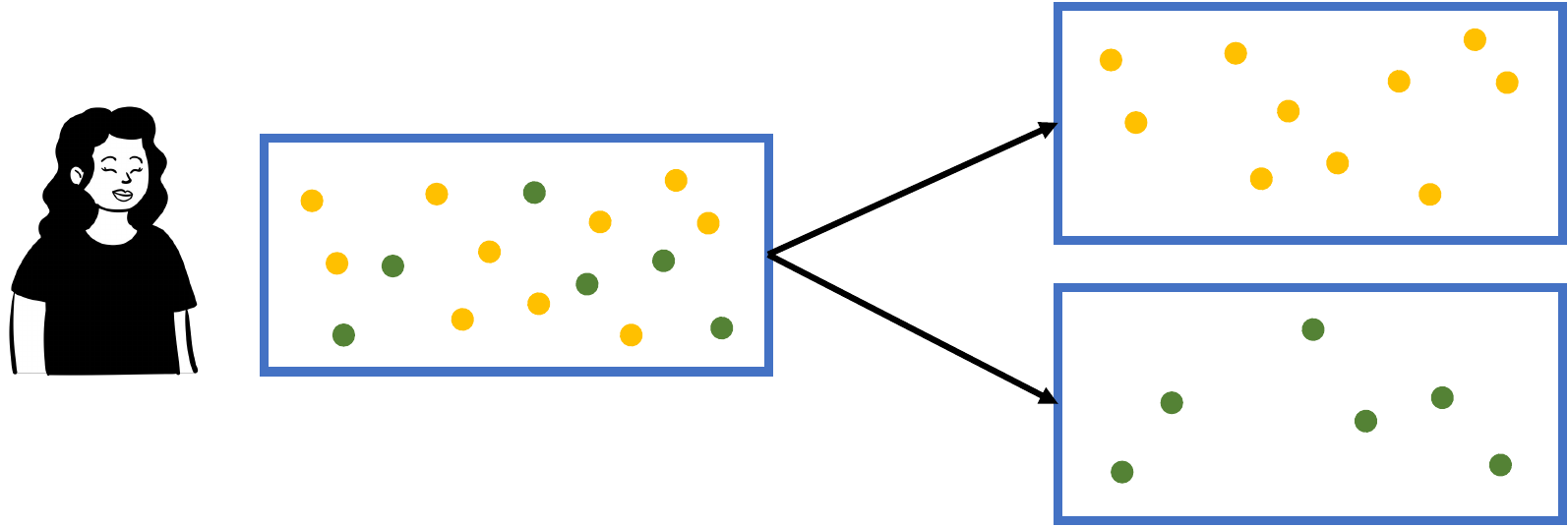}
	\caption{An observer gains predictive advantage by selecting sub-ensembles for which the outcome statistics differ from the predictions on the total ensemble. A physical theory may be considered empirically complete when such a procedure is impossible.}
	\label{fig:empirical-decomposition}
\end{figure}

The example shows how a theory, making probabilistic predictions, can be replaced by an improved theory with refined predictions such that indeterminacy can be reduced (see Figure \ref{fig:empirical-decomposition}). Thus, the old theory may be deemed empirically incomplete. The new theory will add variables to the description and can thus explain what caused the distribution to disperse --- the existence of distinguishable gases. One may ponder if a similar description is conceivable for quantum phenomena. If not, quantum indeterminacy is more fundamental than classical indeterminacy. 

If quantum mechanics isn't empirically complete, non-quantum properties could exist, making predictive advantage possible. More specifically, this may mean that the standard Born rule applies whenever the wave function is used to make predictions. But whenever the post-quantum variables are invoked, the new predictions render it invalid. 

From a purely statistical point of view, any non-extremal probability distribution can be trivially written as a convex combination of distinct probability distributions. In that sense, no prediction-making algorithm of any theory would be empirically complete. However --- and that is one of the crucial questions of this thesis ---  the properties of sub-ensembles in the decomposition may turn out unphysical and contradict other methodological, well-established, or reasonable scientific principles. Thus, a significant task is to characterise empirical extensions. Moreover, it will be essential to seek an explanation for why predictive advantage isn't observed in practice if empirical extensions exist in principle.

One can formulate the notion of empirical completeness in terms of ensemble averages (e.g. for instance, along the lines of what Brown et al.'s `statistical completeness' captures \citep{Brown-et-al-physical-complete}). Suppose theory $T$ assigns a physical state $\lambda\times\mu_Q$ to target system and measurement setup. Imagine a large ensemble $E$ of identical systems, i.e. all prepared in the same initial state $\lambda\times\mu_Q$ according to $T$. Thus, the ensemble is homogeneous. The prediction algorithm of the theory associates to the states the familiar probabilities $p(x)=p(x|\lambda\times\mu_Q)$ for the outcomes of observable $Q$. Brown et al. then define two different averages. The average value $\langle Q\rangle$ (expectation value) of the results we obtain upon \textit{actual} measurement, and  $\langle Q\rangle_T$ for the average value (expectation value) derived from the \textit{predicted} probabilities $p(x)$. Next, decompose $E$ into smaller sub-ensembles $E_1, E_2, ..., E_n$ and evaluate the corresponding expectation values. If the theory is empirically accurate, we expect the two quantities to coincide almost surely for very large ensembles $E$. According to Brown et al., statistical completeness is then a question of whether $\langle Q\rangle$ equals $\langle Q\rangle_T$ for each sub-ensemble. \newpage

\begin{definition}{\textbf{Statistical Completeness (Brown et al.).}}
	Theory $T$ is statistically complete if and only if for each \textit{conceivable sub-ensemble} $E'$ of $E$ that is `pure' in $T$, $\langle Q\rangle$ coincides with $\langle Q\rangle_T$ with probability one (in the limit as the number of systems in $E$ approaches infinity). 
\end{definition} Here, Brown et al.'s notion of a pure ensemble is identical to the notion of a pure or homogeneous ensemble in von Neumann's sense \citep{von2018mathematical}. 

But how is `conceivable ensemble' to be interpreted here? More importantly, how are the sub-ensembles divided? Is it purely theoretical by, say, introducing states finer-grained than those of theory $T$, or is it meant experimentally, i.e. a procedure in the laboratory that distinguishes them? In other words, it remains to be seen whether the ensemble ought to be decomposable operationally (through manifest variables) or purely metaphysically (through specifying further non-manifest variables). From the context in which Brown et al. use the definition, it becomes clear that statistical completeness is broadly construed and doesn't specify \textit{how} the ensembles are decomposed. Thus, it follows from the definition that a theory $T$ is statistically complete when there exists no other theory --- even in principle --- with a more refined description producing different outcome statistics for the sub-ensembles $E_1, E_2, ..., E_n$. In particular, there exists no more information or physical differences in the initial states for making predictions than the statistical algorithm of $T$ entails. A statistically complete theory implies that every system in the homogeneous ensemble is physically the same with respect to measurement outcomes. 

It follows from the definitions that metaphysical completeness is equivalent to statistical completeness. To see this, recall that for a metaphysically complete theory, there exists no extension such that $p(x|\lambda^1\times\mu_Q^1)=\int p(x|\lambda^2\times\mu_Q^2)\rho(\lambda^2|\lambda^1)d\lambda^2$, where we have used the familiar stipulations. If the theories are empirically accurate, their predictions for $\langle Q\rangle_T$ are the same. But since $T_2$ is not an extension of $T_1$, the individual probabilities of each element in the ensemble are equal, i.e. $p(x|\lambda^1\times\mu_Q^1)=p(x|\lambda^2\times\mu_Q^2)$. Therefore, for any conceivable ensemble $E$, we have that $\langle Q\rangle=\langle Q\rangle_{T_1}$ in the limit where the number of systems in $E$ approaches infinity. Moreover, when a theory $T_1$ is statistically complete, each conceivable ensemble homogeneous in $T_1$ leads to the same average $\langle Q\rangle$. As a result, there cannot exist a metaphysical extension of $T_1$, for otherwise, we could find ensembles with deviating predicted averages from $\langle Q\rangle_{T_1}$. In sum, a theory is metaphysically complete if and only if it is statistically complete.  

The concept of a statistically complete or metaphysically complete theory might make perfect sense in capturing a notion of completeness and arguably is useful in particular for the dedicated realist who wants to describe the underlying nature of reality irrespective of what can and cannot be observed. But to answer the question of \textit{empirical} completeness, this definition gets us into trouble were we to equate the two: \textit{If a theory makes probabilistic predictions, one can always trivially find a statistically complete theory in their sense.} In fact, one can always conceive of a fully \textit{deterministic} theory.\footnote{As we’ll see, de Broglie-Bohm theory is an elegant example of a deterministic and statistically complete theory of quantum mechanics.}

Here is a cheap way of constructing a deterministic, hence also statistically complete extension for \textit{any} physical theory that makes probabilistic predictions. Simply by adding a sufficient number of non-manifest variables and cooking up sufficiently complex physical laws, \textit{any} observation in the manifest configuration is describable by a deterministic theory. Consider the whole of the manifest domain, i.e. its complete set of configurations $\mathcal{M}$. And suppose that this set is parametrised by some variable $t$ over time. For every instant $t$, there exists a map $t \mapsto \mathcal{M}_t$. The amount of information that each set of configurations could hypothetically represent can, in principle, be very limited and constrained. For example, the same simple configuration could occur more than once in the `history' of $\mathcal{M}$ spanned by the parameter $t$. In particular, it could be the case that the information content is so limited that no unique map from one configuration to the next, i.e. $\mathcal{M}_t\mapsto\mathcal{M}_{t'}$ could possibly exist. Future configurations could be vastly underdetermined by present configurations. However, and this is the point of this exercise, every $M_t$ can be thought of as corresponding to a different non-manifest state (or set of states) $\lambda_t$, i.e. for any pair $t\neq t' : \lambda_t\neq\lambda_{t'}$. And for such a space, it is always mathematically possible to find a function that uniquely maps one state to the other. Thus, although in this case, the manifest domain is \textit{unpredictable}, the non-manifest domain may describe a deterministic and hence metaphysically complete world. No theory can therefore be provably irreducibly random in the metaphysical sense without in some way constraining the non-manifest domain (cf. also Bricmont's views on determinism in \citealt[Chapter~3.4.1]{bricmont2016making}). Theoretical and metaphysical virtues like simplicity and usefulness may rule out such constructions, but they are always possible in principle. At any rate, deterministic completions of this kind aren't relevant for any claim about the questions of this thesis, namely the \textit{empirical} completeness problem of quantum theory. Nothing would be gained for the \textit{predictability} by completing it \textit{metaphysically}.  

Although very artificial, such a model is always complete and therefore trivialises statistical completeness. The statistical completeness of a theory should therefore invoke at least some assumptions to restrict possible alternative theories that are supposedly more complete. Against the backdrop of this finding, I'm not interested in metaphysical completions since they have little to say about a theory's empirical completeness. Moreover, the concrete example of pilot wave theories shows that the converse is false: empirically complete theories need not be metaphysically complete. Nevertheless, a proof of metaphysical completeness would be stronger than empirical completeness since a metaphysically complete theory must also be empirically complete. For if no variables exist that would even \textit{theoretically} imply predictive advantage, how should the \textit{empirical} predictions be improved? Thus, it is expected that the assumptions leading to a proof of metaphysical completeness are substantial restrictions on such variables' properties. 

Therefore, the definition of statistical completeness must be tweaked to account for the notion I’m interested in here. What’s important for a reasonable account of empirical completeness is that the ensembles are to be distinguished experimentally, i.e. in terms of the configurations in the manifest domain. Is it possible to conceive of a theory whose empirical content permits a pre-selection of sub-ensembles such that $\langle Q\rangle$ and $\langle Q\rangle_T$ don't coincide? If not, then $T$ would indeed be empirically complete. This allows us to formalise empirical completeness in the style of statistical completeness.

\begin{definition}{\textbf{Empirical completeness 2.}}
	\label{def:empirical completeness 2}
	Theory $T$ is empirically complete if and only if for each manifest set of configurations corresponding to some prepared sub-ensemble $E'$ of $E$ that is homogeneous in $T$, $\langle Q\rangle$ coincides with $\langle Q\rangle_T$ with probability one (in the limit as the number of systems in $E$ approaches infinity). 
\end{definition} 

Thus, empirical completeness also accounts for the preparability of physical states that would lead to an average $\langle Q\rangle$, deviant from $\langle Q\rangle_T$.\footnote{A brief note on post-selection: Of course, one could always \textit{a posteriori} divide the large ensemble $E$ into smaller ensembles to make the corresponding averages differ. After all, if a theory makes probabilistic predictions, it must lead to a distribution of measurement results. And we could use the post-measurement results to simply post-select and group them into sub-ensembles such that conditionalised on these sub-ensembles, the averages $\langle Q\rangle$ and $\langle Q\rangle_T$ arbitrarily deviate from one another. This is cheating. Such a move evades prediction altogether and is thus rather a matter of retrodiction. Moreover, the idea falls short of all metaphysical requirements of theory selection and is experimentally and ontologically rather useless. But anyway, such a theory can, at best, only be a metaphysical, not an empirical, extension since prediction-making cannot involve future outcomes.} 

Definitions \ref{def:empirical completeness 1} and \ref{def:empirical completeness 2} of empirical completeness are equivalent: Let theory $T_1$ be empirically complete according to Definition \ref{def:empirical completeness 1} and $y_1$ be the manifest configuration selecting the fixed state of an ensemble of systems. Then, there exists no other theory $T_2$ with manifest configurations $y_2$ such that the predictions of $T_1$ could be refined, i.e. $p(x|y_1)=p(x|y_2)$, for all possible states the system can possess, where $x$ the outcomes of the measurement. But since no theory exists making deviating predictions for some configurations $y_2$, this means $\langle Q\rangle=\langle Q\rangle_{T_1}$. No manifest variables are conceivable on the basis of which we could do so. Therefore, theory $T_1$ is empirically complete according to Definition \ref{def:empirical completeness 2}. The converse follows from the fact that when $T_1$ is empirically complete by \ref{def:empirical completeness 2}, i.e. it isn't possible to find manifest sets of configurations such that $\langle Q\rangle\neq\langle Q\rangle_T$, then there exists no theory $T_2$ and configurations $y_2$ to select the big ensemble into sub-ensembles. This implies that the predictions of theory $T_2$  must coincide with theory $T_1$'s predictions for all configurations $y_2$ in order to reproduce the averages. In sum, Definition \ref{def:empirical completeness 1} and \ref{def:empirical completeness 2} are equivalent. 

In brief, the quest for empirical completeness relative to the standard theory of quantum mechanics comes down to assessing the following:  Suppose a physical system is described by quantum theory by the quantum state $\rho$. Does a theory $T_0$ exist permitting a preparation $y_0$ of some physical state $\lambda\times\mu_Q$ such that $p(x|\rho)\neq p(x|\rho, y_0)$? And if not, what are, if any, the physical principles ruling this out?

\section[Standard Quantum Theory and Quantum Predictions]{Standard Quantum Theory and Quantum Predictions}
\sectionmark{Standard Quantum Theory}
\label{section:quantum theory and quantum predictions}

I will explain here how prediction-making in quantum mechanics maps to the framework introduced above (or at least one way of understanding how this can work for quantum theory). I shall call this, henceforth, standard quantum theory or standard quantum mechanics. For all practical purposes, it is clear what the quantum predictions for the outcomes of measurements are. However, as we shall later see in Section \ref{section:predictability and measurement problem}, quantum mechanics' connection to the empirical results is a subtle business. The common use of the theory is not unambiguous and draws a notoriously murky picture of reality. This may be expected since it could be argued that the theory's framework isn't sufficiently filled in to explain \textit{what} is actually observed in a measurement. Moreover, the lack of a satisfactory solution to its notorious measurement problem isn't helping for a clear-cut account of prediction making. 

In the prediction-making algorithm of standard quantum theory, the possible physical states $\lambda$ and $\mu$ are normalised vectors in complex Hilbert spaces (alternatively, a similar account can be given involving mixed states). The observable $Q$ — the apparatus is set to measure — is determined by the details of the physical interaction with the target system. That is, the dynamical evolution as per an interaction Hamiltonian $H_{int}$ defines the measurement type performed. When we write $\mu_Q$ for the state of the apparatus, we mean that the interaction procedure is of the sort such that we call it a measurement of the observable feature $Q$. Therefore, the choice of the observable also specifies the coupling between target system and apparatus. Let $\lambda:=\ket{\psi} \in \mathcal{H}_S$ and $\mu_Q:= \ket{\phi} \in \mathcal{H}_A$ be the states of target system and apparatus with corresponding Hilbert spaces $\mathcal{H}_S$, $\mathcal{H}_A$ (the initial states are, once more, picked based on manifest configurations). The compound state of the target system and apparatus is postulated to be the tensor product $\ket{\psi}\otimes\ket{\phi}\in \mathcal{H}_S\otimes\mathcal{H}_A$, and its dynamics are dictated by the linear Schrödinger equation. The initial states and the Schrödinger evolution determine the possible measurement outcomes after the coupling time $\Delta t$. As outlined above, the purpose of the physical states is to assign a probability to each outcome $x$ of any measurement, i.e. $p(x) = f_x(\lambda\times\mu_Q)$, where $f_x$ are some probability maps from the state space to the probability space.\footnote{For simplicity, I restrict the present discussion to finite-dimensional Hilbert spaces and discrete outcomes.} 

Given a pointer observable $P_x$ --- represented by a projection-valued measure (PVM) --- with manifest outcome $x$, and given the details of the coupling, there will be a corresponding POVM $E_x$ on the target system. By Gleason's theorem, the probability measures on POVMs are given by quantum states \citep{Gleason-theorem}. 

Moreover, the operator’s properties are restricted by the fact that they need to reproduce the properties of normalised probabilities. But how do we find the relevant operators to obtain the outcome statistics of the apparatus and, in turn, define the observable $Q$? 

We stipulate that the statistics of the post-measurement configuration of the pointer uniquely define the measured observable. We assume that the pointer variable is represented by a set of (smeared) projection operators $P_x$ for every outcome $x$ onto localised configurations of the pointer. Whenever outcome $x$ obtains, the pointer must reside inside the corresponding configuration of the operator $P_x$. Thus, we obtain the operators $E_x$ through the following defining condition — the Born rule: 

\begin{equation}
	p(x)=\bra{\psi}E_x\ket{\psi} := \bra{\psi}\otimes\bra{\phi}U^{\dagger}(\mathds{1}\otimes P_x )U\ket{\psi}\otimes\ket{\phi},
\end{equation} where $U=U(t_0, t_0+\Delta t)=e^{-\frac{i}{\hbar}(H_S+H_A+H_{int})}$, and $H_S$, $H_A$, and $H_{int}$ are the free Hamiltonians and interaction Hamiltonian of the target system and apparatus, respectively. 

The Hamiltonian functions are found through quantising the familiar Hamiltonians of classical physics. Note also that the pointer variable $x$ of the apparatus is a manifest configuration, i.e. a position variable. 

I want to emphasise that something needs to be said about the post-measurement state of the system. Indeed, making such an assumption is unnecessary since making predictions via the Born rule suffices to derive the relevant outcome statistics. That is, we can abstain from introducing a projection postulate (a `collapse' postulate). This is possible since the initial states are assumed as primitive in the description. However, there remains to be a question about how we arrive at the initial states. To this, I shall turn later.

The most general framework that the standard account of quantum mechanics offers to describe a measurement is `positive operator-valued measurements' (POVMs). The unitary Schrödinger evolution of system-cum-apparatus naturally leads to a set of self-adjoint operators that satisfy

\begin{equation}
	\mymathbb{0}\leq E_x \leq\mathds{1}, \text{and} \sum\limits_x E_x =\mathds{1},
\end{equation} by virtue of probability theory. 

As an example of an interaction that usually counts as a measurement of some observable $Q$ (rather than, e.g. a free evolution or other non-measurement-like interactions), I shall mention the familiar von Neumann scheme of measurement:

Consider the state space of the target system and apparatus to be $\mathcal{H_S}=\mathcal{H_A}={L}^2(\mathbb{R})$. Assume further that the total Hamiltonian is $H=H_S+H_A+H_{int}$, i.e. the free Hamiltonians of target system and apparatus and interaction term 

\begin{equation}
	H_{int} := g(t) \lambda Q\otimes P_A,
\end{equation} where $g(t)$ is function restricting the coupling to a limited time interval $\Delta t$, with $\int\limits_{t_0}^{t_0+\Delta t}g(t) dt=1$, and $\lambda$ the coupling strength. Furthermore, let $Q$ be the canonical position operator and $P_A$ the canonical momentum operator. If we consider the coupling interval very small and $\lambda$ very big, we are dealing with an impulsive interaction that is short and strong. Therefore, the interaction term in the Hamiltonian dominates the free evolution, and we can approximate $H\approx H_{int}$. Let the total system’s initial state be $\ket{\Psi(x,y,0)}=\ket{\psi}\otimes\ket{\phi}=\int \psi(x,0)\phi(y,0)\ket{x}\otimes\ket{y}dx dy$.

The outcomes of the apparatus are pointer positions, and we, therefore, take $P_x=\ket{x}\bra{x}$ as the (sharp) apparatus observable, projecting onto a tiny region of configurations $y$ in which the pointer resides after the measurement. For example, $\ket{x}$ could be chosen to be a very narrowly peaked function with small compact support.  

The probability of obtaining outcome $x$ is found by computing

\begin{equation}
	\label{POVM from  interaction}
	p(x)=\bra{\psi}\otimes\bra{\phi}U^{\dagger}\mathds{1}\otimes P_x U \ket{\psi}\otimes\ket{\phi} \overset{!}{=} \bra{\psi}E_x \ket{\psi}.
\end{equation}

As introduced above, this equation lets us read off the corresponding POVM operators $E_x$. The unitary reads $U=U(t_0,t_0+\Delta t)=e^{-\frac{i}{\hbar}\int\limits_{t_0}^{t+\Delta t}\lambda g(t)Q\otimes P_A dt}$. Working out the calculation in Equation \ref{POVM from  interaction} yields

\begin{equation}
	E_x = \int |\phi(y-\lambda q)|^2\chi_x(y) \ket{q}\bra{q} dq dy,
\end{equation} and the outcome probabilities are thus

\begin{equation}
	p(x) = \int |\psi(q)|^2|\phi(y-\lambda q)|^2\chi_x(y) \ket{q}\bra{q} dq dy.
\end{equation}

Formally, we can identify three axioms in the preceding discussion that define the basic components of quantum theory:

\begin{itemize}
	
	\item[(1)] \textbf{Kinematics.} The physical state of a quantum system is represented by a mathematical state vector $\ket{\psi}$ in a Hilbert space $\mathcal{H}$ over the complex numbers. If the state isn't pure, more generally, we assign a density operator $\rho\in\mathcal{S}(\mathcal{H})$ as the (mixed) state of the system, where $\mathcal{S}(\mathcal{H})$ is a convex subset of the self-adjoint trace-class operators on $\mathcal{H}$. \\
	
	\item[(2)]  \textbf{Dynamics.} The quantum states $\ket{\psi}$ evolve in time according to the Schrödinger equation:
	\begin{equation}
		i \hbar \partial_t \psi(t)=\hat{H}\psi(t),
	\end{equation} where $\hat{H}$ is the (essentially) self-adjoint Hamiltonian of the system corresponding to the quantised classical energy. For open systems involving generic states $\rho$, the dynamics require a more complicated master equation such as the Liouville-von Neumann or Lindblad equation.\\
	
	\item[(3)] \textbf{Epistemology.} To each physical feature $Q$ is associated a POVM with self-adjoint operators $E_x$ (the observables). The index $x$ is the pointer configuration and corresponds to the possible outcomes of the apparatus for measurement of $Q$. The probability of finding outcome $x$ in an experiment is obtained by 
	\begin{equation}
		p(x):= \bra{\psi}E_x\ket{\psi}=\tr(E_x \ket{\psi}\bra{\psi}).
	\end{equation} If the state is mixed, we have 
	\begin{equation}
		p(x):= \tr(E_x \rho).
	\end{equation} 
	This probability assignment is standardly called the Born rule for generalised measurements. \\
\end{itemize}

There are good arguments for why only those three postulates are necessary and unproblematic in a minimal account of quantum mechanics (see, e.g. \citealt{Wallace-orthodox-quantum}). For example, state reduction postulates or some \textit{`collapse'} are often not used in practice; measurements are usually disturbing such that \textit{repeatability} doesn't apply (violation of the eigenvalue-eigenvector link), and measurement is usually continuous and non-sharp which corroborates the use of POVMs instead of projections as measurement operators. It can be argued that such additional ingredients are often part of a proposed interpretation of quantum mechanics. The rest of my discussion will proceed without them, as they are unnecessary for the results. What is important for the present discussion is that these are the minimal assumptions necessary for discussing the empirical completeness problem and predictive advantage relative to quantum theory. 

\subsection{Quantum State Preparation}
\label{Section:quantum state preparation}

So far, we have treated the initial quantum states prior to the coupling as primitive. In particular, we have assumed that the target system’s state on which we perform the measurement as \textit{a priori}. We saw how the state is measured in an experiment and how the quantum states connect to the empirical results --- the macroscopic configurations in the manifest domain. But the question remains how the target system is prepared in a given state? After all, to get the prediction off the ground, we need to know what states we are dealing with in the lab. In other words, how do we find the initial state of the target system from the manifest degrees of freedom?  

One natural strategy for accommodating the issue is to use what we know from quantum dynamics and add to the picture conditionalising on the macroscopic degrees of freedom. Hence we introduce a selection criterion based on the manifest configurations. Thus, some manipulation of manifest configuration will have to signify what quantum state the target system and apparatus possess. 
Imagine a source that produces a probabilistic mixture of known pure states describable by a density operator $\rho$. We assume this can be justified by physical reasoning about the details of the source and (quantum) statistical mechanics. We could either proceed with a statistical ensemble $\rho$ of pure quantum states or further divide the mixture into pure sub-ensembles: The states may then be selected into sub-ensembles by classical means and macroscopic pointers. For example, we may produce a beam of spin particles in some classical mixture of spin-up and spin-down states in some appropriate basis. A subsequent measurement coupling splits the beam by a magnetic field. One part of the beam is discarded, and the remaining beam, corresponding to spin-up, is used for further interference experiments. Hence, by this strategy, we have prepared pure spin-up states. In a nutshell, quantum state preparation involves finding a suitable source which we know generates the desired states we wish to prepare with some non-vanishing probability and which are subsequently selected on the basis of macroscopic configurations. Note that those probabilities (with which the states occur) are subjective probabilities, i.e. in the sense of uncertainty in the presence of ignorance of the observer — as opposed to quantum probabilities derived from pure states. In the latter case, we would run into trouble if we applied the same selection procedures since we wouldn't observe interference effects (cf. \citealt{Wallace-orthodox-quantum}). The individual parts of the conditionalised ensembles are treated classically and do not interact. They are therefore said to have \textit{decohered}. Furthermore, the selection process leaves behind the records that are used to distinguish the states (beams), and if we average over them, we will recover the original statistical ensemble $\rho$.   

So we need to decide when a quantum state should be presumed to be either (fully) mixed or pure. The former is permissible of a classical treatment in terms of classical probabilities, whereas the latter isn't. And it is, of course, the notorious measurement problem that I have just alluded to. One can’t just assume that the initial states are primitive. They need to be prepared through some selection process, which requires making assumptions about post-conditionalised states. The measurement problem precisely comprises the issue of dealing with definite classical or indefinite quantum states. But without stipulating that macroscopic degrees of freedom (the configurations in the manifest domain) decide what state was prepared, prediction in quantum mechanics would be impossible. Remarkably, the measurement problem enters right at the beginning, i.e. when it has to be decided what quantum states one is dealing with. 


\section{The Quantum Decorrelation Principle}
\label{section: quantum decorrelation}

\subsection{Pure Quantum States}

\begin{figure}[h]
	\centering
	\includegraphics[width=0.85\linewidth]{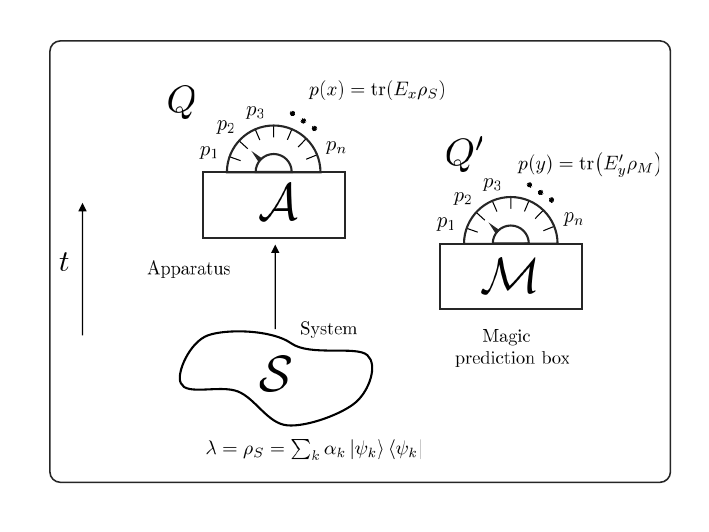}
	\caption{A magic prediction box refines the quantum probabilities for outcomes $x$ of a target system by introducing stronger correlations to additional manifest configurations $y$.}
	\label{fig:magic-prediction-box-quantum}
\end{figure}

I first observe that pure state assignment in quantum theory leads to decorrelation of target system and its environment: Assume the target system is described by the \textit{pure} quantum state $\rho_S:=\ket{\psi}\bra{\psi}$ and the apparatus is set to measure an observable $Q$ represented by the POVM $\{E_x\}$ with manifest outcomes $x$. A second apparatus --- which we shall call the magic prediction box --- is described by the quantum state $\rho_M$ and yields manifest outcomes $y$ (the corresponding observable $Q'$ be represented by the POVM $\{E_y\}$). The joint quantum state of the compound system of target system and magic prediction box thus is $\rho\in\mathcal{S}(\mathcal{H_S}\otimes\mathcal{H_M})$ --- uniquely defining the states $\rho_S=\tr_M(\rho)$ and $\rho_M=\tr_S(\rho)$ of the subsystems. Figure \ref{fig:magic-prediction-box-quantum} depicts the situation. I first treat the case where the manifest outcomes $(x,y)$ are produced by bi-partite systems. I shall turn later to the case of a single system on which different measurements are performed. However, when there is one single system that produces both the outcomes of the primary measurement apparatus and the magic prediction box, two quantum interactions with the target system are required. But then, we would need an account for the post-measurement states after the first interaction. I stayed away from this in my introduction to standard quantum mechanics. As I emphasised, the post-measurement states are usually unknown and often destroyed in the process. Hence, the interesting case is bi-partite systems arising as the joint states of target system and the states of the magic prediction box. In other words, the quantum state space of target system and its environment are distinct Hilbert spaces. 

The general form of a density operator of the bi-partite system can be expressed as

$\rho=\sum_{k}\alpha_k \left(\sum_{r} c_{rk}\ket{\psi_{rk}}\otimes\ket{\psi'_{rk}}\right)\left(\sum_{l} \bar{c}_{lk}\bra{\psi_{lk}}\otimes\bra{\psi'_{lk}}\right)$, where $\alpha_k$ are probabilities, and $\sum_{l}|c_{lk}|^2=1$ are normalised coefficients of the Schmidt decomposition for each entangled state indexed by $k$ (note that this decomposition isn't unique since the total state is mixed). Since the Schmidt states of the secondary system are orthogonal, we find the partial trace and thereby the reduced state of the target system to be $\rho_S=\tr_M(\rho)=\sum_{k}\alpha_k  |c_{r_kk}|^2\ket{\psi_{r_kk}}\bra{\psi_{r_kk}}$, for some numbers $r_k$. But $\rho_S=\ket{\psi}\bra{\psi}$, and as the spectral decomposition is unique, the states $\ket{\psi'_{rk}}$ of this expansion must be the same for all $k,r_k$, and  $\sum_k\alpha_k  |c_{r_kk}|^2=1$. As a result, $\rho$ is separable and we can rewrite $\rho=\sum_{k}\alpha_k \left(\sum_{r} c_{rk}\ket{\psi}\otimes\ket{\psi'_{rk}}\right)\left(\sum_{l} \bar{c}_{lk}\bra{\psi}\otimes\bra{\psi'_{lk}}\right)=\ket{\psi}\bra{\psi}\otimes\left(\sum_{rlk}\alpha_k c_{rk}\bar{c}_{lk}\ket{\psi'_{rk}}\bra{\psi'_{lk}}\right)=\rho_S\otimes\rho_M$. But this immediately implies the statistical independence of magic prediction box and target system: 
\begin{align}
	p(x,y)&= \tr(E_x\otimes E'_{y}\rho)\\ \nonumber
	&=\tr(E_x\ket{\psi}\bra{\psi}\otimes E'_{y}\rho_M)\\ \nonumber
	&=\tr(E_x\ket{\psi}\bra{\psi}) \tr(E'_{y}\rho_M)\\\nonumber
	&=p(x)\cdot p(y).
\end{align}

I will henceforth dub this the quantum decorrelation principle. 

\begin{definition}{\textbf{Quantum decorrelation principle (QDP)}}
	\label{def:quantum decorrelation principle}
	Whenever a pure state describes a quantum system, the outcomes of any measurement are statistically independent of its environment. Therefore, the maximum information encoded in its environment is the target system's pure quantum state. 
\end{definition}

The QDP asserts that whenever a pure state is assigned, according to quantum theory, the uncertainty in the statistics of measurement outcomes is exactly given by the uncertainty contained in the pure quantum state by virtue of Born's rule. In other words, the most accurate prediction possible in quantum theory for measurements of pure states are the Born probabilities. And no records that exist in the environment of the system can improve this prediction. Conversely, whenever manifest configurations exist in the environment --- correlated to the outcome configurations of the target system --- the target system's state must be mixed. This shows that insofar as quantum predictions are unambiguous,  no empirical extension can exist reproducing all predictions of quantum mechanics. Therefore, an alternative theory with predictive advantage will necessarily have to violate the quantum decorrelation principle in some regime. Hence, if there is to be a departure from quantum predictiveness, there has to exist non-quantumness somewhere. And to reproduce the quantum predictions of a pure state, it follows that the non-quantumness will be in the correlations between target system and its environment. Bear in mind that the QDP is a trivial proposition in probability theory: whenever there exist non-trivial correlations with the environment, conditionalising on the additionally correlated variables leads to a decomposition that isn't pure. The important point, however, is that if quantum mechanics isn't empirically complete, then there exist stronger correlations although a quantum state may be pure according to the theory. Thus, the QDP would fail in a theory with predictive advantage over standard quantum theory.  

But how surprising would it be to find such `non-quantumness'? Moreover, what is the nature of theory violation really in this case? From a purely epistemic point of view, it may be argued to be rather innocuous: Let's go back to classical physics and the thermodynamic example of two distinguishable gases in a box. There, a better theory exists with predictive advantage to distinguish two sorts of gases, i.e. nitrogen and oxygen, in a box where they are mixed and brought to thermodynamic equilibrium with total Maxwell-Boltzmann distribution $f_{MB}(v)$. Conditionalised on the variables of the new theory, it predicts two distinct statistical distributions of the velocities for the gases. It is thus conceivable to select sub-ensembles such that the mean velocities for nitrogen and oxygen differ from the mean velocity of the total mixed ensemble. Following a similar description for these thermodynamic states as for quantum states, this can be expressed in the language used before: According to the empirically incomplete theory, target system and environment factorise into independent distributions of the form 

\begin{equation}
	p(v,v')= f_{MB}(v)\cdot g_E(v'), 
\end{equation}   where $g_E(v')$ some thermodynamic velocity distribution of the environmental state. Thus, because the theory doesn't contain the relevant manifest variables to distinguish nitrogen from oxygen, its probabilistic predictions are such that target system and environment are decorrelated whenever the state $f_{MB}(v)$ was prepared. The theory must predict that conditioned on any variables in the environment, the same distribution $f_{MB}(v)$ obtains; thus, target system and environment satisfy a decorrelation principle similar to quantum theory. But according to the theory with predictive advantage, the variables lead to refined predictions and stronger correlations. The point is that decorrelation principles arise from probability theory and the possible variables used to make predictions. 

The difference that seems to set apart the quantum decorrelation principle and the classical decorrelation principle is this: In the example of the classical thermodynamic system, the correlation between target system and environment yielding predictive advantage can be established without changing any of the future physical behaviour of the target system. That is, the discovery of the new theory invoking variables used to update the old predictions has no impact on all the possible measurement outcomes of the target system. Thus, predictive advantage can be viewed as an epistemic update of the predicted probability density.

In contrast, quantum mechanics predicts that the only way to establish stronger correlations to a target system when a pure state is assigned is to create entanglement with its environment. But as a consequence, the physical behaviour of the system will necessarily deviate from a non-entangled system possessing the original state (see the analysis of entangled quantum states below). Thus, entanglement produces operational disturbance in the outcome probabilities of the target system.  

The quantum decorrelation principle is, therefore, a stronger statement about predictability, but at the same time, expected to exist in one way or the other from an epistemic point of view. So in what sense would post-quantum theories with predictive advantage have to be \textit{non-quantum} then? 
The answer may turn out to be: In the simple probabilistic and, therefore, innocuous way. Born rule predictions imply decorrelation in the total distribution of target system and environment when pure states are assigned. But variables exist, leading to predictive advantage and novel correlations, incompatible with the Born prediction. Nevertheless, this follows for variables not present in the quantum framework; therefore, the formalism can't claim their relevance for prediction making. 

To summarise, because the assignment of a quantum state directly implies an assignment of a probability distribution, an empirical extension would have predictive advantage whenever variables exist based on which the probabilities could be updated. In this sense, the Born rule and hence the quantum decorrelation principle are violated.

\subsection{Mixed Quantum States}

For mixed states $\rho_S$, the situation is different. Let's begin with non-entangled states, i.e. the total state of target system and magic prediction box is separable and can be written as $\rho= \sum_{k} \alpha_k \rho_k \otimes \rho_k'$, where $\rho_k$, $\rho_k'$ are density operators on the subsystems and $\alpha_k$ are probabilities. Without loss of generality, we can assume that the operators are pure states. Otherwise, expand them and recollect the terms. Thus, one has

\begin{equation}
	\rho=\sum\limits_k\alpha_k\ket{\psi_k}\bra{\psi_k}\otimes\ket{\psi_k'}\bra{\psi_k'}, 
\end{equation} for some $\psi_k\in\mathcal{H_S},\psi_k'\in\mathcal{H_M}$.

As before, the marginal states of the subsystems are $\rho_S=\tr_M(\rho)$ and $\rho_M=\tr_S(\rho)$. But $\rho_S$ generally is a mixed state now. By calculating the joint probabilities, we can see that they don't factorise. That is, target system and magic prediction box generically are not statistically independent:

\begin{align}
	\label{eqn:separable state}
	p(x,y)&= \tr(E_x\otimes E'_{y}\rho)\\ \nonumber
	&=\tr(\sum_{k} \alpha_k E_x\ket{\psi_k}\bra{\psi_k}\otimes E'_{y}\ket{\psi_k'}\bra{\psi_k'})\\ \nonumber
	&=\sum_{k}\alpha_k \bra{\psi_k}E_x\ket{\psi_k} \bra{\psi_k'}E'_{y}\ket{\psi_k'}.\\\nonumber
	&=\sum_{k} \alpha_k p_k(x)\cdot p_k'(y), 
\end{align} were I defined $p_k(x)=\bra{\psi_k}E_x\ket{\psi_k}$ and $p_k'(y)$ analogously. We can recover the marginal probabilities $p(x), p(y)$ of the subsystem's outcomes by computing the partial trace or, more easily, by directly summing over the joint probabilities $p(x,y)$:

\begin{align}
	\label{eqn:probability i}
	p(x)=\sum_{y} p(x,y) &=\sum_{y}\sum_{k} \alpha_k \bra{\psi_k}E_x\ket{\psi_k} \bra{\psi_k'}E'_{y}\ket{\psi_k'}\\ \nonumber
	&= \sum_{k} \alpha_k \bra{\psi_k}E_x\ket{\psi_k} \bra{\psi_k'}\sum_{y}E'_{y}\ket{\psi_k'}\\ \nonumber
	&=\sum_{k} \alpha_k \bra{\psi_k}E_i\ket{\psi_k}\\\nonumber
	&=\tr(E_x\rho_S).
\end{align} The second last equality holds since the POVM operators sum up to the identity and the states $\psi_k'$ are normalised. Analogously, we obtain $p(y)=\sum_{k} \alpha_k \bra{\psi_k'}E'_{y}\ket{\psi_k'}$. 

In contrast to pure states, the manifest variables $y$ of the secondary system clearly can lead to a predictive advantage. That is, the conditional probability of the outcomes $x$ do, in general, depend on the outcomes $y$:

\begin{equation}
	\label{eqn:conditional probabilities}
	p(x|y)= \frac{\sum_{k}\alpha_k \bra{\psi_k}E_x\ket{\psi_k} \bra{\psi_k'}E'_{y}\ket{\psi_k'}}{\sum_{k} \alpha_k \bra{\psi_k'}E'_{y}\ket{\psi_k'}}=\frac{\sum_k\alpha_k p_k(x)p_k'(y)}{\sum_k \alpha_kp_k'(y)}\neq p(x).
\end{equation} Thus, we find that the quantum prediction algorithm based on mixed states is strictly not empirically complete in the sense of Definition \ref{def:empirical completeness 1}. We can find an empirical extension with manifest configurations $y$ conditioned on which we obtain predictive advantage over the prediction based on the mixed state $\rho_S$ alone. On the face of it, we can interpret the separable state $\rho$ as a probability distribution over uncorrelated pure states for each $k$. In operational terms the ensemble giving rise to $\rho$ can be viewed as the mixture of individual states $\ket{\psi_k}\bra{\psi_k}\otimes\ket{\psi_k'}\bra{\psi_k'}$ each prepared with probability $\alpha_k$. Since each ensemble element describes a pure state, we get from Equation \ref{eqn:separable state} that target system and magic prediction box are uncorrelated for fixed $k$, for all POVMs (according to the decorrelation of pure states). 

Suppose now that the POVM of the magic prediction box are projections onto the $\ket{\psi'}$ states, i.e. $\{E'_{y}\}=\{\ket{\psi_{y}'}\bra{\psi_{y}'}\}$. Assume that the $\ket{\psi'}$ states are orthogonal. The conditional probability for $x$ then reduces to 
\begin{equation}
	\label{eqn:improved probabilities}
	p(x|y)= p_k(x)\mid_{k=x}=\bra{\psi_{y}}E_x\ket{\psi_{y}},
\end{equation} which is a fine-graining of the the outcome probabilities $p(x)$ in Equation \ref{eqn:probability i}. Thus, the magic prediction box could be said to possess a classical epistemic correlation to the preparation of the target system. When on top of that, the POVM of the primary apparatus are projectors, the prediction may even be deterministic since $p(x|y)  \in \{0,1\}$. In a way, this isn't surprising, for mixed states are considered an epistemic state of incomplete knowledge. If the state is pure, however, the knowledge is complete. 

If we fix the state of the target system $\rho_S$ and the target POVM, we can ask what the best strategy is for the magic prediction box to improve the Born rule prediction of $p(x)$. What is the optimal choice of POVM and states, and are the improved probabilities in Equation \ref{eqn:improved probabilities} optimal? The question arises since for every $y$, we still obtain a distribution of conditional probabilities $p(x|y)$ for the outcomes $x$. Hence, how informative can $p(x|y)$ be conditioned on $y$?   

In general, when the POVMs are not projections and the states $\ket{\psi'_k}$ are not orthogonal, the conditional probabilities are a coarse-graining of the probabilities $p_k(x)$ we obtained for orthogonal states and projection POVMs (cf. the general form \ref{eqn:conditional probabilities}). Thus, for a fixed POVM, the previously outlined prediction procedure is optimal when the states of the magic prediction box are orthogonal. To formalise and quantify this intuition, one could use the quantum information-theoretic mutual information between the distributions in $x$ and $y$. We then apply the data processing inequality, which states that coarse-graining cannot increase the mutual information between the two distributions. So the best correlation we get for the probabilities is the one in \ref{eqn:improved probabilities}. 

In other words, the outcome probabilities for the target system are updated according to the following rule

\begin{equation}
	p(x)= \sum\limits_{k}\alpha_k \bra{\psi_{k}}E_x\ket{\psi_{k}}\mapsto \bra{\psi_{y}}E_x\ket{\psi_{y}}=p(x|y),
\end{equation} where $y$ is the outcome in the environment for some projective POVM. 

Even though a magic prediction box sometimes cannot be set up to measure the right POVM to obtain these correlations since insufficient information is given as to how the overall state was prepared, the crucial point is that, in principle, there could exist records in the target system's environment employable for predictive advantage. Moreover, I gave a concrete realisation of an actual preparation for which this can occur. An interesting question is whether the presence of a mixed quantum state for the target system always implies the existence of records in the environment usable for predictive advantage. If not, can records like these be created by some measurement procedure? In the case of pure states, we got a negative answer. For mixed states, predictive advantage can, in principle, exist, e.g. when there is a classical epistemic correlation. As we saw in Section \ref{Section:quantum state preparation} the standard method of selecting pure states from a mixed ensemble can leave traces in the environment that precisely are of this sort of correlation. That is, every time the source creates a pure state in the separable mixture $\rho=\sum\limits_k\alpha_k\ket{\psi_k}\bra{\psi_k}$, there is a recording of which one was prepared, i.e. the index $k$ is known. However, there is no reason to believe that such a recording always exists for mixed states. We can conceive of preparation procedures for mixed states without classical epistemic traces from a selection process.

Let's summarise the observations. In the interesting case of a pure quantum state, the prediction algorithm of quantum mechanics implies that there cannot be magic prediction boxes with predictive advantage over quantum theory. This decorrelation principle mainly derives as a consequence of the tensor product structure of bi-partite quantum systems, the use of POVMs, and the linearity of the trace operator to derive outcome probabilities. For separable mixed states, matters are slightly more complicated, but any prediction improving correlation can be understood as a classical epistemic uncertainty in the quantum density operator. Then manifest configurations can exist on top of the ones indicating the prepared state $\rho_S$ that produce predictive advantage over the predictions from $\rho_S$ alone. That is, correlations can exist such that the uncertainty of the outcomes on $p(x)=\tr(E_x\rho_S)$ can be `improved' up to the probabilities that are obtained from conditioning on the preparation, i.e. $p_k(x)$ when the POVM of the environment are a particular set of projectors. But not by more since, for all other POVMs, the predictions are mixtures of these probabilities. Thus, mixed states effectively reduce to a pure state for every individual preparation since, conditioned on $k$, the target system's state is pure. The answer to the question of whether there are any circumstances in which the predictions we can make about the target system can be improved with respect to what would be given by the Born rule for the state of that system is \textit{no} if the states are pure and \textit{yes} if the states are mixed. But the improvement we get is no more than what we would get if the state had been pure in the first place on every run of the experiment. In this sense, a mixed state could be argued to do always worse than the prediction of some pure state does. But in the mixed state case, the correlations can be used to improve the predictions up to a pure state case. In the next section I will restate that entangled quantum states allow for even stronger correlations. 

\subsection{Entangled Quantum States}

Entangled quantum states of bi-partite systems are quantum density operators that aren't separable and generically have the form  \begin{equation}
	\rho=\sum_{k}\alpha_k \left(\sum_{r} c_{rk}\ket{\psi_{rk}}\otimes\ket{\psi'_{rk}}\right)\left(\sum_{l}\bar{c}_{lk}\bra{\psi_{lk}}\otimes\bra{\psi'_{lk}}\right),
\end{equation} where for each index $k$ the entangled state is expressed in a Schmidt basis. As before, we denote with $\rho_S=\tr_M(\rho), \rho_M=\tr_S(\rho)$ the reduced states of target system and magic prediction box. From the previous considerations on when the target system state is assumed pure, it follows that $\rho_S$ must be mixed. Otherwise, the joint state $\rho$ would have to be separable. But for an entangled (pure) state of the composite system, the mixed state $\rho_S$ cannot be viewed as a mixture being realised as an ensemble of pure states, each prepared with some probability $\alpha_k$ --- often called an improper mixture. Analogously to a separable state, there can exist records indicating the individual preparation $k$. I showed before that this leads to some predictive advantage for the outcome probabilities and can be seen as a classical epistemic correlation of the target system to its environment. I have described that the predictive advantage is at least the improvement gained from conditioning on the preparation $k$.

Therefore, we may assume $\alpha_k=0$ for all but one $k$ without loss of generality. 
Thus, the total entangled state is assumed pure and reads $\rho=\left(\sum_{r} c_{r}\ket{\psi_{r}}\otimes\ket{\psi'_{r}}\right)\left(\sum_{l}\bar{c}_{l}\bra{\psi_{l}}\otimes\bra{\psi'_{l}}\right)$, with real numbers $\sum_{r}|c_r|^2=1$. For the reduced state of the target system, we thus obtain $\rho_S=\sum_{r}|c_r|^2\ket{\psi_{r}}\bra{\psi_{r}}$. In contrast to pure separable states, however, $\rho_S$ can still be mixed and exhibit correlations to the magic prediction box $\rho_M$. 

The conditional probability for outcome $x$ of the target system given outcome $y$ reads

\begin{equation}
	\label{eqn:conditional probabilities entangled}
	p(x|y)= \frac{\sum_{r,l}c_r \bar{c}_l \bra{\psi_l}E_x\ket{\psi_r} \bra{\psi_l'}E'_{y}\ket{\psi_r'}}{\sum_{r} |c_r|^2 \bra{\psi_r'}E'_{y}\ket{\psi_r'}},
\end{equation} which looks similar to the conditional probability we obtained for a separable mixed state (Equation \ref{eqn:conditional probabilities}). There it was shown that the best predictive advantage is reached for projective POVMs of the environment and that the target system has a projective POVM such that the correlation is deterministic. One immediate observation is that for projective POVMs on the basis states of the environment, the updated probability is like in the classical separable case, i.e.

\begin{equation}
	p(x)= \sum\limits_{r}|c_r|^2 \bra{\psi_{r}}E_x\ket{\psi_{r}}\mapsto \bra{\psi_{y}}E_x\ket{\psi_{y}}=p(x|y),
\end{equation} But as opposed to before this holds for the same pure state $\rho$ that is prepared for target system and environment in every run, whereas the correlations for the mixed state arise from the preparation of a mixture of pure states. 

By rewriting the conditional probability \ref{eqn:conditional probabilities entangled} it can be seen that the entangled correlations are stronger than those from a separable mixed state:

\begin{equation}
	p(x|y)= p(x|y)_{sep}+\frac{\sum_{r\neq l}c_r \bar{c}_l \bra{\psi_l}E_x\ket{\psi_r} \bra{\psi_l'}E'_{y}\ket{\psi_r'}}{\sum_{r} |c_r|^2 \bra{\psi_r'}E'_{y}\ket{\psi_r'}},
\end{equation} where $p(x|y)_{sep}$ is the conditional probability derived from the separable state $\rho=\sum_k |c_k|^2\ket{\psi_k}\bra{\psi_k}\otimes\ket{\psi_k'}\bra{\psi_k'}$. Note that when the POVM of the environment are projective operators on the Schmidt basis states, the second (entangled) part of the equation vanishes such that the familiar correlation of the separable state is recovered. But moreover, it is now possible that a deterministic prediction is possible for \textit{all} projective POVMs on the target system. For instance, as is well known for a maximally entangled state, if the measurement bases of target system and environment are identical, the outcomes will be perfectly correlated. That's a feature a separable mixed state cannot have. Another crucial difference is that the target system's pure state is now not pre-determined by the source but by the respective outcomes in the environment. 

A separable mixed state is, in general, empirically distinct from an entangled state in terms of the overall joint statistics. More importantly, however, the reduced state of the target system is identical in both cases. Therefore, entangled correlations are, strictly speaking, a predictive advantage over the correlations of separable mixed states. But similarly to the separable mixed state case, the predictions cannot do better than the one from a pure state, i.e. the one we get from conditionalising on each outcome in the environment. But in the entangled case, the conditionalised pure states are different from the pure states one gets in a mixture. One could argue that the predictive power depends on the quantum state assigned, and that perhaps the uncertainty about the outcomes of a pure state could be explained by the presence of subjective ignorance. That is, the system could be described by a mixed or entangled state for which stronger correlations to the environment do exist. 

Having said that, the quantum behaviour of a system described by a pure state \textit{cannot} in general be reproduced by a separable or entangled system-cum-environment state. This is relevant for the empirical completeness problem insofar as for a given target system, one could argue that its outcomes can be produced by either pure, mixed, or entangled joint states but where each state implies different correlations to the environment and, therefore, different predictive advantage. But since these cases are empirically distinct, the existence of predictive advantage for, e.g. pure states cannot be explained by restoring to alternative quantum state assignment. 

To summarise, quantum theory offers three different ways in which particular measurement statistics of the target system can arise. Either it is described by a pure quantum state yielding outcome probabilities $p(x)=\tr(E_x\ket{\psi}\bra{\psi})$ for arbitrary POVMs $E_x$. But then the environment is decorrelated to all outcomes since only then a pure state can be assigned as we saw above. Or, the target system is described by a mixed state $\rho_S=\sum_{k}\alpha_k\ket{\psi_k}\bra{\psi_k}$. Such a state can arise from a separable or entangled joint state with the environment. Concerning the state $\rho_S$, we saw that both separable and entangled states, in a sense, lead to predictive advantage, where the latter exhibit stronger correlations to the target system's outcomes. But both of these cases cannot reproduce the statistics coming from a pure state assignment for all observables. The existence of a pure state implies empirically distinct phenomena  (e.g. some interference effects etc.). Therefore, the crucial question for the present purposes is \textit{why}, to recover those phenomena and to assign a pure state, the environment has to be decorrelated from the target system. This indicates the old intuition only when `we don't look' a physical system exhibits non-classical behaviour.     

Based on the considerations above, we can qualify the statement that quantum mechanics is a probabilistic theory:  Pure state assignment to a target quantum system generically leads to indeterminate outcomes for all except one POVM measurement (when the POVMs are projectors on the relevant basis states of the density operator). On the other hand, when mixed states describe the target system, the predictions can often be refined and deterministic. In the case of a separable joint state of target system and environment, the statistics sometimes arises from an epistemic uncertainty for which classical correlations may exist with environmental configurations. In the case of an entangled state, the target system's state is mixed but cannot be interpreted as an epistemic ensemble of states since the joint state is still pure (improper mixture). And then again, the correlations to the environment can lead to predictive advantage over the predictions based on the reduced density operator of the target system alone. But since the density operator formalism, unitary evolution, and POVM measurements exhaust quantum prediction making, the pure states are the relevant states to consider for empirical completeness: The Hilbert space representation of a pure state is incompatible with a mixed state representation. This is because the former implies decorrelation to the environment, whereas the latter doesn't. But suppose now that we don't know what quantum state (pure or mixed) was prepared for the target system. Then the question arises whether quantum theory could allow every pure state to find a mixed state exhibiting prediction-improving correlations. But pure states are not only theoretically different from mixed states but also empirically: \textit{It is impossible to recover all the quantum probabilities predicted by a pure state with a mixed separable or entangled state. } 

\section[Ambiguities in Predictability]{Ambiguities in Predictability and the Measurement Problem}
\sectionmark{Ambiguities in Predictability}
\label{section:predictability and measurement problem}

\subsection{Internal vs. External Predictions}

Despite the standard quantum predictions permitting no pre-determination of all measurement outcomes, it would be at least desirable that the quantum description uniquely fixes a probability distribution for each observable. But even this much is not clear. Although for a given quantum state, the theory \textit{does} uniquely determine the outcome probabilities of arbitrary observables, in certain circumstances, ambiguities arise as to what quantum state was prepared and thus unclear what to substitute into the Born rule. Standard quantum mechanics, as introduced in Section \ref{section:quantum theory and quantum predictions}, is sufficient to make predictions for all practical purposes where states are assigned unequivocally, but isn't a framework filled-in enough for unique predictions in all conceivable situations.

In turn, the ambiguities for prediction-making in standard quantum theory may have implications for the empirical completeness problem. Accommodating predictive advantage may thus be intimately bound up with resolutions to the measurement problem. In this section, I shall clarify some options for disambiguating quantum predictions and discuss their implications for the theory's empirical completeness. The clarifying considerations on what I shall henceforth refer to as \textit{internal observation} or \textit{internal prediction} apply to quantum theory irrespectively of its interpretation. 

In the framework introduced, the problem takes the following form. Let's recall the minimal ingredients necessary to make quantum predictions and look at a simple example. Consider a set $S$ of manifest configurations representing the quantum target system. That is, a lab, black box, or some subset of the manifest domain whose configuration signifies the preparation of some (pure) quantum state $\ket{\psi}$. Consider an apparatus consisting of a set $M$ of manifest configurations associated with some apparatus states. Hence, the (non-manifest) quantum states of target system and apparatus are $\lambda:=\ket{\psi}\in\mathcal{H_S}=\mathbb{C}^2$, and $\mu_Q:=\ket{\phi}\in\mathcal{H_M}=\mathbb{C}^2$, respectively.\footnote{For simplicity, all systems are qubits, i.e. two-dimensional Hilbert spaces.} All configurations other than those belonging to $S$ and $M$ are considered to belong to the environment $E$. The joint state of target system, apparatus, and environment are states in the tensor product $\mathcal{H_S}\otimes\mathcal{H_M}\otimes\mathcal{H_E}$, with $\mu_E:=\ket{\varphi}\in\mathcal{H_E}=\mathbb{C}^2$ the states signified by manifest configurations in the environment. 

Target system and apparatus are assumed to interact via some Hamiltonian $H_{int}$ by virtue of which an observable $Q$ is defined (the measurement operators of the pointer will again be assumed to be projections). Suppose further that the apparatus configurations consist of two regions $A$ and $B$, such that whenever all its manifest configurations end up in one of those regions, this is taken as a record of the binary outcome of a measurement with observable $Q$. It is assumed that the apparatus configurations arrange like so: If the target system indicates a preparation of the pure state $\ket{\psi_A}$, the configurations settle in the region $A$ surely, and when state $\ket{\psi_B}$ was prepared, they settle in the region $B$ surely. By the postulate of the Born probability and the assumption that the apparatus results are associated with some projection operators $P_i$, the corresponding outcome probabilities read

\begin{align}
	p(A)&:=\bra{\psi_A}\otimes\bra{\phi}U^{\dagger}\mathds{1}\otimes P_A U \ket{\psi_A}\otimes\ket{\phi} \overset{!}{=}1, \\
	p(B)&:=\bra{\psi_B}\otimes\bra{\phi}U^{\dagger}\mathds{1}\otimes P_B U \ket{\psi_B}\otimes\ket{\phi} \overset{!}{=}1, \nonumber
\end{align} for some projectors $P_x=P_A, P_B$, and unitary operator $U$ (cf. Section \ref{section:quantum theory and quantum predictions}). This relationship can be established if the projection operators satisfy $P_x\ket{\phi_y}=\delta_{xy}\ket{\phi_y} $ with $x,y \in \{A,B\}$, and if for the unitary 

\begin{align}
	U: &\ket{\psi_A}\otimes\ket{\phi} \mapsto \ket{\psi_A}\otimes\ket{\phi_A} \\
	&\ket{\psi_B}\otimes\ket{\phi} \mapsto \ket{\psi_B}\otimes\ket{\phi_B}. 
\end{align} Thus, where the apparatus configuration ends up can be predicted with certainty depending on which of those two states was prepared. 

\begin{figure}[h]
	\centering
	\includegraphics[width=1\linewidth]{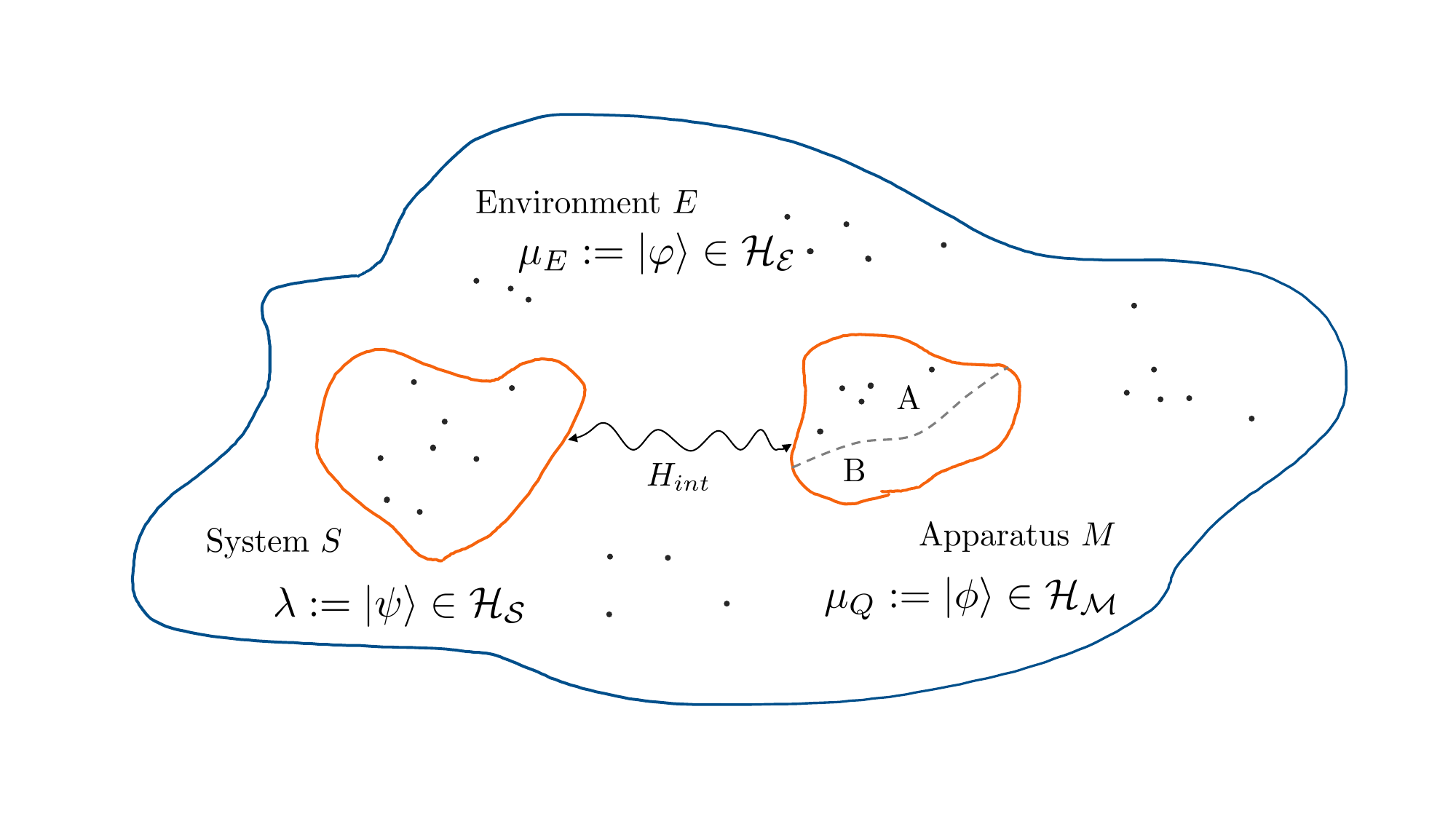}
	\caption{A subset of the manifest domain is split into sets of manifest configurations representing a quantum target system $S$, an apparatus $M$, and the environment $E$. The configurational degrees of freedom related to the target system signify the preparation of a non-manifest quantum state $\lambda:=\ket{\psi}\in\mathcal{H_S}=\mathbb{C}^2$, which interact according to the quantum postulates with the apparatus states $\mu_Q:=\ket{\phi}\in\mathcal{H_M}=\mathbb{C}^2$. The type of interaction and states defines the measured observable $Q$. After the interaction, the apparatus either shows the manifest result that all its configurations end up in region $A$ or $B$. A further interaction of the environment with states  $\mu_E:=\ket{\varphi}\in\mathcal{H_E}=\mathbb{C}^2$ on the joint system of target and apparatus yields manifest configurations in $E$. In terms of predictability, quantum theory justifies the assignment of apparently conflicting predictions: According to the definite post-measurement configuration $A$, $B$, quantum theory should allow a deterministic prediction if the measurement is repeated (if repeated measurements are to give the same results). But according to the manifest configurations possibly existent in the environment, both outcomes should occur with equal probability (if the overall state evolves unitarily).}
	\label{fig:internal-observation}
\end{figure}

But suppose that the initial target system was prepared in the pure state $\frac{1}{\sqrt{2}}(\ket{\psi_A}+\ket{\psi_B})\in\mathcal{H_S}$. The Schrödinger evolution with unitary $U$ implies 

\begin{equation}
	\label{eqn:superposition state ambiguities}
	\frac{1}{\sqrt{2}}(\ket{\psi_A}+\ket{\psi_B})\otimes\ket{\phi}\mapsto 	\frac{1}{\sqrt{2}}(\ket{\psi_A}\otimes\ket{\phi_A} +\ket{\psi_B}\otimes\ket{\phi_B}), 
\end{equation} and the Born probabilities in this case yield $p(A)=p(B)=\frac{1}{2}$.  After the interaction, the apparatus configuration will still show an outcome represented by configurations either in region $A$ or $B$, but the actual result wasn't determined by the quantum state or any of the manifest configurations. Furthermore, nothing else in the environment could have possibly been employed for a predictive advantage. The QDP dictates for pure system states that $p(x,y)=p(x)p(y)$, for $x$ the apparatus outcome configurations and $y$ the configurations of the environment. The environment cannot contain more detailed information about the measurement results (see Section \ref{section: quantum decorrelation}, and Figure \ref{fig:internal-observation}). 

Assume in the following the apparatus to show configuration $A$. The next step is to ask what would be expected to happen if the environment is subject to some subsequent interaction with the joint system of target system and apparatus; what can be predicted about the environment's future manifest configurations? 

This defines the internal vs external distinction. That is, relative to the environment, an \textit{external} prediction is made about the outcome probabilities of measurements on target system and apparatus. However, relative to the apparatus, an \textit{internal} prediction is made about the same measurements. The question is whether the internal prediction differs from the external one by having access to the definite configuration $A$ of the apparatus. 

In order to answer this question, we need to know what state describes the joint system after the interaction of target and apparatus.\footnote{Up to this point, the assumption was that the minimal account of quantum mechanics and Born probabilities apply. All preparable Hilbert states of the target system evolve unitarily, and the standard quantum predictions are supposed to be correct. And, of course, the vast empirical evidence supports those assumptions.} 

But this poses a problem. There exists no agreed-upon rule as to what the state of a quantum system is supposed to be after such measurement interaction. At the very least, the minimal uncontroversial parts of quantum theory I gave don't contain such a rule. The standard quantum framework needs to be completed when it comes to assigning post-measurement states.\footnote{In fact, the controversy about any measurement update rule or state `reduction' was the primary motivation to leave it out of the minimal account in Section \ref{section:quantum theory and quantum predictions}.} Moreover, it is unclear what role the definite manifest configuration $A$ is supposed to play in deciding what state to use to make future predictions.

Nevertheless, further supplementing probability rules are required to address this case of what I henceforth term \textit{internal observation}. Otherwise, the ambiguities in prediction-making remain present in quantum theory. But depending on any such additional law, the predictions of an internal observer may differ. What's more, the tension between unitary processes and `collapse', i.e. the quantum measurement problem, is only part of the issue. So a resolution to the ambiguities in quantum predictability needs to say a good deal more about \textit{what} the probability rules are that apply.        

Here is the problem. Depending on how an improved quantum theory pans out, internal observation can be used to justify the assignment of incompatible quantum states. If quantum state updating were considered a reasonable assumption, one manifest configuration (the outcome $A$ of the primary measurement) justifies the post-measurement state to be $\ket{\psi_A}\otimes\ket{\phi_A}$. On the other hand, the manifest configuration signifying the preparation of the superposition state together with the unitary dynamics would justify the assignment of state $\frac{1}{\sqrt{2}}(\ket{\psi_A}\otimes\ket{\phi_A} +\ket{\psi_B}\otimes\ket{\phi_B})$ to the joint system. Moreover, superposition measurements by the environment on the joint system could further be used to confirm the preparation of such a superposition state. 


This ambiguity is often framed as an inconsistency in the standard quantum framework that needs to be resolved in order to make sense of what the wave function could represent and whether that representation is physical and a complete description. If it is physical and complete, the problem arises of how the appearance of definite outcomes is to be explained for generic superposition states (cf. \citealt{Wigner-measurement-problem,Hermann1940-naturphilosophische-Uberlegungen,albert1995further,Brukner-measurement-problem}).\footnote{The traditional approach has it that a quantum system has a physical property (i.e. a definite manifest configuration) if and only if its state is an \textit{eigenstate} of the corresponding measurement operator. That is, the so-called eigenvector-eigenvalue link demands that the apparatus can indicate a definite configuration ($A$ or $B$) if the state is an eigenstate, i.e. either $\ket{\psi_A}\otimes\ket{\phi_A}$, or $\ket{\psi_B})\otimes\ket{\phi_B}$. That would imply that after the interaction, when the target system's initial state is as in Equation \ref{eqn:superposition state ambiguities}, the apparatus \textit{cannot} indicate a determinate manifest configuration. And the question is why this link is necessary, to begin with. The story often goes like this. If the post-interaction state is an eigenstate, the predictions for future results show the expected behaviour: If the apparatus indicates $A$ the joint state according to this rule must be $\ket{\psi_A}\otimes\ket{\phi_A}$ and a second measurement (e.g. an interaction with the environment on that state in the appropriate basis) will show the same result with certainty. Likewise, for configuration $B$. But, of course, this is akin to some state reduction rule and would prevent the assignment of superposition states on the joint system based on the unitary evolution.}

For the purposes of this thesis, I am not concerned with any attempts to solve the measurement problem. Nevertheless, what is important here is its relevance for prediction-making and, consequently the empirical completeness problem. Whatever is needed to disambiguate the actual predictions in the puzzling case described is something over and above the minimal quantum postulates introduced in Section \ref{section:quantum theory and quantum predictions}.  

The main issue here amounts to the following: If the statistics of future measurements results on the target system plus apparatus are to comply with the standard predictions on superposition states, what role (if any) can the manifest apparatus configuration play in improving that prediction? 


The analysis motivates the introduction of a \textit{relative} prediction and a distinction of the \textit{internal} from the \textit{external} point of view. I shall introduce these two notions and clarify their consequences by applying the previous analysis to Wigner's friend-type scenarios. 

\subsection{Predictive Advantage for Wigner's Friend?}

The ambiguities in quantum predictions often make an appearance in Wigner's friend-type Gedanken experiments and have led many authors to evince (more or less contentious) deficiencies and to derive no-go theorems (see, for example, \citealt{frauchiger2018quantum, Cavalcanti-local-friendliness-no-go, Healey-limits-of-objectivity, Brukner-persistent-reality-of-Wigner}). I shall develop the idea of internal predictions in the previous section in an attempt to disambiguate quantum predictions by applying it to the familiar Wigner's friend Gedanken experiment. There, Wigner and his friend serve as an instance of the external and internal observer, respectively. I will also outline and assess some of the assumptions the friend would have to make to arrive at a sufficient account of quantum predictions. These assumptions need to reach beyond a mere resolution of the measurement problem. 

It is then found that certain scenarios exist for which Wigner's friend, i.e. the internal observer, has (deterministic) predictive advantage over the probabilities that Wigner would assign according to standard quantum theory. On the face of it, this could be employed to construct empirical extensions since the quantum predictions differ and can be refined relative to different observers (i.e. the internal versus external perspective). Hence, the empirical completeness problem would be contingent on the observer making the prediction and how the modified quantum theory will accommodate internal predictions. But I shall also show that for this class of extended quantum theories, the hypothetical predictive advantage is undermined by a fundamental limit to single out empirically the `correct' empirical extension among a multitude of possibilities. In all other cases, the friend either has no predictive advantage due to probabilistic internal predictions or predictive advantage isn't present since a violation of unitary probabilities, similar to dynamical collapse theories, modifies quantum predictability.  

\begin{figure}[h]
	\centering
	\includegraphics[width=1\linewidth]{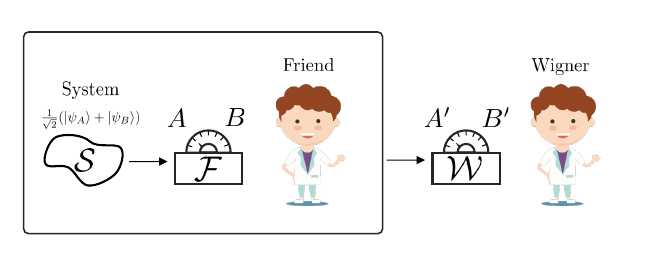}
	\caption{Wigner's friend measures a target system prepared in a superposition state $\frac{1}{\sqrt{2}}(\ket{\psi_A}+\ket{\psi_B})$. He obtains outcomes represented by the manifest configurations $A$ or $B$. Based on the initial state and this outcome, he makes predictions about Wigner's measurement performed in an arbitrary (orthogonal) basis  $\{\ket{\Psi_W}, \ket{\Phi_W}\}$, where $\ket{\Psi_W}=\alpha \ket{\psi_A}\otimes\ket{\phi_A}+\beta \ket{\psi_B}\otimes\ket{\phi_B}$. The friend's predictions for the outcomes of Wigner's external measurement may differ depending on how standard quantum theory is modified. In some cases, the former is uncertain of the result of his measurement, whereas his friend can predict what he will see with certainty. As the `internal' observer, Wigner's friend, therefore, has predictive advantage.}
	\label{fig:wigners-friend}
\end{figure}

To discuss internal prediction for Wigner's friend, suppose the apparatus $M$ represents an observer (the friend with a measurement apparatus) performing measurements on the qubit target system $S$ with manifest outcomes $A, B$ as introduced above. Suppose further that another observer is represented by the environment $E$ (Wigner) able to perform measurements in any desired basis on the compound quantum system of the friend with his apparatus and the target system. After the measurement of Wigner's friend, the friend finds the result either to be the manifest configuration $A$ or $B$. Given the initial state \ref{eqn:superposition state ambiguities}, both occur with probability $p(A)=p(B)=\frac{1}{2}$. Imagine that Wigner can then choose to perform further measurements on the joint system in any (orthogonal) basis $\{\ket{\Psi_W}, \ket{\Psi_W^{\perp}}\}$. One of the basis states is assumed to be of the form $\ket{\Psi_W}=\alpha \ket{\psi_A}\otimes\ket{\phi_A}+\beta \ket{\psi_B}\otimes\ket{\phi_B}.$  Let's further assume that Wigner's outcomes are represented by manifest configurations $A'$ if $\ket{\Psi_W}$ was prepared, and $B'$ if $\ket{\Phi_W}$ was prepared, as shown in Figure \ref{fig:wigners-friend}. Both Wigner and his friend can be asked to make a prediction --- based on the empirical data they have gained so far --- about the outcome probabilities of Wigner's measurement. As before, the puzzle on prediction making is that based on standard quantum theory alone, the friend is not in a position to make a unique prediction, and neither is Wigner. In particular, the friend may wonder whether his apparatus's manifest configuration is relevant to his predictions.  

At this point, the standard quantum formalism is, at best, agnostic and, at worst, ambiguous about the conclusions of the two observers. The friend has a manifest outcome (i.e. $A$ or $B$) at his disposal, and he could argue that further predictions ought to be based on some updated state (which, according to the familiar rules of state preparation, is assumed to be one of the corresponding basis states, see Section \ref{section:quantum theory and quantum predictions}). But Wigner, on the other hand, may assume that the joint system is described by state \ref{eqn:superposition state ambiguities} (as from his perspective, the joint system evolved unitarily). From Wigner's perspective, there shouldn't be a reason to believe that the minimal quantum theory cannot be used to predict his measurement. After all, according to the theory, qualitatively, there is no distinction between a single target system and the joint system of Wigner's friend together with the target system. Hence, Wigner would be in a position to apply the Born rule on the states of the joint system. Even worse, the friend also would have all reason to argue that Wigner's account is right since the friend knows the total state of the joint system, and Wigner could, in principle, communicate to the friend the superposed state of the joint system through a measurement (which is compatible with all assumptions of quantum theory). 

The familiar problem is that both accounts can only be right at one time as empirically relevant contradictions arise when applying the Born rule to both states since the statistics are simply different.\footnote{More strikingly, this can also be seen in the one-shot gedankenexperiment by \citealt{frauchiger2018quantum}. But note that the latter can be argued as amounting to a subjective collapse model. There the state update is subjective and consequently leads to the contradiction in the theorem. Wigner bases his prediction on the total entangled state, whereas the friend uses an updated state based on his manifest result (see, e.g. \citealt{lazarovici2019quantumconsistency}).} As a result, the questions to be answered are: What \textit{is} the \textit{state} or variables based on which a prediction can be made after the initial interaction? And secondly: What \textit{are} the \textit{predictions} following that state if future measurements are performed? The first question arises due to the measurement problem, whereas the second is related to predictability. A modified quantum theory must respond to both questions. Thus, additional assumptions need to be invoked.
Most importantly, a claim has to be made about the friend's manifest result's relevance to Wigner's future measurement result. Should the internal observer's manifest result have predictive significance for the external observer's results over and above the total wave function, the quantum decorrelation principle needs qualification. 

Usually, quantum prediction making doesn't present a problem since, for a particular observer, quantum theory assigns unambiguous probability distributions for the outcomes of arbitrary observables. But in the case at hand, the nesting of observers doesn't allow the familiar split between target system and environment. Thus the situation introduces a kind of `self-measurement' previously not conceived of when applying the standard quantum formalism. 

Several options have been explored for how to modify the standard framework, which (often only implicitly) include some account of internal predictions. It would be helpful to understand possible answers from a more general point of view, which I attempt to provide here. However, it isn't the purpose of this section to discuss the relevance of various no-go theorems. Instead, since I'm concerned with quantum predictability, it is to discuss the different modifications' implications on the empirical completeness problem. Some attempts exist in the literature to explicate what further assumptions would involve, often presented alongside no-go theorems or incompatibility arguments. 

One natural proposal to (partly) accommodate the apparent ambiguity is an assumption on the `intersubjectivity'  between different observers (see, for example, \citealt{Adlam-intersubjectivity}). That is, if Wigner's measurement happens to be in the friend's basis, i.e. $\alpha=1, \beta=0$ such that $\ket{\Psi_W}= \ket{\psi_A}\otimes\ket{\phi_A}$, then the measurement results are assumed to be compatible: Wigner obtains $A'$ if the friend's result was $A$ and $B'$ if the result was $B$. In other words, Wigner essentially `checks' the friend's reading, and it seems plausible that Wigner's result equals what the friend observed. Occasionally this assumption is called `shared facts' \citep{Adlam-intersubjectivity}. Moreover, it also follows from the so-called `absoluteness of observed events' (cf. \citealt{Cavalcanti-local-friendliness-no-go}), and `observer-independent facts' (cf. \citealt{Brukner-persistent-reality-of-Wigner}).\footnote{Note that this kind of intersubjectivity is distinct from the `repeatability' of quantum measurements. If the same measurement is performed twice, the results are in agreement. Moreover, it also implies that if Wigner performed the same measurement as the friend \textit{directly} on the target system, their values would come out identical.  
This consistency is a feature of the empirical content of the traditional theory and is experimentally well-confirmed.}  

One salient feature, however, is that since all these assumptions apply for a single choice of measurement only (i.e. `checking the friend's reading'), they prove insufficient to describe internal observation fully. Wigner can perform measurements in an arbitrary basis, and so the same question arises again for all other measurement choices: How should the friend's predictions be based on his outcome? A good deal more needs to be said in order to account for internal predictions in their generality. I shall remark on this in further detail below.

Of course, there exists the familiar alternative that blocks the problems of internal predictions right from the beginning: an appeal to a state reduction upon measurement of the friend. The intersubjectivity then is implied by the fact that the friend's outcomes determine the post-measurement state by reducing it to either $\ket{\psi_A}\otimes\ket{\phi_A}$ when $A$ obtains, or $\ket{\psi_B}\otimes\ket{\phi_B}$ when $B$ obtains. If a state-update rule is to be invoked a la `collapse' of some sort --- in whatever regime such a collapse might occur --- the empirical export of the modified quantum theory becomes falsifiable. For there will be situations in which Wigner's predictions deviate from the predictions that standard quantum theory gives based on the total entangled state of the target system and the friend. Thus, in the regime where the quantum state needs updating when target system and the friend interact, the standard formalism becomes provably wrong (cf. also generalised Born rules introduced by \citealt{Baumann2021generalized}).

In the context of predictive advantage, it thus proves useful to distinguish two cases: Either the modified quantum theory allows Wigner to apply the standard Born algorithm for the state he obtains during the unitary interaction. Or, upon measurement of the friend or any other mechanism, the modified theory requires Wigner to use an updated state to make predictions.\footnote{Note that the updated state, in general, needn't be (as is standardly the case in dynamic collapse models) one of the basis states. So there might exist intermediate resolutions in which further predictions are based on partially `collapsed' quantum states or a different state altogether.} An appeal to state reduction would ultimately be empirically falsifiable since quantum probabilities are incorrect in some regime. At any rate, the friend would have no predictive advantage if state reduction occurred. In this case, the internal observer never obtains a manifest measurement outcome \textit{while} some external observer would assign superposition states to this internal variable. Thus the quantum decorrelation principle trivially applies since all observers agree on the quantum state describing the system. 

I shall focus on the latter (more interesting) case where the total state remains entangled. This then opens up the possibility of a modification while recovering the quantum probabilities. Wigner's use of the Born rule upon measurements on the total state is thus warranted. It will also be clear that this hypothetical situation introduces a case where predictions have to be made for simultaneous measurement of `incompatible'  (i.e. non-commuting) quantum observables. Contingent on the state and prediction is to be made in a completed quantum theory, the post-measurement states of Wigner's friend sometimes even allow deterministic predictions for these observables.

\begin{figure}[h]
	\centering
	\includegraphics[width=\linewidth]{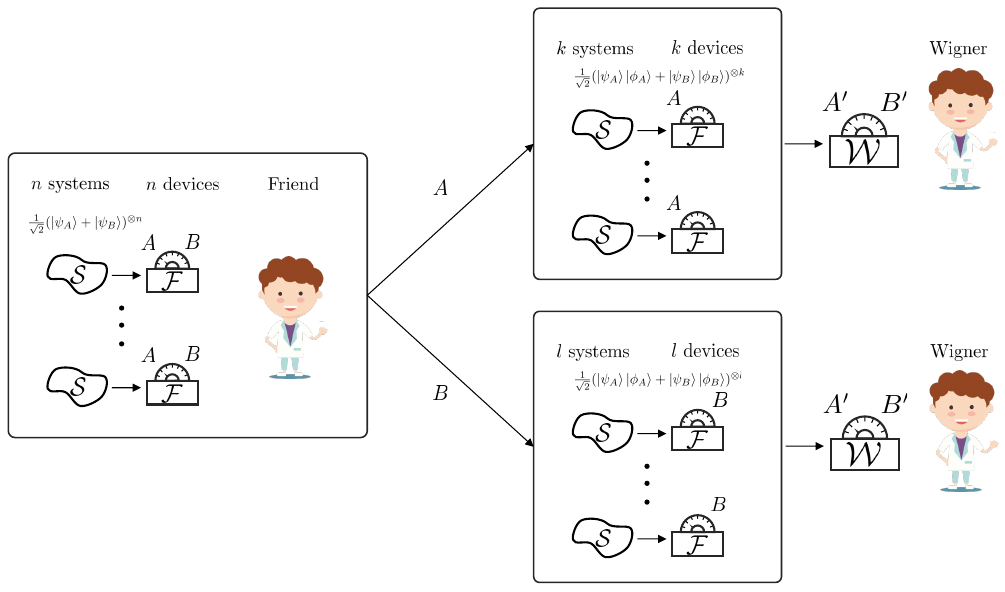}
	\caption{The Wigner's friend type experiment is repeated to collect statistics on a hypothetical predictive advantage. Each incoming spin particle is coupled to a pointer device (suppose the friend has $n$ of them) to obtain a manifest outcome $A$ or $B$. The standard Wigner's friend scenario from before is thus repeated with $n$ target systems. Based on the friend's outcomes, however, he then sorts the ensemble of joint entangled systems, i.e. spin particle plus pointer device, into the ones that yielded configuration $A$ (of which, say, $k$ like that exist) and the ones that yielded $B$ (of which $l=n-k$ exist). Both sub-ensembles are then passed on to Wigner, who performs measurements in the familiar basis  $\{\ket{\Psi_W}, \ket{\Psi_W^{\perp}}\}$ with $\ket{\Psi_W}=\ket{\psi_A}\otimes\ket{\phi_A}$. The corresponding manifest results are $A'$ and $B'$, respectively. Are the friend's predictions different from Wigner's based on the manifest configuration of each measurement device?}
	\label{fig:wigners-friend-predictive-advantage}
\end{figure}

Let's see how the friend could try to select sub-ensembles to gain predictive advantage, and then see what certain assumptions would imply about his abilities. For this, consider the friend to measure and decompose a large number of $n$ identically prepared spin states according to Definition \ref{def:empirical extension} (see Figure \ref{fig:empirical-decomposition} and \ref{fig:wigners-friend-predictive-advantage}). More concretely, each incoming spin particle is coupled to a pointer device (suppose the friend has $n$ of them) to obtain a manifest outcome, i.e. $A$ or $B$. The familiar scenario from before is thus repeated with $n$ target systems. Based on the friend's outcomes, however, he then sorts the ensemble of joint entangled systems, i.e. spin particle plus pointer device, into the ones that yielded configuration $A$ (of which, say, $k$ exist) and the ones that yielded $B$ (of which $l=n-k$ exist).\footnote{Obviously, $k$ and $l$ approach $\frac{n}{2}$ in the limit of large $n$.} Both sub-ensembles are then passed on to Wigner, who performs measurements in a `record checking' basis $\{\ket{\Psi_W}, \ket{\Psi_W^{\perp}}\}$ with $\ket{\Psi_W}=\ket{\psi_A}\otimes\ket{\phi_A}$,  and  $\ket{\Psi_W^{\perp}}$ some orthogonal basis state.

The crucial question is whether each pointer's manifest configurations ($A$ or $B$) have relevance to improving Wigner's predictions over and above Wigner's use of the Born rule. 


Employing intersubjectivity, it can be argued that Wigner's results ought to be consistent with the friend's records, that is, identical. The friend would then be able to predict with \textit{certainty} Wigner's outcome! Thus, providing a deterministic predictive advantage over Wigner's probabilities obtained through the Born rule. Modifying quantum theory by including intersubjectivity leads to a (deterministic) empirical extension of standard quantum theory for at least certain measurements. 

Interestingly, on intersubjectivity, the internal prediction assigns determinate outcomes on two non-commuting quantum observables: The friend knows both the result of the observable when Wigner `checks the reading' and the result of the observable when Wigner checks the total entangled state. But those two measurements are represented by incompatible bases and can't usually be measured simultaneously. 

But if Wigner is supposed to recover the quantum probabilities upon measurement on the entangled joint state, intersubjectivity gets us into trouble with locality. From the perspective of Wigner, assuming the friend observes definite manifest configurations is tantamount to a `hidden variable' theory and thus subject to Bell non-locality. In an extended version of the internal observer case involving two distant copies of Wigner's friend scenarios, the manifest outcome configurations of the friends have to depend on Wigner's measurement settings. Barring the familiar caveats, this was explicitly shown in \citep{Cavalcanti-local-friendliness-no-go}.\footnote{Problems also arise when combining Wigner's friend with the GHZ scenario (cf., for instance, \citealt{WignerGHZ}).} Let's dwell on this more in the present case. If intersubjectivity were to apply, Wigner should observe outcomes identical to what the friend observed. Imagine a different experiment now where Wigner sets up a Bell experiment on the joint states of target spin systems and pointer devices that the friend sorted. He separates each spin particle from the device and sends each pair through Stern-Gerlach apparatuses. That is, he performs measurements on the target systems and devices separately. As usual, to violate Bell inequalities, non-locality will be present, in particular for the manifest configurations of the pointer. 

But suppose Wigner only uses the pairs of the $A$-ensemble. Then the quantum correlations imply that for this set of states, signal locality can be violated. The intuitive argument is this: Since the spins in the ensemble have all the same definite value, they act as a deterministic hidden variable relative to Wigner. Only the $A$-ensemble is used in the experiment, so intersubjectivity implies that all results show `spin up' for both systems corresponding to the state  $\ket{\psi_A}\otimes\ket{\phi_A}$. But if Wigner performs a Bell experiment, at least some predictions conditioned on this hidden variable will involve non-local effects between the outcomes of target system and pointer device (due to Bell's theorem). Thus, how Wigner sets the apparatuses determines the outcome statistics on the other site. This would conflict with signal-locality because we are talking about manifest configurations that are affected. Therefore, intersubjectivity must fail. In the case of de Broglie-Bohm theory this can be seen immediately by using two measurement settings only and conditioning on the Bohmian particle configuration. A detailed account is given in Section \ref{Section:maximally entangled states dBBT advantage}. 



However, although intersubjectivity may have to be abandoned for deterministic predictions, the friend could still gain a probabilistic predictive advantage. That is, for each basis, the friend assigns a probability to Wigner's outcomes $A'$ and $B'$ as a function of his manifest result $A$ and $B$. More formally, let $\ket{\Psi}=\frac{1}{\sqrt{2}}(\ket{\psi_A}\otimes\ket{\phi_A}+\ket{\psi_B}\otimes\ket{\phi_B})$ be the total state of the joint system after the initial interaction, and $\{\ket{\Psi_W},\ket{\Psi_W^{\perp}}\}$ Wigner's basis with $\ket{\Psi_W}=\alpha\ket{\psi_A}\otimes\ket{\phi_A}+\beta\ket{\psi_B}\otimes\ket{\phi_B}$ as before. Then the friend's predictions provide an empirical extension for Wigner's probabilities if
\begin{align}
	p(A')&= |\bra{\Psi_W}\ket{\Psi}|^2= \sum_{z=A, B} p(A'|z)\mu(z)= \frac{1}{2}(p(A'|A)+p(A'|B)), \\ \nonumber
	p(B')&= \left|\bra{\Psi_W^{\perp}}\ket{\Psi}\right|^2= \sum_{z=A, B} p(B'|z)\mu(z)= \frac{1}{2}(p(B'|A)+p(B'|B)).
\end{align} The extension is non-trivial if the friend's probabilities deviate from the quantum prediction for at least one configuration.\footnote{Note that $A$ and $B$ occur with equal probability by construction.}  

What can be said about the probabilities in this extension? An analogous argument can be run by looking at the predictions' compatibility with signal-locality. This time it's more complicated, however. It has to be checked whether any probability assignment satisfies signal-locality by examining all possible measurement scenarios Wigner can come up with. And perhaps there exist probability distributions for which the non-locality washes out just to the right extent to give signal-locality. The friend then acquires non-deterministic predictive advantage. This is indeed so. I treat in more detail this possibility in Chapter \ref{section:Non-locality and quantum predictability}.   Jumping ahead, the conclusion is this. For maximally entangled states like the joint state in the Wigner's friend scenario, a predictive advantage by the friend is impossible given signal-locality. However, predictive advantage can exist without violating signal-locality for a class of non-maximally entangled states of target system and friend.

Where does this leave us regarding the empirical completeness problem? The question is whether the friend's abilities present a genuine construction of an empirical extension. Extended quantum theories exist which  --- in certain special cases, i.e. nested measurements with sufficient macroscopic coherence --- have predictive advantage over the standard theory on the assumption of intersubjectivity. Although, those conflict with signal-locality. Nevertheless, non-deterministic predictive advantage is possible if the joint state of the target system and the friend's pointer device are not maximally entangled. Then the question arises whether there are other means to show that non-maximally entangled states still lead to a violation of signal-locality. I explore these questions on either side in Chapter \ref{section:Non-locality and quantum predictability}.

\chapter{Pilot Wave Theory and Quantum Predictive Advantage}
\chaptermark{Pilot Wave Theory and Predictability}
\label{section:de Broglie-Bohm theory and predictive advantage} 
 
Here I analyse this question of empirical completeness in the case of de Broglie and Bohm's pilot wave theory. Since the Bohmian formalism inherits enough of the quantum formalism, the theory can be shown not to have a predictive advantage. I discuss the precise conditions in the theory that lead to predictive equivalence. Furthermore, I discuss the possibility of predictive advantage when these are violated: the existence of non-equilibrium Bohmian matter, the possibility of `internal prediction' utilising the apparent ambiguity in the quantum prediction-making algorithm, and the case when no `effective' Bohmian wave function exists. I study the relationship between Bohmian predictive advantage and the possibility of signal in the case of pairs of particles. Finally, I comment on the proofs by \citeauthor{Valentini-H-theorem-II} that Bohmian non-equilibrium distributions are necessary and sufficient for signalling in pilot wave theory. For non-maximally entangled states signal-local predictive advantage does indeed exist. 

\section{Pilot Wave Theory}

Whatever one's views regarding the status of de Broglie-Bohm theory as a viable solution to the measurement problem, its virtues in assessing the empirical completeness problem are immediate. At the very least it serves as a concretely worked out instance of a metaphysical and empirical extension, reproducing quantum statistics within the equilibrium regime. In the context of non-equilibrium, Bohmian commitments prove useful to grasp what sort of thing a post-quantum theory with predictive advantage could look like. At its best, the theory serves as an inspiration to identify generic features of post-quantum theories for general claims about empirical completeness. Moreover, pilot wave theory provides a model of quantum mechanics with lessons for predictive advantage from non-locality and signal-locality. 

One could argue that in contrast to standard quantum theory, de Broglie-Bohm theory draws a clearer picture of reality since it provides a framework for both the manifest and non-manifest domain. The theory tells a story about how the manifest configurations come about epistemically but also postulates what precisely comprises the ontology in the non-manifest domain and how it connects to the manifest configurations through dynamical laws. Similar to the standard quantum mechanical story outlined in Section \ref{section:quantum theory and quantum predictions} the target system and apparatus states $\lambda\times\mu_Q$ in the Bohmian theory contain normalised vectors ${\psi}\otimes{\phi}\in \mathcal{H}_S\otimes\mathcal{H}_A$ in complex Hilbert spaces with all the same structure as introduced before, and $Q$ indexing the measured observable and measurement interaction.\footnote{Technically, this assumes that before the measurement interaction a (pure) quantum state can be assigned to the target system. In the Bohmian framework, this means there exists an effective wave function for the system. If such doesn't exist, the universal Bohmian wave function needs to be invoked (for details, see below). Moreover, to be more precise, the wave functions used in pilot wave theory are square integrable functions over $3N$-dimensional configuration space, i.e.  $\psi \in L^2(\mathbb{R}^{3N})$ for an $N$-particle system. Therefore, strictly speaking, the `wave functions' in a Bohmian universe are not Hilbert space vectors (which are equivalence classes rather than wave functions). This indeed introduces subtle qualifications to the empirical equivalence of Bohmian predictions and standard quantum predictions, cf. \citealt{arageorgis2017bohmian}.}  Besides the wave functions, the Bohmian states comprise positions, which at the same time are the fundamental ontology of the theory, i.e. corpuscles with definite configurations at all times. 

A generic initial state for target system and apparatus is thus $\lambda\times\mu_Q= (\psi,q)\times (\phi,q')$. Just like in standard quantum mechanics, the states of compound systems are tensor product states of the individual systems and evolve according to the Schrödinger equation. The position variables, however, evolve according to a non-linear guiding equation. 
In de Broglie-Bohm theory, the probability assignments for the outcomes $x$ of a measurement are deterministic functions of the initial state such that

\begin{equation}
	p(x|\lambda\times\mu_Q)=p(x|(\psi,q)\times (\phi,q'))=p(x|(\psi\otimes\phi)\times(q,q'))\in\{0,1\}.  
\end{equation}

Consistency with standard quantum mechanics requires that in the limit of very large ensembles of states, the \textit{standard} Bohmian probabilities match the standard quantum prediction, i.e.

\begin{equation}
	p(x|\psi)=\int p(x|(\psi,q)\times (\phi,q'))\rho(q,q')dqdq',
\end{equation} for all effective initial apparatus states $\phi$, where $\rho(q,q')$ is the probability distribution over the initial particle configurations $q,q'$ for a given preparation and observable $Q$.  An account of the theory's postulates is given in Appendix \ref{appendix:de Broglie-Bohm theory}, to which I refer for all details.

Across time, the particles follow deterministic trajectories. Like a `pilot wave', the quantum mechanical wave function guides them along those paths. Assuming a particular initial distribution of the particles, one recovers the empirical content of standard quantum theory. 

\section{Bohmian Predictions}
\label{Section:Bohmian predictions}

In de Broglie-Bohm theory, the set of particle configurations in 3-dimensional space constitutes the ontological sector of the non-manifest domain and the Schrödinger and guiding equation the nomological domain. Hence, postulating the existence of definite particle positions in the Bohmian universe is a natural choice to reflect the nature of the empirical content in the manifest domain. Since all the manifest empirical data is encoded in configurational facts in 3-dimensional space, a formulation of the non-manifest domain in terms of particles naturally accounts for our phenomenological experience. Therefore, a pointer device in the Bohmian universe is made out of a subset of Bohmian particle configurations. Furthermore, and crucially the assumption is made that whenever a pointer device records an outcome, i.e., an orientation or position in the manifest domain, it is postulated to signify the very ontological configuration of the pointer itself. In other words, all the spatial degrees of freedom of observable objects like ink on paper, dots on screens, and positions of pointers do precisely coincide with --- usually a collection of --- Bohmian configurations in the non-manifest domain. 

Because of this, since the standard Bohmian theory postulates the quantum equilibrium hypothesis \ref{eqn:QEH} for the distribution of particle configurations, it straightforwardly follows that all manifest configurations exhibit distributions according to the Born statistics. Moreover, the choice of the guiding equation \ref{eqn:guiding equation} consistently ensures that Bohmian configurations satisfy Born rule behaviour at all times, i.e. the Born distribution \ref{eqn:Born rule} is preserved. Notice again that the continuity equation empirically underdetermines the guiding equation (cf. Chapter \ref{section:Weak measurements}).

As a result, whenever a target system possesses a pure quantum wave function, all configurations (non-manifest as well as manifest) must be distributed according to quantum equilibrium. This means that for a given system, the possible states of system-cum-environment may imply restrictions to how strongly the Bohmian configuration of the environment is correlated to the measurement results of an apparatus.  

The concept of the effective wave function is directly linked to Bohmian predictability. Whenever target system and environment obey \ref{eqn:effective wave function}, not only the target system can be said to possess quantum state $\psi$, but also measurement results on the target system must be statistically independent of the configurations in the environment. This holds since all outcomes will be determined by the product state $\psi(x)\phi(y)$ only, which predicts no correlation between configurations $x$ and $y$. Notice that whenever an effective wave function exists, it coincides with the standard quantum state that we would assign.\footnote{Note, however, that after a preparation, the system has an effective wave function, but which effective wave function it has depends on the manifest configuration of the apparatus that has prepared the system. Whenever a target system possesses an effective wave function, it can be treated as a pure state.} This may also be argued to be a plausible posit as an assumption on a necessary initial condition for all practical purposes we face.  
Since all correlations between a target system and its environment are encoded in the quantum wave function states in both the Bohmian universe and the standard quantum theory, the question of the predictive advantage of de Broglie-Bohm theory translates into an investigation of the possible correlations in standard quantum theory.   

As a result, if we can show that whenever a (pure) quantum state is assigned, the environment is decorrelated from the target system, this implies that the Bohmian theory cannot have predictive advantage over standard quantum theory (at least for pure states). Thus, it remains to show that in standard quantum mechanics, pre-measurement correlations between target system and environment imply that the target system can't possess a pure state, which is what was shown in Section \ref{section: quantum decorrelation} with the quantum decorrelation principle.  

Let's summarise. The Bohmian description inherits all of the quantum state structure. The outcome of a measurement in the theory is deterministically determined by the target system's initial configuration and its wave function through both the guiding equation and Schrödinger's equation. In this sense, no Born rule is involved in the theory's microscopic predictions. But if on top of the particle dynamics, quantum equilibrium is assumed, the Bohmian statistics follow Born rule behaviour, and the quantum decorrelation principle applies. For this, the choice of the guidance equation needs to preserve the Born probabilities and the non-manifest Bohmian configurations are assumed to coincide with the manifest configurations, i.e. the apparatus outcomes signify actual Bohmian configurations. The quantum decorrelation principle for de Broglie-Bohm theory thus follows from the conjunction of the following assumptions: That an effective Bohmian wave function, i.e. pure state exists for the target system ($Psi$), the assumption of quantum equilibrium ($QEH$), the dynamics is given by a guidance equation consistent with quantum probabilities ($GEQ$), and that measurement results, i.e. manifest configurations signify Bohmian (non-manifest) configurations ($Re$).  

In short,

\begin{equation}
	\label{eqn:QDP for de Broglie-Bohm theory}
	Psi \wedge  QEH \wedge GEQ \wedge Re \rightarrow QDP
\end{equation}

This set of conditions leads to the conclusion that de Broglie-Bohm theory does not have a predictive advantage over quantum probabilities. It can then be used to discuss the possibility of empirical incompleteness depending on which assumptions are violated. Predictive advantage due to violation of equilibrium is investigated in Sections \ref{section:Non-equilibrium predictive advantage with signalling}. In Section \ref{section:internal observation} I discuss the Bohmian prediction making for cases where the quantum measurement problem is relevant.


\section[Pilot Wave Theory as a Merely Metaphysical Extension of Standard Quantum Mechanics]{Pilot Wave Theory as a Merely Metaphysical Extension of Standard Quantum Mechanics} \sectionmark{Pilot Wave Theory as a Metaphysical Extension}
\label{section:pilot wave as metaphysical extension}

From the definitions of de Broglie-Bohm theory, we can readily see that the Bohmian states introduce a metaphysical extension of the standard quantum framework. Mathematically, we have

\begin{equation}
	\bra{\psi}E_x\ket{\psi}= p(x|\psi\otimes\phi)=\int p(x|(\psi,q)\times(\phi,q'))\mu(q,q')dqdq',
\end{equation} where $E_x$ the POVM operators for measurement outcomes $x$ and $\psi, \phi, q, q'$ the familiar quantum and Bohmian states, and $\mu(q,q')$ an initial distribution of particle configurations consistent with quantum equilibrium. Thus, the Bohmian theory is a metaphysical extension of quantum theory by Definition \ref{def:metaphysical extension}. Since an effective wave function $\psi$ for the target system is assumed to exist, the Bohmian probabilities for position outcomes $x$ are independent of the apparatus state $(\phi,q')$.

Furthermore, the ideas in Section \ref{Section:Bohmian predictions} show that the predictions of standard quantum mechanics limit the theory's prediction-making algorithm. Even though the Bohmian physical state contains more information about future measurement results than the standard quantum state, this information cannot be manifest and is thus of no use for increased predictability. I conclude that standard equilibrium de Broglie-Bohm theory is a \textit{merely} metaphysical extension of quantum mechanics and thus cannot have predictive advantage. In the Bohmian universe, all manifest configurations signify non-manifest (sets of) particle configurations. Moreover, all manifest configurations are distributed according to the quantum equilibrium hypothesis, which is preserved by the Bohmian dynamics. Thus, de Broglie-Bohm theory says that the predictions that can be made on the basis of manifest configurations are just the standard quantum predictions. All the possible manifest configurations in a Bohmian universe in equilibrium cannot be used to produce predictive advantage over quantum predictions. More formally, in terms of our definition of empirical completeness: de Broglie-Bohm theory cannot be an empirical extension of standard quantum mechanics since, for all outcome probabilities derived from the prediction-making algorithms of the two theories, we have

\begin{equation}
	p_{Q}(x|y_1)=p_{B}(x|y_2, y_1), 
\end{equation} and the extension condition trivialises to 

\begin{equation}
	p_{Q}(x|y_1)=\int p_{B}(x|y_2)\rho(y_2|y_1)dy_2= p_{B}(x|y_2)\int \rho(y_2|y_1)dy_2= p_{B}(x|y_2, y_1),
\end{equation} for all POVMs and outcomes $x$, where $y_1$ are all the manifest variables representing the target system state according to standard quantum mechanics, and $y_2$ all the manifest variables representing the same state according to de Broglie-Bohm theory. Hence, standard de Broglie-Bohm theory is not an empirical extension of standard quantum mechanics. 

For pure states, we have in particular 

\begin{align}
	p_{B}(x|y_2) &=\frac{1}{p(y_2)}\tr(E_x \ket{\Psi(x,y_2)}\bra{\Psi(x,y_2)})\\ \nonumber
	&=\frac{1}{p(y_2)}\tr(E_x \ket{\psi(x)\Phi(y_2)}\bra{\psi(x)\Phi(y_2)})\\ \nonumber 
	&=\tr(E_x \ket{\psi(x)}\bra{\psi(x)})=p_{Q}(x),
\end{align} where $\Psi(x,y_2)$ is the Bohmian conditional wave function given apparatus configuration $y_2$ and the third equality holds because the effective wave function $\psi$ is assumed to exist. We have also used the definition of conditional probabilities, i.e. $p(x|y_2)=\frac{p(x,y_2)}{p(y_2)}$.\footnote{The point of predictive equivalence between standard quantum mechanics and de Broglie-Bohm theory is commonly made in standard textbooks. For example, Dürr, Goldstein, and Zangh\'i refer to the impossibility of predictive advantage for pure states in de Broglie-Bohm theory as \textit{absolute uncertainty} where it is expressed as 	
\begin{equation}
		\mathbb{P}(X_t\in dx|Y_t)=\mathbb{P}(X_t\in dx|\psi_t)=|\psi_t(x)|^2,
\end{equation}(cf. \citealt{durr1992quantum}).}

In the following sections, I will discuss the implications for predictive advantage when either of the assumptions is violated. More concretely, I address non-equilibrium (violation of $QEH$), and the absence of an effective wave function (violation of $Psi$) which both give rise to predictive advantage through circumvention of the quantum decorrelation principle. A predictive advantage could also be gained by altering the guiding equation ($GEQ$) or abandoning correspondence of manifest and Bohmian configurations ($Re$). I shall comment on this briefly in the conclusions.

\section{Quantum Non-Equilibrium and Bohmian Predictive Advantage}
\label{non-equilibrium}

One of the crucial prerequisites for the impossibility of predictive advantage in de Broglie-Bohm theory was the assumption of quantum equilibrium. However, predictive advantage could be feasible without an equilibrium distribution in the initial configurations. In fact, I shall give two simple Bohmian models in which violation of quantum equilibrium leads to empirical extensions and, therefore, predictive advantage. 

Since outcomes of experiments are derived from the deterministic history of particles, the initial conditions qua equilibrium hypothesis play an important role in deriving the theory's predictions. 

I stated that quantum equilibrium is preserved which follows from the postulates of de Broglie-Bohm theory. In a Bohmian universe, if the configurations are in equilibrium at some point in time, they are in equilibrium at all times (cf. also \cite{durr2012quantum} for proof). This is a direct consequence of the continuity equation for the Bohmian probabilities obtained from the Schrödinger equation. Since the Bohmian probability density is in quantum equilibrium, i.e. $\rho(\bm{Q},t):=|\Psi(\bm{Q},t)|^2$, and the probability current due to the guidance equation is $\bm{j}=|\Psi(\bm{Q},t)|^2\bm{v}^{\Psi}$, the Schrödinger evolution for the wave function implies 

\begin{equation}
	\label{eqn:Bohmian continuity equation}
	\partial_t\rho+\nabla\bm{j}=0. 
\end{equation} 

Hence, the postulated Bohmian probability current $\bm{j}$ results in a probability density that preserves its form $\rho(\bm{Q},t):=|\Psi(\bm{Q},t)|^2$ at all times. 

As a result, no manipulation, preparation, or measurement in the equilibrium Bohmian universe will ever cause distributions in non-equilibrium. Thus, non-equilibrium matter cannot be created from equilibrium matter.\footnote{There is one caveat here, however. As I indicated before, whenever de Broglie-Bohm theory assigns an effective wave function to an individual system, it coincides with the standard quantum wave function. Moreover, it then satisfies the Schrödinger equation. But what if the system doesn't possess an effective wave function, i.e. when its conditional wave function instead doesn't satisfy the Schrödinger equation? Then, the assumptions mentioned above leading to preserved equilibrium don't apply.}

Therefore, the Bohmian states need to be out of equilibrium for predictive advantage. For these purposes, I shall look at the example of non-equilibrium spin measurements in a Stern-Gerlach apparatus. I will first treat the simple case of a \textit{maximally entangled} two-particle system with initial Bohmian configurations out of equilibrium. This will allow for deterministic empirical extensions, as one would expect. But the correlations established by the prediction improving manifest variables will have to exhibit signalling. That is, there is not only predictive advantage for the outcomes of the two-particle system, but it is also possible that one observer can communicate superluminally signals to the other. This well known feature is expected since pilot wave theory is deterministic and non-local. 

Interestingly, however, I will also describe an alternative deterministic hidden variable description of the two-particle system establishing predictive advantage \textit{without} the presence of signalling correlations. Thus, a deterministic signal-local theory exists with predictive advantage over standard quantum mechanics. Furthermore, such an empirical extension can be constructed for \textit{non-maximally entangled} states. This conforms with the results studied in Chapter \ref{section:Non-locality and quantum predictability}, where generalised Bell inequalities are used to constrain the extent to which a theory can be local, showing that non-maximally entangled states can exhibit local parts.

\section{Bohmian Non-Equilibrium Predictive Advantage}
\label{section:Non-equilibrium predictive advantage with signalling}

I shall now outline a concrete example for predictive advantage due to non-equilibrium for spin-$\frac{1}{2}$ particles.\footnote{For a thorough treatment of spin in pilot wave theory, see, for instance, \citealt{Norsen-spin} whom I closely follow here.} Imagine a Stern-Gerlach apparatus where neutral spin-$\frac{1}{2}$ particles with mass $m$ are emitted in a localised beam of finite width. The description is restricted to two spatial dimensions, the direction $\hat{y}$ in which the beam is emitted and the transverse direction of deflection $\hat{z}$. The particles are prepared in the initial state

\begin{equation}
	\label{eqn:initial bohmian spin state}
	\ket{\psi(y,z)}= \frac{1}{\sqrt{2}}e^{ik_y y}(\ket{0}+\ket{1}), 
\end{equation} where $k_y$ is the wave vector pointing in the $y$-direction, i.e. the beam. For simplicity, we assume the wave function of the spatial degrees of freedom to be a plane wave inside the finite beam of the source, which vanishes outside. Furthermore, the spin-up and spin-down states in $\psi$ are supposed to align with the $z$-direction of the experimental setup, such that $\sigma_z\ket{0}=-1$, and $\sigma_z\ket{1}=1$, where $\sigma_z$ is the Pauli $z$-spin operator. We first compute the Schrödinger evolution $i\hbar\partial_t\psi=\hat{H}\psi$ of the particle's wave function for the Hamiltonian 

\begin{equation}
	\hat{H}=-\frac{\hbar^2}{2m}(\partial_y^2+\partial_z^2)+\mu B z \delta(y)\sigma_z,
\end{equation} That is, suppose an interaction with an inhomogeneous magnetic field of magnitude $B$ aligned with $z$, turned on very briefly when the particles pass through the magnets at $y=0$. By solving Schrödinger's equation, we find the wave function of the system for $y>0$:

\begin{equation}
	\ket{\psi(y.z)}=\frac{1}{\sqrt{2}}e^{ik_y'y}(e^{ik_z z}\ket{0}+e^{-ik_z z}\ket{1}),  
\end{equation} again omitting normalisation in the spatial degrees of freedom. The magnets produce two separated beams propagating in opposite directions in the $z$-plane. The interaction has also slightly changed the wave vector in the $y$-direction to $k_y'$. As for the $y$-waves, we assume that the plane waves in $z$ have only finite widths. Hence, they overlap only in a finite region of space and then are fully separated. 

The experiment has two outcomes as a result, and the interaction setup thus implements a projective measurement. For the spin measurement in $z$-direction, the POVM has exactly two operators. They are $E_0=\frac{1}{2}(\mathds{1}+\sigma_z), E_1=\frac{1}{2}(\mathds{1}-\sigma_z)$, corresponding to a spin-up and spin-down projection, respectively. The quantum probabilities for the two outcomes are, therefore, $p(x)=\bra{\psi}E_x\ket{\psi}=\frac{1}{2}$. 

Suppose now the Bohmian configurations are in quantum equilibrium. For the initial wave function, this means they are equally distributed within the particle beam. The guidance equation can then be integrated to compute individual particle trajectories. This, in turn, can be used to obtain the number of particles that are deflected in either the up or down beam. Recall that the Bohmian guidance equation $v^{\psi}=\frac{\hbar}{m}\Im \frac{\psi^*\nabla\psi}{\psi*\psi}$ is also defined for spinor-valued wave functions. Thus, before the beams are fully separated, the Bohmian velocities are 

\begin{equation}
	\label{eqn:velocity z measurement overlap}
	v^{\psi}=\frac{\hbar k_y}{m} \hat{y},
\end{equation} i.e. particles moving in straight lines in the $\hat{y}$-direction, and after the separation 

\begin{equation}
	\label{eqn: velocity z measurement separated}
	v^{\psi}=\frac{\hbar k'_y}{m} \hat{y}\pm\frac{\hbar k_z}{m}\hat{z}, 
\end{equation} where the sign depends on whether the particle ends up in the upper or lower branch of the beam (which, in turn, is decided by the initial configuration of the Bohmian particle). Note also that within the overlap region, the velocity in $\hat{z}$ remains unaffected. Since exactly half of the particles start in the upper half of the initial beam, the velocities show that exactly half of them will yield a spin-up result ending up in the upper branch, and half of them will yield a spin-up result ending up in the lower branch, as expected. 

The analysis above can be generalised to spin measurements along arbitrary directions, i.e. when the POVM is $E_x=\frac{1}{2}(\mathds{1}+(-1)^x \hat{n}\vec{\sigma})$: The corresponding interaction for this POVM is obtained by rotating and changing the polarity of the magnets to measure in the desired new spin axis $\hat{n}=(\sin\vartheta\cos\varphi, \sin\vartheta\sin\varphi, \cos\vartheta)$.\footnote{Note that turning the Stern-Gerlach apparatus upside down doesn't change the outcomes since flipping both the gradient and the polarity of the magnetic field cancels out the effect on the particles. Thus, in the experiment the measurement settings $\vartheta$ is realised by adapting either the direction of the polarity or gradient.} Assume for simplicity that the measurement of spin lies in the plane orthogonal to the propagation of the beam and that we only rotate the magnets around the propagation of the initial beam, i.e. the $\hat{y}$-axis, by some angle $\vartheta$, i.e. $\varphi=0$. Thus, the direction of measurement simplifies to $\hat{n}=(\sin\vartheta, 0, \cos\vartheta)$, and the corresponding POVM is given by $E_x=\frac{1}{2}(\mathds{1}+(-1)^x\sigma_n)$, where 

\begin{equation}
	\sigma_n:= \hat{n}\vec{\sigma}= 
	\begin{pmatrix}
		\cos\vartheta & \sin\vartheta  \\
		\ \sin\vartheta  & -\cos\vartheta \\
		
	\end{pmatrix}.
\end{equation}

A spin measurement of the initial state (\ref{eqn:initial bohmian spin state}) in this basis yields probabilities $p(x)=\bra{\psi}E_x\ket{\psi}=\frac{1}{2}(1+(-1)^x\sin\vartheta)$. The eigenvectors of the projectors $E_x$ are $\ket{n_0}=(\cos\frac{\vartheta}{2}, \sin\frac{\vartheta}{2})$, and $\ket{n_1}=(\sin\frac{\vartheta}{2}, -\cos\frac{\vartheta}{2})$. By rewriting Equation \ref{eqn:initial bohmian spin state} in this basis and repeating the analysis above, the wave function for $y>0$ is 

\begin{equation}
	\ket{\psi(y.z)}=\frac{1}{\sqrt{2}}e^{ik_y'y}\left[e^{ik_n \hat{n}}(\cos\frac{\vartheta}{2}+\sin\frac{\vartheta}{2})\ket{n_0}+e^{-ik_n \hat{n}}(\sin\frac{\vartheta}{2}-\cos\frac{\vartheta}{2})\ket{n_1}\right],  
\end{equation} where $k_n$ the wave vector of the beams along $\hat{n}$. As opposed to the previous measurement in $\hat{z}$, the Bohmian velocities in the overlap region are now affected so that after the magnetic interaction we have in this region

\begin{equation}
	\label{eqn:v theta 1}
	v^{\psi}=\frac{\hbar k'_y}{m} \hat{y}+\sin\vartheta\frac{\hbar k_n}{m}\hat{n},
\end{equation} Outside the overlap, when the beams have separated again, we still recover

\begin{equation}
	\label{eqn:v theta 2}
	v^{\psi}=\frac{\hbar k'_y}{m} \hat{y}\pm\frac{\hbar k_n}{m}\hat{n},
\end{equation} as before. But since the trajectories are tilted --- proportional to $\sin\vartheta$ --- while passing the overlap, the number of particles ending up in the upper or lower branch of the separated beam depends on the measurement angle $\vartheta$. For instance, if the magnetic axis is chosen to measure in the diagonal basis, i.e. $\vartheta=\frac{\pi}{2}$, all particles are deflected into the upper ($\hat{n}$-direction) beam according to Equations \ref{eqn:v theta 1} and \ref{eqn:v theta 2}. This makes sense as the initial state $\frac{1}{\sqrt{2}}(\ket{0}+\ket{1})$ is an eigenstate of the diagonal basis. In general, the space of initial configurations divides into two fractions: The upper fraction of height $h_0=\frac{1}{2}(1+\sin\vartheta)$ which is deflected up, and the lower fraction of height $1-h_0=\frac{1}{2}(1-\sin\vartheta)$ which is deflected down (see also the following sections and Figure \ref{fig:Stern-Gerlach experiment}). These coincide with the corresponding outcome probabilities as expected. 

\subsubsection{Bohmian Extensions}

Recalling the quantum decorrelation principle, since the Bohmian quantum states in the experiment are assumed pure and in equilibrium, no system in the environment can be correlated to the outcome statistics better than the Born rule predicts. Conversely, if stronger correlations exist to the environment, the quantum states cannot be pure unless Bohmian equilibrium is violated. This allows conceiving of a magic prediction box whose outcomes indicate information about the initial Bohmian configuration of the spin particles in every run of the experiment. If the total state of spin system and prediction box is $\rho=\rho_S\otimes\rho_M=\ket{\psi}\bra{\psi}\otimes\ket{\phi}\bra{\phi}$, one then has $p(x,y)\neq p(x)p(y)$ in non-equilibrium, violating the Born rule applied to the total state. The outcome probabilities on the individual subsystems, however, still satisfy the quantum probabilities, i.e. $p(x)=\tr(\rho_S E_x)$ for the target system, and $p(y)=\tr(\rho_M E'_{y})$ for the prediction box. In brief, the target system and magic prediction box both yield quantum probabilities individually, but are correlated \textit{non}-quantum mechanically.   

For predictive advantage, imagine a non-equilibrium source that produces an ensemble of states $\rho$, i.e. the initial quantum spin states (\ref{eqn:initial bohmian spin state}) with Bohmian configurations equally distributed within the beam and a pure state for the magic prediction box but with initial Bohmian configurations mapping identically to the spin particles' configurations. Since the Bohmian state uniquely fixes the spin measurement results for all Stern-Gerlach POVMs, the outcomes of the magic prediction box --- e.g. if a position measurement is performed on it --- contain all the information to deterministically predict the spin outcomes in any measurement basis. Thus, this type of non-equilibrium on the larger system implies $p(x|y)\in\{0,1\}$. The correlations are much stronger than needed for deterministic predictions in the Stern-Gerlach experiment since each half of the particle beam produces a single spin-up or spin-down result. It suffices to suppose that the magic prediction box produces binary outputs referring to whether the target system particles are located in either the upper or lower half of the initial beam (with respect to the measurement axis $\hat{n}$), as I will show below. These deviations from the Bohmian equilibrium hypothesis can lead to predictive advantage. 

I wish to emphasise that the mere existence of a non-equilibrium distribution of a quantum target system is insufficient for predictive advantage. Nothing has been said about how the predictions are made by virtue of manifest configurations in its environment. In order for predictive advantage to exist, a violation of the quantum decorrelation principle through Bohmian non-equilibrium of the \textit{compound} state of target system and prediction box is necessary. As a result, by Definition \ref{def:empirical completeness 1} the non-equilibrium Bohmian theory is an empirical extension of standard quantum mechanics like so: The manifest outcome configurations, say $z$, of the magic prediction box directly signify the full Bohmian configuration of the target bi-partite system. The measurement axis of the primary and secondary system be $a, b$ and are fully specified by some angles $\vartheta_1$ and $\vartheta_2$. Formally, the empirical extension for a maximally entangled bi-partite 2-qubit state $\rho=\ket{\Psi}\bra{\Psi}$ (Equation \ref{eqn:initial bohmian bipartite spin state}) reads
\begin{equation}
	p(x,y|a,b,\rho)= \int p(x,y|a,b,z)\mu(z)dz,
\end{equation} where $p(x,y|a,b,\rho)=\tr(E_x^a\otimes E_y^b\rho)\neq p(x,y|a,b,z)\in\{0,1\} $ are the quantum predictions and the ones of the empirical extension for the outcomes of the primary and secondary particles conditioned on $z$. The inequality holds for at least one $z$. The outcomes $(x,y)$ of the target system are thus in quantum equilibrium whereas the correlation of the target system with the configurations $z$ of the magic prediction box demonstrates the existence of quantum non-equilibrium. Otherwise, the $z$ variables couldn't contain more information for the outcomes of measurements than the preparation of the state $\ket{\Psi}$. 
Whenever a pair of particles is created, the magic prediction box indicates one of the manifest configurations $z$ with probability $\mu(z)$ and corresponding deterministic predictions $p(x,y|a,b,z)$. 

Although the initial Bohmian configurations could have many possible values, as I shall show it will suffice to consider certain sets of configurations $z$ for the empirical extension since the measurement outcomes are already fully determined by in what half of the beam the particles reside.

\subsection{Maximally-Entangled States and Bohmian Predictive Advantage}
\label{Section:maximally entangled states dBBT advantage}

A magic prediction box of the kind introduced will always lead to the possibility of signalling if the state is maximally entangled. To see this consider two Stern-Gerlach apparatuses measuring bi-partite spin particles. Suppose the source emits two beams propagating in opposite directions --- one primary beam of particles towards $\hat{y}$ and another secondary beam in the opposite direction towards $-\hat{y}$. The party to which signals are transmitted is assumed to measure in the fixed basis aligned with the $\hat{z}$-axis. The other party possesses the magic prediction box and measures the primary particles in any desired axis $\hat{n}$. The two individual experiments are space-like separated. Furthermore, the initial quantum state of the compound system be maximally entangled, i.e.

\begin{equation}
	\label{eqn:initial bohmian bipartite spin state}
	\ket{\Psi(y_1,z_1,y_2,z_2)}= \frac{1}{\sqrt{2}}e^{ik_y y_1}e^{-ik_y y_2}(\ket{00}+\ket{11}). 
\end{equation} 

As discussed for single particle systems and using the same definitions, the Bohmian velocities are obtained in exactly the same way. By rewriting the initial state in the basis of the $\hat{n}$-basis for the primary measurement and solving the Schrödinger evolution, the quantum state after the primary particle's interaction with the Stern-Gerlach magnets reads (omitting normalisation in the spatial degrees of freedom):

\begin{dmath}
	\label{eqn:initial bohmian bipartite spin state after interaction}
	\ket{\Psi(y_1,z_1,y_2,z_2)}= \frac{1}{\sqrt{2}}e^{ik_y y_1}e^{-ik_y y_2}\left[e^{ik_n \hat{n}}\ket{n_0}\otimes(\cos\frac{\vartheta}{2}\ket{0}+\sin\frac{\vartheta}{2}\ket{1}) 
	+e^{-ik_n \hat{n}}\ket{n_1}\otimes(\sin\frac{\vartheta}{2}\ket{0}-\cos\frac{\vartheta}{2}\ket{1})\right].  
\end{dmath} 

From which the outcome probabilities are obtained by tracing over the opposite particle or directly by 

\begin{equation}
	p(x)=\bra{\Psi}E_x\otimes\mathds{1}\ket{\Psi}=p(y)=\bra{\Psi}\mathds{1}\otimes E'_{y}\ket{\Psi}=\frac{1}{2}.
\end{equation} This shows that for a maximally entangled state, the probabilities are uniform for any choice of POVM, i.e. axis $\hat{n}$. Moreover, for the Bohmian velocities of the primary particle, we recover the familiar quantities for an ordinary $\hat{z}$-spin measurement, i.e. Equation \ref{eqn:velocity z measurement overlap} and Equation \ref{eqn: velocity z measurement separated}  for the overlap and separated beam region, with $\hat{z}$ replaced by $\hat{n}$. Regardless of what axis $\hat{n}$ is measured, half of the particles are deflected into the up beam and half into the down beam. 

But the relevant branch of the total wave function guiding the secondary particle now depends on the primary particle's trajectory. Assuming the primary particle is measured first if it ends up in the upper beam, the down term in the total wave function vanishes since this branch has support only for the upper beam. More concretely, in the overlap region of the two beams for the secondary particle, its velocities are

\begin{equation}
	\label{eqn:velocity maximally entangled overlap}
	v^{\Psi}=-\frac{\hbar k'_y}{m} \hat{y}\pm\cos\vartheta\frac{\hbar k_z}{m}\hat{z},
\end{equation} where the sign is determined by the spin result of the primary apparatus. That is, $+$ for $\ket{n_0}$, and $-$ for $\ket{n_1}$. As before in the single-particle case, when the sub-beams have fully separated, the trajectories continue with velocity

\begin{equation}
	\label{eqn:velocity maximally entangled separated}
	v^{\Psi}=-\frac{\hbar k'_y}{m} \hat{y}\pm\frac{\hbar k_z}{m}\hat{z}.
\end{equation} The sign is plus if the particle ends up in the upper half of the beam, i.e. $z>0$ after the overlap, and minus otherwise. For example, when the primary system is measured in the $\hat{z}$-basis as well, i.e $\vartheta=0$, the two velocities in Equation \ref{eqn:velocity maximally entangled overlap} and Equation \ref{eqn:velocity maximally entangled separated} coincide. Thus, whenever the primary particle is deflected upwards, the secondary particle is deflected upwards, irrespective of its initial configuration. Likewise, if the primary particle is deflected downwards, the secondary particle is deflected downwards. 

\begin{figure}[h]
	\centering
	\includegraphics[width=1\linewidth]{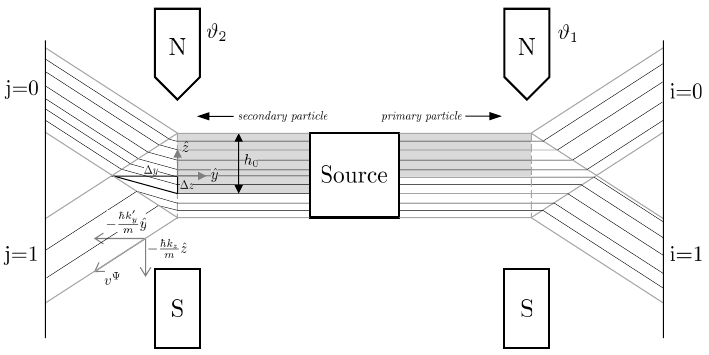}
	\caption{Two Stern-Gerlach experiments measuring a system of two spin-$\frac{1}{2}$ particles, emitted in an entangled state from a source in the centre of the experiment. The measurement settings $\vartheta_1$ and $\vartheta_2$ of the two magnets determine the outcome probabilities and hence correlations between primary and secondary particles. The wave branches of the quantum state overlap until they reach the magnets, where they are separated. The thin lines depict the Bohmian trajectories of the particles as they pass through the experiment. Within the overlap region where the separation happens, the slope $\frac{\Delta z}{\Delta y}$ of the trajectory is determined by both the measurement settings of the primary and secondary apparatus. The shaded area with height $h_0$ shows the fraction of secondary particles deflected into the up beam, which in turn is interpreted as outcome $0$, or spin up. Particles deflected downwards are interpreted as showing outcome $1$, or spin down. For clarity, the rotation of the magnets around the $\hat{y}$ axis depending on the setting $\vartheta_1$ and $\vartheta_2$  isn't represented in the figure.} 
	\label{fig:Stern-Gerlach experiment}
\end{figure}

Generally, the fraction of secondary particles being deflected up or down depends on the primary setting $\vartheta$. To quantify this fraction, the slope $\frac{v_z}{v_y}$ needs to be computed (see Figure \ref{fig:Stern-Gerlach experiment}). The height $h_0$ within the beam for which particles yield the up result can then be related to this slope. Assuming the total height of the beam to be of unit length $1$, one has

\begin{equation}
	\frac{v_z}{v_y} = \frac{\Delta z}{\Delta y}= \frac{h_0-\frac{1}{2}}{\frac{1}{2}\frac{k_y'}{k_z}},
\end{equation} where $v_z, v_y$ the velocity components of the secondary particle in the overlap area. The length $\Delta y={\frac{1}{2}\frac{k_y'}{k_z}}$ is determined by the velocity in the separated beams only. Hence, for the height we obtain

\begin{equation}
	h_0=\frac{1}{2}\left(\frac{k_y'}{k_z}\frac{v_z}{v_y}+1\right).
\end{equation}

For the simple case discussed, i.e. $\vartheta_2=0, \vartheta_1:=\vartheta$ this reads $h_0=\frac{1}{2}(1+\cos\vartheta)=\cos^2\frac{\vartheta}{2}$  if $n_0$ occurs for the primary system, and $h_0=\frac{1}{2}(1-\cos\vartheta)=\sin^2\frac{\vartheta}{2}$ if $n_1$ occurs for the primary system.

The primary observer can now employ the magic prediction box to send superluminal signals like so: In every run of the experiment, she reads off from the prediction box whether the particle was emitted in the upper or lower half of the beam. Every time it shows \textit{up}, she measures in the setting where $\vartheta=0$, and every time it shows \textit{down}, she performs a measurement in the setting $\vartheta=\pi$ by either flipping the gradient or polarity of the magnetic field. As a result, \text{all} secondary particles --- no matter where they are initially located in the opposite beam --- are shunted into the up sub-beam. For the sign in Equation \ref{eqn:velocity maximally entangled overlap} will always be positive. The secondary observer will thus repeatedly observe the same result. Conversely, she will repeatedly observe the \textit{opposite} result when the procedure is reversed. That is, the primary apparatus performs a $\vartheta=\pi$ measurement when the magic prediction box indicates \textit{up} configurations and $\vartheta=0$ when \textit{down}. All secondary particles are then shunted in the opposite direction. The resulting outcome string can be interpreted as the transmission of one bit of information, e.g. $0$ for the up and $1$ for the down string. Note also that any other setting ($\vartheta\neq 0, \pi$) will lead to more uncertainty in the outcomes of both the primary and secondary observer. 

In general, for any pair of arbitrary settings $(a=\vartheta_1,b=\vartheta_2)$ there are four possible measurement results $(x,y)\in\{0,1\}^2$ with corresponding quantum probabilities and deterministic Bohmian predictions. By rewriting the spin part of the maximally entangled state (\ref{eqn:initial bohmian bipartite spin state after interaction}) in the measurement basis $(a,b)$ it can be seen that it only depends on the difference of the measurement angles $\vartheta_1-\vartheta_2$:

\begin{dmath}
	\label{eqn:maximally entangled state in arbitrary basis}
	\ket{\Psi}= \frac{1}{\sqrt{2}}\left[\cos(\frac{\vartheta_1}{2}-\frac{\vartheta_2}{2})\ket{n_0}\otimes\ket{n_0'}-\sin(\frac{\vartheta_1}{2}-\frac{\vartheta_2}{2})\ket{n_0}\otimes\ket{n_1'}+\sin(\frac{\vartheta_1}{2}-\frac{\vartheta_2}{2})\ket{n_1}\otimes\ket{n_0'}+\cos(\frac{\vartheta_1}{2}-\frac{\vartheta_2}{2})\ket{n_1}\otimes\ket{n_1'}\right].
\end{dmath}

Without loss of generality, one of the angles can therefore be set to zero, i.e. $\vartheta_2:=0$. The quantum probability functions hence are recovered to be $p(x,y)=\frac{1}{2}\cos^2(\frac{\vartheta_1}{2})$ if $x=y$ and $p(x,y)=\frac{1}{2}\sin^2(\frac{\vartheta_1}{2})$ otherwise, as before. The marginal probabilities $p(x)=\sum_y p(x,y)=\frac{1}{2}$ and $p(y)=\sum_x p(x,y)=\frac{1}{2}$ are clearly independent of the measurement angles and thus signal-local. 

Conditioned on the additional variables of the Bohmian extension, the predictions differ, however. Since pilot wave theory comes with concrete dynamics (the guiding equation), the equations can be integrated to quantify the theory's non-locality. This, in turn, allows one to obtain a direct analysis of when predictive advantage is possible while obeying the signal-locality constraint. By doing this, four domains of configurations are identified, uniquely determining the four possible measurement results $(x,y)\in\{0,1\}^2$: 

\begin{equation}
	h_0^{n_0}=\cos^2\frac{\vartheta_1}{2},
\end{equation} for the primary outcome $n_0$, and

\begin{equation}
	h_0^{n_1}=\sin^2\frac{\vartheta_1}{2},
\end{equation} for the primary outcome $n_1$. 

\begin{figure}[h]
	\centering
	\includegraphics[width=1\linewidth]{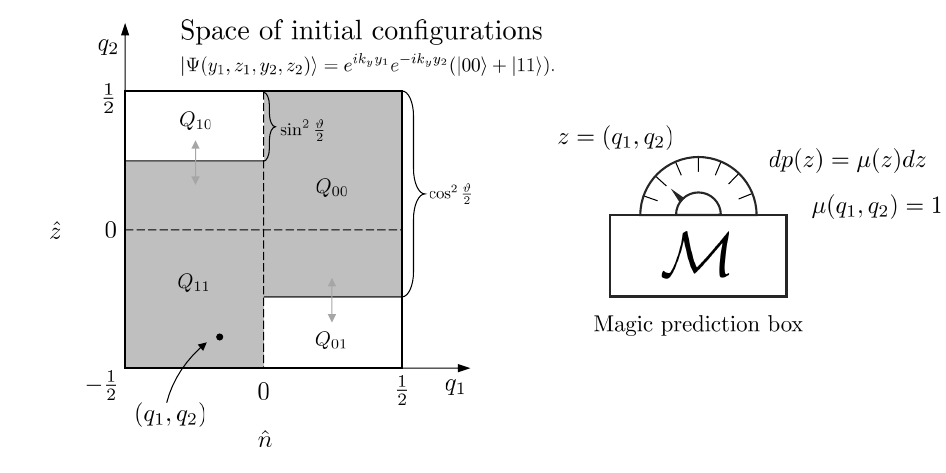}
	\caption{The figure shows the deterministic predictions of a magic prediction box with manifest variables $z=(q_1,q_2)$ corresponding to the initial Bohmian configurations for a maximally entangled quantum state of two spin-$\frac{1}{2}$ particles. The size of the shaded areas depends on the measurement parameter $\vartheta$, i.e. their heights are $\cos^2\frac{\vartheta}{2}$ and $\sin^2\frac{\vartheta}{2}$, respectively. The quadrangle depicts the space of all possible initial Bohmian configurations of the two particles within the two beams of the source ($q_1$ the initial height in the beam along measurement axis $\hat{n}$ for the primary system, and $q_2$ the initial height in the beam along measurement axis $\hat{z}$ for the secondary system). The primary measurement axis is tilted by an angle $\vartheta$ with respect to the $\hat{z}$-axis. Every point in the domain of configurations lies within one of the four areas $Q_{xy}$ for which one of the four deterministic outcomes occurs. For simplicity, the heights of the beams are assumed to be $1$. By varying the measurement axis the whole configuration space is surveyed. As a result, for any point $(q_1,q_2)$, two distinct outcomes can occur depending on the chosen angle $\vartheta$. An empirical extension of this sort is thus impossible if signal-locality is to be respected.} 
	\label{fig:bohmian-predictions-maximally-entangled-state}
\end{figure}

Figure \ref{fig:bohmian-predictions-maximally-entangled-state} illustrates the situation. By varying the measurement setting of the primary apparatus, the whole space of configurations is surveyed such that signal-locality can be violated for all settings of the secondary apparatus. Thus, for maximally entangled states (\ref{eqn:initial bohmian bipartite spin state}), non-equilibrium predictive advantage isn't possible without signalling. 

In order to avoid signalling, therefore, the statistics of the secondary observer must not depend in any way on the primary observer's measurement settings. But this will be impossible for any empirical extension of the maximally entangled two-particle state: In light of Bell's theorem, the non-locality in pilot wave theory is a universal feature in all hidden variable theories. Thus, the existence of fully deterministic predictive advantage implies the ability to signal in all deterministic hidden variable theories when the magic prediction box has access to the relevant variables (see Chapter \ref{section:Non-locality and quantum predictability} for details). As it turns out, though, there exist quantum states leading to non-equilibrium predictive advantage \textit{without} signalling for particular deterministic models. To this, I shall turn in Chapter \ref{section:Non-locality and quantum predictability}. Nevertheless, all bi-partite states in a Bohmian universe are signal-non-local for all pairs of particles. To this, I turn in the following. But signal-local non-trivial non-equilibrium distributions for non-maximally entangled states could still lead to predictive advantage in pilot wave theory.

\subsection{Non-maximally Entangled States and Bohmian Predictive Advantage}

For maximally entangled states, I have shown how the existence of a magic prediction box would imply signalling if deterministic predictions could be made. In this section, I shall analyse non-maximally entangled states. For this case, too, any pair of Bohmian particles turns out to be non-local. Thus, as for maximally-entangled states, deterministic Bohmian empirical extensions don't exist without signalling.

However, as will be seen in Chapter \ref{section:Non-locality and quantum predictability}, for non-maximally entangled pure states, in general, one \textit{can} find a \textit{signal-local} magic prediction box empirically distinct from quantum probabilities for more general deterministic hidden variable theories. I shall discuss in Section \ref{section: generalised non-equilibrium} below how this observation qualifies Valentini's proof that `non-equilibrium distributions are sufficient and necessary to imply instantaneous signalling' \citep{Valentini-H-theorem-II}. 

Imagine like before the same magic prediction box for making deterministic predictions conditioned on the variable $z$ for a general class of non-maximally entangled states


\begin{equation}
	\label{eqn:initial bohmian bipartite spin state non-maximally entangled}
	\ket{\Psi(y_1,z_1,y_2,z_2)}= e^{ik_y y_1}e^{-ik_y y_2}(\alpha\ket{00'}+\sqrt{1-\alpha^2}\ket{11'}),  
\end{equation} where $\{\ket{00'}, \ket{11'}\}$ the Schmidt basis of the bi-partite system. The marginal quantum probabilities then are

\begin{align}
	p(0)=&\bra{\Psi}E_0\otimes\mathds{1}\ket{\Psi}=\bra{\Psi}\mathds{1}\otimes E'_{0}\ket{\Psi}=\alpha^2, \\
	p(1)=&\bra{\Psi}E_1\otimes\mathds{1}\ket{\Psi}=\bra{\Psi}\mathds{1}\otimes E'_{1}\ket{\Psi}=1-\alpha^2,
\end{align} when both POVMs are measurements aligned with the Schmidt basis.

For arbitrary measurement angles $\vartheta_1, \vartheta_2$ (as in Figure \ref{fig:Stern-Gerlach experiment}) of primary and secondary Stern-Gerlach devices, respectively, rewrite the non-maximally entangled spin state to obtain

\begin{dmath}
	\label{eqn:non-maximally entangled state in arbitrary basis}
	\ket{\Psi}= \left(\alpha\cos\frac{\vartheta_1}{2}\cos\frac{\vartheta_2}{2}+\sqrt{1-\alpha^2}\sin\frac{\vartheta_1}{2}\sin\frac{\vartheta_2}{2}\right)\ket{n_0}\otimes\ket{n_0'}-\left(\alpha\cos\frac{\vartheta_1}{2}\sin\frac{\vartheta_2}{2}-\sqrt{1-\alpha^2}\sin\frac{\vartheta_1}{2}\cos\frac{\vartheta_2}{2}\right)\ket{n_0}\otimes\ket{n_1'}+\left(\alpha\sin\frac{\vartheta_1}{2}\cos\frac{\vartheta_2}{2}-\sqrt{1-\alpha^2}\cos\frac{\vartheta_1}{2}\sin\frac{\vartheta_2}{2}\right)\ket{n_1}\otimes\ket{n_0'}+\left(\alpha\sin\frac{\vartheta_1}{2}\sin\frac{\vartheta_2}{2}+\sqrt{1-\alpha^2}\cos\frac{\vartheta_1}{2}\cos\frac{\vartheta_2}{2}\right)\ket{n_1}\otimes\ket{n_1'}.
\end{dmath}

This time the amplitudes aren't symmetric in the measurement parameters as was the case for the maximally entangled state, i.e. do not only depend on the difference of the angles but on both individually. 

As before, the fraction of the height that is deflected upwards can be computed as a function of the measurement angles (these are just the conditional probabilities as was shown before). First, consider the conditional outcome for when the primary apparatus shows the result $\ket{n_0}$:  

Performing the analogous calculations as in the previous section, for the velocity we get
\begin{equation}
	\label{eqn:velocity non-maximally entangled overlap}
	v^{\Psi}=-\frac{\hbar k'_y}{m} \hat{y}+\frac{1}{N_0}\left((\alpha^2-\sin^2\frac{\vartheta_1}{2})\cos\vartheta_2+\alpha\sqrt{1-\alpha^2}\sin\vartheta_1\sin\vartheta_2\right)\frac{\hbar k_z}{m}\hat{z},
\end{equation}

and for the height $h_0^{n_0}$

\begin{equation}
	h_0^{n_0}=\frac{1}{N_0}\left(\alpha\cos\frac{\vartheta_1}{2}\cos\frac{\vartheta_2}{2}+\sqrt{1-\alpha^2}\sin\frac{\vartheta_1}{2}\sin\frac{\vartheta_2}{2}\right)^2,
\end{equation} where we defined $N_0=\alpha^2\cos^2\frac{\vartheta_1}{2}+(1-\alpha^2)\sin^2\frac{\vartheta_1}{2}$.

And similarly for $\ket{n_1}$:

\begin{dmath}
	v^{\Psi}=-\frac{\hbar k'_y}{m} \hat{y}+\left[(\alpha^2-\cos^2\frac{\vartheta_1}{2})\cos\vartheta_2-\alpha\sqrt{1-\alpha^2}\sin\vartheta_1\sin\vartheta_2\right]\frac{\hbar k_z}{m}\hat{z},
\end{dmath}

and for the height $h_0^{n_1}:$

\begin{equation}
	h_0^{n_1}=\frac{1}{N_1}\left(\alpha\sin\frac{\vartheta_1}{2}\cos\frac{\vartheta_2}{2}-\sqrt{1-\alpha^2}\cos\frac{\vartheta_1}{2}\sin\frac{\vartheta_2}{2}\right)^2,
\end{equation} where we defined $N_1=\alpha^2\sin^2\frac{\vartheta_1}{2}+(1-\alpha^2)\cos^2\frac{\vartheta_1}{2}$.

The equations reduce to the quantities from above for the simple case when the secondary apparatus performs a $\hat{z}$-measurement, i.e. $\vartheta_2=0$, and when the state is maximally entangled. It is not difficult to see that again the conditional probabilities, i.e. the heights $h_0^{n_0}$, $h_0^{n_1}$ survey the full configurations space since $h_0^{n_0}, h_0^{n_1}\in\left[0,1\right]$ (cf. the analogous situation in Figure \ref{fig:bohmian-predictions-maximally-entangled-state} for maximally-entangled states just with a different dependency on the measurement angles). Thus, for arbitrary combinations of measurement settings $\vartheta_2$, it is, like before, always possible to find a setting of the primary apparatus for which the outcome changes. Hence, the prediction violates signal-locality. 

It can also be seen that the corresponding heights simply represent the conditional probabilities for the outcomes of the secondary observer depending on the primary outcome. We can check whether the conditional probabilities match the quantum probabilities and are signal-local. As opposed to the maximally entangled state, however, due to the lack of symmetry in the initial particle distribution (i.e. weighted by the entanglement coefficient $\alpha$), the fractions of primary particles that are deflected up and down are not uniform and depend on the measurement angle $\vartheta_1$. More concretely, according to the equilibrium distribution, a fraction of $\alpha^2$ of pairs are emitted in the upper half, and a fraction $1-\alpha^2$ on the lower half of the beam. The number of primary particles shunted up is

\begin{equation}
	N_0=\alpha^2\cos^2\frac{\vartheta_1}{2}+(1-\alpha^2)\sin^2\frac{\vartheta_1}{2},
\end{equation} and 

\begin{equation}
	N_1=\alpha^2\sin^2\frac{\vartheta_1}{2}+(1-\alpha^2)\cos^2\frac{\vartheta_1}{2},
\end{equation}

shunted down, respectively. These quantities equal the normalisation constants $N_0, N_1$ above. The marginal probabilities for the outcomes of the secondary can then be computed, e.g. for the up outcome, 

\begin{equation}
	p(0|\theta_2)=N_0 h_0^{n_0}+N_1 h_0^{n_1}=\alpha^2\cos^2\frac{\vartheta_2}{2}+(1-\alpha^2)\sin^2\frac{\vartheta_2}{2},
\end{equation} which is independent of the primary measurement angle and thus signal-local as expected. A similar relationship holds for the down outcome. 

In sum, for both maximally and non-maximally entangled states, every pair of Bohmian trajectories is non-local in the sense that for every initial configuration, there exist measurement settings for which the corresponding outcomes are different. That is, if a magic prediction box were to encode information about the exact Bohmian configuration, it inevitably could be employed for signalling for any pair of particles.

This may not come as a big surprise since pilot wave theory is non-local, but it shows the interesting feature: No subset of particle pairs is local (no matter the degree of entanglement of the quantum state). In light of the findings of the following sections, pilot wave theory is thus more non-local than required to account for quantum probabilities. Interestingly, deterministic empirical extensions exist for non-maximally entangled states that conform to signal-locality. For instance, Elitzur, Popescu, and Rohrlich devise a concrete example where a convex combination of a local (and deterministic) and non-local correlations reproduces the quantum probabilities for the non-maximally entangled states in Equation \ref{eqn:initial bohmian bipartite spin state non-maximally entangled} (see \citealt{ELITZUR199225}). 

Notice also that the possibility of using Bohmian trajectories to signal is due to the fact that for every single initial configuration, there always exists \textit{some} set of measurement settings for which the corresponding outcomes differ. Thus, the answer may look different if the space of possible measurement bases were restricted to a finite number. 

The number of settings plays a crucial role in determining whether signal-local extensions are possible. For example, suppose the setup is as in the standard Bell scenario with only two different fixed settings for the primary and secondary observer. Then for a given prediction of the magic prediction box and secondary setting, it may not be possible to tune the primary system to the setting for which the Bohmian trajectories lead to different outcomes, thus not being able to signal. In other words, not the whole square of initial configurations in Figure \ref{fig:bohmian-predictions-maximally-entangled-state} is accessible by tuning the primary setting $\vartheta_1$ if the set of possible measurement angles is restricted to only two. But clearly, if a general claim about predictive advantage is to be made, then the observer has to be thought of as potentially having access to all means of signalling and, thus, all settings in an experiment. A concrete example of a signal-local predictive advantage for the standard Bell scenario with two measurement settings is a convex combination of PR-boxes. 

In Westman’s approach to non-locality in pilot wave theory, an upper and lower bound on the degree of non-locality to violate Bell inequalities is derived based on Hardy bounds (cf. \citealt{Westman-transition-sets,HARDY-bounds}). It is shown that for two settings, the Bohmian theory is more non-local than minimally required and less non-local than maximally allowed to reproduce the quantum predictions. This, however, may not be surprising given that pilot wave theory reproduces \textit{all} quantum probabilities for arbitrarily many measurements. It was expected that an alternative hidden variable theory with less non-locality could reproduce the two-setting Bell experiment.   

This proves crucial in generalising the relationship between signal-locality and empirical completeness (Chapter \ref{section:Non-locality and quantum predictability}). For the results only hold in the limit of arbitrarily many measurement bases. Then, strictly speaking, signal-local empirical extensions do exist for any fixed finite set of measurements.

\section[Non-Equilibrium and Probabilistic Bohmian Advantage]{Non-Equilibrium and Probabilistic Bohmian Advantage} \sectionmark{Bohmian Predictive Advantage} 
\sectionmark{Bohmian Predictive Advantage} 

\label{section: generalised non-equilibrium}

The previous analysis shows that for both the maximally and non-maximally entangled states, every single initial configuration of both particles leads to final configurations, i.e. the measurement results, that depend on both the primary and secondary settings. Thus, for all entangled pure states, every single particle trajectory is, in fact, non-local. Furthermore, for all outcomes of the secondary observer, there exists no subset of initial configurations which isn't affected by the remote setting of the primary observer. In other words, for every Bohmian initial configuration, one can find a pair of measurement settings such that their respective outcomes differ. Therefore, no Bohmian predictive advantage exists, which is signal-local \textit{and} deterministic for any choice of settings. 

However, the result does not suggest that there can’t exist a signal-local \textit{probabilistic} Bohmian predictive advantage. It doesn't rule out that there could be subsets of Bohmian initial distributions \textit{out of equilibrium}, which are still invariant under variation of the primary setting.  Individual trajectories may be non-locally affected by a change of measurement basis, but there might be a signal-local probability distribution that is invariant under non-local distant measurements. Hence, if one does a partial coarse-graining the non-locality disappears. 

Let’s dwell a little more on this idea. Recall the definition of an empirical extension for bi-partite systems, i.e. the decomposition of predictions

\begin{equation}
p(x,y|a,b,\rho)=\int p(x,y|a,b,z)\mu(z)dz.
\end{equation} If $z=(q_1,q_2)$ the Bohmian initial configurations, then each corresponding prediction $p(x,y|a,b,z)\in\left\{0,1\right\}$ is deterministic and non-local, i.e. also signal-non-local. 

But suppose that the outcome variable $z$ of the magic prediction box indicates some probability preparation $\rho(q_1,q_2)$ on the configuration space of the initial Bohmian configurations (i.e. on the square in Figure \ref{fig:bohmian-predictions-maximally-entangled-state}). Since the Bohmian dynamics are deterministic, this gives a unique outcome distribution $p(x,y|a,b,z)$, which is the prediction for $z$. Note that $\rho(q_1,q_2)$ doesn't per se signify a Bohmian non-equilibrium preparation. By construction, it says that whenever some $z$ obtains, the magic prediction box indicates that the current pair of particles can be thought of as belonging to the preparation $\rho(q_1,q_2)$. Nevertheless, although when averaging over $z$, the Born probabilities are recovered, for each $z$, the distribution $\rho(q_1,q_2)$ can also be seen as Bohmian non-equilibrium preparation. 

The empirical extension is designed to recover the quantum probabilities on average as stipulated in the equation above. The initial configurations of the pairs of particles are in Bohmian equilibrium. Nevertheless, predictive advantage only exists when the QDP (cf. Definition \ref{def:quantum decorrelation principle}) is violated. Thus, Bohmian non-equilibrium has to come in somewhere. But as discussed to violate QDP means the correlations between the magic prediction box and target system are non-quantum, i.e. cannot be described by standard quantum probabilities and assignment of a pure state. As a result, Bohmian non-equilibrium only shows up in the total distribution of the system plus the magic prediction box. 

Let the four different regions in Figure \ref{fig:bohmian-predictions-maximally-entangled-state} leading to the four possible outcomes $(x,y)\in \left\{0,1\right\}^2$ be labelled with $Q_{xy}$. As can be read from the figure, the outcome probabilities can be written as (note, as before, due to symmetry, one measurement setting is set to 0)

\begin{equation}
p(x,y|a,b,z)=p(x,y|\theta,z)=\int_{Q_{xy}(\vartheta)}\rho_z(q_1,q_2)dq_1 dq_2, 
\end{equation} with the marginal probabilities for the secondary observer

\begin{equation}
p(y|\vartheta,z)=\sum_x p(x,y|\vartheta,z)=\sum_x \int_{Q_{xy}(\vartheta)}\rho_z(q_1,q_2)dq_1 dq_2. 
\end{equation}  The question now is whether a \textit{signal-local} (i.e. $p(y|\vartheta,z)=p(y|z)$) $\rho(q_1,q_2)$ exists that is \textit{non-trivial} (i.e. $\rho_z(q_1,q_2)\neq const.$). If yes, signal-local Bohmian predictive advantage is possible. 

There is a trivial instance doing the job: the quantum equilibrium distribution (as it satisfies signal-locality). Indeed, in equilibrium $\rho(q_1,q_2)=1$ and 

\begin{equation}
p(y|\vartheta,z)=\sum_x \int_{Q_{xy}(\vartheta)}dq_1 dq_2= \sum_x |Q_{xy}(\vartheta)|=\frac{1}{2}(\sin^2\frac{\vartheta}{2}+\cos^2\frac{\vartheta}{2})=\frac{1}{2}
\end{equation} since the four regions are disjoint, for all settings $\vartheta$. 

So the question here is whether the flat quantum equilibrium distribution is the only one. Valentini's proof that `non-equilibrium distributions are sufficient and necessary to imply instantaneous signalling' shows precisely this and generalises it to deterministic hidden variable theories \cite{Valentini-H-theorem-II, VALENTINI2002273}. It is proven that whenever an entangled Bohmian bi-partite system is out of quantum equilibrium, the setup can be used to send superluminally signals.  

However, his proof is not as general as one would hope. A closer look at his analysis reveals that in the proof, only singlet states are considered, i.e. maximally entangled states. It is thus still to be found whether the proof goes through for arbitrarily entangled bi-partite states. A signal-local empirical extension for non-maximally entangled states would thus not be in tension with Valentini's claims. This would be compatible with the observation that for non-maximally entangled states, signal-local non-equilibrium and, thus, predictive advantage is possible.   

As mentioned, deterministic signal-local empirical extensions exist for non-maximally entangled states (cf. Chapter \ref{section:Non-locality and quantum predictability} for details). But it leaves open the question of whether there exists a non-trivial \textit{Bohmian} non-equilibrium distribution that is signal-local.

\section{Internal Bohmian Predictions}
\label{section:internal observation}
Here I will apply the considerations of Section \ref{section:predictability and measurement problem} on internal predictions to pilot wave theory. That is, I explicate how the Bohmian framework addresses the ambiguities in quantum prediction making. 

As described, in the Bohmian account, the quantum decorrelation result is valid under the assumptions of equilibrium distributions, the coincidence of manifest configurations and non-manifest configurations, and a probability preserving dynamics, i.e. guiding equation (see Equation \ref{eqn:QDP for de Broglie-Bohm theory}). But crucially, it was also assumed that an \textit{effective} wave function exists, which turns out to be identical to the standard quantum wave function. That is, the manifest record in the environment signifies the preparation of a quantum state for a separate target system. When an effective Bohmian wave function exists, the predictions are identical to the standard quantum account. But in the case of internal predictions, a scenario was devised in which the manifest record denotes a quantum state and is also a degree of freedom to which this state is assigned. It may then happen that some configuration encodes information about both the wave function and the initial Bohmian configuration. Thus, leading to predictive advantage. The non-existence of an effective wave function does present a case where the quantum decorrelation principle is violated. 

However, from the previous considerations in Section \ref{section:pilot wave as metaphysical extension} it became clear that the theory satisfies the quantum decorrelation principle in quantum equilibrium. A Bohmian universe starting in quantum equilibrium stays in equilibrium due to its preservation under time evolution and continuity of quantum probabilities along Bohmian trajectories. This immediately ensures that for Bohmian observers assigning an effective wave function to a subsystem, the outcomes are distributed according to the Born rule. Thus, whatever the manipulations and details of an internal observer could be, the quantum decorrelation principle will prevent the creation of Bohmian non-equilibrium with respect to the environments relative to which an effective wave function exists. Superficially, this means that if they exist, internal predictions cannot be represented manifestly in the environment; otherwise, the quantum decorrelation principle would be violated. Therefore, either an internal observer isn't able to achieve this dynamically, or if internal predictive advantage is indeed possible, to separate the relevant sub-ensembles to produce a genuine empirical extension or a combination of both. It would be interesting to work out the details of how an internal observer fails to achieve this in a Bohmian universe. I do not succeed in this in the current section, but I shall make some preliminary remarks.    

Recall the setup of Section \ref{section:predictability and measurement problem} on ambiguities in quantum prediction making. The friend measures a target system prepared in $\frac{1}{\sqrt{2}}(\ket{\psi_A}+\ket{\psi_B})$. An outcome is obtained represented by the manifest configurations $A$ or $B$. Based on the initial state and this configuration, an `internal' prediction is made on Wigner's outcomes who performs a subsequent measurement in an arbitrary (orthogonal) basis  $\{\ket{\Psi_W}, \ket{\Phi_W}\}$, where $\ket{\Psi_W}=\alpha \ket{\psi_A}\otimes\ket{\phi_A}+\beta \ket{\psi_B}\otimes\ket{\phi_B}$.

The relevant question then was whether the friend's record has any significance for predicting the probabilities of Wigner's outcomes. De Broglie-Bohm theory must have a clear answer to this question since the Bohmian framework is complete in that it tells an unambiguous story about how manifest measurement outcomes relate to the variables of the non-manifest domain. 

Applying the considerations mutatis mutandis to pilot wave theory, the Bohmian commitments imply that the manifest configuration the friend has access to signifies the existence of a (set of) Bohmian particles located in $A$ or $B$. Moreover, a configurational record must also be accessible to the friend and Wigner representing the prepared state of the target system. In sum, after the friend's interaction with the target system, Wigner's prediction is based on applying the Born rule to the quantum state $\ket{\Psi}$, whereas the friend's prediction involves the quantum state plus some additional configuration $q_F$ denoting the outcome on the target system, i.e. the total Bohmian state $\left(\Psi, q_F\right)$, where $q_F\in\left\{A,B\right\}$.    

After the interaction, the joint system is described by the maximally entangled bi-partite state $\ket{\Psi}=\frac{1}{\sqrt{2}}(\ket{\psi_A}\otimes\ket{\phi_A}+\ket{\psi_B}\otimes\ket{\phi_B})$, and Wigner measures in the bases defined before. 

The compound system of friend and target system thus resembles the Bohmian experiment of a spin-$\frac{1}{2}$ particle studied in Chapter \ref{section:de Broglie-Bohm theory and predictive advantage}. To model the Wigner's friend type experiment in pilot wave theory, it can therefore be assumed that Wigner performs a Stern-Gerlach measurement of spin on the friend, and that their outcomes play the same role as the Bohmian configurations in Section \ref{Section:maximally entangled states dBBT advantage}. By defining $\alpha:=\sin\frac{\theta}{2}, \beta:=\cos\frac{\theta}{2}$ the identical situation as in the Bohmian treatment of spin is recovered. That is, Wigner's measurement is a POVM with elements $E_x=\frac{1}{2}(\mathds{1}+(-1)^x \hat{n}\vec{\sigma})$, and outcome probabilities  $p(x)=\bra{\psi}E_x\ket{\psi}=\frac{1}{2}(1+(-1)^x\sin\vartheta)$, with Wigner's outcomes $x\in\left\{A',B'\right\}$. Since Wigner has information about the quantum state, only these probabilities are his best prediction. However, since the friend has access to the Bohmian configuration too, he can make a deterministic prediction: A fraction of $\frac{1}{2}(1+\sin\vartheta)$ initial Bohmian configurations yield outcome $A'$, and a fraction of $\frac{1}{2}(1-\sin\vartheta)$ yields outcome $B'$.\footnote{Incidentally, this shows that the result of the friend's device cannot be a single configuration, for the deterministic outcome could otherwise not be encoded in the spatial degree of freedom of the particle.} 

But as a result, if there were agreement on the validity of the guidance equation, the friend, in fact, is in a position to make a deterministic prediction about Wigner's outcomes, whereas Wigner can only predict the standard quantum probabilities for arbitrary measurements. Thus, at least so it seems, having predictive advantage over standard quantum theory. The Bohmian predictions moreover entail that the theory satisfies `intersubjectivity' as introduced in Section \ref{section:predictability and measurement problem}. If Wigner measures in the `record checking' basis (i.e. $\alpha=1$), then his outcomes coincide with the friend's records. The friend can, therefore, indeed predict Wigner's result with certainty. Likewise, for arbitrary bases, the friend can also employ the Bohmian guidance equation and work out the deterministic outcome of Wigner's measurements. Since the friend is an internal observer, the possibility for predictive advantage relative to Wigner does not per se lead to inconsistencies.    

The relevant question is, however, whether, based on his records, Wigner's friend can select out an ensemble of identical states into sub-ensembles for predictive advantage according to the definition of an empirical extension (cf. Definitions \ref{def:empirical completeness 1} and \ref{def:empirical completeness 2} as well as Figures \ref{fig:empirical-decomposition} and \ref{fig:wigners-friend-predictive-advantage}). First, recall that in quantum equilibrium, the Bohmian theory is signal-local on the manifest domain. From the preceding considerations, it then must follow that some mechanism in theory prevents the friend from signalling even when a deterministic prediction about Wigner's outcomes can be made. For otherwise, in bi-partite experiments, the possibility of deterministic predictions was shown to violate signal-locality for all initial Bohmian configurations. Moreover, since quantum equilibrium is preserved under unitary evolution, it will also have to be the case that the friend's records cannot be employed to create quantum \textit{non}-equilibrium out of a Bohmian universe in equilibrium. This points in the direction that the friend's record may not be useful for constructing an empirical extension. As alluded to before, for any external observer relative to which an effective wave function exists, in particular, Wigner, the predictive advantage of the friend, including his memory, can never be made manifest in the environment and thus be communicated. When Wigner performs a measurement on the friend, the friend's records and thus memory will be erased. For in equilibrium de Broglie-Bohm theory satisfies the quantum decorrelation principle. Therefore, the friend's alleged predictive advantage is only available to him and can't be used to construct a genuine empirical extension.\footnote{Notice that this, in turn, also spoils any hopes to dispel the vast underdetermination of the Bohmian trajectories. No empirical statistics could ever be gathered to single out one version of the guidance equation over another.} The details of how exactly the internal states are disturbed depending on Wigner's measurement choice I leave to future work. 

\section{Conclusion}

I discussed the case of pilot wave theory regarding predictive advantage. I expounded how --- in equilibrium --- the theory is merely a metaphysical extension of standard quantum mechanics. Thus, the Bohmian framework entails that no manifest information can, in principle, be present in the environment of a target system such that measurement uncertainty would be reduced with respect to the quantum state. This is a consequence of the fact that the quantum probabilities are recovered from 1) the existence of an effective Bohmian wave function, 2) initial Bohmian configurations being in quantum equilibrium, 3) the probability preserving guidance equation, and 4) the coincidence between manifest and non-manifest Bohmian configurations.  

For predictive advantage to exist in a Bohmian universe, the quantum decorrelation principle has to be violated. I showed how deterministic empirical extensions can be constructed by drawing on the details of Bohmian configurations out of equilibrium. Violation of the quantum decorrelation principle in pilot wave theory can be achieved by considering non-equilibrium Bohmian distributions, where the quantum probabilities are recovered for the target system but not for the joint system of target and environment. I showed that every single pair of Bohmian particles in a Stern-Gerlach experiment is non-local and depends on the settings of both primary and secondary observers. This holds for arbitrary bi-partite states, both maximally and non-maximally entangled. Thus, no deterministic Bohmian predictive advantage can exist without violation of signal-locality. 

By looking at non-deterministic predictions through non-equilibrium preparations restricted only by the assumption of signal-locality, Valentini's proofs imply that no out-of-equilibrium predictive advantage is possible for maximally entangled bi-partite spin-$\frac{1}{2}$ systems. However, signal-locality is generally not sufficient to rule out predictive advantage, as the models mentioned show. The predictive advantage is possible without violating signal-locality (see \citealt{ELITZUR199225}, \citealt{Ghirardi-Romano-inequivalent}) for at least some subsets of predictions (see the explicit toy theory by Elitzur, Popescu, and Rohrlich in Section \ref{section:partly deterministic extensions}). I suspect that by coarse-graining the deterministic Bohmian empirical extension discussed here, a similar conclusion can be reached for pilot wave theory for non-maximally entangled states. That is, Bohmian (non-deterministic) predictive advantage should in principle exist for this case. Regarding deterministic predictive advantage for non-maximally entangled states, pilot wave theory is special since the theory doesn't allow local sets of Bohmian configurations (which is what the models, as mentioned earlier, do establish). 

I point out that Valentini's claims thus have to be qualified that all non-equilibrium Bohmian distributions give rise to signalling (\citealt{VALENTINI2002273}, \citealt{Valentini-H-theorem-II}). In his proof, it is tacitly assumed that the state is maximally entangled. In line with Valentini's idea, nevertheless, \textit{deterministic} Bohmian predictive advantage through non-equilibrium leads to signalling for arbitrary entangled states as outlined above. 

It would also be interesting to investigate predictive advantage through violation of 3) and 4), when the guidance equation is modified or the relationship between non-manifest Bohmian configurations and manifest observable configurations is given up. In the latter case, it would mean that, at least sometimes, the positions of Bohmian particles don't coincide with the observable manifest configurations. The former would require an alternative guiding equation not conforming with the continuity equation and thus not preserving quantum probabilities. The latter, for example, could be conceived of as becoming directly aware of Bohmian positions without using a measurement such that the Bohmian positions are not constrained by the quantum decorrelation principle. In both cases, it seems difficult to argue in what regime the relevant violation should occur, but further work is necessary.  

\chapter[Quantum Non-Locality and Predictability]{Quantum Non-Locality and Predictability}
\chaptermark{Non-Locality and Predictability}
\label{section:Non-locality and quantum predictability}
 
The idea to connect locality with quantum completeness dates back to Einstein, Podolsky and Rosen, who argued for the incompleteness of quantum theory based on the existence of entangled states and their non-local properties \citep{einstein1935can}. If quantum measurement outcomes aren't further pre-determined by more than the probabilistic quantum prediction, they must be non-local to recover the perfect correlations of entangled states. Supplementing the quantum description with so-called hidden variables could then perhaps restore determinism, thus completing the theory to void what is often called `action at a distance'. But by Bell's theorem, any such completions must be non-local too. 

As shown in Chapter \ref{section:de Broglie-Bohm theory and predictive advantage}, the concrete case of pilot wave theory has proven useful to assess the connection between the theory's predictability and non-locality. Pilot wave theory reproduces the quantum probabilities and its dynamics are explicitly non-local. On the other hand, arbitrary empirical extensions  --- as introduced in \ref{eqn:general bi-partite empirical extension}, are defined by a set of probability distributions \textit{without} a dynamics. The additional variables $z$ are not concretely defined nor have dynamics. That is, it will be more difficult to make claims about the degree of non-locality of empirical extensions and thus has to be implicit. Therefore, a natural strategy to express the amount of non-locality thereof is by resorting to Bell-type inequalities.  

In this chapter, I shall examine some well-known attempts to generalise the results of Chapter \ref{section:de Broglie-Bohm theory and predictive advantage} on pilot wave theory: First, no signal-local empirical extension can be fully deterministic. That is, at least some measurement outcomes must be indeterminate if signal-locality is satisfied. Secondly, no signal-local empirical extension exists for the case of maximally entangled states. This implies that quantum predictions are empirically complete for this set of quantum states. Thirdly, the predictions for non-maximally entangled states permit signal-local --- even partly deterministic --- decompositions. However, by coupling the bi-partite system to another system, the results by \citeauthor{colbeck2011no} establish empirical completeness for this case too under the assumption of parameter independence. Finally, I will comment on the fact that parameter independence is insufficient for a general proof of quantum empirical completeness (including all types of systems such as single target systems). Nevertheless, these theorems, in particular the one by Colbeck and Renner, may be reformulated to give a strong proof of empirical completeness on the grounds of signal-locality. 


\section{Quantum Predictions for Bi-Partite Systems}

I start with a characterisation of the space of density operator states for bi-partite 2-dimensional Hilbert spaces over the complex numbers to study the empirical completeness problem for this case. Physical systems that are described by this space are, for instance, pairs of spin-$\frac{1}{2}$ particles or quantum bits.\footnote{Although more general proofs exist for the claims I will be making for arbitrary high dimensions, I will be concerned only with the simpler case of a pair of 2-dimensional quantum systems.} Furthermore, I classify the space of all possible generalised measurements, i.e. POVMs on that space. This will serve as the most generic and simplest description available for studying bi-partite quantum systems, their possible correlations and, in particular, empirical completeness. I will then look at theories that preserve the quantum mechanical predictions for these classes of systems but may invoke hypothetical variables leading to predictive advantage as introduced in Chapter \ref{section:the empirical completeness problem}. I consider empirical extensions of quantum theory and investigate their compatibility with familiar locality assumptions such as signal-locality. Thus, the claims made in the case of non-equilibrium pilot wave theory generalise to arbitrary empirical extensions. 

Recall the three basic quantum postulates from Section \ref{section:quantum theory and quantum predictions} formulated on the manifest domain. Applied to the pair of quantum systems, here they are: 

\begin{itemize}
	
	\item The physical state of the quantum system is represented by a density operator $\rho\in\mathcal{S}(\mathcal{H}\otimes\mathcal{H})$ as the (mixed) state of the system, where $\mathcal{S}(\mathcal{H}\otimes\mathcal{H})$ is a convex subset of the self-adjoint trace-class operators on $\mathcal{H}=\mathbb{C}^2\otimes\mathbb{C}^2$.  
	
	\item To each physical feature $Q$ is associated a POVM with self-adjoint operators $E_i$ (the observables). The index $i$ is the manifest pointer configuration and corresponds to the possible outcomes of the apparatus for measurement of $Q$. The probability of finding outcome $(i,j)$ in an experiment for the pair of systems is obtained by the Born rule
	\begin{equation}
		p(i,j):= \tr(E_i \otimes E_j \rho),
	\end{equation} since we consider two spatially separated non-interacting systems, i.e. the total Hamiltonian is the sum of two commuting Hamiltonians for the individual systems. Thus, all measurements will have the form $E_i \otimes E_j$. 
\end{itemize}

Moreover, generic states $\rho$ evolve in time according to the Schrödinger evolution, or the more general Liouville-von Neumann equation for open systems. The possible measurements are determined by the quantum interaction with the measurement apparatus and need to satisfy the probability requirements dictated by the Born rule: 

\begin{equation}
	\label{POVM definition}
	\mymathbb{0}\leq E_i \leq\mathds{1}, \text{and} \sum\limits_i E_i =\mathds{1},
\end{equation} by virtue of probability theory. 

Since the Hilbert space of the individual systems is $\mathbb{C}^2$, all operators on that space can be represented by $2\times 2$ matrices. Thus, the self-adjoint operators are elements in $\spn\{\sigma_0,\sigma_x, \sigma_y, \sigma_z\}$, where $\sigma_0=\mathds{1}$ and $\sigma_x, \sigma_y, \sigma_z$ the familiar traceless Pauli matrices. Every POVM element can therefore be written as 
\begin{equation}
	E_i= \frac{1}{2}(\alpha_0^i\mathds{1}+\bm{\alpha}^i\cdot\bm{\sigma}),
\end{equation} where we defined $\bm{\alpha}^i=(\alpha_x^i,\alpha_y^i,\alpha_z^i)$ and $\bm{\sigma}=(\sigma_x, \sigma_y, \sigma_z)$. The factor $\frac{1}{2}$ is a useful convention. Each operator $E_i$ is positive semi-definite if and only if $\alpha_0^i\geq \alpha_z^i$ and $|\alpha_0^i|^2\geq |\bm{\alpha}^i|^2$. By using the defining property \ref{POVM definition}, the real constants are further restricted to $\sum_i \alpha_0^i=2$ and $0\leq\alpha_0^i\leq 2$.  If the POVM is a projective measure, i.e. $E_i^2=E_i$, then by comparing the constants in this equation and using $\tr(E_i^2)=\tr(E_i)$, we get two possibilities:  Either $\alpha^i_0=2, |\bm{\alpha}^i|^2=0$, which corresponds to the trivial measurement with a single operator $E_0=\mathds{1}$. Or, $\alpha^i_0=1, |\bm{\alpha}^i|^2=1$. In this case the POVM has exactly two elements as $\sum_i \tr(E_i)=2$. Analogously, one can study a representation of the quantum states $\rho_S$ on each subsystem with Hilbert space $\mathbb{C}^2$. Generic mixed states are density operators, i.e. $2\times2$ positive self-adjoint matrix operators that have trace $1$. Thus, they are represented by the same operator space as POVMs with the further requirement of the trace. That is, 

\begin{equation}
	\rho_S= \frac{1}{2}(\mathds{1}+\bm{n}^i\cdot\bm{\sigma}_i),
\end{equation} with $\bm{n}=(n_x,n_y,n_z)$, $|\bm{n}|^2\leq 1$. The density operator is pure if its spectral decomposition has only one eigenvalue, or, equivalently $\tr(\rho_S^2)=1$, i.e.  $|\bm{n}|^2=1$. For the fully mixed state, one has $\bm{n}=\bm{0}$. 

More concretely, as in the Bohmian case, the states and measurements I'm interested in for the present considerations are the following: The system is described by bi-partite (pure) quantum states in their Schmidt basis parametrised by some (real) parameter $\alpha$, i.e.

\begin{equation}
	\label{eqn:arbitrary entangled bipartite state}
	\ket{\Psi}=\alpha\ket{00'}+\sqrt{1-\alpha^2}\ket{11'}, 
\end{equation} and corresponding density operator $\rho=\ket{\Psi}\bra{\Psi}$. A subsystem's state is then recovered by the partial trace $\rho_S=\tr_{S'}(\rho)$. The measurements are POVMs $E_i\otimes E'_j$ on the bi-partite state, parametrised by the angles $\vartheta_1, \vartheta_2$.\footnote{Notice also that the joint POVM corresponds to local observables since the measurements are non-entangled.} Again, for simplicity, the basis states of the two sub-systems are assumed to coincide such that both angles are rotations around the $\hat{y}$ spin axis (cf. the analysis in Chapter \ref{section:de Broglie-Bohm theory and predictive advantage}). Formally, this means we have $E_i=\frac{1}{2}(\mathds{1}+(-1)^i\sigma_n)$, where 

\begin{equation}
	\sigma_n:= \hat{n}\vec{\sigma}= 
	\begin{pmatrix}
		\cos\vartheta & \sin\vartheta  \\
		\ \sin\vartheta  & -\cos\vartheta \\
		
	\end{pmatrix}, 
\end{equation} for the measurement axis $\hat{n}=(\sin\vartheta, 0, \cos\vartheta)$.

\section{Post-Quantum Theories and Predictability}

I shall now investigate what happens if we only keep the empirical content of quantum theory, i.e. the possible correlations, and drop its prediction-making algorithm in terms of preparations and measurements of quantum states.  

For clarity, let's denote the manifest outcomes of measurements on the primary and secondary systems as $x$ and $y$ configurations and the possible observables, i.e. interactions and settings, as $a$ and $b$. The set of all quantum correlations is then 

\begin{equation}
	p(x,y|a,b,\rho)=\tr(E_x^a\otimes E_y^b \rho),
\end{equation} with the POVMs and states as defined before. 

If the outcomes were completely \textit{unrestricted}, they would only need to satisfy a probability requirement, i.e.

\begin{equation}
	\sum_{x,y} p(x,y|a,b)=1,
\end{equation} for all settings $a,b$. Such unrestricted distribution could, in principle, show all sorts of unphysical dependencies on the settings $a,b$, such as signalling. When imposing \textit{signal-locality} the correlations further satisfy
\begin{align}
	p(x|a,b)&=p(x|a),\\
	p(y|a,b)&=p(y|b), 
\end{align} where the marginals $p(x|a,b)=\sum_y p(x,y|a,b)$ and $p(y|a,b)=\sum_x p(x,y|a,b)$. I neglect to index the state $\rho$ when it is obvious which one is being used for making predictions. 

\section{Empirical Extensions for Bi-Partite Systems}

By the definitions of Chapter \ref{section:the empirical completeness problem}, an empirical extension for a bi-partite system is then a theory $T$ where the \textit{same} system is described by some states represented by configurations $z\in \mathcal{M}$, and it holds that $p(x,y|a,b,z)\neq p(x,y|a,b,\rho)$ for some $x,y$, and it reproduces the quantum probabilities for the bi-partite system, i.e.
\begin{equation}
	\label{eqn:general bi-partite empirical extension}
	p(x,y|a,b,\rho)=\int p(x,y|a,b,z)\mu(z)d z,
\end{equation} where $\mu(z)$ is some probability distribution over the target system states of theory $T$. This can be seen as a decomposition of the quantum probabilities indexed by the variable $z$, and the question is what their properties are. Based on how the probabilities decompose, sub-ensembles could be created for any $z$ predictively inequivalent to the Born statistics. Note that $T$ could be viewed as a special so-called hidden variable theory. The additional variables $z\in\mathcal{M}$ are manifest configurations and, therefore, not hidden. But they do (partly) complete the quantum state in the sense of adding accessible properties that improve the theory's predictions.  Strictly speaking, this also doesn't make it an ontological model as commonly understood since the ontic states are not \textit{a priori} assumed to be accessible. Moreover, I don't assume that the $z$-variables contain all the information about the physical state prepared as is standardly supposed in ontological models (cf. \citealt{Spekkens-ontological-models}). The manifest configurations leading to predictive advantage may, in principle, reflect only very little information about hypothetical physical states with predictive advantage. If no empirical extension of this sort exists, the quantum description is called empirically complete. Conversely, if an empirical extension exists, we say that theory $T$ has a predictive advantage over quantum theory. 

Alternatively, this can be thought of as a magic prediction box subject to a description by this empirical extension such that it produces manifest outcomes $z$ conditioned on which the predictions differ from the quantum probabilities. In other words, the distribution $p(x,y|a,b,z)$ introduces a decomposition of the quantum probabilities $p(x,y|a,b,\rho)$. For any $z$, a new prediction is obtained that differs from the quantum probabilities for at least one value of $z$. By Definition \ref{def:empirical completeness 2} in Chapter \ref{section:the empirical completeness problem} this means that conditioning on $z$ realises a selection of sub-ensembles of states for which the corresponding outcome averages $\langle Q\rangle_T$ are distinct from the quantum averages $\tr(E_Q\rho)$ for at least some measurement $E_Q:=E_x^a\otimes E_y^b$.

In order to avoid signalling the predicted probabilities of theory $T$ have to satisfy 

\begin{align}
	\label{eqn:signal locality}
	p(x|a,b,z)&=p(x|a,z)\\ \nonumber
	p(y|a,b,z)&=p(y|b,z), 
\end{align} Note that these requirements do not express the standard notion of parameter independence. The variables $z$ are directly observable manifest configurations and can therefore be used to select and prepare ensembles displaying non-quantum behaviour. In this sense, the difference between locality in terms of parameter independence and signal-locality is the status of the variables or states employed for prediction making. Here I am concerned with variables in the manifest domain. Thus, Equation \ref{eqn:signal locality} effectively is a signal-locality assumption. But whenever no assumption is made about the status of the relevant variables, it may express parameter independence. In Chapter \ref{section:de Broglie-Bohm theory and predictive advantage} my analysis of Bohmian predictions shows that the theory implies no predictive advantage while being parameter \textit{dependent}. As a result, in such a case, this means the variables featured in the hypothetical prediction-making algorithm were shown to be inaccessible. The significance of distinguishing signal-locality and parameter independence will appear again when assessing the relevance of the theorem by \citeauthor{colbeck2011no} for a general proof of empirical completeness.   

In the next sections, I examine the potential predictive advantage of theory $T$ when subject to signal-locality. We find that, generally, predictive advantage exists for non-maximally entangled states, whereas it doesn't for maximally entangled states.  

\sectionmark{No Fully Deterministic Extensions}
\section[No Fully Deterministic Signal-Local Empirical Extensions]{No Fully Deterministic Signal-Local Empirical Extensions} \sectionmark{No Fully Deterministic Extensions}
\label{section:no fully deterministic extensions}

Most of the early ideas on completeness of quantum theory were concerned with the possibility of \textit{complete} pre-determination and \textit{deterministic} predictability of quantum phenomena. But more generally, an empirical extension may be probabilistic and still entail predictive advantage over standard quantum predictions.   

The case of deterministic predictive advantage in pilot wave theory allows for a simple generalisation. 
Let $p(x,y|a,b,z)$ again be the predictions of an empirical extension. A deterministic prediction entails that the outcomes $x,y$ are functions of the settings $a,b$ and extension variable $z$. That is, $(x,y)=(f(a,b,z), g(a,b,z))$, and so 

\begin{align}
	\label{eqn:deterministic predictions}
	p(x,y|a,b,z)&=\delta_{(x,y),((f(a,b,z),g(a,b,z)))}\\ \nonumber
	&=\delta_{(x),f(a,b,z)}\delta_{(y),g(a,b,z)}\\ \nonumber
	&=p(x|a,b,z)p(y|a,b,z). 
\end{align} But when on top of that, the probabilities are to satisfy signal-locality, then 
\begin{equation}
	\label{eqn:signal-local deterministic predicitons}
	p(x,y|a,b,z)=p(x|a,b,z)p(y|a,b,z)=p(x|a,z)p(y|b,z). 
\end{equation} Thus, the predictions describe uncorrelated and signal-local correlations that neither depend on the outcomes of the other site nor its settings. But such cannot reproduce all quantum correlations due to Bell's theorem. Therefore, $p(x,y|a,b,z)$ cannot be (fully) deterministic (cf. also \citealt{PR-boxes}, \citealt{Masanes}, and \citealt{Cavalcanti-predictability}).\footnote{As is well-known, in the context of Bell non-locality, since deterministic hidden variable theories cannot violate outcome independence they cannot reproduce all quantum predictions if they are also assumed to be parameter independent.} 

Thus, we conclude that any empirical extension of quantum mechanics necessarily makes indeterministic predictions for at least some settings and outcomes. This is true in any world where correlations violate Bell inequalities, and signalling is assumed impossible. Since both maximally and non-maximally entangled states violate a Bell inequality, no empirical extension can be fully deterministic. It is still possible that theories with deterministic predictive advantage exist for at least \textit{some} measurement outcomes. For instance, the model by Elitzur et al. can introduce an empirical extension for which the outcomes of the local ensemble can be predicted deterministically \citep{ELITZUR199225}. 

The natural question arises what the minimally non-local partly deterministic extensions could be, and so it would be the aim to find a lower bound. In the following sections, I show that for the maximally entangled state, it's zero for all empirical extensions whereas, for non-maximally entangled states, it isn't. 

Then the question remains if the unavoidable empirical randomness is quantum, i.e. reproducing the standard quantum predictions. The next sections show that non-trivial indeterministic empirical extensions exist in certain cases. Hence, other physical principles are required to explain why quantum theory would be empirically complete. 

\section[Partly Deterministic Signal-Local Empirical Extensions]{Partly Deterministic Signal-Local Empirical Extensions}
\sectionmark{Partly Deterministic Extensions}
\label{section:partly deterministic extensions}

The mere existence of some non-locality in quantum correlations shows the impossibility of signal-local empirical extensions that are \textit{fully} deterministic. In other words, no empirical extension exists for which all outcomes are predicted with certainty. A natural next step is to study whether there perhaps exist \textit{partly} deterministic signal-local extensions. That is, most predictions are probabilistic whilst \textit{some} are deterministic.

To achieve this, the quantum predictions are split into signal-local deterministic and non-local parts. Formally, a sufficient condition for a signal-local (partly) deterministic predictive advantage to exist is, if the quantum correlation \ref{eqn:general bi-partite empirical extension} can be written as a decomposition of the following form

\begin{equation}
	p(x,y|a,b,\rho)=p\cdot p_L(x,y|a,b)+(1-p)\cdot p_{NL}(x,y|a,b),
\end{equation} where $p_L(x,y|a,b)$ the signal-local deterministic correlation, and $p_{NL}(x,y|a,b)$ the non-local indeterministic correlation. The idea to study local parts of quantum correlations and rewrite them this way was first proposed by \cite{ELITZUR199225}. The decomposition can be represented as an empirical extension as follows
\begin{align}
	p(x,y|a,b,\rho)&=\int p(x,y|a,b,z)\mu(z)d z\\ \nonumber
	&=p\int p_L(x,y|a,b,z)\mu_L(z)dz\\ \nonumber
	&+(1-p)\int p_{NL}(x,y|a,b,z)\mu_{NL}(z)dz,
\end{align} That is, the manifest variables of the magic prediction box are prepared with probability $p\cdot \mu_L(z)dz$ for the signal-local deterministic predictions and with probability $(1-p)\cdot \mu_{NL}(z)dz$ for all other signal-local but indeterministic predictions. Notice that $\mu_L(z),\mu_{NL}(z)$ are disjoint. If $p$ is non-vanishing and such a model can be shown to exist, then signal-local deterministic predictive advantage is conceivable.  

By Equations \ref{eqn:deterministic predictions} and \ref{eqn:signal-local deterministic predicitons} every probability distribution in the first term is equivalent to 

\begin{equation}
	p_L(x,y|a,b,z)=p_L(x|a,z)p_L(y|b,z). 
\end{equation} Since the second term is signal-local but non-local on the manifest level, the probabilities don't factorise for either manifest or non-manifest variables. 

What can be inferred about the existence of the deterministic signal-local part? To study the properties of the decomposition, let's introduce a generalised Bell inequality for arbitrary numbers of measurements. In order to do this let $x,y \in \{0,1\}$ be the possible manifest outcomes of $N$ different measurements with primary settings $a:=\vartheta_k$, $k=1, ..., N$ and secondary settings $b:=\vartheta_l$, $l=1,..., N$ and consider the measure

\begin{align}
	I_N:= &\langle|x_{\vartheta_1}-y_{\vartheta_1}|\rangle+\langle|y_{\vartheta_1}-x_{\vartheta_2}|\rangle
	+ \langle|x_{\vartheta_2}-y_{\vartheta_2}|\rangle\\
	&+ ... + \langle|x_{\vartheta_N}-y_{\vartheta_N}|\rangle+\langle|y_{\vartheta_N}-x_{\vartheta_1}-1|\rangle,\nonumber
\end{align} where the expectation value is defined as $\langle A\rangle:=\sum_{i=0}^{1} i p(A=i)=p(A=1)$. By defining $x_{\vartheta_{N+1}}:=x_{\vartheta_1}+1 ~(\bmod~ 2)$ it can be written as $I_N=\sum\limits_{n=1}^N  (\langle|x_{\vartheta_n}-y_{\vartheta_n}|\rangle+\langle|y_{\vartheta_n}-x_{\vartheta_{n+1}}|\rangle)$. Up to minor details this expression corresponds to the so-called chained Bell inequalities first introduced by \citep{BRAUNSTEIN-caves-chained-bell-inequalities}. See in particular the generalisation to arbitrary dimensions in \citep{Barrt-Kent-Pironio-local-decomposition} who I closely follow in the subsequent presentation. This is the generalised Bell inequality.\footnote{For $N=2$ it coincides with the familiar CHSH inequality.}  

For the signal-local deterministic term, one has $I_N(L)\geq 1$. To see this, observe that 
\begin{align}
	&|x_{\vartheta_1}-y_{\vartheta_1}|+|y_{\vartheta_1}-x_{\vartheta_2}|
	+ |x_{\vartheta_2}-y_{\vartheta_2}|+ ... + |x_{\vartheta_N}-y_{\vartheta_N}|+|y_{\vartheta_N}-x_{\vartheta_1}-1|\geq \\ &|x_{\vartheta_1}-y_{\vartheta_1}+y_{\vartheta_1}-x_{\vartheta_2}
	+x_{\vartheta_2}-y_{\vartheta_2}+ ... + x_{\vartheta_N}-y_{\vartheta_N}+y_{\vartheta_N}-x_{\vartheta_1}-1|=1. \nonumber
\end{align} Moreover, each term in $I_N$ is positive and, therefore, $I_N(NL)\geq 0$. In turn, this can be used to bound the quantity $p$, i.e. since the Bell measure is a sum of probability, we can write  $I_N(QM)=pI_N(L)+(1-p)I_N(NL)$. From which, it follows that 

\begin{equation}
	p\leq I_N(QM).
\end{equation} (cf. also Barrett et al. for arbitrary dimensions). 

I now compute the value of $I_N$ for arbitrary states and will show that it vanishes for a special set of measurement settings in the case of maximally-entangled states. For the expectation values in the sum, only those quantities are relevant for which $|x_{\vartheta_k}-y_{\vartheta_l}|=1$. Since all outcomes can be $0$ or $1$, this is satisfied whenever the outcomes are unequal. There are two cases for which this is the case, and the corresponding measurements are $E_1^{\vartheta_k}\otimes E_0^{\vartheta_l}$ and $E_0^{\vartheta_k}\otimes E_1^{\vartheta_l}$. Each term (except the last) in the sum is therefore given by

\begin{equation}
	\langle|x_{\vartheta_k}-y_{\vartheta_l}|\rangle=p(|x_{\vartheta_k}-y_{\vartheta_l}|=1)=\tr(E_1^{\vartheta_k}\otimes E_0^{\vartheta_l}\rho)+\tr(E_0^{\vartheta_k}\otimes E_1^{\vartheta_l}\rho). 
\end{equation} This holds for all terms but the last, where the argument is $1$ if and only if both outcomes are equal. For the latter case, the relevant probability is thus

\begin{equation}
	\langle|x_{\vartheta_k}-y_{\vartheta_l}-1|\rangle=p(|x_{\vartheta_k}-y_{\vartheta_l}-1|=1)=\tr(E_0^{\vartheta_k}\otimes E_0^{\vartheta_l}\rho)+\tr(E_1^{\vartheta_k}\otimes E_1^{\vartheta_l}\rho). 
\end{equation}  A simple calculation for when $\rho$ is maximally entangled (i.e. $\alpha=\frac{1}{\sqrt{2}}$) yields that all these terms are equal to $\sin^2(\frac{\vartheta_k}{2}-\frac{\vartheta_l}{2})$, and the last term is equal to $\cos^2(\frac{\vartheta_k}{2}-\frac{\vartheta_l}{2})$.\footnote{This can also be read off directly from the quantum state (\ref{eqn:maximally entangled state in arbitrary basis}) by squaring and summing the relevant amplitudes.} 

We can now choose a set of measurement angles for which the internal $\left[0,\pi\right]$ is divided into $N$ equally spaced settings for both the primary and secondary measurement. That is, define, for example, the primary setting as 

\begin{equation}
	\vartheta_k:=\frac{\pi}{N}(k-1), k=1,..., N,
\end{equation} and the secondary setting as 

\begin{equation}
	\vartheta_l:=\frac{\pi}{N}(l-\frac{1}{2}), l=1,..., N.
\end{equation} With these definitions each term in $I_N$ is identical and reads $\sin^2\frac{\pi}{4N}$. Hence, there are $2N$ equal terms in the sum and 

\begin{equation}
	I_N= 2N\sin^2\frac{\pi}{4N}.
\end{equation} Clearly, this implies that in the limit of infinite measurement settings 
\begin{equation}
	\lim\limits_{N\rightarrow \infty}I_N=0, 
\end{equation} (cf. \citealt{Barrt-Kent-Pironio-local-decomposition}). It immediately follows that $p=0$ in the decomposition of quantum probabilities. Thus, no (partly) deterministic empirical extension can exist.\footnote{Strictly speaking, it only means that if $p$ vanishes, every individual preparation of the entangled systems is non-local. But it doesn't show that there can't exist a non-local preparation for which some predictions are deterministic for a proper subset of measurement settings (cf. \citealt{Cabello-model-Barret} for a model like that). However, the result from Barrett et al. presented in the next section also rules out this possibility.} This is in line with the analysis in the previous chapter and Elitzur, Popescu, and Rohrlich's observation that \textit{every} pair of particles in a maximally entangled state is, in fact, non-local.

Note the relevance of an infinite number of settings again. In the finite case, signal-local empirical extensions exist for maximally entangled states since $I_N$ doesn't vanish. For instance, in the familiar Bell experiment where $N=2$, the correlations can be reproduced by a convex combination of PR-boxes (see \citealt{PR-boxes} and \citealt{Barret-PR-boxes}). Another model that simulates maximal quantum entanglement as convex combinations of local correlations and non-local correlations was proposed in \citealt{Gisin-simulating-wihth-PR}.

For arbitrary entangled states, we arrive at a different conclusion, however. For each term except the last the analogous computation gives  $\sin^2(\frac{\vartheta_k}{2}-\frac{\vartheta_l}{2})-(\alpha\sqrt{1-\alpha^2}-\frac{1}{2})\sin\vartheta_k\sin\vartheta_l$, and the last term $\cos^2(\frac{\vartheta_1}{2}-\frac{\vartheta_N}{2})+(\alpha\sqrt{1-\alpha^2}-\frac{1}{2})\sin\vartheta_1\sin\vartheta_N$. To compute $I_N$ define for convenience the primary and secondary measurement angle as $\vartheta_n, \vartheta'_n$, and $\vartheta_{N+1}:=\vartheta_1-\pi$. Summing up gives the generalised Bell measure for non-maximally entangled states, i.e.

\begin{align}
	I_N&=\sum\limits_{n=1}^N  (\langle|x_{\vartheta_n}-y_{\vartheta_n}|\rangle+\langle|y_{\vartheta_n}-x_{\vartheta_{n+1}}|\rangle)\\
	&=\sum_{n=1}^{N}\left[\sin^2(\frac{\vartheta_n}{2}-\frac{\vartheta_n'}{2})+\sin^2(\frac{\vartheta'_n}{2}-\frac{\vartheta_{n+1}}{2})\right]\\ \nonumber
	&-(\alpha\sqrt{1-\alpha^2}-\frac{1}{2})\sum_{n=1}^{N}\left[\sin\vartheta_n\sin\vartheta_n'+\sin\vartheta_n'\sin\vartheta_{n+1}\right]\nonumber
\end{align} which reduces to the maximally entangled case, i.e. for $\alpha=\frac{1}{\sqrt{2}}$. The first sum equals the generalised Bell measure for maximally entangled states and vanishes in the limit of arbitrarily many settings chosen as before. As opposed to maximally entangled states, it is not clear here, however, what set of measurements minimises the expression. Thus, to find the most partly deterministic predictive advantage (and hence the maximum value for $p$), it would be necessary to find

\begin{equation}
	p\leq \min\limits_{N, \vartheta_1, ..., \vartheta_N,\vartheta'_1, ... ,\vartheta_N} I_N,
\end{equation} where the minimisation happens over all possible sets of $2N$ measurement angles $\vartheta_{n}, \vartheta'_{n}$ and $N\in\mathbb{N}$.\footnote{For the set of measurements picked to minimise the Bell inequality for the maximally entangled state a calculation shows $I_N=2N\sin^2\frac{\pi}{4N}+N\cos\frac{\pi}{2N}$, which in fact diverges for $N\rightarrow \infty$.} 

It is rather difficult to compute the bound for $p$, but generally, it is non-zero. Hence, partly deterministic empirical extensions for non-maximally entangled spin-$\frac{1}{2}$ pairs could, in principle, be possible. Indeed, to show that a concretely worked out empirical extension exists, I shall mention the model by \cite{ELITZUR199225}. They construct a model that works for $\alpha\geq\sqrt{1-\alpha^2}$, and applied in the context used here, it can be expressed as follows. 

\begin{equation}
	p(x|\vartheta,z)= 
	\begin{cases}
		1-x ;& 0\leq \vartheta < \frac{\pi}{2}\\
		\frac{1}{2};  &\vartheta = \frac{\pi}{2}\\
		x;  & \frac{\pi}{2}<\vartheta \leq \pi,
	\end{cases}
\end{equation} where $\vartheta$ is the relative angle of the measurement axis as defined before, and the outcome $x=0,1$. In this case the $p=\frac{1}{2}(\frac{1}{2}-\alpha\sqrt{1-\alpha^2})$.\footnote{The small detail that the predictions aren't deterministic for the single angle $\vartheta=\frac{\pi}{2}$ shouldn't bother us too much.} The same definitions apply to the outcome $y$ and secondary setting $\vartheta'$. This is an (almost fully) deterministic model which is signal-local, i.e. $p_L(x,y|\vartheta,\vartheta',z)=p(x|\vartheta, z)p(y|\vartheta', z)\in\{0,1\}$ for all settings $\vartheta$ but one. 


To conclude, signal-local (partly) deterministic empirical extensions of the form introduced here do, indeed, exist for non-maximally entangled bi-partite states. In this approach, signal-locality alone isn't a sufficient assumption to arrive at the empirical completeness of quantum probabilities. 

\section{Signal-Local Empirical Extensions}

The use of generalised Bell inequalities led to the result that deterministic predictive advantage is bound by locality for generic quantum states. In other words, the more entangled a bi-partite quantum system is, the less deterministic predictive advantage is possible. In the case of maximally entangled states, there exists none. In the previous section, the decomposition of quantum correlations into signal-local deterministic and non-local components was shown to be non-trivial in the case of non-maximally entangled states. This allows partly deterministic predictive advantage. Moreover, a concrete model can be provided for such empirical extensions. 

The questions we are left with are 1) Can signal-local \textit{indeterministic} empirical extensions as well be ruled out in general for maximally entangled systems, and 2) Does there exist an alternative approach towards empirical completeness based on signal-locality for non-maximally entangled systems? 

Indeed, the answer to both questions seems to be yes. First, a theorem by \citeauthor{Barrt-Kent-Pironio-local-decomposition} establishes that any signal-local decomposition of quantum probabilities is trivial for maximally entangled states, i.e. all predictions have to coincide with the Born probabilities. This, in fact, shows that no predictive advantage (of any kind) can exist for such systems. Secondly, the results by \citeauthor{colbeck2011no} on the impossibility of `extensions with improved predictive power' for non-maximally entangled states utilising stronger assumptions can be re-interpreted as a proof for empirical completeness from signal-locality alone. I shall turn to the former in this section and the latter in the next.

\subsection{Empirical Completeness for Maximally Entangled Systems}

Barrett et al.'s crucial idea is to employ the locality measure $I_N$ to bound individual outcome probabilities of signal-local correlations. They show that for every decomposition of quantum probabilities, all outcomes are undetermined and equally likely.  

With the standard definitions, it works like so. Let $p(x,y|a,b,z)$ be the predictions of a hypothetical empirical extension, i.e. $\int p(x,y|a,b,z)\mu(z)$. Since we require them to be signal-local, the marginals $p(x|a,z)=p(x|a,b,z)=\sum_{y}p(x,y|a,b,z), \forall b$ and $p(y|b,z)=p(y|a,b,z)=\sum_{x}p(x,y|a,b,z), \forall a$ are well defined for all outcomes $x, y$, settings $a, b$, and preparations $z$. The aim is to prove that $|p(x|a,z)-p(1-x|a,z)|=0$, and likewise $|p(y|b,z)-p(1-y|b,z)|=0$ for all measurement settings $a, b$. As a result, all outcomes are equiprobable. Due to the probability requirement $\sum_x p(x)=\sum_y p(y)=1$, they further must equal $p(x)=p(y)=\frac{1}{2}$ if all are to be identical, and thus coincide with the corresponding quantum probabilities for maximally entangled states.   

Recall that for the averages it holds that $\langle |A|\rangle=p(|A|=1)=1-p(|A|=0)$. Therefore, the generalised Bell measure reads

\begin{align}
	I_N&=\sum\limits_{n=1}^N  (\langle|x_{\vartheta_n}-y_{\vartheta_n}|\rangle+\langle|y_{\vartheta_n}-x_{\vartheta_{n+1}}|\rangle)\\  \nonumber
	&=\sum_{n=1}^{N}\left[p(x_{\vartheta_n}\neq y_{\vartheta_n})+p(y_{\vartheta_n}\neq x_{\vartheta_{n+1}})\right]\\ \nonumber &=2N-\sum_{n=1}^{N} \left[p(x_{\vartheta_n}=y_{\vartheta_n})+p(y_{\vartheta_n}=x_{\vartheta_{n+1}})\right].
\end{align}

Since the marginal probabilities are sums of the joint probabilities $p(x,y|a,b,z)$ we have $p(x_{\vartheta_k}=y_{\vartheta_l})=\sum_{w=0}^{1}p(x_{\vartheta_k}=w, y_{\vartheta_l}=w)\leq\min(p(x_{\vartheta_k}=q),p(y_{\vartheta_l}=q)+\min(1-p(x_{\vartheta_k}=q),1-p(y_{\vartheta_l}=q)$ for any choice of outcomes $q=0,1$. By using absolute values to compute the minimum, i.e. $\min(x,y)=\frac{1}{2}(x+y)-\frac{1}{2}|x-y|$, this gives  

\begin{equation}
	p(x_{\vartheta_k}=y_{\vartheta_l})\leq 1-|p(x_{\vartheta_k}=q)-p(y_{\vartheta_l}=q)|.
\end{equation} Substituting into the Bell measure and defining arbitrary values $q_n$ then gives

\begin{align}
	I_N&\geq \sum_{n=1}^{N} \left[|p(x_{\vartheta_n}=q_n)-p(y_{\vartheta_n}=q_n)+|p(y_{\vartheta_n}=q_n)-p(x_{\vartheta_{n+1}}=q_n)|\right]\\ \nonumber
	&\geq \sum_{n=1}^{N} |p(x_{\vartheta_n}=q_n)-p(x_{\vartheta_{n+1}}=q_n)|,
\end{align} and by using the triangle inequality in the second step. 

We are free to choose the values for $q_n$. Defining $q_1=...=q_{k-1}=x$ and $q_k=...=q_N=1-x$ for some index $k$ and once again employing the triangle inequality yields

\begin{equation}
	I_N\geq |p(x_{\vartheta_1}=x)-p(x_{\vartheta_{k}}=x)+p(x_{\vartheta_k}=1-x)-p(x_{\vartheta_{N+1}}=1-x)|. 
\end{equation} By definition  $x_{\vartheta_{N+1}}:=x_{\vartheta_1}+1(\mod 2)$ and so 

\begin{equation}
	I_N\geq |p(x_{\vartheta_{k}}=x)-p(x_{\vartheta_k}=1-x)|= |p(x|a=\vartheta_k,z)-p(1-x|a={\vartheta_k},z)|
\end{equation} for any choice of setting $\vartheta_{k}$. The signal-local empirical extension has to recover the quantum probabilities on average, and the proof is concluded by 

\begin{align}
	&\int|p(x|a=\vartheta_k,z)-p(1-x|a={\vartheta_k},z)|\mu(z)dz\\ \nonumber
	&\leq \int I_N \mu(z)dz= I_N(QM)\rightarrow 0 ~\text{as}~ N\rightarrow \infty. 
\end{align}

The derivation is symmetric in the outcomes and settings of the two parties; hence, it equally holds for the probabilities for $y=0,1$. Thus, any prediction is independent of the preparation $z$. The result implies the quantum decorrelation principle for maximally entangled bi-partite systems and any hypothetical magic prediction box, i.e. $p(x,y|a,b,z)=p(x,y|a,b)p(z)$ for any manifest variable $z$.\footnote{For the result to go through it also needs to be assumed that the measurement settings can be chosen independently of the variables for predictive advantage, i.e. what is often dubbed \textit{settings independence}; cf. also Section \ref{section:empirical completeness for non-maximally entangled systems}} In the context of pilot wave theory it moreover implies Valentini's theorem on singlet states ruling out the existence of signal-local non-equilibrium Bohmian preparations \citep{VALENTINI2002273, Valentini-H-theorem-II}. Furthermore, the upper bound for predictive advantage for arbitrarily entangled systems quantifies the claim that in hidden variable theories, signal-locality implies unpredictability (cf. \cite{VALENTINI2002273, Cavalcanti-predictability}).  


\subsection{Empirical Completeness for Non-Maximally Entangled Systems}
\label{section:empirical completeness for non-maximally entangled systems}
The discussion so far has led to the conclusion that quantum theory is empirically complete for maximally entangled bi-partite states. That is, no predictive advantage can exist, and the quantum decorrelation principle must hold in any empirical extension if it's to be signal-local. 

Now to the problem of empirical completeness of non-maximally entangled states. There exists, in fact, a theorem that claims to have established exactly this. The results by Colbeck and Renner purportedly show that \textit{`no extension of quantum theory can have improved predictive power'.} for arbitrary quantum states and even individual systems \citep{colbeck2011no}. If the claim holds up to scrutiny in the general framework presented, it would be a robust result on the empirical completeness problem. 

Of course, the crux of the matter lies in the details of the assumptions used. The proof has been analysed and criticised, e.g. by \cite{HERMENS2020121, Leifer2014extended, LEEGWATER201618, Landsman, Ghirardi-on-Colbeck-Renner}. The criticism is mainly about Colbeck and Renner's assumptions implying parameter independence and thus making it a much weaker claim; the status of `free will' and its relationship to the central premise; moreover, lacking the foundation for addressing individual systems for which even stronger assumptions seem to be necessary. Importantly, Hermens and Landsman close some technical gaps in the proof for non-maximally entangled states. Together with those improvements, it makes Colbeck and Renner's proof a valid and sound derivation for predictive completeness of quantum probabilities in these cases. 

The Colbeck-Renner theorem proceeds on two basic assumptions. In prose, 

\begin{enumerate}
	\item[(FR)] Measurement settings can be chosen freely: `[...] the input, A, of a measurement process can be chosen such that it is uncorrelated with certain other spacetime random variables, namely all those whose coordinates lie outside the future lightcone of the coordinates of A.'
	\item[(QM)] Measurement outcomes obey quantum statistics, and all processes within quantum theory can be considered as unitary evolutions if one takes into account the environment.\footnote{They add that this needs only hold for `microscopic processes on short timescales.'} 
\end{enumerate}

The main result is then stated as the quantum probabilities being `the most accurate prediction'  of measurement outcomes. That is, for any system measured with setting $a$ and outcomes $x$, no additional information exists improving the quantum probability prediction $p(x|a)$. At this point, Colbeck and Renner mention that such additional information provided by a hypothetical `extension' of the theory is \textit{accessible} at any time, which in the present context means it must be a manifest configuration. 

The proof consists of three components. First, by invoking chained Bell inequalities and (FR) they show that quantum predictions cannot be improved for maximally-entangled states. This is essentially a reworking of Barrett et al.'s proof --- a version of which I gave above.\footnote{In particular, Colbeck and Renner quantify the correlations of quantum outcomes by the measure (adapting to the present definitions)\begin{equation}
		I_N:=p(x=y|a=\vartheta_0,b=\vartheta_{2N-1})+\sum\limits_{|j-k|=1}p(x\neq y|a=\vartheta_j,b=\vartheta_k),
	\end{equation} where the measurement settings $\vartheta_j, \vartheta_k, j\in \left\{0,2, ..., 2N-2\right\}, k\in\left\{1,3, ..., 2N-1\right\}$ are defined similarly as before to give $N$ equally distributed measurement angles for both sites. This measure corresponds to the Bell inequality measure $I_N$ introduced in the preceding analysis of Barrett et al.'s proof. In a similar fashion, they then show that this quantity restricts the outcome probabilities of bi-partite systems. Notice again that as is the case for Barrett et al.'s proof and standard Bell inequalities alike, the results by Colbeck-Renner also require an assumption on settings independence.} The second step contains the novel idea of appending an initially non-maximally entangled system to another quantum system which is evolved unitarily to a maximally entangled state to which the proof of the first part is applied.\footnote{Incidentally, \citeauthor{Ghirardi-Romano-inequivalent} argue that Colbeck and Renner's proof must be flawed by constructing models for non-maximally entangled states which are empirically inequivalent to standard quantum predictions (cf. also the previous section on this valid possibility and the concrete model of \cite{ELITZUR199225}). But these concerns, I would argue, are unwarranted. For non-maximally entangled states, Colbeck and Renner invoke an extra step by coupling the system to a further pair of systems and then show that it would lead to signalling. Thus, the proof, in this case, is indirect, and they must therefore be aware that non-maximally entangled states are special in that sense.} Thirdly, Colbeck and Renner argue that every individual quantum system can be viewed as part of a bi-partite entangled system, thereby concluding the proof for \textit{arbitrary} quantum systems. Regarding individual quantum systems, an important shortcoming of the proof was identified by \citeauthor{HERMENS2020121}. He highlights that further strong assumptions are needed to arrive at the claim for individual systems. Specifically, it has to be assumed that the process coupling of such a system and unitarily evolving it to the entangled bi-partite case is non-disturbing on the variables relevant for predictive advantage. But this seems to be generally violated (see, for instance, \citealp{maroney2017measurements}).

I will focus solely on the aspects relevant to the proof on non-maximally entangled states and argue that the result may be interpreted as an empirical completeness proof for signal-local extensions. But first, a few general comments. 

Unfortunately, Colbeck and Renner do not address unambiguously in what sense predictive advantage and `extensions' of quantum theory are to be understood. That is, they don't deal with the quantum decorrelation principle or point out in what regime if any, quantum predictions have to be violated. On the face of it, based on the two assumptions, their claim on empirical completeness appears even paradoxical. Recall the quantum decorrelation principle saying that a target system decouples from its environment whenever a pure state is assigned. If `no extension of quantum theory can have improved predictive power' \textit{while} preserving the standard quantum predictions, it is unclear what such an `extension' is supposed to look like. After all, any theory reproducing the quantum probabilities will also have to contain the decorrelation property. Even worse, assumption (QM) explicitly refers to the environment. But of course, if the environment contained hypothetical configurations leading to predictive advantage, it cannot be part of a quantum description (for otherwise, no pure state can be assigned to the target system). As I described earlier, a genuine empirical extension necessarily involves non-quantum variables for which the standard quantum predictions are violated in some regimes. An empirical extension cannot be constructed without breaking the QDP. Under the assumption that (QM) holds universally, the Colbeck-Renner theorem, strictly speaking, could be a result of the variables in the \textit{non}-manifest domain only. That is, the improved determination of measurement results on the level of some \textit{uncontrollable} (i.e. in general non-manifest) parameters. However, this is undoubtedly not what Colbeck and Renner are getting at given their statements about the information provided by possible extensions being accessible.

In the same vein, the theorem doesn't distinguish between a metaphysical and empirical extension. This highly confuses the analysis, for it is hence not entirely clear if the result pertains to manifest or non-manifest variables. But the implications for the empirical completeness problem crucially hinge on what the `prediction' improving variables represent (cf. Chapter \ref{section:the empirical completeness problem}). Thus, again it's necessary to delineate the manifest from the non-manifest in the context of this theorem to assess its relevance to predictive advantage. Contingent on what the salient variables signify, different conclusions may be drawn. I shall turn to this in the next section. I comment on the implications of treating the relevant extending variables $z$ as being either ontic (non-manifest), or manifest configurations. In light of the former option, the Colbeck-Renner theorem reads as a statement on metaphysical completeness based on Bell's parameter independence. In contrast, I argue that on the latter account, it's a result on empirical completeness based on the weaker assumption of signal-locality. This strengthens the theorem and contrasts the views commonly held in the literature and by the authors themselves. 


\subsection{Signal-Locality, Parameter Independence, and `Free Choice'}

Consider, as before, a bi-partite quantum system with outcomes $x,y$ and settings $a,b$, respectively, and a magic prediction box providing hypothetical predictive advantage though manifest variables $z$. In Colbeck and Renner's approach, it is furthermore assumed that the box takes some settings $c$. 

Formally, with these definitions, (FR) can be expressed as the assumption

\begin{equation}
	p(a|b,y,c,z)=p(a).
\end{equation} According to the authors, the measurement choice is thus independent of any other influences and can, therefore, be considered a `free choice'. By symmetry, the same is assumed for all choices of $b$ and $c$, i.e. that $p(b|a,x,c,z)=p(b)$ and $p(c|a,x,b,y)=p(c)$. 

It is then shown that (FR) leads to what they call `non-signalling' constraints. Again adopting the notation of this thesis, these are

\begin{align}
	\label{eqn:Colbeck Renner no-signalling consraints}
	p(x,y|a,b,c)&= p(x,y|a,b), \\ \nonumber
	p(x,z|a,b,c)&= p(x,z|a,c), \\ \nonumber
	p(z,y|a,b,c)&= p(z,y|c,b). 
\end{align} 

In conjunction, these three constraints imply the relevant locality condition on which the proof rests. By straightforward manipulations of conditional probabilities from the second and third conditions, we get

\begin{align}
	p(x|a,b,c,z)&= p(x|a,c,z), \\ \nonumber
	p(y|a,b,c,z)&= p(y|c,b,z). 
\end{align} 

This is daringly close to signal-locality as encountered above (when the irrelevant setting $c$ is ignored).\footnote{As it seems to me the reason for introducing the variable $c$, i.e. a setting for the magic prediction box, is to achieve symmetry in the assumptions on the outcomes and settings of all systems involved.} So does this show that the Colbeck-Renner theorem can be seen as deriving from signal-locality alone? Basically, yes, but matters are slightly more complicated, and a little more needs to be said.  

In the accounts of Hermens and Leifer, the theorem is restated in the context of ontological models, providing an unambiguous underpinning of the result. There the `free-choice' assumption (FR) straightforwardly translates to parameter independence (and settings independence), as seen below. This is implied by the fact that the relevant variables enter the story as the \textit{full} ontic (non-manifest) state $z=\lambda$ of a preparation determining all outcome probabilities of arbitrary measurements. This seems to be the view as standardly conceived (see, for instance,  \cite{Leifer2014extended,Ghirardi-on-Colbeck-Renner,HERMENS2020121,LEEGWATER201618,Landsman}). Indeed, since the non-manifest domain is essentially metaphysically unrestricted (deterministic metaphysical extensions are always possible), strong assumptions will be necessary to prove the impossibility of metaphysical and --- as a consequence --- empirical extensions. In this case, it is realised by assuming parameter independence on the non-manifest domain. However, what is of interest in the context of \textit{empirical} completeness is deriving quantum probabilities from signal-locality. Colbeck and Renner claim in their paper's supplementary information that the `non-signalling' constraints (\ref{eqn:Colbeck Renner no-signalling consraints}) imply parameter independence as understood in Bell's sense.

Moreover, they purport their assumptions to rule out de Broglie-Bohm theory. But this can only be true if (FR) does imply parameter independence which is violated by the Bohmian theory. It's opaque to me why this should be true given the fact that de Broglie-Bohm theory is signal-local and parameter \textit{dependent} (cf. Chapter \ref{section:de Broglie-Bohm theory and predictive advantage} on the inaccessibility of Bohmian configurations). As repeatedly highlighted in this thesis, signal-locality --- a property of the manifest domain --- is compatible with pilot wave theories being non-local, i.e. parameter dependent --- a property of the non-manifest domain. 
In light of their latter statement, the authors must refer to $z$ as non-manifest variables, i.e. the $\lambda$ states of a hidden variables theory.

On the other hand, as indicated in how their assumptions and main claim are stated, it seems fair to say that Colbeck and Renner are concerned with accessible manifest configurations $z$ conditioned on which quantum predictions are hypothetically improved. In other words, all variables are supposed to only refer to `directly observable objects' which can be determined operationally by clocks and rods. This, I would argue, entails that their assumption on locality, or `free choice', is a condition on the manifest domain and can thus, if anything, amount to signal-locality, but certainly not parameter independence as understood in the context of hidden variable theories. 

It needs to be clarified which way Colbeck and Renner's approach is to be understood. The challenge, therefore, is to assess the relationship between signal-locality and Colbeck and Renner's `free choice' assumption (FR). If signal-locality, as defined in the present context, suffices for the result, it would establish a general proof of empirical completeness for arbitrary bi-partite states. 

If $z$ are supposed to represent non-manifest variables $\lambda$, i.e. hidden variables supplementing the quantum state to determine the measurement outcomes. Then, the familiar notion of Bell's parameter independence is recovered. That is,

\begin{align}
	\label{eqn:parameter independence}
	p(x|a,\lambda)&=p(x|a,b, \lambda)\\ \nonumber
	p(y|b,\lambda)&=p(y|a,b, \lambda).
\end{align}

Furthermore, the standard notion of `settings independence' is recovered (in the literature, also often called `no conspiracy', `free will', or `measurement independence'):

\begin{align}
	p(a|b, \lambda)&=p(a)\\ \nonumber
	p(b|a,\lambda)&=p(b).
\end{align}

But claiming that de Broglie-Bohm theory violates the `free choice' (FR) assumption sits uncomfortably with the fact that in the theory, measurement settings can, of course, be chosen freely. In equilibrium, the Bohmian configurations $\lambda$ are statistically independent of the measurement settings $a, b$. Moreover, the theory is parameter dependent, i.e. violates Equations \ref{eqn:parameter independence}, but satisfies signal-locality, i.e. satisfies the equations when $\lambda$ is replaced by $z$. It's, therefore, not clear why Colbeck and Renner think pilot wave theory should be incompatible with their assumptions. 

Indeed, as I would argue, the extending variables need not be the complete ontic states $\lambda$. Instead, they can be conceived of as manifest variables signifying whatever is accessible of the ontic state whilst violating signal-locality. Put differently, any $z$ may indicate a signal-local preparation of the bi-partite system, which doesn't rule out that each of these preparations may furthermore be determined by additional ontic states $\lambda$. They can thus be understood as a coarse-graining of some non-manifest initial states $\lambda$, which in turn can be highly non-local. 

It's worth mentioning that such an approach is akin to Bohmian non-equilibrium predictive advantage. That is, the system is prepared in a non-equilibrium distribution of initial configurations, which can lead to signal-local correlations in the case of maximally-entangled states. The variables $z$ of a magic prediction box may represent non-quantum preparations recovering Born probabilities on average. The analogue of the course-graining in a Bohmian world is signal-local non-equilibrium distributions over initial particle configurations. A similar idea is captured in Ghirardi and Romano's approach, which resembles the idea that the $z$ variables don't necessarily refer to the full ontic states \citep{Ghirardi-Romano-inequivalent}. In the present terminology, they essentially consider deterministic hidden variable theories whereby the ontic states are split into two parts, i.e. $\lambda =\left(\mu,\tau\right)$. The preparation of the quantum state is then given by some distribution $\rho(\lambda)=\rho(\mu|\tau)\rho(\tau)$, and the quantum averages are reproduced by assuming $p_Q(x,y)=\int p(x,y|\lambda)\rho(\lambda)d\lambda$. Due to quantum non-locality, the outcome correlations conditioned on the ontic state $\lambda$ will have to be parameter dependent. But on the intermediate level, averaging may wash out the non-locality. That is, $\int p(x,y|\mu,\tau)\rho(\mu|\tau)d\mu$ would be signal-local for all $\tau$. They then show that such signal-local distributions only exist for non-maximally entangled states. In this sense, the variables $z$ in an empirical extension could be seen as taking the role of $\tau$ in the intermediate distributions, i.e. $p(x,y|z)=\int p(x,y|\mu,z)\rho(\mu|z)d\mu$.

Ironically, this evades their own criticism of the Colbeck-Renner theorem. First, the argument that (FR) amounts to parameter independence doesn't go through if the relevant variables in the proof are interpreted in their way. Moreover, in the case of non-maximally entangled states, they concretely show the existence of models predictively inequivalent to quantum probabilities and use this to corroborate their criticism. However, the authors seem to be oblivious that Colbeck and Renner's proof for non-maximally entangled states is based on coupling the system to another pair of systems and doesn't follow directly. Therefore, their argument does not present a problem with the proof. 

I believe the relevant variables need not denote ontic states in this way. Put more bluntly, whatever the ontic, i.e. non-manifest, states of a quantum system may be (and irrespective of whether this presents a deterministic or indeterministic hidden variable theory), the variables $z$ are simply representing \textit{manifest} configurations in the environment without any \textit{a priori} bearing on non-manifest states.  

As indicated at the beginning of this section, signal-locality is a property of the correlations on the manifest configurations and thus doesn't constrain the behaviour of any non-manifest variables, i.e.

\begin{align}
	p(x|a)&=p(x|a,b)=\sum_y p(x,y|a,b)\\ \nonumber
	p(y|b)&=p(y|a,b)=\sum_x p(x,y|a,b),
\end{align} for all settings $a, b$, and outcomes $x,y$.

Moreover, since all preparations $p(x,y|a,b,z)$ in an empirical extension are conditioned on manifest configurations $z$, those as well satisfy the analogous signal-locality relations (cf. Equations \ref{eqn:signal locality}). Importantly, these signal-locality constraints are not in conflict with the non-locality of the variables in the non-manifest domain as I stated before. By definition, $z$ are variables the theory can bring about as configurations in the manifest domain and are not in one-to-one correspondence with the variables of the non-manifest domain. Notice again that metaphysical extensions need not be empirical extensions, as, for example, was seen in the case of pilot wave theory. For there, the theory constrains under what circumstances the non-manifest Bohmian configurations manifest themselves as manifest outcomes of measurements.   

Let's summarise. If (FR) is supposed to imply parameter independence, Colbeck and Renner's claim about possible extensions is significantly weaker than a general claim on empirical completeness by virtue of signal-locality alone. One assumption is a condition imposed on the non-manifest domain, and the other is imposed on the manifest domain. However, nothing prevents us from seeing the variables for predictive advantage as being manifest configurations (and perhaps also a coarse-graining over non-manifest variables). The primary assumption on this reading amounts to signal-locality, which is the only relevant condition on which the derivation relies. Thus, the Colbeck-Renner theorem seems stronger than expected since parameter independence is just signal-locality if the salient variables $z$ are assumed to be manifest configurations --- at least regarding the empirical completeness of non-maximally entangled systems. 

\subsection{Conclusion}

In this chapter I studied empirical completeness for bi-partite systems. I first observed that any hypothetical predictive advantage could not be fully deterministic due to non-locality and Bell's theorem. On the other hand, signal-locality did not rule out partly deterministic empirical extensions for non-maximally entangled systems. Then, I mentioned a concrete model with predictive advantage for this case. Generalised Bell inequalities proved a strong approach for studying predictive advantage for arbitrary numbers of measurement settings. The results by Barrett et al. presented proof of empirical completeness for maximally entangled states based on signal-locality.

Consequently, neither (partly) deterministic nor indeterministic empirical extensions exist. The clarifications on what the prediction-improving variables mean in the Colbeck-Renner theorem add an interesting twist to the theorem. It may be considered more substantial than it seems, given that, in the relevant sense, the basic assumption on which the proof rests amounts to the weaker notion of signal-locality rather than parameter independence. This, in turn, presents a generalisation of empirical completeness to non-maximally entangled states based on the weaker assumption of signal-locality. No proof for single systems is yet available, however. Thus combining the findings demonstrates signal-locality as a physical principle that explains the irreducible probabilism in quantum theories, at least for bi-partite systems.

\section{Outlook}
\label{section:outlook}
\sectionmark{Outlook}

I shall conclude the study of the empirical completeness problem with some comments and possible avenues for a general proof of the impossibility of predictive advantage. I will lay out some pertinent ideas in a fairly staccato fashion. 

\textit{Single systems.} Given the empirical completeness of bi-partite quantum systems based on signal-locality, one major step towards a universal claim requires proof for a single quantum system. The Colbeck-Renner theorem seems too limited for such an account, but another approach seems more promising. Since signal-locality can be applied only to the correlations of two or more systems, single quantum systems aren't affected by the results mentioned. As Hermens aptly points out, a claim on the completeness of single systems will require further assumptions on the nature of interactions and how systems are compounded. This turns out problematic. Hermens argues that the conditions necessary to prove Colbeck and Renner's claim for single systems may be unwarranted. The derivation relies on assumptions about the non-invasiveness of the relevant variables when coupling a second system to the target system. These seem generally violated (cf., for instance, \citealt{maroney2017measurements}). See also \citealt{LEEGWATER201618} which closes a number of gaps in the proof of Colbeck and Renner. 

One proposal that seems promising to me is moving from signal-locality to non-contextuality. First, locality can be shown to be a special case of non-contextuality, and second, non-contextuality applies to single systems. In the same sense that contextuality proofs generalise non-locality theorems, empirical completeness for bi-partite systems may be generalised to single systems. Resorting to assumptions on contextuality could close the gap and establish a general proof for arbitrary systems.
The work of Chen and Montina seems to present a viable attempt for the sort of thing empirical completeness is after based on contextuality \citep{Chen-Montina-contextuality}. 

\textit{Temporal dynamics and predictability.} Moreover, all considerations so far only addressed probabilities and predictions at a single time. One idea could be resorting to the dynamics of signal-local extensions to see whether this could restrict the possibility of predictive advantage over quantum theory. There may be a dynamical explanation why --- as a kind of equilibrium --- a universe which begins as one in which improved predictability is possible will tend to one in which quantum theory is empirically complete. Another approach could be to investigate the dynamical properties of empirical extensions. Recalling the definition of empirical extensions, predictive advantage should be preserved, i.e.

\begin{equation}
	|\psi(x,t)|^2=p(x|a,\rho(t))= \int p_t(x|a,z)\mu(z)dz,
\end{equation} for all $t$. By unitarily evolving the quantum state $\rho(t)$, this may pose restrictions on how the empirical extension's probabilities $p_t(x|a,z)$ evolve in time. 

In a Valentini-style for argument and resorting to a coarse-graining process, the empirical extensions could perhaps be shown to approach the quantum predictions. This holds, for example, in pilot wave theory where initial non-equilibrium distributions of Bohmian configurations tend to quantum equilibrium over short time spans \citep{valentini2005dynamical}. Moreover, it seems plausible that a similar property could hold for empirical extensions with arbitrary variables. Hence, during the evolution of empirical extensions, statistical mixing over time could wash out predictive advantage such that quantum probabilities are recovered in some equilibrium limit.   

\textit{Measurement disturbance and predictive advantage.} Along similar lines, the dynamics of measurement interactions may also play a crucial role in prediction-making. Perhaps the only way to incorporate measurements into a hypothetical empirical extension in a way that is compatible with quantum predictions may pose theoretical limits to what extent a system needs to be disturbed undergoing interactions. Thus, the dynamics present a trade-off between predictability and invasiveness of the relevant variables involved  --- akin to quantum uncertainty principles. An instance of such could be found in weak measurements. A different idea related to a dynamical explanation of empirical completeness could invoke a postulate on the disturbance of measurements. If it's true that in some sense or the other, any measurement interaction invariable affects relevant features of the system, then predictive advantage may be limited by such a process.

\textit{Weak values and predictive advantage.} The basic intuition behind weak measurements is to gain information about a quantum system without significantly disturbing it (see Chapter \ref{section:Weak measurements}). This can be done at the cost of gaining very little information in every interaction, which is then averaged over many runs of the experiment to obtain the weakly measured observable. May this perhaps allow predictive advantage? After all, the remarkable claim is sometimes made that weak measurements could violate Heisenberg's uncertainty principles (see, for example, in \citealt{Rozema-uncertainty-weak-measurement}). This immediately raises the question of whether weak measurements or versions thereof can be employed for predictive advantage. In Chapter \ref{section:Weak measurements}, I have shown that, at least in the case of pilot wave theory, weak measurements cannot provide any means to gain more information about a Bohmian system than what is already possible with standard quantum measurements. I suspect that a similar claim holds more generally, but further work is needed.  

\chapter*{Conclusions}
\addcontentsline{toc}{chapter}{Conclusions}
\chaptermark{Conclusions}

By providing a suitable framework, in this thesis, I studied two related topics in the foundations of quantum theory. There I introduced the relevant concepts, domains, and variables in a physical theory and for prediction making. The first part was concerned with problems related to the question of what quantum measurements signify. This question mostly appears in the form of quantum `paradoxes'. I showed that a number of these puzzles resolve by delineating the manifest variables of a theory from the non-manifest variables. The discussion started with the allegedly paradoxical claims commonly made on so-called `surreal' observables and `surrealistic' trajectories in pilot wave theory, where detectors are purportedly fooled. This first example served as an instructive case for inconsistent reasoning on what quantum outcomes are supposed to signify. The central insight is that quantum measurement outcomes \textit{a priori} have no bearing on variables in the non-manifest domain.  

I then discussed the `Delayed choice Quantum Erasure' experiment by drawing an analogy with Bell-type measurements and giving an account in standard quantum theory. I show that there is no need to resort to claims about any retrocausal action into the past by clarifying the role of the relevant parts in the experiment involved and giving an account in de Broglie-Bohm theory.

Next, I investigated weak measurements in the context of Bohmian trajectories. I first showed that weak velocity measurements admit a straightforward standard quantum mechanical description. In the second step, it turns out that reconstructing particle trajectories through weak values can provide no empirical argument for Bohmian trajectories and the theory's vastly underdetermined guidance equation. The upshot of the analysis by which the puzzle resolves is that one shouldn't na\"ively identify weak positions with actual Bohmian positions. As a result, presenting yet another instance of the importance of carefully distinguishing the manifest from the non-manifest. 

Moving on to measurements and disturbance, I began with an account of classical physics. It turns out that measurements can be designed not to disturb the system in any way, and thus predictions can be made with arbitrary precision. Regarding quantum theory, I point out that the question of what is or isn't disturbed in a measurement again depends on what the pertinent variables are deemed to represent. Trivially, since quantum states are in the sense epistemic as they represent probability distributions via Born's rule, they must undergo a disturbance or `updating' upon measurement. For ontic states, i.e. non-manifest configurations, the story is more complicated, but similar claims seem to hold. I commented on the potential relevance of measurement disturbance for the empirical completeness problem. 

The other topic of the thesis was the empirical completeness problem of quantum mechanics. That is, whether quantum probabilities are fundamental or could be refined by future theories. In contrast to the questions on the mere existence of `hidden variables' and determinism --- which here is captured by the notion of a metaphysical extension --- the aim was to investigate empirical extensions, i.e. theories with predictive advantage based on accessible information --- the manifest configurations. I started with an account of quantum predictions which led to the introduction of what I dubbed the `quantum decorrelation principle': Whenever a (pure) quantum state is assigned to a target system, the outcomes of arbitrary measurements decouple from all manifest configurations in its environment. Therefore, this principle will have to be violated for an empirical extension to exist. Furthermore, I engaged with ambiguities in quantum prediction making due to the measurement problem and outlined possible approaches to close this gap. 

The discussion on quantum empirical completeness started with carrying out a case study on predictive advantage in pilot wave theory. I clarified under what assumptions the Bohmian framework satisfies the quantum decorrelation principle and concluded that it is merely a metaphysical extension of standard quantum theory. I discussed the various options for when the quantum decorrelation principle is violated. A careful reconstruction of the behaviour of Bohmian particles in a Bell spin measurement reveals that every single pair of particles is non-local both for maximally and non-maximally entangled states. Deterministic predictive advantage, therefore, is impossible.

Moreover, although signal-local Bohmian non-equilibrium distributions can be shown not to exist for maximally entangled states, they might for non-maximally entangled states. There are deterministic hidden variable theories for which such is constructed explicitly. It would be interesting to see whether a straightforward way exists in the approach discussed to rule out a signal-local non-equilibrium predictive advantage for maximally entangled states and explicitly construct a signal-local non-equilibrium predictive advantage for non-maximally entangled states. It would also be interesting to apply the general results on the impossibility of predictive advantage employing generalised Bell-inequalities and to investigate what happens when Bohmian bi-partite states are coupled to further systems which is a crucial step in the general proofs of empirical completeness for bi-partite systems. Moreover, some implications for empirical completeness for single systems may be drawn by looking into the ongoings of coupling systems to single systems in de Broglie-Bohm theory. 

I argued that empirical completeness is established for arbitrary bi-partite quantum states. It was first demonstrated that empirical extensions could not be fully deterministic because of Bell's theorem. However, signal-locality does not rule out partly deterministic empirical extensions for non-maximally entangled systems. The latter case allows explicit constructions of models with predictive advantage. By invoking generalised Bell inequalities for arbitrary numbers of measurement settings, the results by Barrett et al. present a suitable proof of empirical completeness for maximally entangled states. That is, neither (partly) deterministic nor indeterministic empirical extensions exist, conforming with signal-locality. The relevant part of Colbeck and Renner's theorem was then reinterpreted as a generalisation to non-maximally entangled states. This was done by investigating the status of the relevant assumptions on which the proof rests. Against the authors' own beliefs, it was shown to derive from the basic assumption of signal-locality --- a constraint on the manifest domain only. Thus the results demonstrate signal-locality as a physical principle that explains the fundamental chanciness in present and future quantum theories. In turn, this reconciles us to many quantum features as aspects of limits on Nature's predictability. However, no general proof is currently available for arbitrary single systems. I outlined some ideas for future work regarding a general proof of quantum empirical completeness. More concretely, the quantum nature of measurement disturbance may intimately be tied to limits on predictive advantage; there could be a dynamical explanation for why possible empirical extensions --- as a kind of equilibrium --- will tend to a universe in which improved predictability is impossible. Empirical completeness for single systems may be established on the grounds of non-contextuality --- as a generalisation of locality. 

In sum, the contemporary mainstream view that quantum predictions are irreducibly probabilistic, thus, becomes a precise theorem. Consequently, not only can quantum mechanical measurement outcomes not be predicted with certainty, the signal-local predictions of any theory must be exactly the quantum probabilities. It holds as a generic feature of physics, including post-quantum theories. 
Insofar as we hold on to the --- arguably irrevocable --- principle of relativistic locality, it seems fair to say that one more thing we may add to the very few things we can be certain about: \textit{the certainty of uncertainty.}


\appendix
\chapter{de Broglie-Bohm Theory}
\label{appendix:de Broglie-Bohm theory}

\addtocontents{toc}{\protect\setcounter{tocdepth}{-2}}
\section{Postulates}

de Broglie-Bohm theory is best conceived of as an example of what \citet{popper1967quantum} dubbed a `quantum theory without observer' (cf. \citealt{goldstein1998quantum}; \citealt[esp. Section~8]{allori2008common}): it aspires to provide an understanding of quantum phenomena without fundamental recourse to non-objective (i.e. subjective or epistemic) notions. In this vein, these are `[...] some words which, however legitimate and necessary in application, have no place in a formulation with any pretension to physical precision: system, apparatus, environment, microscopic, macroscopic, [...], observable, information, measurement' \citep[p.~215]{bell_against_aspect_2004}.
Such endeavours grew out of the dissatisfaction with influential presentations of QM, notably by von Neumann, Heisenberg and (common readings of) Bohr (see, e.g. \citealt{jammer1974philosophy}; \citealt[Ch.~VIII, IX]{scheibe2006philosophie}; \citealt{cushing1996causal}).  

In de Broglie-Bohm theory the quantum mechanical wave function guides point particles along deterministic paths like a `pilot wave'. Assuming a particular initial distribution of the particles, one recovers the empirical content of QM. 

More precisely, for an $N$-particle system, de Broglie-Bohm theory can be taken to consist of three postulates. (I closely follow \citet{teufel2009bohmian}, to whom I refer for all details.)

\begin{itemize}
	\label{eqn:postulates}
	\item[(1)] In its non-relativistic form, de Broglie-Bohm theory is a theory about (massive, charged, etc.\footnote{For the present purposes, I will elide subtleties concerning the ascription of such intrinsic properties (cf. \citealt{brown1995bohm, brown1996bovine, brown1996cause}). I will also set aside Esfeld's (\citeyear{esfeld2014quantum, esfeld2017minimalist}) `Humeanism without properties' (the ontology and ideology of which is limited to primitively occupied spacetime points and the spatiotemporal relations).}) particles. At all times, they occupy definite positions. 
	
	\item[(2)] The wave function $\Psi\colon\mathbb{R}^{3N}\times\mathbb{R}\to \mathbb{C}$ satisfies the standard $N$-particle Schrödinger Equation in the position representation:
	\begin{equation}
		i\hbar \frac{\partial}{\partial t} \Psi(\bm{Q},t)=\hat{H}\Psi(\bm{Q},t) 
	\end{equation} with the $N$-particle Hamiltonian $\hat{H}=-\sum\limits_{i=1}^{N}\frac{\hbar^2}{2m_i}\Delta + V(\bm{Q},t)$, where $\nabla_i=\frac{\partial}{\partial \bm{Q}_i}$, $i=1,...,N$ acts on the $i$-th position variable $\bm{Q}_i$ and $\bm{Q}:=(\bm{Q}_1,..., \bm{Q}_N)$.
	
	\item[(3)] The continuous evolution of the $i$-th particle's position $\bm{Q}_i(t)\colon \mathbb{R}\to\mathbb{R}^3$ in $3$-dimensional Euclidean space is generated by the flow of the velocity field\footnote{For motivations, see \citealt[Chapter 4]{passon2004bohmsche}.}
	\begin{equation}
		\label{standardguidancelaw}
		v_i^{\Psi}:= \frac{\hbar}{m_i}\Im \frac{\Psi^*\nabla_i \Psi}{\Psi^*\Psi}|_{(\bm{Q}_1(t), ..., \bm{Q}_N(t))}.
	\end{equation} Note that Equation \ref{standardguidancelaw} is also defined for spinor-valued functions $\Psi$, e.g. accounting for spin. 
	That is, the particle position $\bm{Q}_i$ obeys the so-called guidance equation
	\begin{equation}
		\label{eqn:guiding equation}
		m_i\bm{\dot{Q}}_i=v_i^{\Psi}.
	\end{equation}
	For all relevant types of potentials, unique solutions (up to sets of initial conditions of measure zero) have been shown to exist \citep{teufel2005simple}. 
	Notice that $v_i^{\Psi}$ depends on all particle positions simultaneously. This is the origin of de Broglie-Bohm theory's manifest action-at-a-distance in the form of an instantaneous non-locality (see, e.g. \citealt[Section~13]{goldstein-sep-qm-bohm}).
	
	\item[(4)] The wave function induces a natural (and, under suitable assumptions, unique, see \citet{goldstein2007uniqueness}) measure on configuration space, the Born rule measure:
	\begin{equation}
		\label{eqn:Born rule}
		\mathbb{P}^{\Psi}(d^{3N}\bm{Q}):=|\Psi(\bm{Q})|^2d^{3N}\bm{Q}.
	\end{equation}
	It quantifies which (measurable) sets of particle configurations $\mathcal{Q}\subseteq \mathbb{R}^{3N}$ count as large (`typical') --- how common configurations of the type represented by $\bm{Q}$ are amongst all possible configurations. That is:
	\begin{equation}
		\label{eqn:QEH}
		\int_{\mathcal{Q}}d^{3N}\bm{Q}|\Psi(\bm{Q})|^2=1-\varepsilon, \nonumber
	\end{equation} for some small $\varepsilon$ (see \citealt{maudlin2011three, durr2019typicality, dustin2015typicality} for details; cf. \citealt{frigg2009typicality, frigg2011typicality}).\footnote{Typicality raises intriguing questions about whether an appeal to it is explanatory (and if so, in which sense). For a recent account, see \citet{wilhelm2019typical}.} This definition of typicality respects a generalised sense of time independence. A universe typical in this sense is said to be in quantum equilibrium (see \citealt{durr1992quantum} for further details). The continuity equation for $|\Psi(\bm{Q})|^2$ obtained from the Schrödinger Equation implies that a system is in quantum equilibrium at all times, if and only if in equilibrium at \textit{some} point in time. This is called the Quantum Equilibrium Hypothesis (QEH). 
	
\end{itemize}

\textit{Postulating} QEH is not the only option to arrive at the Born rule distributions. For instance, the approaches of  \citet{bohm1953proof},  \citet{valentini1992pilot}, and \citet{valentini2005dynamical} try to derive quantum equilibrium distributions from \textit{dynamical} convergence to equilibrium. The ideas are approximately analogues to the arguments of thermodynamic equilibrium in statistical mechanics. I will discuss the implications of non-equilibrium initial distributions in Section \ref{non-equilibrium}. 

\section{Effective Wave Functions}

Consider now a de Broglie-Bohmian $N$-particle universe, satisfying these four axioms. An $M$-particle subsystem is said to possess an `effective' wave function $\psi$, if the universal wave function (i.e. the wave function of the universe) $\Psi\colon X\times Y \to \mathbb{C}$, with $X$ and $Y$ denoting the configuration space of the subsystem and its environment, respectively, can be decomposed as

\begin{equation}
	\label{eqn:effective wave function}
	\forall (x,y) \in X\times Y\colon \Psi(x,y) = \psi(x)\Phi(y) + \Psi_{\perp}(x,y).
\end{equation} Here $\Phi$ and $\Psi_{\perp}$ have macroscopically disjoint $y$-support and $y \subseteq \supp (\Phi)$.  That is, the configurations in which $\Phi$ and $\Psi_{\perp}$ vanish are macroscopically distinct (e.g. correspond to distinct pointer positions). For negligible interaction with their environment, the effective wave function $\psi(x)$ of subsystems exists and can be shown to satisfy the linear Schrödinger equation. However, in situations where a system doesn't possess an effective wave function, according to de Broglie-Bohm theory, the evolution of its wave function will deviate from the linear Schrödinger evolution: In this case, the Schrödinger evolution of the universal wave function determines the dynamics of the subsystem by its conditional wave function, i.e. $\psi(x):=\Psi(x,y)\mid_{y=y_0}$, which doesn't decompose according to Equation \ref{eqn:effective wave function}. The Bohmian prediction is then still unique when considering the universal system's total wave function and configuration. But then the question arises whether the environment, e.g. measurement apparatus, is a sensible prediction-making device. We may also ask what the predictions \textit{are} in that case and whether they could lead to predictive advantage. It is, therefore, important to investigate the exact nature of state preparation for these cases. To this, I turned in Section \ref{section:internal observation}. 

\bibliography{library}

\end{document}